\documentclass[11pt,a4paper]{article}
\pdfoutput=1

\usepackage{jheppub-nosort,amsthm}

\usepackage{amsmath,amssymb,amsthm,amscd}
\usepackage{physics}
\usepackage{tikz}
\usetikzlibrary{decorations.pathmorphing}
\tikzset{node distance=2cm, auto}
\tikzset{snake it/.style={decorate, decoration=snake}}
\usepackage{epsfig}
\usepackage{psfrag,comment}
\usepackage{graphicx}
\usepackage{caption}
\usepackage{subcaption}
\usepackage{mathrsfs}
\usepackage{xcolor}
\usepackage[normalem]{ulem}
\usepackage{bm}
\usepackage{lscape}
\usepackage{diagbox}
\usepackage{upgreek}

\usepackage{url}
\newcommand\doilink[1]{\href{http://dx.doi.org/#1}{#1}}
\newcommand\arxivlink[1]{\href{http://arxiv.org/abs/#1}{#1}}

\newcommand{\CB}{{\mathcal B}}

\newcommand{\CH}{{\mathcal H}}

\newcommand{\CN}{{\mathcal N}}

\newcommand{\CR}{{\mathcal R}}
\newcommand{\CS}{{\mathcal S}}

\newcommand{\CZ}{{\mathcal Z}}

\def\BN{{\mathbb N}}
\def\BZ{{\mathbb Z}}
\def\BR{{\mathbb R}}
\def\BC{{\mathbb C}}

\def\BS{{\mathbb S}}

\newcommand{\be}{\begin{equation}}
\newcommand{\ee}{\end{equation}}
\newcommand{\ba}{\begin{aligned}}
\newcommand{\ea}{\end{aligned}}
\newcommand{\bea}{\begin{eqnarray}}
\newcommand{\eea}{\end{eqnarray}}
\newcommand{\bean}{\begin{eqnarray*}}
\newcommand{\eean}{\end{eqnarray*}}

\newcommand{\p}{\partial}

\def\r{\right\rangle}

\def\1{\mathbf{1}}
\def\0{|\1\r}

\newcommand{\rme}{{\mathrm{e}}}
\newcommand{\rmi}{{\mathrm{i}}}
\newcommand{\rmd}{{\mathrm{d}}}

\def\XXint#1#2#3{{\setbox0=\hbox{$#1{#2#3}{\int}$}
     \vcenter{\hbox{$#2#3$}}\kern-.5\wd0}}

\newenvironment{rcases}
  {\left.\begin{aligned}}
  {\end{aligned}\right\rbrace}

\newcommand{\PI}{{\textsf{P}$_{\textsf{I}}$}}
\newcommand{\PII}{{\textsf{P}$_{\textsf{II}}$}}

\makeatletter
\newsavebox\myboxA
\newsavebox\myboxB
\newlength\mylenA

\newcommand*\widebar[2][0.75]{%
    \sbox{\myboxA}{$\m@th#2$}%
    \setbox\myboxB\null
    \ht\myboxB=\ht\myboxA%
    \dp\myboxB=\dp\myboxA%
    \wd\myboxB=#1\wd\myboxA
    \sbox\myboxB{$\m@th\overline{\copy\myboxB}$}
    \setlength\mylenA{\the\wd\myboxA}
    \addtolength\mylenA{-\the\wd\myboxB}%
    \ifdim\wd\myboxB<\wd\myboxA%
       \rlap{\hskip 0.8\mylenA\usebox\myboxB}{\usebox\myboxA}%
    \else
        \hskip -0.5\mylenA\rlap{\usebox\myboxA}{\hskip 0.5\mylenA\usebox\myboxB}%
    \fi}
\makeatother

\newdimen\tableauside\tableauside=1.0ex
\newdimen\tableaurule\tableaurule=0.4pt
\newdimen\tableaustep
\def\phantomhrule#1{\hbox{\vbox to0pt{\hrule height\tableaurule width#1\vss}}}
\def\phantomvrule#1{\vbox{\hbox to0pt{\vrule width\tableaurule height#1\hss}}}
\def\sqr{\vbox{%
  \phantomhrule\tableaustep
  \hbox{\phantomvrule\tableaustep\kern\tableaustep\phantomvrule\tableaustep}%
  \hbox{\vbox{\phantomhrule\tableauside}\kern-\tableaurule}}}
\def\squares#1{\hbox{\count0=#1\noindent\loop\sqr
  \advance\count0 by-1 \ifnum\count0>0\repeat}}
\def\tableau#1{\vcenter{\offinterlineskip
  \tableaustep=\tableauside\advance\tableaustep by-\tableaurule
  \kern\normallineskip\hbox
    {\kern\normallineskip\vbox
      {\gettableau#1 0 }%
     \kern\normallineskip\kern\tableaurule}%
  \kern\normallineskip\kern\tableaurule}}
\def\gettableau#1{\ifnum#1=0\let\next=\null\else
\squares{#1}\let\next=\gettableau\fi\next}
\title{Resurgent Stokes Data for Painlev\'e Equations and Two-Dimensional Quantum (Super) Gravity}

\author[a]{Salvatore~Baldino,}
\affiliation[a]{CAMGSD, Departamento de Matem\'atica, Instituto Superior T\'ecnico,\\ Universidade de Lisboa, 1049-001 Lisboa, Portugal\\}
\emailAdd{salvatore.baldino@}

\author[a]{Ricardo~Schiappa,}
\emailAdd{ricardo.schiappa@}

\author[a]{Maximilian~Schwick,}
\emailAdd{maximilian.schwick@}

\author[a]{Roberto~Vega\,}
\emailAdd{roberto.vega@tecnico.ulisboa.pt}

\abstract{
Resurgent-transseries solutions to Painlev\'e equations may be recursively constructed out of these nonlinear differential-equations---but require Stokes data to be globally defined over the complex plane. Stokes data explicitly construct connection-formulae which describe the nonlinear Stokes phenomena associated to these solutions, via implementation of Stokes transitions acting on the transseries. Nonlinear resurgent Stokes data lack, however, a first-principle computational approach, hence are hard to determine generically. In the Painlev\'e~I and Painlev\'e~II contexts, nonlinear Stokes data get further hindered as these equations are resonant, with non-trivial consequences for the interconnections between transseries sectors, bridge equations, and associated Stokes coefficients. In parallel to this, the Painlev\'e~I and Painlev\'e~II equations are string-equations for two-dimensional quantum (super) gravity and minimal string theories, where Stokes data have natural ZZ-brane interpretations. This work computes for the first time the complete, analytical, resurgent Stokes data for the first two Painlev\'e equations, alongside their quantum gravity or minimal string incarnations. The method developed herein, dubbed ``closed-form asymptotics'', makes sole use of resurgent large-order asymptotics of transseries solutions---alongside a careful analysis of the role resonance plays. Given its generality, it may be applicable to other distinct (nonlinear, resonant) problems. Results for analytical Stokes coefficients have natural structures, which are described, and extensive high-precision numerical tests corroborate all analytical predictions. Connection-formulae are explicitly constructed, with rather simple and compact final results encoding the full Stokes data, and further allowing for exact monodromy checks---hence for an analytical proof of our results.
}


\arxivnumber{2203.13726}

\begin{document}

\maketitle

\vfill

\eject

\allowdisplaybreaks

\section{Introduction and Summary}\label{sec:intro}

Over one hundred years ago, Paul Painlev\'e embarked on a quest to find new classes of special  functions, beyond the realms of elliptic and classical-special functions \cite{p02, p06}. It was already well-known at the time that a large number of special functions could be defined via ordinary differential equations (ODEs) (see, \textit{e.g.}, \cite{olbc10}), and that, in almost all such cases, the resulting ODE was \textit{linear}. Painlev\'e's quest hence started off by asking if it could be possible to define \textit{new} special functions---beyond the classical ones---but via generic \textit{nonlinear} ODEs instead?

Such a seemingly simple question opened a century's mathematical Pandora's box. To start, the ability to define new, sensible functions very much depends upon the nature of their would-be singularities. Now, whereas linear ODEs \textit{only} have \textit{fixed}\footnote{A fixed singularity of an ODE is a singularity in its solutions whose location \textit{does not} depend on initial data/ boundary conditions, \textit{i.e.}, a singularity which only depends upon the ODE and not upon any particular solution.} singularities, it turns out that nonlinear ODEs may have \textit{both} fixed and \textit{movable}\footnote{A movable singularity of an ODE is otherwise. It is a singularity in the solution whose location \textit{will depend} upon the initial/boundary conditions selecting that particular solution. It varies as initial/boundary data vary.} singularities. On top of this lies the myriad of possible singularities one may find---at its broadest split, either \textit{single}-valued or \textit{multi}-valued (branch point) singularities. In such wide contexts, and to ensure that the ``nonlinear special-function programme'' would be feasible, the \textit{Painlev\'e property} arises: these are ODEs whose solutions have \textit{no movable multi-valued} singularities. Let us follow Painlev\'e in trying to classify them.

For first-order Painlev\'e-type ODEs there is not much to say. It turns out, one either finds equations that are reducible to linear ODEs, or else equations that may be solved via elliptic functions (all of those are actually deducible from, say, the Weierstrass $\wp$-function; in which case there is a \textit{single} ``new''---well known!---special function at this level). The level of complexity jumps dramatically as soon as one turns to second-order Painlev\'e-type ODEs, of the form
\be
u''(z) = \CR \left( u',u,z \right),
\ee
\noindent
with $\CR$ rational in $u$ and $u'$, and locally analytic in $z$. In a nutshell, Painlev\'e's classification of this type of second-order ODEs \cite{p02, p06} tells us that most ($44$, to be precise) are solvable in terms of previously known functions (\textit{e.g.}, elliptic functions, classical special functions), hence bringing nothing new to the table. But there are $6$ canonical ODEs which require the introduction of \textit{new transcendental\footnote{This just means that their \textit{general} solutions cannot be expressed in terms of previously known functions (\textit{e.g.}, rational functions, exponential functions, elliptic functions, classical special functions, and so on).} functions} in order to describe their \textit{general} solutions---these are the famous six Painlev\'e equations, Painlev\'e~I through Painlev\'e~VI. This set of six Painlev\'e transcendents is the first historical example of ``nonlinear special functions'', which have received a great deal of attention over the past $100$ years. We refer to, \textit{e.g.}, \cite{k94, c03, c06, olbc10, d14, d16, c19}, for introductions, reviews, and references on the above highlights---of what is a very long, rich, and on-going history and literature, hence one which is also too large to review herein. Let us further point-out that taking the programme further, to higher-order equations, is an open on-going research problem. 

\begin{figure}
\centering
     \begin{subfigure}[h]{0.235\textwidth}
         \centering
         \includegraphics[width=\textwidth]{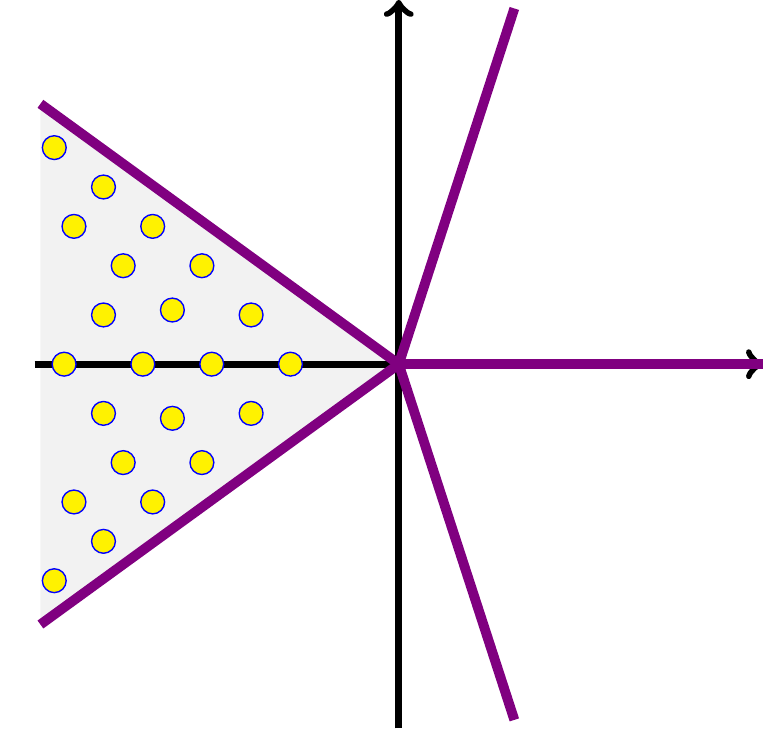}
     \end{subfigure}
\hspace{5mm}
     \begin{subfigure}[h]{0.235\textwidth}
         \centering
         \includegraphics[width=\textwidth]{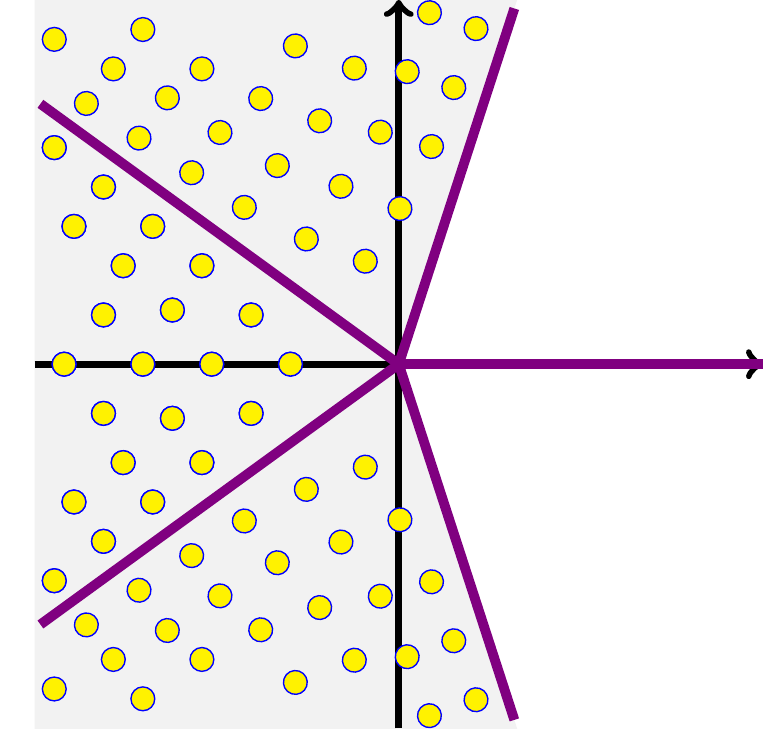}
     \end{subfigure}
\hspace{5mm}
     \begin{subfigure}[h]{0.235\textwidth}
         \centering
         \includegraphics[width=\textwidth]{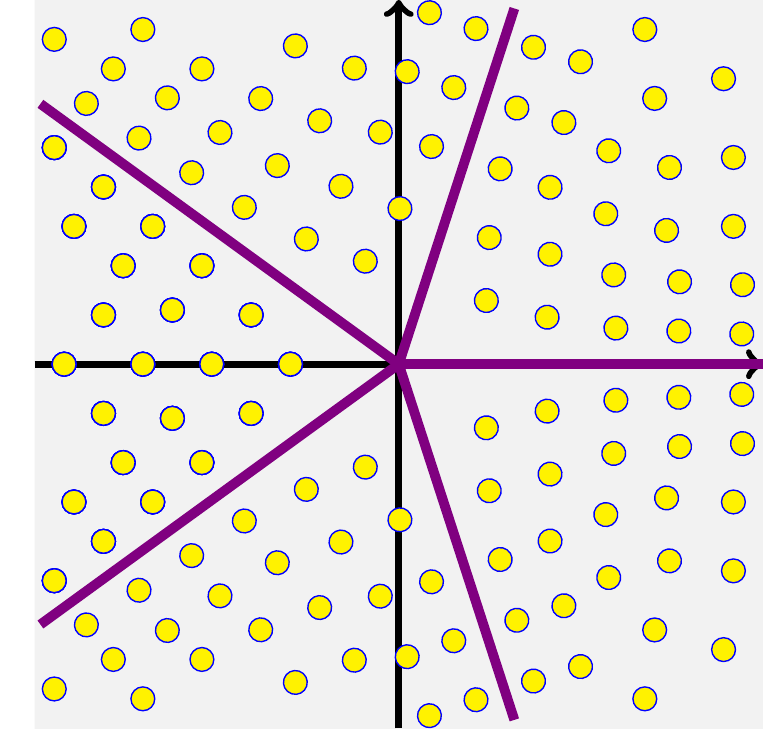}
     \end{subfigure}
\caption{Illustrative plots of the complex $z$-plane for \PI~solutions, alongside its five-fold ``pizza slicing''. From left to right, images plot the possible Boutroux classifications of \PI~solutions, where the yellow circles represent the double-pole movable singularities. In sequence, we plot a \textit{trintronqu\'ee}, a \textit{tronqu\'ee}, and a general solution. See the main text for details.}
\label{fig:P1-singular-Boutroux}
\end{figure}

\begin{figure}
\centering
     \begin{subfigure}[h]{0.235\textwidth}
         \centering
         \includegraphics[width=\textwidth]{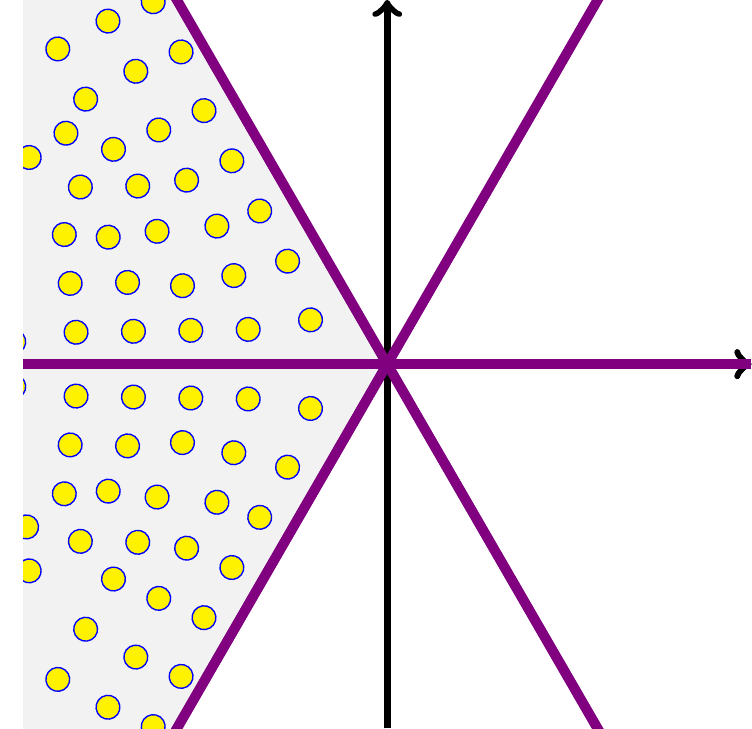}
     \end{subfigure}
\hspace{0mm}
     \begin{subfigure}[h]{0.235\textwidth}
         \centering
         \includegraphics[width=\textwidth]{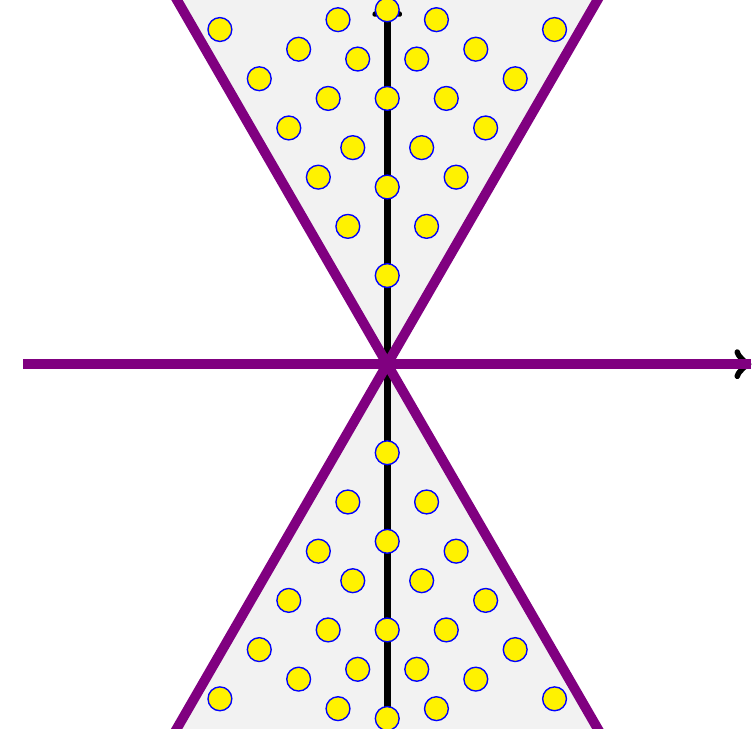}
     \end{subfigure}
\hspace{0mm}
     \begin{subfigure}[h]{0.235\textwidth}
         \centering
         \includegraphics[width=\textwidth]{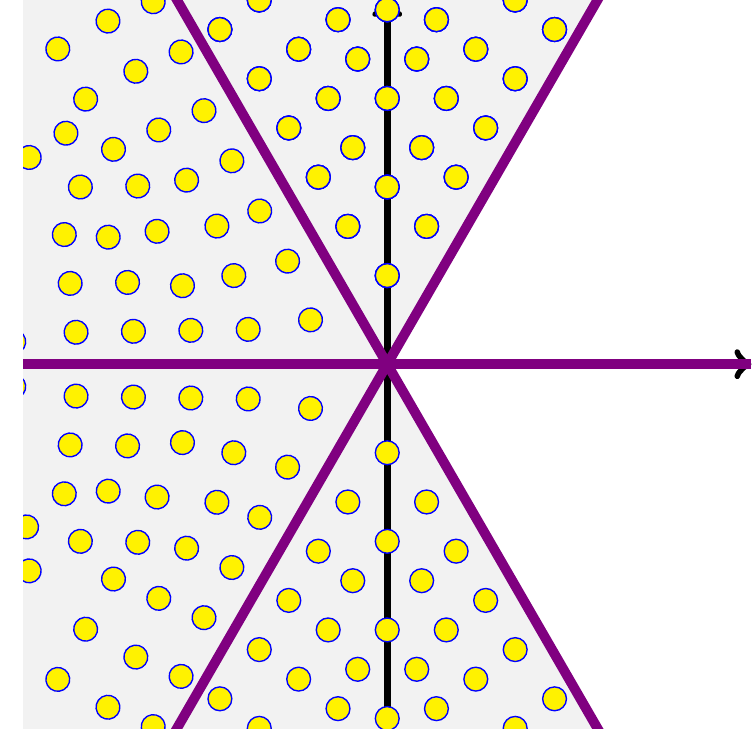}
     \end{subfigure}
\hspace{0mm}
     \begin{subfigure}[h]{0.235\textwidth}
         \centering
         \includegraphics[width=\textwidth]{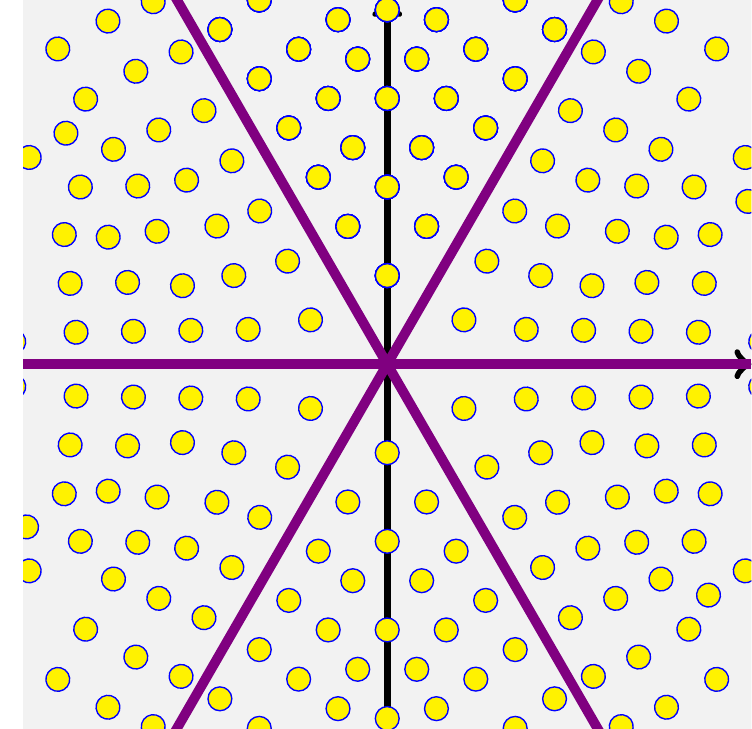}
     \end{subfigure}
\caption{Illustrative plots of the complex $z$-plane for \PII~solutions, alongside its six-fold ``pizza slicing''. From left to right, images illustrate the possible Boutroux classifications of \PII~solutions, where the yellow circles represent the simple-pole movable singularities: a \textit{tritronqu\'ee}, a \textit{bitronqu\'ee}, a \textit{tronqu\'ee}, and a general solution. See the main text for details.}
\label{fig:P2-singular-Boutroux}
\end{figure}

In the present work we shall be interested in the Painlev\'e~I equation (henceforth simply denoted by \PI),
\be
\label{eq:PI-eq}
u_{\text{I}}^2 (z) - \frac{1}{6}\, u_{\text{I}}'' (z) = z,
\ee
\noindent
and in the (homogeneous) Painlev\'e~II equation (similarly, henceforth denoted simply by \PII),
\be
\label{eq:PII-eq}
\mu_{\text{II}}^3 (z) - \frac{1}{2}\, \mu_{\text{II}}'' (z) = z\, \mu_{\text{II}} (z).
\ee
\noindent
As mentioned above, solutions to these equations are transcendental hence hard to simply describe \textit{quantitatively} (but more on this below). However we do know, essentially by definition, that they will have \textit{movable poles}---as we shall review in section~\ref{sec:review}, these are \textit{double}-poles in the case of \PI, and \textit{simple}-poles in the case of \PII. Consequently, precisely because of their (movable) singularity structure, \PI~and \PII~solutions are simple to describe \textit{qualitatively}---as was worked out soon after Painlev\'e's initial results by Boutroux \cite{b13, b14}. In order to swiftly describe Boutroux's classification of Painlev\'e solutions, let us first point out straightforward symmetries of \PI~and \PII~solutions. \PI~\eqref{eq:PI-eq} has a natural $\BZ_5$ symmetry, invariant under
\be
\begin{rcases}
z &\mapsto \kappa\, z \quad \\
u_{\text{I}} &\mapsto \kappa^{-2}\, u_{\text{I}} \quad
\end{rcases} \qquad \text{with} \qquad \kappa^5 = 1.
\ee
\noindent
Hence if $u_{\text{I}}^\star (z)$ is a solution of \PI, so are its above ``five-fold rotations''. It turns out to be convenient to partition the complex $z$-plane for \PI~solutions into \textit{five} radial sectors, as illustrated in figure~\ref{fig:P1-singular-Boutroux}---but this will be properly discussed in subsection~\ref{subsec:directmonodromy}, where this five-fold split actually corresponds to (anti-)Stokes lines for \PI. The double-poles of \PI~solutions (asymptotically) accumulate in each of these sectors. Something similar occurs for \PII. First, \eqref{eq:PII-eq} has definite parity hence is $\BZ_2$-invariant under $\mu_{\text{II}} \mapsto -\mu_{\text{II}}$. Second, akin to what happened for \PI, \PII~has a natural $\BZ_3$ symmetry being further invariant under
\be
\begin{rcases}
z &\mapsto \kappa\, z \quad \\
\mu_{\text{II}} &\mapsto \kappa^{-1}\, \mu_{\text{II}} \quad
\end{rcases} \qquad \text{with} \qquad \kappa^3 = 1.
\ee
\noindent
Hence if $\mu_{\text{II}}^\star (z)$ is a solution of \PII, so are its ``six-fold rotations'' (combined via reflection). It turns out to be convenient to partition the complex $z$-plane for \PII~solutions into \textit{six} radial sectors, as illustrated in figure~\ref{fig:P2-singular-Boutroux}---again, this will be properly discussed in subsection~\ref{subsec:directmonodromy}, where this six-fold split actually corresponds to (anti-)Stokes lines for \PII. The simple-poles of \PII~solutions (asymptotically) accumulate in each sector. With this rough motivation in mind, one qualitatively classifies Painlev\'e solutions depending on which such ``pizza slices'' are populated\footnote{With singularities asymptotically constrained inside each slice, given its boundaries are (anti-)Stokes lines.} with movable singularities, and which ones are singularity-free. Boutroux denoted the different types of solutions as \cite{b13, b14} (see as well, \textit{e.g.}, \cite{k94, dz95, c97, cc97, hl99, jk01, olbc10, fw11, n12, n13, cch13, fw14, n14, d14, d16}): \textit{tritronqu\'ee} (solutions free of poles in $4$ adjacent sectors; ``lattices'' of poles throughout the remaining sectors), \textit{tronqu\'ee} (solutions free of poles in $2$ adjacent sectors; ``lattices'' of poles throughout the remaining ones), and general solutions (all sectors are populated with movable singularities). In the case of \PII~one may also construct the Hastings--McLeod \cite{hm80} or \textit{bitronqu\'ee} solution, which is real and pole-free on the real line. All these different solutions are schematically illustrated in figures~\ref{fig:P1-singular-Boutroux} and~\ref{fig:P2-singular-Boutroux}.

Having qualitatively understood where Painlev\'e movable-poles accumulate, for different solutions, we may now go back and ask if one may actually locate them \textit{quantitatively}---which would greatly amount to fully describing Painlev\'e transcendents. We are interested in the construction of general solutions to either \PI~or \PII~starting off with (inverse) power-series expansions around the (single) irregular point at infinity $z \sim \infty$. Due to the nature of this fixed singularity, these expansions are \textit{asymptotic}, hence require exponential ``beyond-all-orders'' corrections. This is best tackled within the \textit{resurgent transseries} framework \cite{e81, s14, ms16, abs18}, where general transseries for \PI~were constructed in \cite{gikm10, asv11} and for \PII~in \cite{sv13} (see as well, \textit{e.g.}, \cite{y85, k88, jk88, jk92b, k94, dz95, akt96, kt98, jk01, cc01, o05, msw07, m08, msw08, as13, d14, s15, d16})---and which we review in section~\ref{sec:review}. These asymptotic solutions and their associated resurgent transseries start-off at $z \sim +\infty$, implying if one wants to reach any of the pole-populated sector of figures~\ref{fig:P1-singular-Boutroux} and~\ref{fig:P2-singular-Boutroux}---where, recall, radial-sector boundaries correspond to (anti-)Stokes lines---one must necessarily deal with Stokes phenomena \cite{s64}. In the resurgent transseries context for Painlev\'e solutions this amounts to \textit{nonlinear} Stokes phenomena, hence to an \textit{infinite} amount of Stokes data lacking a first-principle computational approach\footnote{Let us further stress that their connection formulae are \textit{transcedental} functions of initial/boundary data---as we shall see, Stokes data turn out to be \textit{zeta-numbers} themselves---hence, in any case, generically hard to compute.}. Prior to this work, only \textit{one} Stokes coefficient was analytically known for each Painlev\'e equation. This is the ``canonical'' coefficient appearing in the leading \textit{perturbative asymptotics} (it is also the single non-trival coefficient in the Riemann--Hilbert formulation; or the coefficient computed from the matrix integral around a non-trivial instanton saddle). For \PI~this number was (Painlev\'e conventions will appear in section~\ref{sec:review} and Stokes data notation in section~\ref{sec:stokes-resonance})
\be
\label{eq:exampleStokesP1}
N_1^{(1)} = - \rmi\, \frac{\sqrt[4]{3}}{2\sqrt\pi},
\ee
\noindent
and for \PII~it was
\be
\label{eq:exampleStokesP2}
N_1^{(1)} = -\rmi\, \frac{1}{\sqrt{2\pi}}.
\ee
\noindent
By now these two numbers have been computed in many different ways---the first of which made use of indirect methods (linearization via Riemann--Hilbert, Lax pairs, or isomonodromic deformations \cite{ifk94, i03}), for both \PI~\cite{k88, kk93, k94, k04a} and \PII~\cite{k94, ik03, k04b}. But the existence of an underlying infinite amount of other Stokes data was only realized later, when addressing the asymptotics of \textit{instanton sectors} in the Painlev\'e transseries in the seminal paper \cite{gikm10}, and then digging deeper into these \textit{two-parameter} transseries structures \cite{asv11, sv13}. In particular, these complete Stokes data have a very clean \textit{raison d'\^etre} within resurgence \cite{e81}, appearing throughout all resurgence relations in-between distinct transseries sectors; see, \textit{e.g.}, \cite{abs18}. However, all these additional data were previously only known \textit{numerically} \cite{gikm10, asv11, sv13}. Many (empirical) relations between these coefficients were also found \cite{gikm10, asv11, sv13, as13}, which led to long lists of numbers begging for an explanation. It is our goal in this paper to compute\footnote{A first---yet not successful---attempt at finding these numbers and their structure was reported in October 2017 in \cite{asv17a, asv17b}. Their correct structure was found later and reported in June 2019 in \cite{s19}, up to a single number, which was then finally reported in February 2021 in \cite{s21} (the contents of this paper).} the complete, analytical, resurgent Stokes data for \PI~and \PII. We wish to tackle this problem as directly as possible\footnote{Other direct methods appearing in the mathematical literature include, \textit{e.g.}, for \PI~\cite{cch12, cch13}.} (\textit{i.e.}, bypassing indirect methods such as Riemann--Hilbert and the like), so that its solution might be applicable to arbitrary Painlev\'e-type ODEs (more below). This is done by introducing a method, which we dub ``closed-form asymptotics'', that solely exploits resurgent large-order asymptotics of the Painlev\'e transseries---hence, hopefully, general enough for applications in broad classes of nonlinear systems. Only then may one write fully general connection-formulae implementing Stokes transitions; hence write transseries solutions in all sectors of the complex $z$-plane (in fact more than one Stokes transition may be required); hence finally compute exact locations of Painlev\'e poles (upon specified initial/boundary data). Note that this final step will still require resummation methods to handle the resulting transseries, an analysis which will be addressed elsewhere. Stokes data thus play a \textit{fundamental role} in the study of resurgent-transseries general-solutions to Painlev\'e equations, and our present results finally close the analyses started-out in \cite{gikm10, asv11, sv13}.

One absolutely remarkable aspect of the \PI~\eqref{eq:PI-eq} and \PII~\eqref{eq:PII-eq} equations---specially in light of their purely  mathematical origin---is their appearance in the nonperturbative study of two-dimensional (2d) quantum gravity and minimal string theory. Specifically, \PI~appears in the framework of 2d quantum gravity \cite{gm90a, ds90, bk90, d90, gm90b}, \PII~appears in the framework of 2d quantum supergravity \cite{ps90a, dss90, ps90b}, and both appear within minimal (super) string theory \cite{kms03b, ss03, ss04b} (see, \textit{e.g.}, \cite{g91, gm93, dgz93, n04, ss04a} for reviews). In particular, \PI~is the string-equation describing the (exact) specific-heat of the lowest multicritical  \textit{hermitian} matrix-model. It is also the simplest minimal string theory (in the conformal background \cite{mss91, ss03}). Similarly, \PII~is the string equation describing the \textit{square-root} of the (exact) specific-heat of the lowest multicritical  \textit{unitary} matrix-model; the simplest minimal superstring theory. Due to their role in nonperturbative quantum gravity and string theory, and also due to their relation to (double-scaled) hermitian/unitary random matrices, there has been long-lasting recurrent physical interest in understanding the multi-instanton content of solutions to these Painlev\'e equations; see, \textit{e.g.}, \cite{gz90a, gz90b, d91, gz91, d92, ez93, akk03, ss03, hhikkmt04, st04, kkm04, iy05, iky05, m06, msw07, m08, msw08, kmr10}. These multi-instanton analyses---describing D-brane exponential-corrections \cite{s90, p94} ``beyond-all-orders'' of the string-theoretic perturbative asymptotic expansion \cite{gp88}---were what later naturally led to the aforementioned resurgent-transseries analyses for both \PI~\cite{gikm10, asv11} and \PII~\cite{m08, sv13}, hence bridging the gap between mathematical and physical interests\footnote{Which exists in other directions; \textit{e.g.}, a reformulation of Painlev\'e connection problems in terms of quantum-mechanical exact-WKB analysis was achieved in \cite{t95, t99, t00} for the case of \PI, with a modern counterpart in \cite{is15}.} in these equations. In particular, it is precisely this string-theoretic connection which sparks our focus of interest solely in these first two Painlev\'e equations (and which does not hold for the remaining ones).

This is actually just the beginning of a fascinating story taking us into the realm of \textit{higher-order} Painlev\'e-type ODEs. Both \PI~and \PII~sit at the bottom of (distinct) hierarchical towers of increasingly-complicated, higher-order nonlinear ODEs, describing the specific-heat of all hermitian/unitary\footnote{To be precise, and as already mentioned for \PII, the unitary case yields the \textit{square-root} of the specific heat.} (respectively) multicritical models with \textit{one}-matrix origin. These are the Korteweg--de~Vries (KdV) hierarchy, arising from \PI~\cite{gd75}, and the modified KdV (mKdV) hierarchy, arising from \PII~\cite{ps90b} (see \cite{gs21} for a discussion in our resurgent transseries and Stokes data contexts). Due to their origin in Painlev\'e-type ODEs, this plethora of multicritical and string-theoretic models share common physical and mathematical properties. All specific-heat transcendents along the KdV (mKdV) hierarchy have fields of movable double (simple) poles \cite{gz90b, ps90b}---akin to what happened for \PI~(\PII)---but now with more intricate would-be Boutroux-type classifications. The corresponding (nonperturbative) string-theoretic partition functions follow from the specific heats; where, in both cases, movable poles translate to simple zeroes of the partition functions (this is briefly illustrated in section~\ref{sec:review}). This implies the Boutroux-type classification is, to some extent, a classification of different string-theoretic phases---hence that accessing them will again require complete, analytical, resurgent Stokes data for all these equations; hence that these data will play a \textit{fundamental role} in the uncovering of the associated string physics (only now our results are opening\footnote{Albeit computing generic resurgent Stokes data for all string equations in the KdV and mKdV hierarchies is likely a daunting endeavor. One first step was taken in \cite{gs21}, computing the ``canonical'' Stokes coefficient for all multicritical and string theoretic models associated to the KdV hierarchy (see below for \PI/\PII).} many new analyses; not closing!).

Having in mind eventually addressing all string theories in the aforementioned (m)KdV hierarchy---obtaining their resurgent-transseries structures alongside their complete Stokes data---then direct methods to compute the latter straight-out of string-equations are of prime relevance (hence our ``closed-form asymptotics'' as already mentioned). But this should be complemented with physically-motivated calculations. Indeed, in the physics literature the most interesting calculations arise from matrix models and minimal strings. Herein, ``canonical'' Stokes data has an eigenvalue-tunneling or ZZ-brane \cite{zz01} one-loop amplitude interpretation \cite{m03, kopss04, emms22} and may be computed directly from the matrix integral, for both \PI~\cite{d92, hhikkmt04, msw07} and \PII~\cite{kkm04, m08} (see as well \cite{ss03, st04, iy05, iky05}). For all multicritical and minimal string theoretic models, these ``canonical'' Stokes coefficients were recently computed in \cite{gs21}. The main question, of course, is how to compute all \textit{other} Stokes data? Hopefully, ``closed-form asymptotics'' is general enough to be applicable to all string-equations along the (m)KdV hierarchy (albeit, as mentioned, this may be a daunting task). Let us stress, however, that our long-term goal is to achieve a \textit{direct} calculation of all resurgent Stokes data out of the matrix model/minimal string ZZ-brane interpretation alone. This would probably greatly illuminate the proper role of Stokes data---but is at this stage impaired by the fact that we do not know which types of ZZ-branes could yield the remaining data (only the ``canonical'' coefficients \cite{gs21}). If this was to be achievable, the whole (m)KdV Stokes data would likely unfold. And if that were to happen, Stokes data for Jackiw--Teitelboim gravity would likely also follow (see \cite{gs21} and references therein).

One final remarkable aspect of the \PI~\eqref{eq:PI-eq} and \PII~\eqref{eq:PII-eq} equations is their relation to gauge theories in \textit{four} dimensions. Building upon \cite{agt09, w09, gil12, gl16} it was shown in \cite{blmst16} that the partition functions of \PI~and \PII~relate to certain (distinct) four-dimensional $\CN = 2$ superconformal gauge theories. Although exploring in any further detail such relation is far from the scope of the present work, there is one particular aspect of relevance to our analysis. Essentially by construction such correspondence yields natural variables parametrizing the space of initial/boundary conditions for our ODEs, and these variables seem to lead to the simplest formulation of connection formulae at the level of the Painlev\'e partition function \cite{lr16}. Interestingly enough, these are the very same variables used in the exact WKB analysis of \cite{t95, t99, t00} (see \cite{kt05} for a review), as shown in \cite{i19a, i19b}. As we shall see in section~\ref{sec:stokes}, the ``resurgence origin'' of this particular parametrization stems from a special property of resurgent-transseries Painlev\'e solutions: they are\footnote{In fact, to a great extent, resonance is also the mathematical reason why there is a correspondence between Painlev\'e and gauge-theory partition functions---see the discussions on ``framings'' throughout this paper.} \textit{resonant}, \textit{i.e.}, instanton actions arise in symmetric pairs (this will be reviewed in section~\ref{sec:review}). Choosing to parametrize the moduli-space of initial/boundary conditions---in other words, of \textit{transseries parameters}---in the natural variables arising from ``factoring out'' resonance, leads to the aforementioned simple formulation of connection formulae and gauge theory relation \cite{blmst16, lr16, i19a, i19b}. This will allow us to reformulate the (complicated) nonlinear Stokes data in a rather simple and compact final package. 

The precise contents of this paper are as described in the following. We begin in section~\ref{sec:review} with a swift overview of resurgent-transseries constructions of Painlev\'e solutions---as constructed in \cite{gikm10, asv11, sv13}. This includes both mathematical constructions, alongside physical interpretations in quantum gravity and minimal strings. In particular, we introduce the concept of ``framing'' in the organization of a transseries, which is directly related to resonance and will play a key role in our subsequent Stokes analysis. This is discussed in section~\ref{sec:stokes-resonance}, where we show how resonant transseries imply specific properties for Stokes data, Borel residues, and connection formulae. Starting in this section we need to assume the reader has some working knowledge of \cite{abs18} in order to proceed. Section~\ref{sec:asymptotics} starts with a somewhat general discussion of resurgent asymptotics in the Painlev\'e context, reviewing large-order asymptotics and its uses in the calculation of Stokes data. It then builds its way to the introduction of ``closed-form asymptotics''. The discussion quickly becomes rather technical, but we made our best effort to keep it as pedagogical as possible. Results for complete, analytical, resurgent Stokes data are in section~\ref{sec:results}. This includes results specifically tailored for the Painlev\'e equations, for their quantum-gravity incarnations, and for their string-theoretic incarnations. The reader interested in results but not in the procedure may jump directly to this section, where all data is presented also with many illustrative examples. As we mentioned above, we expect these results may be generalizable to the full (m)KdV-hierarchy string equations. Further, one thing we know for sure is that they are generalizable to the matrix-model origins of either \PI~or \PII. As explained in \cite{asv11, sv13} Painlev\'e (critical) Stokes data immediately translates to matrix-model (off-critical) Stokes data (and vice-versa). Hence our results immediately yield the complete resurgent Stokes data of the \textit{quartic matrix model} (which would, for instance, immediately become relevant in a two-parameter transseries extension of the analysis in \cite{csv15}). To make sure our results are rock-solid, we performed extensive numerical checks. An overview of all those numerics may be found in section~\ref{sec:numerics}, with further details included in appendix~\ref{app:numerics}. Having computed resurgent Stokes data one may finally discuss the nonlinear Stokes phenomenon, and we construct connection-formulae implementing transseries Stokes transitions in our last section~\ref{sec:stokes}. Upon implementing ``diagonal framing'' at the tau-function/partition-function transseries level, such connection formulae simplify considerably. In particular, the complete (and complicated) nonlinear Stokes data may, in this way, be fully packaged in a rather simple and compact final result---whose non-trivial ``numerology content'' reduces to \eqref{eq:exampleStokesP1} and \eqref{eq:exampleStokesP2}. In particular we implement the direct monodromy calculation at Painlev\'e solutions level, and how to map it to the aforementioned isomonodromy calculation which exists in the literature. Having achieved such calculation and such map is tantamount to a proof of our earlier conjectures, making our analysis come full circle and hence closing our paper.

\section{Painlev\'e Equations and Resurgent Transseries}\label{sec:review}

In order to set the stage, let us begin by addressing resonant resurgent-transseries solutions to the Painlev\'e equations \eqref{eq:PI-eq} and \eqref{eq:PII-eq}, briefly reviewing the results in \cite{gikm10, asv11, sv13}. This should also highlight the need for Stokes data, at both transseries and alien calculus levels---albeit we will come back to alien calculus in section~\ref{sec:stokes-resonance}. At the same time, we also address the role these equations play in 2d quantum (super) gravity and minimal string theory---already mentioned in the introduction. Finally, we discuss transseries ``framing''; rectangular versus diagonal.

\subsection{Painlev\'e~I, 2D Quantum Gravity, and Minimal Strings}\label{subsec:PI2dQG}

Let us begin by addressing \PI~\eqref{eq:PI-eq}, which we repeat herein:
\be
\label{eq:PI-eq-v2}
u_{\text{I}}^2 (z) - \frac{1}{6}\, u_{\text{I}}'' (z) = z.
\ee
\noindent
We are following the conventions in \cite{gikm10, asv11}, associated to a matrix-model origin with \textit{odd} potential \cite{msw07}. If one were to consider an \textit{even}-potential origin instead, we would find the $\frac{1}{3}$ normalization which is also natural in the Gel'fand--Dikii KdV potentials context \cite{gd75} (see \cite{gs21} for results in this latter normalization). In the mathematics literature, one other common normalization is instead \cite{c03, c06}
\be
\tilde u_{\text{I}}'' = 6\, \tilde u_{\text{I}}^2 + \tilde z.
\ee
\noindent
Of course all choices trivially relate to each other. We mostly work with \eqref{eq:PI-eq-v2} as we are building upon \cite{gikm10, asv11}, but on occasion we shall also translate our results to the $(2,3)$ minimal-string normalization \cite{gs21} (where $g_{\text{s}}$ is the string coupling)
\be
\label{eq:PI-(2,3)}
u_{(2,3)}^2 - \frac{1}{3} g_{\text{s}}^2\, u''_{(2,3)} = - \frac{4\sqrt{2}}{3}\, z,
\ee
\noindent
as this is the (KdV) normalization\footnote{See, \textit{e.g.}, \cite{msw07, gs21} for the relation between the \eqref{eq:PI-(2,3)} $\frac{1}{3}$ and the \eqref{eq:PI-eq-v2} $\frac{1}{6}$ normalizations.} which matches against string-theoretic world-sheet calculations once $z$ is tuned to the conformal background \cite{mss91, ss03, gs21} (in this case, this is $z \to - \frac{3}{4\sqrt{2}}$).

The construction of resurgent transseries solutions to nonlinear differential equations \cite{c95, c98}, and in particular of resurgent transseries solutions for our (resonant) Painlev\'e systems, begins with a perturbative solution, say $u_{\text{I,pert}} (z)$, expanded in inverse powers of the variable $z$, around $z \sim \infty$. Such a perturbative expansion with asymptotics $u_{\text{I,pert}} (z) \sim \sqrt{z}$ at infinity is easily obtained as 
\be
\label{eq:PI-asymp}
u_{\text{I,pert}} (z) \simeq \sqrt{z} \left( 1 - \frac{1}{48}\, z^{-\frac{5}{2}} - \frac{49}{4608}\, z^{-5} - \frac{1225}{55296}\, z^{-\frac{15}{2}} - \frac{4412401}{42467328}\, z^{-10} - \cdots \right).
\ee
\noindent
This power-series is \textit{asymptotic}, with perturbative coefficients growing factorially fast, $u^{(0)}_g \sim (2g)!$, in which case nonperturbative instanton-type corrections are needed in order to properly define\footnote{Which is not enough---one very much needs Borel resummations as well, but we leave them for the next section.} a complete \PI~solution. These come in the form of a transseries solution. In the variable\footnote{For the moment, just a convenient variable. Below we show it is in fact the (multicritical) string coupling.} $x = z^{-\frac{5}{4}}$, \PI~admits a one-parameter transseries solution of the form
\be
\label{eq:PI-1PTS}
u_{\text{I}} \left(x;\sigma_1\right) = x^{-\frac{2}{5}} \sum_{n=0}^{+\infty} \sigma_1^n\, \rme^{- \frac{n A}{x}}\, x^{n \beta} \sum_{g=0}^{+\infty} u_g^{(n)}\, x^g.
\ee
\noindent
Here $\sigma_1$ is the transseries parameter, $A$ is the instanton action, $\beta$ some characteristic exponent, and $n$ is the multi-instanton number. Plugging this back into \PI~\cite{gikm10, asv11} recursively determines the transseries perturbative coefficients $u_g^{(n)}$ around the $n$-instanton sector and further fixes
\be
\label{eq:PI-Abeta}
A = \pm \frac{8\sqrt{3}}{5}, \qquad \beta = \frac{1}{2}.
\ee
\noindent
The two signs\footnote{Where the specific \textit{symmetric-pair} solution is an immediate telltale of resonance.} for the instanton action are due to the second-order nature of \PI, and already make clear that a full solution entails constructing a \textit{two}-parameter transseries. This is\footnote{When comparing formulae, keep in mind that the factor $x^{-\frac{2}{5}}$ was factored-out most of the time in \cite{asv11}.} \cite{gikm10, asv11}
\be
\label{eq:PI-2PTS}
u_{\text{I}} \left(x;\sigma_1,\sigma_2\right) = x^{-\frac{2}{5}} \sum_{n=0}^{+\infty} \sum_{m=0}^{+\infty} \sigma_1^n\, \sigma_2^m\, \rme^{- \left( n-m \right) \frac{A}{x}}\, \sum_{k=0}^{k_{nm}} \left( \frac{\log x}{2} \right)^k\, \sum_{g=0}^{+\infty} u_{2g}^{(n,m)[k]}\, x^{g+\beta_{nm}^{[k]}}.
\ee
\noindent
The notation is the same as above, only now with two transseries parameters, and with the added intricacies \cite{asv11}
\be
\label{eq:kbeta-def}
k_{nm} = \min \left( n,m \right) -n\, \delta_{nm}, \qquad \beta_{nm}^{[k]} = \frac{1}{2} \left( n+m \right) - \left\lfloor \frac{1}{2} \left( k + k_{nm} \right) \right\rfloor.
\ee
\noindent
In particular, $\beta_{nm}^{[k]}$ is the starting power of the $(n,m)$ asymptotic series. The coefficients $u_{2g}^{(n,m)[k]}$ are again recursively determined by plugging this \textit{ansatz} into \PI, albeit this is best done by working with a variable $w =\sqrt{x}$ hence the reason why we are now labelling coefficients with a $2g$ subscript (see \cite{asv11} for these details, alongside the full recursion relation). Note how transmonomial $\sim \rme^{-\frac{1}{x}}$ powers are \textit{not} all independent as one roams the $(n,m)$ $\BN^2$ transseries lattice---which implies that the transseries \eqref{eq:PI-2PTS} is \textit{resonant}; see, \textit{e.g.}, \cite{abs18}. It should also be immediately clear that there must be more to the above logarithms than initially meets the eye---after all, we expect Painlev\'e solutions to be meromorphic. Indeed, the ``logarithmic sectors'' in the above transseries are \textit{not} independent of each other; rather they are a ``resonant rearrangement'' of the transseries solution \cite{asv11}, as
\be
u_{g}^{(n,m)[k]} = \frac{1}{k!} \left( \frac{4}{\sqrt{3}} \left( m-n \right) \right)^k u_{g}^{(n-k,m-k)[0]}.
\ee
\noindent
This implies the sum in $k$ may be exactly evaluated, trading logarithms with exponentiation of transseries parameters. One obtains\footnote{Note that when the transseries coefficients have no $[k]$ superscript index, we are simply setting it to $[0]$.} \cite{asv11}:
\be
u_{\text{I}} \left(x;\sigma_1,\sigma_2\right) =  x^{-\frac{2}{5}} \sum_{n=0}^{+\infty} \sum_{m=0}^{+\infty} \sigma_1^n\, \sigma_2^m\, \rme^{- \left( n-m \right) \frac{A}{x}}\, x^{- \frac{2}{\sqrt{3}} \left( n-m \right) \sigma_1 \sigma_2} \sum_{g=0}^{+\infty} u_{2g}^{(n,m)}\, x^{g+\beta_{nm}}.
\ee

Some examples of nonperturbative transseries sectors\footnote{The coefficients we display are defined according to
\be
u_{\text{I}}^{(n,m)} (x) \simeq x^{-\frac{2}{5}} \sum_{g=0}^{+\infty} u_{2g}^{(n,m)}\, x^{g+\beta_{nm}}.
\ee
} in \eqref{eq:PI-2PTS} are \cite{asv11}
\bea
u^{(1,0)}_{\text{I}} (x) &\simeq& x^{\frac{1}{10}} - \frac{5}{64 \sqrt{3}}\, x^{\frac{11}{10}} + \frac{75}{8192}\, x^{\frac{21}{10}} - \frac{341329}{23592960 \sqrt{3}}\, x^{\frac{31}{10}} + \cdots, \\
u^{(2,0)}_{\text{I}} (x) &\simeq& \frac{1}{6}\, x^{\frac{3}{5}} - \frac{55}{576 \sqrt{3}}\, x^{\frac{8}{5}} + \frac{1325}{36864}\, x^{\frac{13}{5}} - \frac{3363653}{53084160 \sqrt{3}}\, x^{\frac{18}{5}} + \cdots, \\
u^{(1,1)}_{\text{I}} (x) &\simeq& - x^{\frac{3}{5}} - \frac{75}{512}\, x^{\frac{13}{5}} - \frac{300713}{1572864}\, x^{\frac{23}{5}} - \frac{4807377125}{7247757312}\, x^{\frac{33}{5}} - \cdots, \\
u^{(2,1)}_{\text{I}} (x) &\simeq& \frac{11}{72}\, x^{\frac{11}{10}} - \frac{985}{4608 \sqrt{3}}\, x^{\frac{21}{10}} + \frac{597575}{15925248}\, x^{\frac{31}{10}} - \frac{660060187}{5096079360 \sqrt{3}}\,  x^{\frac{41}{10}} + \cdots.
\eea
\noindent
The coefficients in all these perturbative expansions also grow factorially fast, turning every transseries sector asymptotic. All sectors, however, still relate to each other via resurgence \cite{e81}, as alien calculus relates distinct transseries sectors to each other by means of resurgence relations whose proportionality factors are Stokes data (see, \textit{e.g.}, \cite{abs18})---more on this in the next section.

The bridge to 2d quantum gravity, also denoted the (hermitian) ``$k=2$ multicritical model'', is rather simple \cite{gm90a, ds90, bk90, d90, gm90b}. The \PI~solution \eqref{eq:PI-eq-v2} describes the \textit{specific-heat} of the simplest multicritical model, where the \textit{string coupling} $g_{\text{s}}$ relates to the $z$ (or $x$) variable as 
\be
g_{\text{s}} = x = z^{-\frac{5}{4}}.
\ee
\noindent
The free energy and partition function of this system follows from its specific heat via the usual
\be
\label{eq:PI-F/Z}
F_{\text{I}}'' (z) = - u_{\text{I}} (z), \qquad Z_{\text{I}} (z) = \exp F_{\text{I}} (z).
\ee
\noindent
From the perturbative specific-heat \eqref{eq:PI-asymp} it is clear that the free energy has the usual string-theoretic genus-expansion (the large $z$ expansion is a small $g_{\text{s}}$ expansion),
\be
F_{\text{I}} (g_{\text{s}}) \simeq - \frac{4}{15}\, \frac{1}{g_{\text{s}}^2} + \frac{1}{60} \log g_{\text{s}} + \frac{7}{5760}\, g_{\text{s}}^2 + \frac{245}{331776}\, g_{\text{s}}^4 + \frac{259553}{159252480}\, g_{\text{s}}^6 + \cdots,
\ee
\noindent
and the exponential transmonomials in \eqref{eq:PI-1PTS} become the usual D-brane weights \cite{s90, p94}
\be
\sim \exp \left(- \frac{1}{g_{\text{s}}} \right)
\ee
\noindent
(which, in this context, correspond to ZZ-brane contributions \cite{ss03, gs21, emms22}).

As already mentioned, Painlev\'e movable-poles translate to simple zeroes of the partition function. This is now simple to verify. Instead of trying to solve \eqref{eq:PI-eq-v2} with an (asymptotic) expansion around the (fixed) irregular point $z \sim \infty$, let us consider instead an expansion around some (movable) singularity, $z_0$. A Laurent-series \textit{ansatz} of arbitrary (negative) degree about $z_0$ in \eqref{eq:PI-eq-v2} immediately yields degree $2$ (the well-known double poles of \PI) and further fixes its structure as
\be
u_{\text{I}} (z) = \frac{1}{\left( z-z_0 \right)^2} + \frac{3 z_0}{5} \left( z-z_0\right)^2 + \left( z-z_0 \right)^3 + \lambda \left( z-z_0 \right)^4 + \frac{3 z_0^2}{25} \left( z-z_0 \right)^6 + \cdots.
\ee
\noindent
The two transseries parameters $\sigma_1$, $\sigma_2$, parametrizing initial/boundary conditions of the second-order ODE \PI, have now been traded by $z_0$ and $\lambda$ (albeit the map in-between them is highly non-trivial). Following \eqref{eq:PI-F/Z} to reach free energy and then partition function immediately yields
\be
Z_{\text{I}} (z) = - \left( z-z_0 \right) + \frac{z_0}{20} \left( z-z_0 \right)^5 + \frac{1}{20} \left( z-z_0 \right)^6 + \frac{\lambda}{30} \left( z-z_0 \right)^7 + \frac{z_0^2}{1120} \left( z-z_0 \right)^9 + \cdots.
\ee
\noindent
This shows how \PI~double-poles became $Z_{\text{I}} (z)$ simple zeroes.

For completeness, let us address the $(2,3)$ minimal-string \eqref{eq:PI-(2,3)}. Without surprise, its perturbative free-energy\footnote{The relation between free-energy and specific-heat gets slightly upgraded to $g_{\text{s}}^2 F_{(2,3)}'' (z) = - \frac{1}{2} u_{(2,3)} (z)$; where the $\frac{1}{2}$ factor arises due to change from the \eqref{eq:PI-eq-v2} $\frac{1}{6}$ to the \eqref{eq:PI-(2,3)} $\frac{1}{3}$ normalization \cite{msw07, gs21}.} has the standard string-theoretic genus-expansion (we have already tuned to the conformal background, hence no-longer any $z$-dependence)
\be
F_{(2,3)} (g_{\text{s}}) \simeq - \frac{3}{80}\, \frac{1}{g_{\text{s}}^2} + \frac{7}{810}\, g_{\text{s}}^2 + \frac{245}{6561}\, g_{\text{s}}^4 + \frac{519106}{885735}\, g_{\text{s}}^6 + \frac{10699640}{531441}\, g_{\text{s}}^8 + \cdots.
\ee
\noindent
The ZZ-brane instanton action is now
\be
A = \pm \frac{3\sqrt{3}}{5},
\ee
\noindent
and a couple of free-energy nonperturbative transseries sectors are
\bea
F_{(2,3)}^{(1,0)} (g_{\text{s}}) &\simeq&  - \frac{1}{3 \sqrt{6}}\, g_{\text{s}}^{\frac{1}{2}} + \frac{37}{216 \sqrt{2}}\, g_{\text{s}}^{\frac{3}{2}} - \frac{6433}{10368 \sqrt{6}}\, g_{\text{s}}^{\frac{5}{2}} + \frac{12741169}{11197440 \sqrt{2}}\, g_{\text{s}}^{\frac{7}{2}} -\cdots, \\ 
F_{(2,3)}^{(2,0)} (g_{\text{s}}) &\simeq& - \frac1{108}\, g_{\text{s}} + \frac{109}{3888 \sqrt{3}}\, g_{\text{s}}^2 - \frac{11179}{279936}\, g_{\text{s}}^3 + \frac{11258183}{50388480 \sqrt{3}}\, g_{\text{s}}^4 - \cdots, \\ 
F_{(2,3)}^{(1,1)} (g_{\text{s}}) &\simeq& \frac{6}{5}\, g_{\text{s}}^{-1} + \frac{5}{36}\, g_{\text{s}} + \frac{15827}{77760}\, g_{\text{s}}^3 + \frac{6630865}{3359232}\, g_{\text{s}}^5 + \cdots, \\ 
F_{(2,3)}^{(2,1)} (g_{\text{s}}) &\simeq& - \frac{71}{81\sqrt6}\, g_{\text{s}}^{\frac{3}{2}} + \frac{2999}{1944 \sqrt{2}}\, g_{\text{s}}^{\frac{5}{2}} - \frac{25073507}{2519424 \sqrt{6}}\, g_{\text{s}}^{\frac{7}{2}} + \cdots. 
\eea

\subsection{Painlev\'e~II and 2D Quantum Supergravity}\label{subsec:PII2dSG}

\PII~\eqref{eq:PII-eq} follows in complete parallel with the last subsection. We first repeat it herein:
\be
\label{eq:PII-eq-v2}
\mu_{\text{II}}^3 (z) - \frac{1}{2}\, \mu_{\text{II}}'' (z) = z\, \mu_{\text{II}} (z).
\ee
\noindent
We are following the conventions\footnote{We are also focusing on the case of vanishing parameter, in order to connect to 2d supergravity in the following.} in \cite{m08, sv13}. Note that this normalization is the natural one in the mKdV hierarchy \cite{ps90a, ps90b}, and there are now no issues of even versus odd matrix-model potentials. In the mathematics literature, one other common normalization is instead \cite{c03, c06}
\be
\tilde \mu_{\text{II}}'' = 2\, \tilde \mu_{\text{II}}^3 + \tilde z\, \tilde \mu_{\text{II}},
\ee
\noindent
which is of course trivially related to ours. As we build upon \cite{sv13}, we always work with \eqref{eq:PII-eq-v2}.

As for \PI, we begin with a perturbative solution with asymptotics $\mu_{\text{II,pert}} (z) \sim \sqrt{z}$ at infinity. This is easily obtained as
\be
\label{eq:PII-asymp}
\mu_{\text{II,pert}} (z) \simeq \sqrt{z} \left( 1 - \frac{1}{16}\, z^{-3} - \frac{73}{512}\, z^{-6} - \frac{10657}{8192}\, z^{-9} - \frac{13912277}{524288}\, z^{-12} - \cdots \right).
\ee
\noindent
Just as for \PI, the above \PII~perturbative power-series is \textit{asymptotic}; a nonperturbative solution is only properly defined via a transseries completion. In the variable\footnote{Again, for now just a convenient variable. Below this will turn out to be the string coupling.} $x = z^{-\frac{3}{2}}$, \PII~admits a one-parameter transseries solution of the form
\be
\label{eq:PII-1PTS}
\mu_{\text{II}} \left(x;\sigma_1\right) = x^{-\frac{1}{3}} \sum_{n=0}^{+\infty} \sigma_1^n\, \rme^{- \frac{n A}{x}}\, x^{n \beta} \sum_{g=0}^{+\infty} u_g^{(n)}\, x^g.
\ee
\noindent
Here $\sigma_1$ is the transseries parameter, $A$ is the instanton action, $\beta$ some characteristic exponent, and $n$ is the multi-instanton number. Inserting this \textit{ansatz} back into \PII~\cite{m08, sv13} recursively determines the transseries perturbative coefficients $u_g^{(n)}$ around the $n$-instanton sector and further fixes
\be
\label{eq:PII-Abeta}
A = \pm \frac{4}{3}, \qquad \beta = \frac{1}{2}.
\ee
\noindent
The two signs for the instanton action are again due to the second-order nature of \PII, and again make clear that a full solution entails constructing a \textit{resonant} \textit{two}-parameter transseries. This is\footnote{When comparing formulae, keep in mind that the factor $x^{-\frac{1}{3}}$ was factored-out most of the time in \cite{sv13}.} \cite{sv13}
\be
\label{eq:PII-2PTS}
\mu_{\text{II}} \left(x;\sigma_1,\sigma_2\right) = x^{-\frac{1}{3}} \sum_{n=0}^{+\infty} \sum_{m=0}^{+\infty} \sigma_1^n\, \sigma_2^m\, \rme^{- \left( n-m \right) \frac{A}{x}}\, \sum_{k=0}^{k_{nm}} \left( \frac{\log x}{2} \right)^k\, \sum_{g=0}^{+\infty} u_{2g}^{(n,m)[k]}\, x^{g+\beta_{nm}^{[k]}}.
\ee
\noindent
This is also pretty much the exact same structure as \eqref{eq:PI-2PTS}, including the definitions \eqref{eq:kbeta-def} (but of course all transseries coefficients are distinct---they are now recursively determined by plugging this \textit{ansatz} into \PII, which is again done working with the variable $w =\sqrt{x}$ hence the reason we again label them with a $2g$ subscript; all details alongside the full recursion may be found in \cite{sv13}). The same holds concerning resonance and the remark on logarithms---and its resolution. Also here the ``logarithm sectors'' in the transseries are not independent of each other; rather they are a hallmark of resonance in this case. One now finds \cite{sv13}
\be
u_{g}^{(n,m)[k]} = \frac{1}{k!} \left(8 \left( m-n \right) \right)^k u_{g}^{(n-k,m-k)[0]}.
\ee
\noindent
Again, this implies the sum in $k$ may be exactly evaluated, once more trading logarithms with exponentiation of transseries parameters. One obtains \cite{sv13}:
\be
\mu_{\text{II}} \left(x;\sigma_1,\sigma_2\right) = x^{-\frac{1}{3}} \sum_{n=0}^{+\infty} \sum_{m=0}^{+\infty} \sigma_1^n\, \sigma_2^m\, \rme^{- \left( n-m \right) \frac{A}{x}}\, x^{- 4 \left( n-m \right) \sigma_1 \sigma_2} \sum_{g=0}^{+\infty} u_{2g}^{(n,m)}\, x^{g+\beta_{nm}}.
\ee
\noindent
Some examples of nonperturbative transseries sectors\footnote{Again, the coefficients we display are defined according to
\be
\mu_{\text{II}}^{(n,m)} (x) \simeq x^{-\frac{1}{3}} \sum_{g=0}^{+\infty} u_{2g}^{(n,m)}\, x^{g+\beta_{nm}}.
\ee} in \eqref{eq:PII-2PTS} are \cite{sv13}
\bea
\mu^{(1,0)}_{\text{II}} (x) &\simeq& x^{\frac{1}{6}} -\frac{17}{96}\, x^{\frac{7}{6}} + \frac{1513}{18432}\, x^{\frac{13}{6}} - \frac{850193}{5308416}\, x^{\frac{19}{6}} + \cdots, \\
\mu^{(2,0)}_{\text{II}} (x) &\simeq& \frac{1}{2}\, x^{\frac{2}{3}} - \frac{41}{96}\, x^{\frac{5}{3}} + \frac{5461}{9216}\, x^{\frac{8}{3}} - \frac{1734407}{1327104}\, x^{\frac{11}{3}} + \cdots, \\
\mu^{(1,1)}_{\text{II}} (x) &\simeq& - 3 x^{\frac{2}{3}} - \frac{291}{128}\, x^{\frac{8}{3}} - \frac{447441}{32768}\, x^{\frac{14}{3}} - \frac{886660431}{4194304}\, x^{\frac{20}{3}} - \cdots, \\
\mu^{(2,1)}_{\text{II}} (x) &\simeq& x^{\frac{7}{6}} - \frac{115}{48}\, x^{\frac{13}{6}} + \frac{30931}{18432}\, x^{\frac{19}{6}} - \frac{4879063}{663552}\, x^{\frac{25}{6}} + \cdots,.
\eea
\noindent
The coefficients in all these perturbative expansions also grow factorially fast, turning every transseries sector asymptotic. All sectors, however, still relate to each other via resurgence \cite{e81}, as alien calculus relates distinct transseries sectors to each other by means of resurgence relations whose proportionality factors are Stokes data (see, \textit{e.g.}, \cite{abs18})---more on this in the next section.

The bridge to 2d quantum supergravity, also denoted\footnote{``Hermitian-$k$'' of the previous subsection, and ``unitary-$k$'' herein, are of course not the same $k$.} the (unitary) ``$k=1$ multicritical model'', is rather simple \cite{ps90a, ps90b, kms03b}. The \PII~solution \eqref{eq:PII-eq-v2} describes the \textit{square-root} of the specific heat of the simplest (unitary) multicritical model. This also explains why we have denoted \PII~solutions as $\mu_{\text{II}}$ rather than $u_{\text{II}}$, as we have left the $u$-variable to precisely denote the specific heat. The string coupling $g_{\text{s}}$ now relates to the $z$ (or $x$) variable as
\be
g_{\text{s}} = x = z^{-\frac{3}{2}}.
\ee
\noindent
The free energy and partition function of this system follows from its specific heat\footnote{Due to the $\mu \mapsto -\mu$ \PII~$\BZ_2$-symmetry, \PII~solutions come in pairs but this is irrelevant for the specific-heat.} via the usual
\be
\label{eq:PII-F/Z}
F_{\text{II}}'' (z) = - u_{\text{II}} (z) = - \mu^2_{\text{II}} (z), \qquad Z_{\text{II}} (z) = \exp F_{\text{II}} (z).
\ee
\noindent
From the perturbative specific-heat \eqref{eq:PII-asymp} it is clear that the free energy has the usual string-theoretic genus-expansion (the large $z$ expansion is a small $g_{\text{s}}$ expansion),
\be
F_{\text{II}} (g_{\text{s}}) \simeq - \frac{1}{6}\, \frac{1}{g_{\text{s}}^2} + \frac{1}{12} \log g_{\text{s}} + \frac{3}{128}\, g_{\text{s}}^2 + \frac{63}{1024}\, g_{\text{s}}^4 + \frac{2407}{4096}\, g_{\text{s}}^6 + \cdots,
\ee
\noindent
and the exponential transmonomials in \eqref{eq:PII-1PTS} become the usual D-brane weights \cite{s90, p94}
\be
\sim \exp \left(- \frac{1}{g_{\text{s}}} \right).
\ee

Also in the present \PII~case Painlev\'e movable-poles will translate to simple zeroes of the partition function. Using the same strategy as for \PI, let us try to solve \eqref{eq:PII-eq-v2} with a Laurent expansion around some (movable) singularity, $z_0$. Such an \textit{ansatz} of arbitrary (negative) degree about $z_0$ in \eqref{eq:PII-eq-v2} immediately yields degree $1$ (the well-known simple\footnote{Double-poles of the corresponding specific heat, as expected from a statistical-mechanical standpoint.} poles of \PII) and further fixes its structure as
\be
\mu_{\text{II}} (z) = \frac{1}{z-z_0} + \frac{z_0}{3} \left( z-z_0 \right) + \frac{1}{2} \left( z-z_0 \right)^2 + \lambda \left( z-z_0 \right)^3 + \frac{z_0}{18} \left( z-z_0 \right)^4 + \cdots.
\ee
\noindent
As for \PI, the two transseries parameters of the second-order ODE have been (non-trivially) traded by $z_0$ and $\lambda$. Following \eqref{eq:PII-F/Z}, it is immediate to reach the free energy, fix an integration constant, and finally obtain the partition function
\be
Z_{\text{II}} (z) = - \left( z-z_0 \right) + \frac{z_0}{3} \left( z-z_0 \right)^3 + \frac{1}{6} \left( z-z_0 \right)^4 + \frac{1}{108} \left( 18 \lambda - 5 z_0^2 \right) \left( z-z_0 \right)^5 - \frac{z_0}{30} \left( z-z_0 \right)^6 + \cdots.
\ee
\noindent
This shows how \PII~simple-poles became $Z_{\text{II}} (z)$ simple zeroes.

\subsection{Transseries Structures: Resonance and Framing}\label{subsec:reson-frame}

What the two previous subsections clearly show is that the \PI~and \PII~cases are extremely similar. In fact we can write them both in one go, which will greatly facilitate the upcoming analysis in our paper---as we shall see also their Stokes structure will be extremely similar.

Both \PI~and \PII~two-parameter transseries, \eqref{eq:PI-2PTS} and \eqref{eq:PII-2PTS}, are pretty much the same and we will write them as (herein $\gamma_{\text{I}} = \frac{2}{5}$ and $\gamma_{\text{II}} = \frac{1}{3}$ ensure the leading $\sim \sqrt{z}$ Painlev\'e behavior)
\be
\label{eq:PI/II-2PTS}
\Phi \left(x;\sigma_1,\sigma_2\right) = x^{-\gamma} \sum_{n=0}^{+\infty} \sum_{m=0}^{+\infty} \sigma_1^n\, \sigma_2^m\, \rme^{- \left( n-m \right) \frac{A}{x}}\, \sum_{k=0}^{k_{nm}} \left( \frac{\log x}{2} \right)^k\, \sum_{g=0}^{+\infty} u_{2g}^{(n,m)[k]}\, x^{g+\beta_{nm}^{[k]}}.
\ee
\noindent
It will be convenient in the following to have this broken down into its constituents. The two-parameter transseries, $\Phi \left(x;\sigma_1,\sigma_2\right)$,
\be
\label{eq:trans_form1}
\Phi \left(x;\sigma_1,\sigma_2\right) = x^{-\gamma} \sum_{n=0}^{+\infty} \sum_{m=0}^{+\infty} \sigma_1^n\, \sigma_2^m\, \rme^{- \left( n-m \right) \frac{A}{x}}\, \Phi_{(n,m)}(x),
\ee
\noindent
is herein split into a sum over its nonperturbative $\Phi_{(n,m)}$ sectors
\be
\label{eq:trans_form2}
\Phi_{(n,m)} (x) = \sum_{k=0}^{k_{nm}} \left(\frac{\log x}2\right)^k \Phi_{(n,m)}^{[k]}(x),
\ee
\noindent
each of which being given by an asymptotic series
\be
\label{eq:trans_form3}
\Phi_{(n,m)}^{[k]} (x) \simeq \sum_{g=0}^{+\infty} u_{2g}^{(n,m)[k]}\, x^{g+\beta_{nm}^{[k]}}.
\ee
\noindent
We had already seen that transseries data \eqref{eq:kbeta-def} was the same for \PI~and \PII. We repeat it herein:
\be
\label{eq:trans_form4}
k_{nm} = \min \left( n,m \right) -n\, \delta_{nm}, \qquad \beta_{nm}^{[k]} = \frac{1}{2} \left( n+m \right) - \left\lfloor \frac{1}{2} \left( k + k_{nm} \right) \right\rfloor.
\ee
\noindent
Finally, logarithmic sectors are not independent, as all coefficients satisfy:
\be
\label{eq:properties_coefs}
u_{2g}^{(n,m)[k]} = \frac{1}{k!}\, \Big( \alpha \left(m-n\right) \Big)^k u_{2g}^{(n-k,m-k)[0]}, \qquad \left(-1\right)^{g+\beta^{[k]}_{nm}}\, u_{2g}^{(n,m)[k]} = \left(-1\right)^{\frac{n+m}{2}}\, u_{2g}^{(m,n)[k]}
\ee
\noindent
(we have also included an existing reflection symmetry valid for all coefficients, and which is the same for \PI~and \PII). The only distinctions we have---besides the obvious transseries coefficients and instanton actions---is the parameter we have denoted by $\alpha$ above:
\be
\alpha_{\text{I}} = \frac{4}{\sqrt{3}}, \qquad \alpha_{\text{II}} = 8.
\ee
\noindent
In particular, the logarithmic $k$-sum in \eqref{eq:PI/II-2PTS} may be evaluated exactly, to
\be
\label{eq:trans_form1-LOGSUM}
\Phi \left(x;\sigma_1,\sigma_2\right) = x^{-\gamma} \sum_{n=0}^{+\infty} \sum_{m=0}^{+\infty} \sigma_1^n\, \sigma_2^m\, \rme^{- \left( n-m \right) \frac{A}{x}}\, x^{- \frac{1}{2} \alpha \left( n-m \right) \sigma_1 \sigma_2}\, \Phi_{(n,m)}^{[0]} (x).
\ee

\begin{figure}
	\centering
	\includegraphics[scale=1.2]{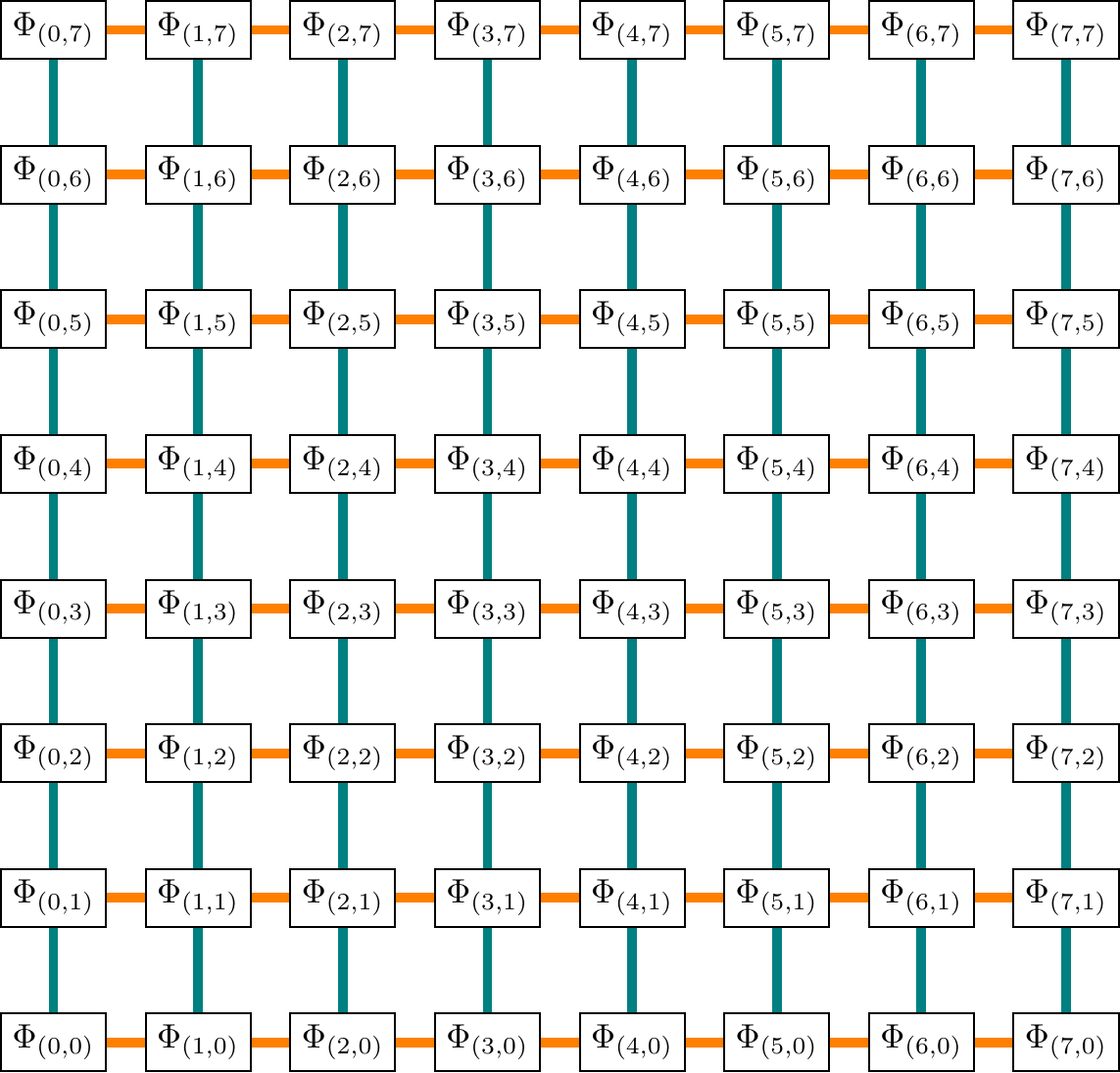}
	\caption{Visual representation of the different sectors building the Painlev\'e transseries \eqref{eq:trans_form1}. As explained in the text, they occupy $(n,m)$ sites on a semi-positive rectangular-lattice, with $n$ ``moving'' along the orange directions and $m$ along the green directions. Compare with figure~\ref{fig:lattice}.}
	\label{fig:lattice-rectangularframing}
\end{figure}

\begin{figure}
	\centering
	\includegraphics[scale=1.2]{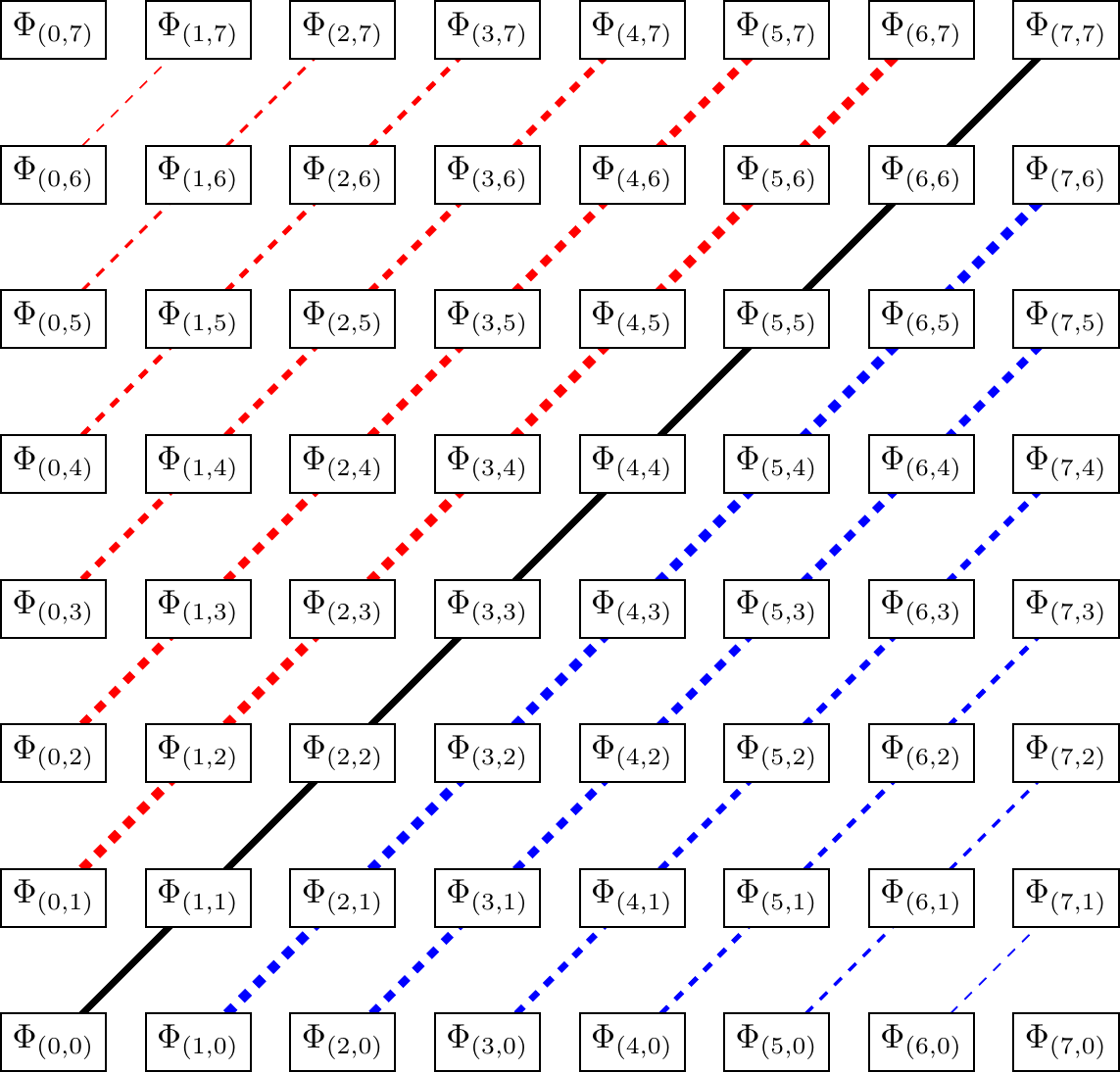}
	\caption{The Painlev\'e transseries, or ``alien lattice'', in diagonal framing \eqref{eq:PI/II-2PTS-diagonal}. In this resonant case, sectors on the main diagonal (black) $\Phi_{(n,n)}$ all have vanishing instanton action, and all sectors along the same diagonal have the same instanton action (always a multiple integer of $A$). In blue we have sectors with positive action, in red we have sectors with negative action. Each line corresponds to a fixed $k$ in equations \eqref{eq:diagonalsummation}-\eqref{eq:diagonalforwardsummation}-\eqref{eq:diagonalbackwardsummation}.}
	\label{fig:lattice}
\end{figure}

To finish setting the stage, let us discuss transseries ``organizations'' in light of resonance (and which will have clear impact in the structure of Stokes data, as we shall see in the following). In fact, by definition, resonance is itself a statement about the transseries organization, \textit{e.g.}, when defined as the existence of \textit{distinct} $(n,m)$ sectors with the \textit{same} transmonomial exponential weight---in our examples, $\sim \exp \left( - \left( n-m \right) \frac{A}{x} \right)$. More generically and more precisely, our two-parameter transseries \eqref{eq:trans_form1} collects nonperturbative sectors $\Phi_{(n,m)}$ labelled on a $\BZ_{\ge0}^2$ semi-positive \textit{rectangular}-lattice as depicted in figure~\ref{fig:lattice-rectangularframing}. These are the sectors which appear in the bridge equations of alien calculus (more in the following), relating distinct sectors to each other via alien derivation and Stokes data---and which further relate to Borel singularities in a natural way \cite{abs18}. On the complex Borel $s$-plane, potential singularities are located at $s = \boldsymbol{\ell} \cdot \boldsymbol{A}$ with $\boldsymbol{\ell} \in \BZ^2$ an integer-valued vector and $\boldsymbol{A} = \left( A, -A \right)$ the pair of Painlev\'e actions. This defines a map projecting the transseries grid into the complex Borel plane as \cite{abs18}
\begin{align}
\mathfrak{P} : \BZ^2 & \to \BC \nonumber \\
\boldsymbol{\ell} & \mapsto \boldsymbol{A} \cdot \boldsymbol{\ell},
\label{eq:map-Z2toC}
\end{align}
\noindent
which is \textit{not} one-to-one once in the resonant case; \textit{i.e.}, $\ker \mathfrak{P} \neq \boldsymbol{0}$. In the present Painlev\'e case, this kernel is generated by the integer multiples of $\boldsymbol{\mathfrak{n}} = \left( 1,1 \right)$. This vector defines the \textit{diagonal} direction of the kernel. In light of this, it is now natural to ask if instead of organizing the transseries in the original $\Phi_{(n,m)}$ ``rectangular framing'' \eqref{eq:trans_form1}, one might instead organize it in ``diagonal framing'', \textit{i.e.}, along the kernel direction; as depicted in figure~\ref{fig:lattice}. Certainly in this case \textit{distinct} transseries sectors will now have \textit{distinct} transmonomial exponential weights---albeit the transseries sectors themselves will be more convoluted. Rewriting \eqref{eq:trans_form1} or \eqref{eq:trans_form1-LOGSUM} in diagonal framing is absolutely straightforward,
\be
\label{eq:PI/II-2PTS-diagonal}
\Phi \left(x;\sigma_1,\sigma_2\right) = \Phi^{(0)} (x,\upmu) + \sum_{k=1}^{+\infty} \sigma_1^{k}\, \rme^{- k\, \frac{A}{x}}\, \Phi^{(k)}_{+} (x,\upmu) + \sum_{k=1}^{+\infty} \sigma_2^{k}\, \rme^{+ k\, \frac{A}{x}}\, \Phi^{(k)}_{-} (x,\upmu).
\ee
\noindent
Herein we have momentarily denoted $\upmu \equiv \sigma_1 \sigma_2$, and introduced the ``new'' sectors
\bea
\label{eq:diagonalsummation}
\Phi^{(0)} (x,\upmu) &:=& x^{-\gamma} \sum_{\ell=0}^{+\infty} \left( \sigma_1 \sigma_2 \right)^{\ell} \Phi_{(\ell,\ell)}^{[0]} (x) = x^{-\gamma} \sum_{\ell=0}^{+\infty} \sum_{g=0}^{+\infty} \left( \sigma_1 \sigma_2 \right)^{\ell} u_{2g}^{(\ell,\ell)}\, x^{g+\beta_{\ell\ell}}, \\
\label{eq:diagonalforwardsummation}
\Phi^{(k)}_{+} (x,\upmu) &:=& x^{-\gamma - k\, \frac{\alpha}{2}\, \sigma_1 \sigma_2} \sum_{\ell=0}^{+\infty} \left( \sigma_1 \sigma_2 \right)^{\ell} \Phi_{(\ell+k,\ell)}^{[0]} (x) = \\
&=& x^{-\gamma - k\, \frac{\alpha}{2}\, \sigma_1 \sigma_2} \sum_{\ell=0}^{+\infty} \sum_{g=0}^{+\infty} \left( \sigma_1 \sigma_2 \right)^{\ell} u_{2g}^{(\ell+k,\ell)}\, x^{g+\beta_{\ell+k,\ell}}, \nonumber \\
\label{eq:diagonalbackwardsummation}
\Phi^{(k)}_{-} (x,\upmu) &:=& x^{-\gamma + k\, \frac{\alpha}{2}\, \sigma_1 \sigma_2} \sum_{\ell=0}^{+\infty} \left( \sigma_1 \sigma_2 \right)^{\ell} \Phi_{(\ell,\ell+k)}^{[0]} (x) = \\
&=& x^{-\gamma + k\, \frac{\alpha}{2}\, \sigma_1 \sigma_2} \sum_{\ell=0}^{+\infty} \sum_{g=0}^{+\infty} \left( \sigma_1 \sigma_2 \right)^{\ell} u_{2g}^{(\ell,\ell+k)}\, x^{g+\beta_{\ell,\ell+k}}.\nonumber
\eea

As we will make clear as we go on, resonance is one of the reasons why it is hard even just trying to \textit{guess} Stokes data. But it will be by tackling resonance head-on, in the resurgent asymptotic relations, that we have managed to bypass it and \textit{analytically} determine these data. This precisely entails looking along the direction of the projection kernel of Stokes data, which was already seen at the level of the transseries itself in the above discussion.

\section{Resurgent Stokes Data in the Resonant Setting}\label{sec:stokes-resonance}

Having made transseries structures clear for both \PI~and \PII~solutions, we still need to properly set-up resurgence and, in sequence, our main players: Stokes data. We refer the reader to the pedagogical introduction in \cite{abs18}, which we lightening review in the following.

Recall that \textit{all} sectors in our transseries are given by asymptotic series \eqref{eq:trans_form3}, with zero radius of convergence, hence requiring Borel resummation in order to yield finite values. This procedure occurs in three steps. First one takes the Borel transform of any asymptotic power-series via $\CB \left[ x^{\alpha+1} \right] (s) := \frac{s^{\alpha}}{\Gamma \left( \alpha+1 \right)}$. This produces a convergent power-series at the origin, which may then be analytically continued throughout the complex $s$-plane to find the function $\CB \left[ \Phi \right] (s)$. Finally, picking a direction $\theta$ of integration on the complex $s$-plane, one obtains the Borel $\theta$-resummation of the asymptotic power-series via Laplace transform
\be
\label{eq:Borel-resum}
\CS_{\theta} \Phi (x) = \int_{0}^{\rme^{\rmi\theta} \infty} \rmd s\, \CB \left[ \Phi \right] (s)\, \rme^{-\frac{s}{x}}.
\ee
\noindent
This simple story turns extremely interesting once one realizes that $\CB \left[ \Phi \right] (s)$ is not entire, and the fundamental role its singularity structure plays. In fact, the above integral \eqref{eq:Borel-resum} will not be defined along rays $\theta_{\text{S}}$ which encounter a singularity of $\CB \left[ \Phi \right] (s)$---these are the \textit{Stokes lines} on the complex Borel plane. In order to describe what happens as the Borel resummation crosses a Stokes line, one first defines \textit{lateral}\footnote{These are always with respect to the Stokes line along $\theta_{\text{S}}$, and we can henceforth drop the ``S'' subscript.} Borel resummations $\CS_{\theta^{\pm}} \Phi$. These two turn out to be related by the action of the \textit{Stokes automorphism} $\underline{\mathfrak{S}}_{\theta}$ \cite{e81},
\be
\label{eq:Stokes-auto}
\CS_{\theta^{+}} = \CS_{\theta^{-}} \circ \underline{\mathfrak{S}}_{\theta}.
\ee
\noindent
Note that, for example, in the simple case of a one-parameter transseries with a Stokes line along $\theta=0$ and a singularity at $A \in \BR^+$---such as, say, \eqref{eq:PI-1PTS} or \eqref{eq:PII-1PTS}---this formulation is elementary and just describes Stokes phenomena as in \cite{s64}. One finds
\be
\label{eq:Stokes-pheno-1pt}
\CS_{+} u \left(x;\sigma\right) = \CS_{-} u \left(x;\sigma+S_1\right),
\ee
\noindent
with $S_1$ the corresponding Stokes coefficient. In full generality, being an automorphism, $\underline{\mathfrak{S}}_{\theta}$ must be the exponential of a derivation---this is the (directional, pointed) \textit{alien derivative} $\underline{\dot{\Delta}}_{\theta}$
\be
\label{eq:Stokes-auto-alien-der}
\underline{\mathfrak{S}}_{\theta} = \exp \underline{\dot{\Delta}}_{\theta}.
\ee
\noindent
The standard alien derivative follows immediately. Let $\left\{ \omega_\theta \right\}$ denote the set of Borel singularities with same argument $\theta$. Then:
\be
\label{eq:dot_al_der}
\underline{\dot{\Delta}}_{\theta} = \sum_{\omega \in \left\{ \omega_\theta \right\}} \rme^{-\frac{\omega}{x}}\, \Delta_{\omega},
\ee
\noindent
with $\Delta_{\omega}$ the standard alien derivation. For our two-parameter transseries \eqref{eq:trans_form1}, with Borel singularities located at $s = \boldsymbol{\ell} \cdot \boldsymbol{A}$ via the projection map \eqref{eq:map-Z2toC}, the action of the alien derivative on a specific transseries sector $\Phi_{\boldsymbol{n}}$, with $\boldsymbol{n} = \left(n,m\right)$, is given by\footnote{This is actually \textit{not} correct, as this expression is only valid for a \textit{non}-resonant transseries. It is nonetheless more pedagogical to start with just this formula, and in any case we will write the correct one right below.}
\be
\label{eq:alien-bridge-equation-NR}
\Delta_{\boldsymbol{\ell} \cdot \boldsymbol{A}} \Phi_{\boldsymbol{n}} = \boldsymbol{S}_{\boldsymbol{\ell}} \cdot \left( \boldsymbol{n} + \boldsymbol{\ell} \right) \Phi_{\boldsymbol{n} + \boldsymbol{\ell}}.
\ee
\noindent
This result sometimes goes by the name of the \textit{bridge equation} \cite{e81}. Herein, $\boldsymbol{S}_{\boldsymbol{\ell}}$ is the two-dimensional Stokes vector associated to the $\boldsymbol{\ell} \cdot \boldsymbol{A}$ Borel singularity or, more precisely, associated to the $\boldsymbol{\ell}$ transseries-lattice site. These are the \textit{Stokes data}, the coefficients we set-out to compute. It turns out \cite{abs18} that they are very much more accessible when working on the complex Borel plane, where the analogue of \eqref{eq:alien-bridge-equation-NR} becomes\footnote{This expression is not a strict equality: it \textit{solely} displays the local, \textit{singular} component of the Borel transform.}
\be
\label{eq:borel-sing-NR}
\CB \left[ \Phi_{\boldsymbol{n}} \right] (s) \Big|_{s = \boldsymbol{\ell} \cdot \boldsymbol{A}} \sim \mathsf{S}_{\boldsymbol{n} \to \boldsymbol{n} + \boldsymbol{\ell}} \times \CB \left[ \Phi_{\boldsymbol{n} + \boldsymbol{\ell}} \right] (s-\boldsymbol{\ell} \cdot \boldsymbol{A})\, \frac{\log \left( s-\boldsymbol{\ell} \cdot \boldsymbol{A} \right)}{2\pi\rmi},
\ee
\noindent
and where the proportionality factors $\mathsf{S}_{\boldsymbol{n} \to \boldsymbol{n} + \boldsymbol{\ell}}$ are the Borel residues. They encode the exact same information as Stokes data (and in fact obviously relate to each other; see \cite{abs18} for many such formulae). In some sense they are the ``unexponentiated'' version of Stokes data, as one may write the action of the Stokes automorphism on a specific transseries sector as\footnote{The infinite sum truncating if we hit the transseries-lattice boundary.} \cite{abs18}
\be
\label{eq:stokes-bridge-equation-NR}
\underline{\mathfrak{S}}_{\theta_{\boldsymbol{\ell}}} \Phi_{\boldsymbol{n}} = \Phi_{\boldsymbol{n}} - \sum_{k=1}^{+\infty} \mathsf{S}_{\boldsymbol{n} \to \boldsymbol{n} + k \boldsymbol{\ell}}\, \rme^{- k \frac{\boldsymbol{\ell} \cdot \boldsymbol{A}}{x}}\, \Phi_{\boldsymbol{n} + k \boldsymbol{\ell}}.
\ee
\noindent
Once all this data is on the table, one may walk the road back to the Stokes automorphism \eqref{eq:Stokes-auto} and finally fully describe the crossing of a Stokes line. This makes it clear how whereas transseries expansions---essentially by construction---immediately represent local solutions to our Painlev\'e equations, they can only be understood as global solutions once Stokes data is known. We will come back to the resulting connection formulae in section~\ref{sec:stokes}. We also refer the interested reader to \cite{abs18} and its references for a detailed exposition of all these concepts.

\begin{figure}
	\centering
	\includegraphics[scale=1.2]{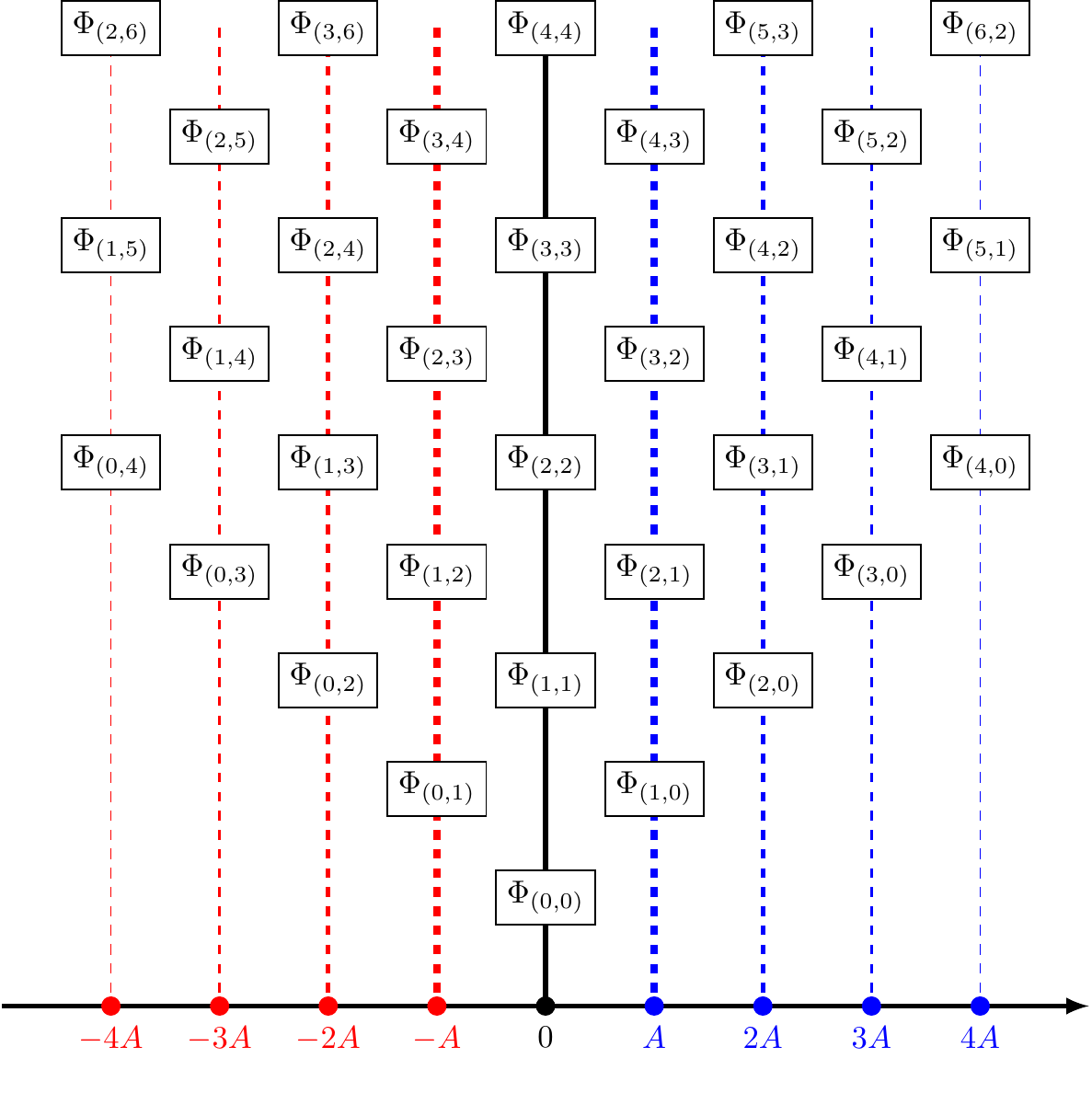}
	\caption{Schematic projection of the transseries ``alien lattice'' into the Borel plane, in the present resonant setting. Multiple nodes in the lattice get projected to the same point, and the singularity structure on the Borel plane becomes more complicated than in the non-resonant setting. Under this projection, all singularities on the Borel plane are located at $nA$, with $n \in \BZ^{\times}$.}
	\label{fig:singular}
\end{figure}

Let us run this story again, but now in our precise Painlev\'e context and being fully explicit on what concerns resonance. Following our discussion in subsection~\ref{subsec:reson-frame}, in the resonant setting the projection map \eqref{eq:map-Z2toC}
\be
\mathfrak{P}: \boldsymbol{\ell} \mapsto \boldsymbol{A} \cdot \boldsymbol{\ell}, \qquad \boldsymbol{\ell} \in \BZ^2,
\ee
\noindent
has a non-trivial kernel, $\ker \mathfrak{P} \neq \boldsymbol{0}$. For our Painlev\'e equations, where the action-vector has the form $\boldsymbol{A} = \left( A, -A \right)$, this is simply
\be
\ker \mathfrak{P} = \left\{ n \left(1,1\right ), n \in \BZ \right\}.
\ee
\noindent
Resonance also plays a distinctive role at the resurgent level, where multiple transseries sectors have the same action, or transmonomial weight, as we illustrated in figure~\ref{fig:lattice}. In fact, in the resonant setting, all transseries sectors of the form $\Phi_{\bm{n}+\bm{k}}$ with $\bm{k} \in \ker\mathfrak P$ will contribute to the very same Borel singularity---which immediately implies the singularity structure cannot possibly be as simple as was illustrated in \eqref{eq:borel-sing-NR} (hence, neither can \eqref{eq:alien-bridge-equation-NR} exactly hold). A simple visualization of this projection is illustrated in figure~\ref{fig:singular}.

\subsection{Setup: Organizing Resonant Stokes Vectors}\label{subsec:reson-stokes}

\begin{figure}
	\centering
	\includegraphics[scale=0.49]{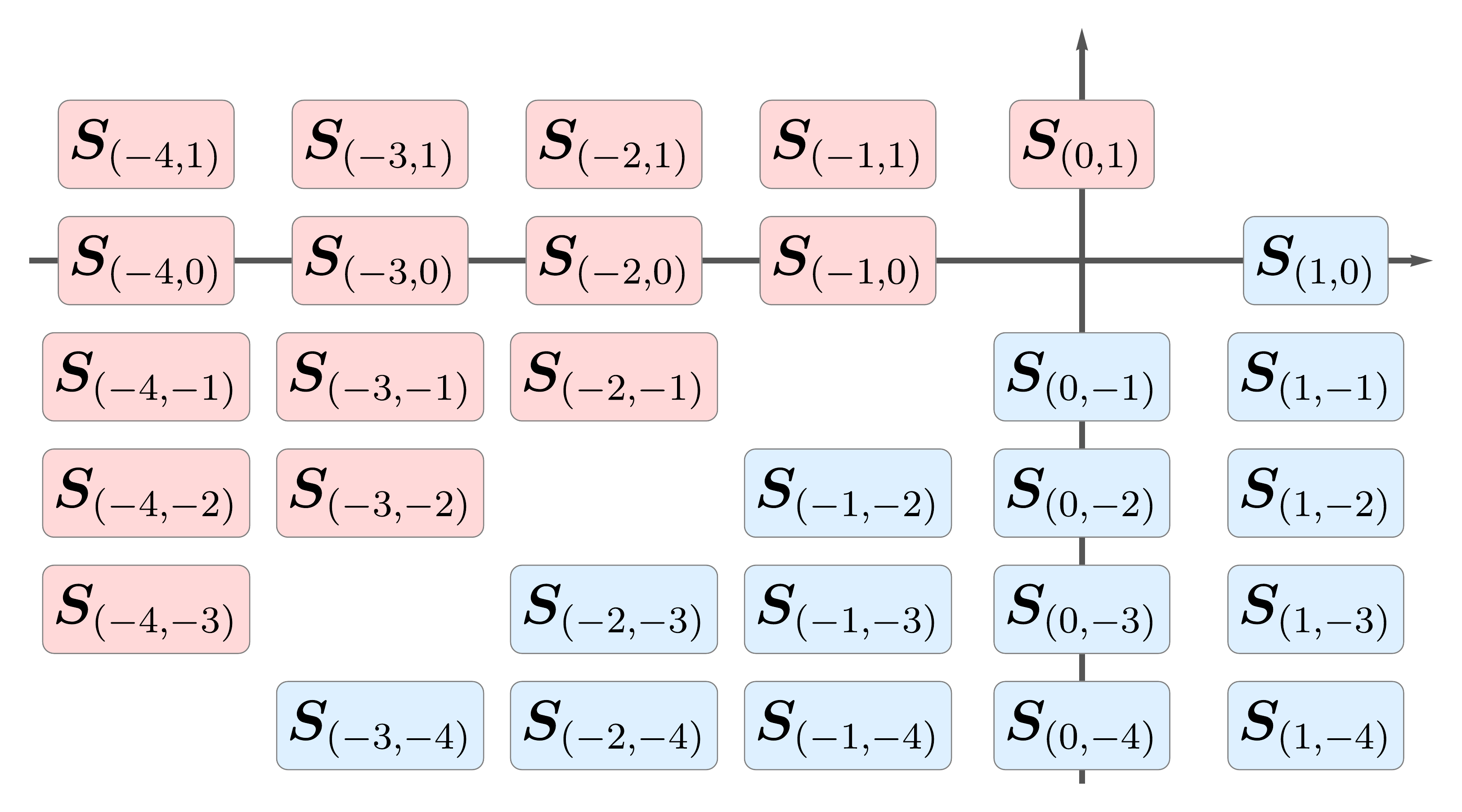}
	\caption{Organization of the Stokes vectors on a two-dimensional lattice. Stokes vectors along the same diagonal are associated to the same instanton-action, \textit{i.e.}, they will appear together upon action of the corresponding alien derivative, \eqref{eq:alien_def_coords_2-a}-\eqref{eq:alien_def_coords_2-b}. This fact, that multiple Stokes vectors are associated to the same instanton-action, is of course a consequence of resonance.}
	\label{fig:lattice_stokes}
\end{figure}

The correction required to make \eqref{eq:alien-bridge-equation-NR} precise follows from looking at figure~\ref{fig:singular} (but see \cite{abs18} for a proper derivation). The alien derivative on a specific sector $\Phi_{\bm{n}}$ now ``sees'' the whole kernel-direction, transseries sectors and Stokes vectors alike,
\be
\label{eq:alien_def_vectors}
\Delta_{\bm{\ell} \cdot \bm{A}} \Phi_{\bm{n}} = \sum_{\bm{p} \in \ker \mathfrak{P}} \bm{S}_{\bm{\ell}+\bm{p}} \cdot \left( \bm{n}+\bm{\ell}+\bm{p} \right) \Phi_{\bm{n}+\bm{\ell}+\bm{p}}.
\ee
\noindent
For our Painlev\'e cases, it will be often convenient to rewrite \eqref{eq:alien_def_vectors} in component notation. Note how the projection map $\mathfrak{P}$ has the same action on every representative of classes in $\BZ^2/\ker\mathfrak{P}$, so it is convenient to choose representatives with one null component. In particular, we will distinguish between \textit{forward} alien derivatives---derivatives $\Delta_{\bm{\ell} \cdot \bm{A}}$ with $\bm{\ell} \cdot \bm{A} > 0$---and \textit{backward} alien derivatives---derivatives $\Delta_{\bm{\ell} \cdot \bm{A}}$ with $\bm{\ell} \cdot \bm{A} < 0$. With $\ell \in \BN^+$, we simply define $\Delta_{\ell A} := \Delta_{(\ell,0) \cdot \bm{A}}$ and $\Delta_{-\ell A} := \Delta_{(0,\ell) \cdot \bm{A}}$. In this case, equation \eqref{eq:alien_def_vectors} becomes
\bea
\label{eq:alien_def_coords_1-a}
\Delta_{\ell A} \Phi_{(n,m)} &=& \sum_{p \in \BZ} \bm{S}_{(\ell+p,p)} \cdot \left[ \begin{array}{c}n+\ell+p\\m+p\end{array} \right] \Phi_{(n+\ell+p,m+p)}, \\
\label{eq:alien_def_coords_1-b}
\Delta_{-\ell A} \Phi_{(n,m)} &=& \sum_{p \in \BZ} \bm{S}_{(p,\ell+p)} \cdot \left[ \begin{array}{c}n+p\\m+\ell+p\end{array} \right] \Phi_{(n+p,m+\ell+p)}.
\eea
\noindent
These formulae may be simplified given the natural resurgence bounds on Stokes vectors \cite{abs18}, namely, $\bm{S}_{(p,q)}$ vanishes if either $p>1$ or $q>1$ or $(p,q)=(1,1)$. Further, another simplification on the infinite-sum arises from the fact that by definition $\Phi_{(n,m)}=0$ for $n<0$ or $m<0$. Finally, it will be useful for the following to perform the substitution $p\to-p-\ell+1$. We may then rewrite \eqref{eq:alien_def_coords_1-a}-\eqref{eq:alien_def_coords_1-b} as
\bea
\label{eq:alien_def_coords_2-a}
\Delta_{\ell A} \Phi_{(n,m)} &=& \sum_{p=0}^{\min(n+1,m+1-\ell)} \bm{S}_{(1-p,1-p-\ell)} \cdot \left[ \begin{array}{c}n+1-p\\m+1-p-\ell\end{array} \right] \Phi_{(n+1-p,m+1-p-\ell)}, \\
\label{eq:alien_def_coords_2-b}
\Delta_{-\ell A} \Phi_{(n,m)} &=& \sum_{p=0}^{\min(n+1-\ell,m+1)} \bm{S}_{(1-p-\ell,1-p)} \cdot \left[ \begin{array}{c}n+1-p-\ell\\m+1-p\end{array} \right] \Phi_{(n+1-p-\ell,m+1-p)}.
\eea
\noindent
These are the alien derivatives we shall use in the following. In accordance with these expressions, let us define a \textit{forward} Stokes vector, a vector $\bm{S}_{(p,q)}$ with $p-q>0$, and a \textit{backward} Stokes vector, a vector $\bm{S}_{(p,q)}$ with $p-q<0$. Note that Stokes vectors are organized on a two-dimensional lattice, almost entirely sitting in the third quadrant of $\BZ^2$ \cite{abs18}---which is illustrated in figure~\ref{fig:lattice_stokes}.

Let us make a remark on conventions. In previous work, notably the one we construct upon \cite{asv11, sv13, as13}, Stokes data notation used different conventions. When comparing with those papers, we find Stokes data denoted by $S^{(a)}_b$ and $\widetilde{S}^{(a)}_b$ therein. Our present work follows in line with the vectorial notation of \cite{abs18}. In order to compare all different conventions, it is enough to compare our bridge equations \eqref{eq:alien_def_coords_2-a}-\eqref{eq:alien_def_coords_2-b} with the corresponding equations in \cite{asv11, sv13, as13}. The resulting map between notations is then:
\be
\label{eq:old_new_not}
\bm{S}_{(1-p,1-p-\ell)} = \left[ \begin{array}{c} S_\ell^{(p)} \\ \widetilde{S}_\ell^{(p-\ell)} \end{array} \right], \qquad \bm{S}_{(1-p-\ell,1-p)} = \left[ \begin{array}{c} S_{-\ell}^{(p-\ell)} \\ \widetilde{S}_{-\ell}^{(p)} \end{array} \right],
\ee
\noindent
for $\ell \in \BN^+$ and $p \in \BN$. Imposing reality of the transseries solution at real, positive $x$ constrains the forward Stokes vectors to be purely imaginary \cite{as13}. This behavior has been checked numerically for both \PI~\cite{asv11} and \PII~\cite{sv13}. For backward Stokes vectors, however, no such reality condition holds---the components of the vectors have been numerically observed to show non-trivial phases. 

\subsection{Setup: Organizing Resonant Borel Residues}\label{subsec:reson-borel}

Having understood resonant Stokes data, let us next address resonant Borel residues. Whereas Stokes data are the building blocks of resurgence as understood via alien calculus \eqref{eq:alien-bridge-equation-NR}, their rearrangement into Borel residues essentially appears everywhere else. To start-off with, when studying Borel singularities as in \eqref{eq:borel-sing-NR}. But also when addressing the resurgent large-order behavior of transseries sectors, all asymptotic formulae explicitly depend on this rearrangement of Stokes data into Borel residues---to the extent that one may think of Borel residues as some sort of ``amplitudes'' measuring the effect of resurgence: \textit{e.g.}, at large-order, the Borel residue $\mathsf{S}_{\boldsymbol{n} \to \boldsymbol{m}}$ measures the influence of the sector $\Phi_{\boldsymbol{m}}$ on the large-order behavior of the $\Phi_{\boldsymbol{n}}$ sector. In light of the alien-derivative ``resonant upgrade'' from \eqref{eq:alien-bridge-equation-NR} to \eqref{eq:alien_def_vectors}, it is now also simple to see how Borel singularities behave under resonance (looking at figure~\ref{fig:singular} or else going back to \cite{abs18}),
\be
\label{eq:borel-sing-R}
\CB \left[ \Phi_{\boldsymbol{n}} \right] (s) \Big|_{s = \boldsymbol{\ell} \cdot \boldsymbol{A}} \sim \sum_{\bm{p} \in \ker\mathfrak{P}} \mathsf{S}_{\boldsymbol{n} \to \boldsymbol{n} + \boldsymbol{\ell} + \bm{p}} \times \CB \left[ \Phi_{\boldsymbol{n} + \boldsymbol{\ell} + \bm{p}} \right] (s-\boldsymbol{\ell} \cdot \boldsymbol{A})\, \frac{\log \left( s-\boldsymbol{\ell} \cdot \boldsymbol{A} \right)}{2\pi\rmi}.
\ee

As already mentioned with \eqref{eq:stokes-bridge-equation-NR}, Borel residues are the relevant combinations when spelling out the action of the Stokes automorphism \eqref{eq:Stokes-auto-alien-der} on specific transseries sectors (and, eventually, we will see how they build-up connection formulae). For our Painlev\'e problems, with instanton actions $\pm A$, the only non-trivial Stokes automorphisms are $\underline{\mathfrak{S}}_0$ and $\underline{\mathfrak{S}}_\pi$. They can be expressed in terms of the Borel residues as
\bea
\label{eq:borel_def-a}
\underline{\mathfrak{S}}_0 \Phi_{(n,m)} &=& \Phi_{(n,m)} - \sum_{\ell=1}^{+\infty} \rme^{-\ell \frac{A}{x}} \sum_{p \in \BZ} \mathsf{S}_{(n,m)\to(n+\ell-p,m-p)}\, \Phi_{(n+\ell-p,m-p)}, \\
\label{eq:borel_def-b}
\underline{\mathfrak{S}}_\pi \Phi_{(n,m)} &=& \Phi_{(n,m)} - \sum_{\ell=1}^{+\infty} \rme^{+\ell \frac{A}{x}} \sum_{p \in \BZ} \mathsf{S}_{(n,m)\to(n-p,m+\ell-p)}\, \Phi_{(n-p,m+\ell-p)}.
\eea
\noindent
As always, we implicitly define $\Phi_{(a,b)}$ to vanish if $a<0$ or $b<0$. Via \eqref{eq:borel_def-a}-\eqref{eq:borel_def-b} above, we may now split the Borel residues in two classes, much like we did for the alien derivatives and Stokes vectors in \eqref{eq:alien_def_coords_2-a}-\eqref{eq:alien_def_coords_2-b}. Denote $\mathsf{S}_{(n,m)\to(a,b)}$ as a \textit{forward} Borel residue if it appears in the action of $\underline{\mathfrak{S}}_0$ upon $\Phi_{(n,m)}$; and as a \textit{backward} Borel residue if it instead appears in the action of $\underline{\mathfrak{S}}_\pi$ upon $\Phi_{(n,m)}$. It is immediate to see that a Borel residue $\mathsf S_{(n,m)\to(a,b)}$ is forward if and only if $n-m<a-b$, and it is backward if and only if $n-m>a-b$.

\begin{figure}
	\centering
	\includegraphics[scale=1.1]{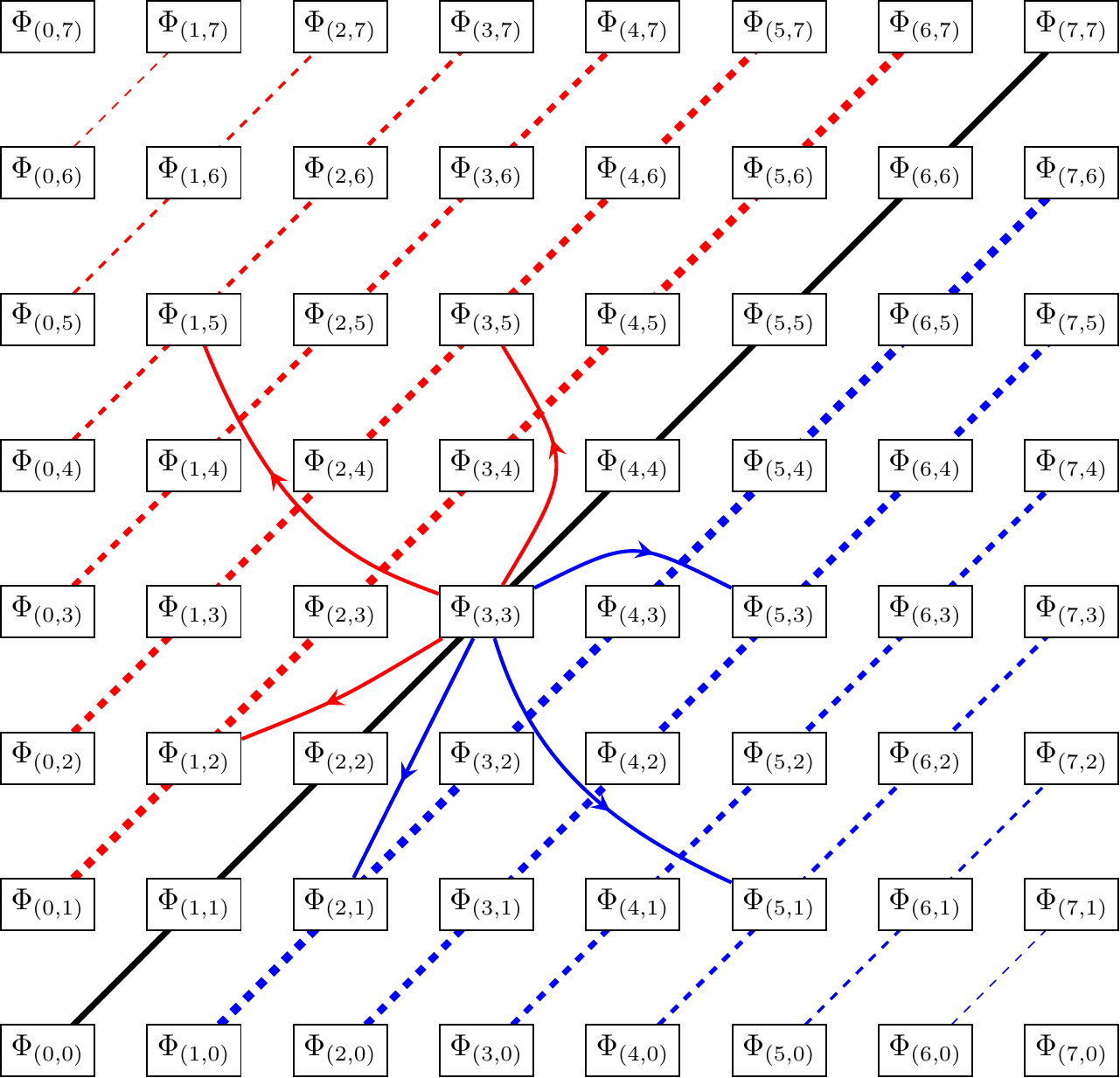}
	\caption{Organization of the Borel residues on the transseries lattice. Being complex-numbers (not vectors) in-between two distinct nodes, one convenient way to visualize Borel residues is by interpreting them as ``amplitudes'' for the resurgence of the linked nodes. In the plot, we have illustrated various sectors which resurge upon the asymptotic behavior of the $\Phi_{(3,3)}$ sector (their thickness illustrates the strength of their contribution). We have represented with blue arrows the Borel residues of positive action, and with red arrows the Borel residues of negative action.}
	\label{fig:borel_org}
\end{figure}

Borel residues are slightly more difficult to display in a graphical representation as compared to Stokes vectors---whereas the latter essentially depend on a lattice site and are immediate to organize as in figure~\ref{fig:lattice_stokes}, the former depend on ``starting'' and ``ending'' lattice nodes. As such, one convenient way to represent them on the two-dimensional transseries ``alien lattice'' with sectors $\Phi_{(n,m)}$, is with arrows in-between the joined sectors. We plot one such visualization in figure~\ref{fig:borel_org}.

\subsection{Relating Borel Residues and Stokes Vectors}\label{subsec:stok_to_borel}

Borel residues and Stokes vectors obviously encode the exact same information, in which case we may determine ones from the others---see \cite{abs18} for many explicit such formulae in generic cases. Let us make these relations precise in the present Painlev\'e example. In fact it will turn out that there is a ``minimal set'' of Borel residues out from which all Stokes data may be constructed. Their relation stems from Stokes data being an ``exponentiated'' version of Borel residues; via the Stokes automorphisms \eqref{eq:Stokes-auto-alien-der} which are now:
\bea
\label{eq:stok_to_borel_formula-a}
\underline{\mathfrak{S}}_0 &=& \exp \underline{\dot{\Delta}}_0 = \exp \left( \sum_{\ell=1}^{+\infty} \rme^{-\ell\frac{A}{x}}\, \Delta_{\ell A} \right), \\
\label{eq:stok_to_borel_formula-b}
\underline{\mathfrak{S}}_\pi &=& \exp \underline{\dot{\Delta}}_\pi = \exp \left( \sum_{\ell=1}^{+\infty} \rme^{+\ell\frac{A}{x}}\, \Delta_{-\ell A} \right).
\eea
\noindent
The explicit relation between Stokes vectors and Borel residues is immediately obtained by simply applying \eqref{eq:stok_to_borel_formula-a}-\eqref{eq:stok_to_borel_formula-b} to any sector, expanding the exponential (using definitions \eqref{eq:alien_def_coords_2-a}-\eqref{eq:alien_def_coords_2-b} to recursively compute multiple alien derivatives), and then collecting the resulting terms so as to fit them appropriately in \eqref{eq:borel_def-a}-\eqref{eq:borel_def-b}. There are three important properties in this relation:
\begin{itemize}
\item In light of our ``forward-backward'' definitions, any forward Borel residue will only be a combination of forward Stokes vectors; and any backward Borel residue will only be a combination of backward Stokes vectors (actually, hence those definitions).
\item The bounds on the sums in the alien derivatives \eqref{eq:alien_def_coords_2-a}-\eqref{eq:alien_def_coords_2-b} also translate to bounds in the Borel residue formulae. For example, when considering forward alien derivatives \eqref{eq:alien_def_coords_2-a} on $\Phi_{(n,m)}$, one obtains linear combinations of the sectors $\Phi_{(n+\ell-p,m-p)}$, with $p \geq 0$. On Borel residues, this translates to $\mathsf S_{(n,m)\to(n+\ell-p,m-p)}$ and $\mathsf S_{(n,m)\to(n-p,m+\ell-p)}$ vanishing if $p<0$. It is then convenient to rewrite \eqref{eq:borel_def-a}-\eqref{eq:borel_def-b} with such explicit bounds, as
\bea
\label{eq:borel_def-2-a}
\underline{\mathfrak{S}}_0 \Phi_{(n,m)} &=& \Phi_{(n,m)} - \sum_{\ell=1}^{+\infty} \rme^{-\ell \frac{A}{x}} \sum_{p=0}^{\min(n+\ell,m)} \mathsf{S}_{(n,m)\to(n+\ell-p,m-p)}\, \Phi_{(n+\ell-p,m-p)}, \\
\label{eq:borel_def-2-b}
\underline{\mathfrak{S}}_\pi \Phi_{(n,m)} &=& \Phi_{(n,m)} - \sum_{\ell=1}^{+\infty} \rme^{+\ell \frac{A}{x}} \sum_{p=0}^{\min(n,m+\ell)} \mathsf{S}_{(n,m)\to(n-p,m+\ell-p)}\, \Phi_{(n-p,m+\ell-p)}.
\eea
\noindent
These constraints on Borel residues are illustrated in figure~\ref{fig:constraints_borel}.
%
\begin{figure}
	\centering
	\includegraphics[scale=1.23]{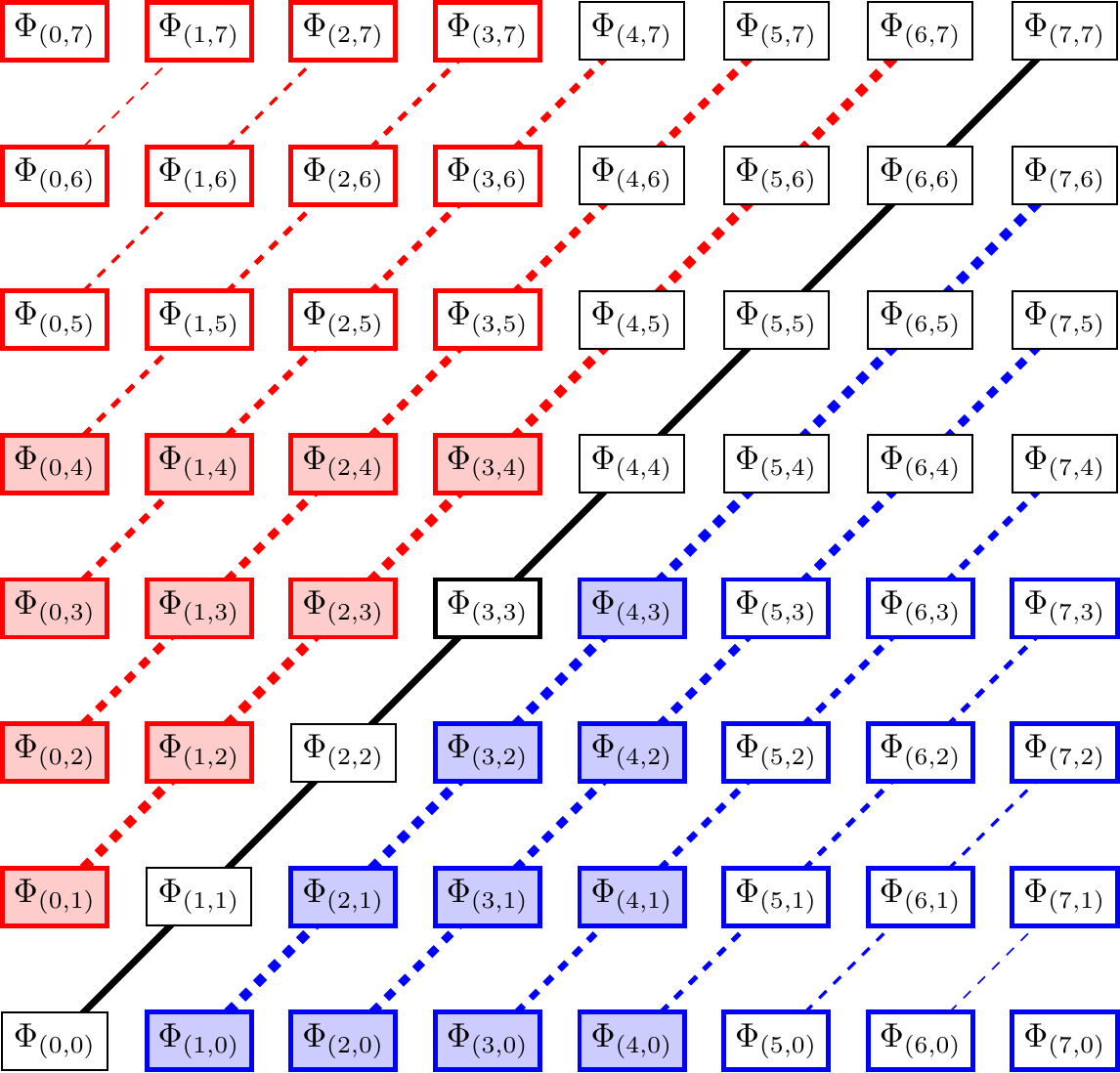}
	\caption{Illustration of the constraints on Borel residues. As in figure~\ref{fig:borel_org}, let us focus on the $\Phi_{(3,3)}$ sector, which is framed in black in the plot. From this sector, the non-vanishing forward Borel residues are those connecting $\Phi_{(3,3)}$ to all sectors \textit{framed} in blue---but where \textit{only} the Borel residues connecting to sectors \textit{colored} in blue are actually needed to construct all other Borel residues $\mathsf{S}_{(3,3)\to(p+s,p)}$. An analogous situation holds for the backward residues (now in red).}
	\label{fig:constraints_borel}
\end{figure}
%
\item There is a ``minimal set'' of Borel residues which yields \textit{all} Stokes data (and, conversely, all other Borel residues). To characterize this minimal set, let us focus on forward Stokes data---the same discussion holds for backward Stokes data. The relation between Borel residues and Stokes data implies that every forward Borel residue will be of the form
\be
\label{eq:rest_borel-a}
\mathsf{S}_{(n,m)\to(n+\ell-p,m-p)} = - \bm{S}_{(\ell-p,-p)} \cdot \left[ \begin{array}{c}n+\ell-p\\m-p\end{array} \right] + \mathsf{R}_{(n+\ell-p,m-p)}^{(n,m)},
\ee
\noindent
where the remainder term $\mathsf{R}_{(n+\ell-p,m-p)}^{(n,m)}$ is a linear combination of products of forward Stokes vectors $\bm{S}_{(t-q,-q)}$, with $t<\ell$ and $q\leq p$, and all the actions of the Stokes vectors in a given term of the product will sum up to $\ell$. By working inductively on $\ell$, we can deduce that all Stokes data $\bm{S}_{(t-q,-q)}$ with $t<\ell$ and $q\leq p$ are known. Then, all remainder terms are known, and we write instead
\be
\label{eq:rest_borel-b}
\bm{S}_{(\ell-p,-p)} \cdot \left[ \begin{array}{c}n+\ell-p\\m-p\end{array} \right] = \mathsf{R}_{(n+\ell-p,m-p)}^{(n,m)} - \mathsf{S}_{(n,m)\to(n+\ell-p,m-p)},
\ee
\noindent
where in this rewrite the right-hand side is fully known: the $\mathsf{R}_{(n+\ell-p,m-p)}^{(n,m)}$ are known by the inductive hypothesis, and the Borel residues with action $\ell$ have been obtained using some numerical procedure (more below). Next, consider two copies of the previous equation, where we choose $n=m=p$ and $n=m=p+1$,
\be
\begin{cases}
\, \bm{S}_{(\ell-p,-p)} \cdot \left[ \begin{array}{c}\ell\\0\end{array} \right] = \mathsf{R}_{(\ell,0)}^{(p,p)} - \mathsf{S}_{(p,p)\to(\ell,0)}, \\
\\
\, \bm{S}_{(\ell-p,-p)} \cdot \left[ \begin{array}{c}\ell+1\\1\end{array} \right] = \mathsf{R}_{(\ell+1,1)}^{(p+1,p+1)} - \mathsf{S}_{(p+1,p+1)\to(\ell+1,1)},
\end{cases}
\ee
\noindent
and rearrange these two equations as a matrix equation:
\be
\left[ \begin{array}{cc}\ell&\,\,\,0\\\ell+1&\,\,\,1\end{array} \right] \bm{S}_{(\ell-p,-p)} = \left[ \begin{array}{c}\mathsf{R}_{(\ell,0)}^{(p,p)} - \mathsf{S}_{(p,p)\to(\ell,0)}\\ \mathsf{R}_{(\ell+1,1)}^{(p+1,p+1)} - \mathsf{S}_{(p+1,p+1)\to(\ell+1,1)}\end{array} \right].
\ee
\noindent
The Stokes vector $\bm{S}_{(\ell-p,-p)}$ finally follows from matrix inversion (with determinant $\ell$, hence always invertible) acting on the right-hand side which is known. In conclusion, the set of Borel residues
\be
\label{eq:sufficientsetofborels}
\left. \left\{ \mathsf{S}_{(q,q)\to(t,0)},\, \mathsf{S}_{(q+1,q+1)\to(t+1,1)} \,\,\, \right| \,\,\, 1 \leq t < \ell,\, 0 \leq q \leq p \right\}
\ee
\noindent
is sufficient to construct all remainder terms $\mathsf{R}_{(\ell,0)}^{(p,p)}$ and $\mathsf{R}_{(\ell+1,1)}^{(p+1,p+1)}$; hence, alongside the Borel residues $\mathsf{S}_{(p,p)\to(\ell,0)}$ and $\mathsf{S}_{(p+1,p+1)\to(\ell+1,1)}$, they are sufficient to construct all Stokes vectors $\bm{S}_{(\ell-p,-p)}$. A completely analogous result holds for the backward direction: remainder terms for $\bm{S}_{(-p,\ell-p)}$ are now constructed from the set
\be
\left. \left\{ \mathsf{S}_{(q,q)\to(0,t)},\, \mathsf{S}_{(q+1,q+1)\to(1,t+1)} \,\,\, \right| \,\,\, 1 \leq t < \ell,\, 0 \leq q \leq p \right\},
\ee
\noindent
with the vector itself obtained with the added knowledge of $\mathsf{S}_{(p,p)\to(0,\ell)}$ and $\mathsf{S}_{(p+1,p+1)\to(1,\ell+1)}$.
\end{itemize}
\noindent
Then, for all purposes and from now on, we may solely focus on Borel residues which start at transseries sectors of vanishing instanton-action (the main diagonal). Everything else follows. In addition, we will later find a relation between forward and backward Borel residues, that allows us to obtain one set from the other: this relation will hence allow us to solely focus on an even smaller subset of the data, the set of forward Borel residues which start at diagonal sectors.

\subsection{The General Structure of Stokes Vectors}\label{subsec:vec_struct}

It turns out that not only the Borel residue story is simpler than it seems at first sight---as we have just seen in the previous subsection---but also Stokes vectors have a simpler structure than what it seems at first sight. In order to understand this, we need to make use of two facts:
\begin{itemize}
\item In \cite{asv11}, for \PI, and in \cite{sv13}, for \PII, numerical observations suggested that the following relation between vectorial-components of Stokes vectors should hold:
\be
\label{eq:structure_1}
\boldsymbol{S}^{(2)}_{(1-p,-p)} = - \frac{p}{p+1}\, \boldsymbol{S}^{(1)}_{(1-p,-p)}.
\ee
\noindent
All our present additional numerical data confirms this relation, in fact with improved accuracy, in which case we shall assume it to be true for arbitrary $p$.
\item In \cite{abs18}, albeit in the non-resonant setting, the following fact was shown. In order for the commutator of two alien derivatives, say $\Delta_{\boldsymbol{n} \cdot \boldsymbol{A}}$ and $\Delta_{\boldsymbol{m} \cdot \boldsymbol{A}}$, to still result\footnote{With the obvious exception of the commutator between $\Delta_{\boldsymbol{n} \cdot \boldsymbol{A}}$ and $\Delta_{-\boldsymbol{n} \cdot \boldsymbol{A}}$.} in an alien derivative, $\Delta_{\left( \boldsymbol{n}+\boldsymbol{m} \right) \cdot \boldsymbol{A}}$, then their corresponding Stokes vectors must verify the following necessary and sufficient proportionality relation:
\be
\label{eq:585}
\boldsymbol{S}_{\boldsymbol{n}+\boldsymbol{m}} \propto \left( \boldsymbol{S}_{\boldsymbol{n}} \cdot \boldsymbol{m} \right) \boldsymbol{S}_{\boldsymbol{m}} - \left( \boldsymbol{S}_{\boldsymbol{m}} \cdot \boldsymbol{n} \right) \boldsymbol{S}_{\boldsymbol{n}}.
\ee
\noindent
We have numerically verified this relation to hold, with great accuracy, for all our (resonant!) Painlev\'e data. As such, we shall assume it to be true for arbitrary Stokes vectors (this will later be checked in section~\ref{sec:numerics}, more specifically in tables~\ref{tab:pIstruct} and~\ref{tab:pIIstruct} therein). For the moment, we focus on its consequences for generic Stokes data.
\end{itemize}

\begin{figure}
	\centering
	\includegraphics[scale=0.35]{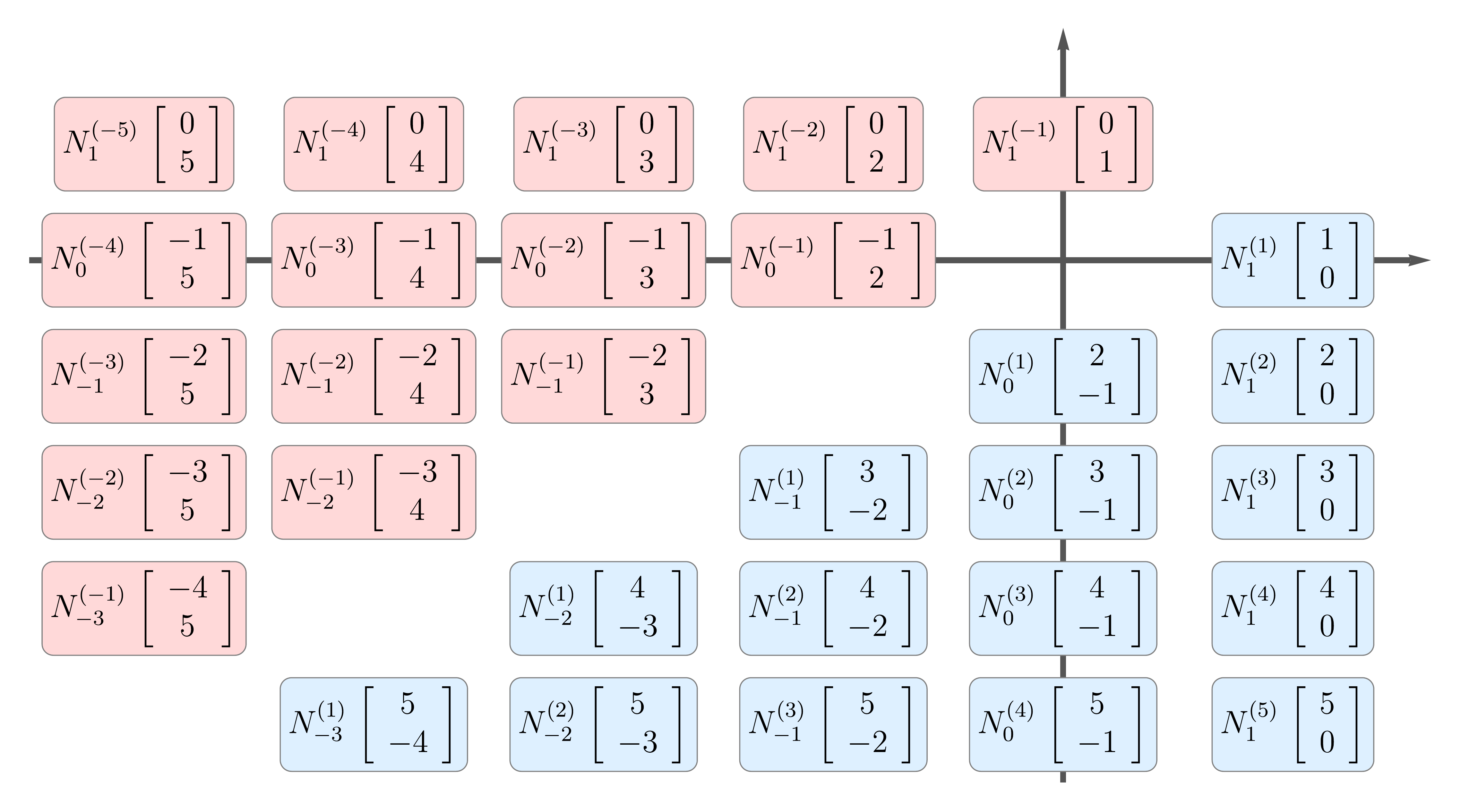}
	\caption{Vectorial structure of Stokes data (upgrading figure~\ref{fig:lattice_stokes}). In the proportionality factors $N_{p}^{(\ell)}$ of \eqref{eq:vec_struct-a}-\eqref{eq:vec_struct-b} the superscript indicates which diagonal we are on; whereas the subscript indicates the ``depth'' along the selected diagonal. For forward vectors, $p$ corresponds to the $x$-coordinate on the grid, while for backward it corresponds to the $y$-coordinate.}
	\label{fig:vec_struct}
\end{figure}

Combining \eqref{eq:structure_1} and \eqref{eq:585}, we obtain the following general structure of Stokes vectors:
\bea
\label{eq:vec_struct-a}
\boldsymbol{S}_{(1-p,1-p-\ell)} &=& N_{1-p}^{(\ell)}\, \left[ \begin{array}{c}p+\ell\\-p\end{array} \right], \\
\label{eq:vec_struct-b}
\boldsymbol{S}_{(1-p-\ell,1-p)} &=& N_{1-p}^{(-\ell)}\, \left[ \begin{array}{c}-p\\p+\ell\end{array} \right].
\eea
\noindent
In these expressions, $\ell$ and $p$ are integers with\footnote{From now on, whenever we write Stokes vectors as $\boldsymbol{S}_{(1-p,1-p-\ell)}$ and $\boldsymbol{S}_{(1-p-\ell,1-p)}$ we are implicitly assuming the bounds $\ell > 0$ and $p \geq 0$ for integer $\ell$ and $p$.} $\ell > 0$ and $p \geq 0$. This resulting vector structure is illustrated in figure~\ref{fig:vec_struct}---and is in fact rather simple: we have reduced our unknowns to the proportionality factors $N_{1-\ell}^{(\ell)}$ and $N_{1-\ell}^{(-\ell)}$ (this also finally explains the notation used earlier in the introduction, for \eqref{eq:exampleStokesP1} and \eqref{eq:exampleStokesP2}). This structure also has a consequence for the computation of Borel residues: instead of needing two Borel residues in order to construct a single Stokes vector, we now only need a single residue. In particular---and for numerical convenience as will be explained in more detail in section~\ref{sec:numerics}---we shall always compute the Borel residues $\mathsf{S}_{(n,n)\to(\ell,0)}$ numerically, and then compute all others by reconstruction via their relation with Stokes data.

\subsection{Stokes Transitions as Flows on Moduli Space}\label{subsec:stokes-trans}

Up to now we have discussed the action of either alien derivative or Stokes automorphism \eqref{eq:Stokes-auto-alien-der} upon specific transseries sectors $\Phi_{\boldsymbol{n}}$; \textit{e.g.}, \eqref{eq:alien-bridge-equation-NR} and \eqref{eq:stokes-bridge-equation-NR}, respectively. In order to understand the Stokes transitions or connection formulae\footnote{Hence making Stokes phenomenon fully explicit; for instance as in \eqref{eq:Stokes-pheno-1pt}.} associated to Stokes automorphisms, one first needs to rewrite these formulae as acting on the \textit{full} two-parameter transseries \eqref{eq:trans_form1}. For what concerns $\Delta_{\omega}$, this is in fact the original way to write the bridge equation \cite{e81}
\be
\label{eq:bridge-original}
\dot{\Delta}_{\boldsymbol{\ell} \cdot \boldsymbol{A}} \Phi = \boldsymbol{S}_{\boldsymbol{\ell}} (\boldsymbol{\sigma}) \cdot \frac{\partial \Phi}{\partial\boldsymbol{\sigma}}.
\ee
\noindent
The proportionality vector on the right-hand side is dictated by Stokes data, \textit{i.e.}, its $\boldsymbol{\sigma} = \left( \sigma_1,\sigma_2 \right)$ dependence is fixed (see below). The great advantage of writing the bridge equation like this is that the Stokes automorphism immediately becomes a flow on the space of transseries parameters, yielding connection formulae. Introducing
\be
\underline{\boldsymbol{S}}_{\theta}(\boldsymbol{\sigma}) \equiv \sum_{\boldsymbol{\ell} \in\left\{ \boldsymbol{\ell}_\theta \right\}}\boldsymbol{S}_{\boldsymbol{\ell}}(\boldsymbol{\sigma}),
\ee
\noindent
essentially along the same lines as in \eqref{eq:dot_al_der} but where we now use $\left\{\boldsymbol{\ell}_\theta\right\}$ to denote the set of Borel singularities with same argument $\theta$, \textit{i.e.}, the set of lattice\footnote{In the resonant setting there exists different $\boldsymbol{\ell}_1$ and $\boldsymbol{\ell}_2$ such that $\boldsymbol{\ell}_1 \cdot  \bm{A} = \boldsymbol{\ell}_2 \cdot \bm{A}$. In this case, the set $\left\{ \bm{\ell}_\theta \right\}$ is defined as a quotient set with the identification that two vectors are in the same class if they are projected to the same number. Sums over such a set will require a choice of representative in the class. In all computations that follow, the choice of representative is non influential.} vectors $\boldsymbol{\ell}$ with $\arg \left( \boldsymbol{\ell} \cdot \boldsymbol{A} \right) = \theta$; then appropriately using the bridge equation \eqref{eq:bridge-original} in the Stokes automorphisms \eqref{eq:Stokes-auto-alien-der} yields
\be
\label{eq:Stokes-flow-simplified}
\underline{\mathfrak{S}}_{\theta} \Phi \left(x; \boldsymbol{\sigma}\right) = \rme^{\underline{\boldsymbol{S}}_{\theta} (\boldsymbol{\sigma}) \cdot \frac{\partial}{\partial\boldsymbol{\sigma}}}\, \Phi \left(x; \boldsymbol{\sigma}\right) \equiv \Phi \left(x; \underline{\pmb{\BS}}_{\theta} (\boldsymbol{\sigma}) \right),
\ee
\noindent
where $\boldsymbol{\sigma} \mapsto \underline{\pmb{\BS}}_{\theta} (\boldsymbol{\sigma})$ is the automorphism generated by the two-parameter flow of the vector field $\underline{\boldsymbol{S}}_{\theta} (\boldsymbol{\sigma}) \cdot \frac{\partial}{\partial\boldsymbol{\sigma}}$ and it explicitly yields the connection formula associated to the Stokes transition \cite{as13}. The trivial one-parameter example associated to \eqref{eq:Stokes-pheno-1pt} is
\be
\underline{\mathfrak{S}}_{0} \Phi \left(x; \sigma\right) = \rme^{S_1 \frac{\partial}{\partial\sigma}}\, \Phi \left(x; \sigma\right) = \Phi \left(x; \underline{\BS}_0 (\sigma) \right) \equiv \Phi \left(x; \sigma+S_1\right). 
\ee
\noindent
Of course for our Painlev\'e problems, transseries solutions \eqref{eq:trans_form1} are parametrized by $\left( \sigma_1, \sigma_2 \right) \in \BC \times \BC$, hence the corresponding Stokes automorphisms acting on \eqref{eq:trans_form1} will likely yield more complicated Stokes transitions than this. Iteration of these ``jumps'', occurring at Stokes lines, lays down a trajectory in $\boldsymbol{\sigma}$-space which allows us to go \textit{anywhere} on the Painlev\'e complex plane, hence turning our local transseries solution into a full-fledged \textit{global} solution. Note, however, that our solutions still obey the Painlev\'e property, hence this hopping trajectory eventually returns to its starting point. This implies we could have started \textit{anywhere} on this ``closed-loop'' trajectory, and we would still be describing the very \textit{same} solution. In other words, the moduli-space of initial-data or boundary-conditions is not truly parametrized by $(\sigma_1,\sigma_2)$ living in $\BC \times \BC$, but requires being modded out by this equivalence relation---originated by the Stokes automorphism---which we have just described. This is an intricate discussion, out of the scope of this work, which has seen very interesting geometrical understanding in, \textit{e.g.}, \cite{g76, o79}.

Let us explicitly formalize the aforementioned concepts in our (resonant) Painlev\'e contexts. First, rewrite the two-parameter transseries \eqref{eq:trans_form1} in slightly more compact vector notation,
\be
\label{eq:full_ts_vec_notation}
\Phi \left( x; \boldsymbol{\sigma} \right) = \sum_{\boldsymbol{n}\in\BN_0^2} \boldsymbol{\sigma}^{\boldsymbol{n}}\, \rme^{- \frac{\boldsymbol{n} \cdot \boldsymbol{A}}{x}}\, \Phi_{\boldsymbol{n}}(x),
\ee
\noindent
where we are using multi-index notation; \textit{e.g.}, $\bm{\sigma}^{\bm{n}}$ is defined as $\bm{\sigma}^{\bm{n}} \equiv \prod_{i=1}^2 \sigma_i^{n_i}$ (basically, this just so that the upcoming derivation is equally valid for arbitrary $k$-parameter transseries). Given $\boldsymbol{A} = \left( A, -A \right)$, the (singular) Stokes lines \eqref{eq:Stokes-auto} are at $\theta_{\text{S}} = 0,\pi$. For the purpose of describing the Stokes-automorphism flow, it is also convenient to introduce arbitrary powers of \eqref{eq:Stokes-auto-alien-der} \cite{as13}
\be
\label{eq:stok_al_der}
\underline{\mathfrak{S}}_{\theta}^{\uptau}= \exp \left( \uptau\, \underline{\dot\Delta}_{\theta} \right).
\ee
\noindent
This is simply computed from $\underline{\mathfrak{S}}_{\theta}$ by replacing every Stokes vector $\boldsymbol{S}_{\boldsymbol{\ell}}$ with $\uptau\, \boldsymbol{S}_{\boldsymbol{\ell}}$. Its corresponding Borel residues (computed from ``$\uptau\, \boldsymbol{S}_{\boldsymbol{\ell}}$ Stokes data'') will be denoted by $\mathsf{S}^{(\uptau)}_{\bm{n} \to \bm{m}}$. Let us next, in sequence, compute the alien derivative as in \eqref{eq:bridge-original} and Stokes automorphism as in \eqref{eq:stok_al_der}, applied to the full Painlev\'e transseries, in order to produce connection formulae---which is to say, find the vectorial functions $\underline{\pmb{\BS}}_{0,\pi} (\boldsymbol{\sigma})$. To run the calculation generically, let us further denote by $\left\{ \boldsymbol{\ell}_{\theta} \right\}$ the subset of vectors $\boldsymbol{\ell} \in \BZ^2$ producing Borel singularities $\boldsymbol{\ell}_\theta \cdot \boldsymbol{A}$ with same argument $\theta$ (the sum over such singularities translates to a sum over such subset). The calculation then proceeds
along the following steps:
\begin{itemize}
\item The directional, pointed alien derivative \eqref{eq:dot_al_der} acts on transseries sectors via \eqref{eq:alien_def_vectors}, as
\be
\underline{\dot\Delta}_{\theta} \Phi_{\bm{n}} = \sum_{\bm{\ell} \in \left\{ \bm{\ell}_{\theta} \right\}} \sum_{\bm{p} \in \ker \mathfrak{P}} \rme^{-\frac{\bm{\ell} \cdot \bm{A}}{x}}\, \bm{S}_{\bm{\ell}+\bm{p}} \cdot \left( \bm{n}+\bm{\ell}+\bm{p} \right) \Phi_{\bm{n}+\bm{\ell}+\bm{p}}.
\ee
\noindent
Its action on the full transseries \eqref{eq:full_ts_vec_notation} can then be written as
\be
\label{eq:action_pointed}
\underline{\dot\Delta}_{\theta} \Phi \left( x; \boldsymbol{\sigma} \right) = \sum_{\bm{\ell} \in \left\{ \bm{\ell}_{\theta} \right\}} \sum_{\bm{p} \in \ker \mathfrak{P}} \sum_{\boldsymbol{n}\in\BN_0^2} \boldsymbol{\sigma}^{\boldsymbol{n}}\, \rme^{- \frac{\left( \boldsymbol{n}+\bm{\ell}+\bm{p} \right) \cdot \boldsymbol{A}}{x}}\, \bm{S}_{\bm{\ell}+\bm{p}} \cdot \left( \bm{n}+\bm{\ell}+\bm{p} \right) \Phi_{\bm{n}+\bm{\ell}+\bm{p}} (x).
\ee
\noindent
We added the term $\bm{p} \cdot \bm{A}$ in the exponent, changing nothing as $\bm{p} \in \ker \mathfrak{P}$. The simple shift $\bm{n}+\bm{\ell}+\bm{p} \to \bm{n}$ in the sums then yields
\be
\label{eq:ac_pointed}
\underline{\dot\Delta}_{\theta} \Phi \left( x; \boldsymbol{\sigma} \right) = \sum_{\boldsymbol{n}\in\BN_0^2} \sum_{\bm{\ell} \in \left\{ \bm{\ell}_{\theta} \right\}} \sum_{\bm{p} \in \ker \mathfrak{P}} \boldsymbol{\sigma}^{\boldsymbol{n}-\bm{\ell}-\bm{p}}\, \rme^{- \frac{\boldsymbol{n} \cdot \boldsymbol{A}}{x}}\, \bm{S}_{\bm{\ell}+\bm{p}} \cdot \bm{n}\, \Phi_{\bm{n}} (x).
\ee
\noindent
If one now defines
\be
\label{eq:def_pointed}
\underline{\boldsymbol{S}}_{\theta} (\boldsymbol{\sigma}) := \sum_{\bm{\ell} \in \left\{ \bm{\ell}_{\theta} \right\}} \sum_{\bm{p} \in \ker \mathfrak{P}} \boldsymbol{\sigma}^{-\bm{\ell}-\bm{p}} \left[ \begin{array}{c} \sigma_1\, \bm{S}_{\bm{\ell}+\bm{p}}^{(1)} \\ \sigma_2\, \bm{S}_{\bm{\ell}+\bm{p}}^{(2)} \end{array} \right]
\ee
\noindent
then \eqref{eq:ac_pointed} immediately becomes\footnote{This result is a particular case of the more general statement proven in \cite{s14}.}
\be
\label{eq:bridge-directional}
\underline{\dot\Delta}_{\theta} \Phi = \underline{\boldsymbol{S}}_{\theta} (\boldsymbol{\sigma}) \cdot \frac{\partial \Phi}{\partial\boldsymbol{\sigma}},
\ee
\noindent
leading to both the (directional) bridge equation \eqref{eq:bridge-original} and the vector field $\underline{\boldsymbol{S}}_{\theta} (\boldsymbol{\sigma})$ generating the flow \eqref{eq:Stokes-flow-simplified}. The above equality of course shows how $\underline{\dot\Delta}_{\theta}$ is indeed a vector field on the space of transseries parameters, and, furthermore, \eqref{eq:def_pointed} explicitly shows how $\underline{\boldsymbol{S}}_{\theta} (\boldsymbol{\sigma})$ is solely dictated by the Stokes data with fixed $\bm{\sigma}$-dependence. Conversely, via \eqref{eq:def_pointed}, one may also think of $\underline{\boldsymbol{S}}_{\theta} (\boldsymbol{\sigma})$ as the generating function\footnote{For example, in the case of a (non-resonant) one-parameter transseries of the sort \eqref{eq:PI-1PTS}-\eqref{eq:PII-1PTS} one finds
\bea
\underline{\boldsymbol{S}}_{0}^{(1)} (\sigma) &=& \bm{S}_{1}^{(1)}, \\
\underline{\boldsymbol{S}}_{\pi}^{(1)} (\sigma) &=& \bm{S}_{-1}^{(1)}\, \sigma^{2} + \bm{S}_{-2}^{(1)}\, \sigma^{3} + \bm{S}_{-3}^{(1)}\, \sigma^{4} + \bm{S}_{-4}^{(1)}\, \sigma^{5} + \cdots.
\eea
} of full Stokes data.
\item The Stokes automorphism \eqref{eq:stok_al_der} acts on transseries sectors via \eqref{eq:borel_def-a}-\eqref{eq:borel_def-b}, as
\be
\underline{\mathfrak{S}}_{\theta}^{\uptau}\, \Phi_{\bm{n}} = \Phi_{\bm{n}} - \sum_{\bm{\ell} \in \left\{ \bm{\ell}_{\theta} \right\}} \rme^{-\frac{\bm{\ell} \cdot \bm{A}}{x}} \sum_{\bm{p} \in \ker \mathfrak{P}} \mathsf{S}^{(\uptau)}_{\bm{n} \to \bm{n}+\bm{\ell}+\bm{p}}\, \Phi_{\bm{n}+\bm{\ell}+\bm{p}}.
\ee
\noindent
Its action on the full transseries \eqref{eq:full_ts_vec_notation} can then be written as
\be
\underline{\mathfrak{S}}_{\theta}^{\uptau}\, \Phi \left( x; \boldsymbol{\sigma} \right) = \Phi \left( x; \boldsymbol{\sigma} \right) - \sum_{\bm{\ell} \in \left\{ \bm{\ell}_{\theta} \right\}} \sum_{\bm{p} \in \ker \mathfrak{P}} \sum_{\boldsymbol{n}\in\BN_0^2} \boldsymbol{\sigma}^{\boldsymbol{n}}\, \rme^{- \frac{\left( \boldsymbol{n}+\bm{\ell}+\bm{p} \right) \cdot \boldsymbol{A}}{x}}\, \mathsf{S}^{(\uptau)}_{\bm{n} \to \bm{n}+\bm{\ell}+\bm{p}}\, \Phi_{\bm{n}+\bm{\ell}+\bm{p}} (x).
\ee
\noindent
Performing the same shift in the sums as before yields
\be
\label{eq:trans_on_pars}
\underline{\mathfrak{S}}_{\theta}^{\uptau}\, \Phi \left( x; \boldsymbol{\sigma} \right) = \sum_{\boldsymbol{n}\in\BN_0^2} \left( \boldsymbol{\sigma}^{\boldsymbol{n}} - \sum_{\bm{\ell} \in \left\{ \bm{\ell}_{\theta} \right\}} \sum_{\bm{p} \in \ker \mathfrak{P}} \mathsf{S}^{(\uptau)}_{\boldsymbol{n}-\bm{\ell}-\bm{p} \to \bm{n}}\, \boldsymbol{\sigma}^{\boldsymbol{n}-\bm{\ell}-\bm{p}} \right) \rme^{- \frac{\boldsymbol{n} \cdot \boldsymbol{A}}{x}}\, \Phi_{\bm{n}} (x).
\ee
\noindent
Via the bridge equation \eqref{eq:bridge-directional}, the action of the Stokes automorphism \eqref{eq:stok_al_der} on the full transseries \eqref{eq:full_ts_vec_notation} may also be written as
\be
\label{eq:Stokes-flow-simplified-TAU}
\underline{\mathfrak{S}}_{\theta}^{\uptau}\, \Phi \left( x; \boldsymbol{\sigma} \right) = \rme^{\uptau\, \underline{\boldsymbol{S}}_{\theta} (\boldsymbol{\sigma}) \cdot \frac{\partial}{\partial\boldsymbol{\sigma}}}\, \Phi \left( x; \boldsymbol{\sigma} \right) \equiv \Phi \left(x; \underline{\pmb{\BS}}_{\theta}^{(\uptau)} (\boldsymbol{\sigma}) \right),
\ee
\noindent
which is basically the ``$\uptau$-version'' of \eqref{eq:Stokes-flow-simplified}. Matching of \eqref{eq:trans_on_pars} and \eqref{eq:Stokes-flow-simplified-TAU} implies\footnote{One technical assumption is further required: that if $\Phi_{\bm{n}}(x)$ and $\Phi_{\bm{m}}(x)$ have the same action in the exponential transmonomial, \textit{i.e.}, $\bm{n}-\bm{m} \in \ker \mathfrak{P}$, then $\Phi_{\bm{n}}(x)$ has the same asymptotic behavior as $\Phi_{\bm{m}}(x)$ (up to a non-zero constant) when $x\to0$ if and only if $\bm{n} = \bm{m}$. In both \PI~and \PII~cases this condition holds as the asymptotic behavior of the sectors is given by
\begin{align*}
\Phi_{(p+\ell,p)}(x)\sim u_0^{(p+s,p)[k]}x^{\frac{\ell}{2}}(\log x)^p.\label{eq:asy_beh}
\end{align*}
\noindent
Two sectors $\Phi_{(p+\ell,p)}$ and $\Phi_{(q+\ell,q)}$ have the same asymptotic behavior only if $p=q$. For the sectors of negative action, the same asymptotic behavior holds.}
\be
\label{eq:consistency}
\left( \underline{\pmb{\mathbb{S}}}_{\theta}^{(\uptau)} (\bm{\sigma}) \right)^{\bm{n}} = \bm{\sigma}^{\bm{n}} - \sum_{\bm{\ell} \in \left\{ \bm{\ell}_{\theta} \right\}} \sum_{\bm{p} \in \ker \mathfrak{P}} \mathsf{S}^{(\uptau)}_{\boldsymbol{n}-\bm{\ell}-\bm{p} \to \bm{n}}\, \boldsymbol{\sigma}^{\boldsymbol{n}-\bm{\ell}-\bm{p}}.
\ee
\noindent
Determining the automorphism generated by the Stokes flow now amounts to finding the vector function $\underline{\pmb{\mathbb{S}}}_{\theta}^{(\uptau)} (\bm{\sigma}) : \BC^2 \to \BC^2$ satisfying \eqref{eq:consistency}---which explicitly shows how $\underline{\pmb{\mathbb{S}}}_{\theta}^{(\uptau)} (\bm{\sigma})$ is solely dictated by Borel residues data, with fixed $\bm{\sigma}$-dependence. Conversely, at $\uptau = 1$, one may also think of $\underline{\pmb{\mathbb{S}}}_{\theta} (\bm{\sigma})$ as the generating function\footnote{For example, in the case of a (non-resonant) one-parameter transseries of the sort \eqref{eq:PI-1PTS}-\eqref{eq:PII-1PTS} one finds
\bea
\underline{\BS}_{0} (\sigma) &=& \sigma - \mathsf{S}_{0 \to 1}, \\
\underline{\BS}_{\pi} (\sigma) &=& \sigma - \mathsf{S}_{2 \to 1}\, \sigma^{2} - \mathsf{S}_{3 \to 1}\, \sigma^{3} - \mathsf{S}_{4 \to 1}\, \sigma^{4} - \mathsf{S}_{5 \to 1}\, \sigma^{5} - \cdots.
\eea
} of full Borel residues data.
\end{itemize}

In section~\ref{sec:results}, we shall present closed-form formulae for \eqref{eq:def_pointed} in the context of both \PI~and \PII~solutions alongside their associated (string theoretic) free energies---hence closed-form formulae generating all their Stokes data. One might then be tempted to use a similar procedure (say, via \eqref{eq:consistency}) to compute $\underline{\pmb{\mathbb{S}}}^{(\uptau)}_{\theta}( \bm{\sigma})$. For example, the $i$th component of $\underline{\pmb{\mathbb{S}}}^{(\uptau)}_{\theta} (\bm{\sigma})$ can be obtained by picking $\bm{n} = \bm{e}_i$ in \eqref{eq:consistency},
\be
\label{eq:transitions}
\left( \underline{\pmb{\mathbb{S}}}_{\theta}^{(\uptau)} (\bm{\sigma}) \right)^{\bm{e}_i} = \bm{\sigma}^{\bm{e}_i} - \sum_{\bm{\ell} \in \left\{ \bm{\ell}_{\theta} \right\}} \sum_{\bm{p} \in \ker \mathfrak{P}} \mathsf{S}^{(\uptau)}_{\bm{e}_i-\bm{\ell}-\bm{p} \to \bm{e}_i}\, \boldsymbol{\sigma}^{\boldsymbol{e}_i-\bm{\ell}-\bm{p}}.
\ee
\noindent
In practice, however, this computation is hard to perform due to the intricate relations between Borel residues and Stokes vectors. An alternative path is to find $\underline{\pmb{\mathbb{S}}}_{\theta} (\bm{\sigma})$ by integration of $\underline{\boldsymbol{S}}_\theta (\boldsymbol{\sigma})$, which is done by solving the system of differential equations
\bea
\label{eq:diff_eq-a}
\underline{\boldsymbol{S}}_\theta \left(\underline{\pmb{\mathbb{S}}}_{\theta}^{(\uptau)} (\bm{\sigma}) \right) &=& \frac{\rmd}{\rmd\uptau}\, \underline{\pmb{\mathbb{S}}}_{\theta}^{(\uptau)} (\bm{\sigma}), \\
\label{eq:diff_eq-b}
\underline{\pmb{\mathbb{S}}}_{\theta}^{(0)} (\bm{\sigma}) &=& \bm{\sigma}.
\eea
\noindent
Once the vector functions $\underline{\pmb{\mathbb{S}}}_{\theta}^{(\uptau)} (\bm{\sigma})$ are obtained, \eqref{eq:consistency} finally indicates how to use them to get generating functions for the Borel residues (as mentioned above), and can be used as a confirmation that the transitions are the correct ones. In particular, evaluation at $\uptau=1$ will by definition give the functions $\underline{\pmb{\mathbb{S}}}_{\theta} (\bm{\sigma})$. Later, in section~\ref{sec:stokes}, we shall use this approach to present closed-form formulae for \eqref{eq:transitions}, again in the contexts of both \PI~and \PII---hence closed-form formulae generating all Borel residues. We will also see how there is an appropriate choice of coordinates which rewrites these functions almost trivially as simple shifts.

\section{From Large-Order Asymptotics to Closed-Form Asymptotics}\label{sec:asymptotics}

Having understood Stokes data, Borel residues, the relevance of their generating functions and how they reorganize themselves into the Stokes automorphism---which in some sense is what one really needs to compute in order to access them all---, we may turn to the actual calculations. Due to the prominent role Stokes data or Borel residues play in large-order asymptotics---see \cite{abs18} for generics and \cite{gikm10, asv11, sv13} for Painlev\'e---this is where we start: how resurgence dictates the asymptotic growth of the coefficients $u_g^{(n,m)[k]}$, in terms of the other sectors and weighted by the Borel residues. Large-order asymptotics was extensively used in \cite{asv11, sv13} to compute Stokes data numerically, which we now build and improve upon. Subsection~\ref{subsec:largeorder-asymp} discusses a method based on large-order analysis, which computes arbitrary Stokes data in a systematic way. Subsection~\ref{subsec:closedform-asymp} then presents the method of ``closed-form asymptotics'', a procedure which we use to conjecture closed-form expressions for the Stokes data and that will be the basis for our closed-form results---later presented in the following section~\ref{sec:results}.

\subsection{Large-Order Asymptotics: Review and Upgrades}\label{subsec:largeorder-asymp}

Resurgent large-order asymptotics is a computational technique which relates the (asymptotic) growth of the coefficients in the transseries sectors to each other \cite{abs18}. It makes the consequences of resurgence explicit in relating different sectors, and allows access (in principle) to all Stokes data---albeit in a numerical form. This technique has been largely used in the literature; the interested reader may refer to \cite{gikm10, asv11, sv13} for previous applications to Painlev\'e equations. The exposition in this paper, however, will follow the guidelines and notations in \cite{abs18}, adapting it to the resonance setting and exploring the consequences of the symmetries of \PI~and \PII~problems.

Large-order asymptotics is based on the Cauchy theorem. Defining the \textit{discontinuity} operator across a Stokes line,
\be
\operatorname{Disc}_\theta := \1 - \underline{\mathfrak{S}}_\theta,
\ee
\noindent
we observe that the only angles across which $\operatorname{Disc}_\theta$ acts non-trivially are $\theta=0,\pi$. Then, the Cauchy theorem\footnote{Note how, in principle, there should be a term in the Cauchy theorem containing an integration around the singularity at $x=0$. See \cite{bw73, cs78} for a discussion on why this term does not contribute in the \PI~and \PII~cases.} can be written as\footnote{Throughout, the symbol $\simeq$ denotes an \textit{asymptotic} equality defined in the following way: a function $f(x)$ is said to be asymptotic to a formal power-series in $x$ with coefficients $c_n$ in the limit $x\to0$ if, for every $N\in\BN$,
\be
f(x) - \sum_{n=0}^N c_n x^n = o \left(x^{N+1}\right).
\ee
\noindent
Two functions are asymptotic to each other if they are asymptotic to the same formal power-series. Note that a function can be asymptotic to a power-series with zero radius of convergence: this is what happens in the non-trivial resurgence examples, where the sectors in the transseries are represented by formal power-series. An analogous definition holds for functions that are asymptotic to power-series in $x^{-1}$ in the $x\to{+\infty}$ limit.}
\be
\Phi \left( x; \boldsymbol{\sigma} \right) \simeq - \frac{1}{2\pi\rmi} \int_0^{+\infty} \rmd w\, \frac{\operatorname{Disc}_0 \Phi \left( w; \boldsymbol{\sigma} \right)}{w-x} - \frac1{2\pi\rmi} \int_0^{-\infty} \rmd w\, \frac{\operatorname{Disc}_\pi \Phi \left( w; \boldsymbol{\sigma} \right)}{w-x}.
\ee
\noindent
Expanding the transseries in powers of $\bm{\sigma}$, we find the analogous statement for  transseries sectors
\be
\Phi_{(n,m)}(x) \simeq - \frac1{2\pi\rmi}\int_0^{+\infty} \rmd w\, \frac{\operatorname{Disc}_0 \Phi_{(n,m)}(w)}{w-x} - \frac1{2\pi\rmi}\int_0^{-\infty} \rmd w\, \frac{\operatorname{Disc}_\pi\Phi_{(n,m)}(w)}{w-x}.
\ee
\noindent
Setting $n=m$ one gets relations for the diagonal sectors, which, as discussed in subsection~\ref{subsec:stok_to_borel}, will yield all the necessary Borel residues needed to compute arbitrary Stokes data. Using \eqref{eq:borel_def-2-a}-\eqref{eq:borel_def-2-b} we obtain (after a convenient translation $p\to n-p$)
\bea
\Phi_{(n,n)}(x) &\simeq& - \frac{1}{2\pi\rmi} \sum_{\ell=1}^{+\infty} \sum_{p=0}^{n} \mathsf{S}_{(n,n)\to(p+\ell,p)} \int_0^{+\infty} \rmd w\, \rme^{-\ell\frac{A}{w}}\, \frac{\Phi_{(p+\ell,p)}(w)}{w-x} - \nonumber \\
&&
\label{eq:asy_series}
- \frac{1}{2\pi\rmi} \sum_{\ell=1}^{+\infty} \sum_{p=0}^n \mathsf{S}_{(n,n)\to(p,p+\ell)} \int_0^{-\infty} \rmd w\, \rme^{\ell\frac{A}{w}}\, \frac{\Phi_{(p,p+\ell)}(w)}{w-x}.
\eea
\noindent
Now use \eqref{eq:trans_form2}-\eqref{eq:trans_form3}-\eqref{eq:trans_form4} to convert the asymptotic equality between formal power-series into an asymptotic equality between power-series \textit{coefficients}. Here we use that $\beta_{(n,n)}^{[0]}=n$, alongside the integrals:
\bea
\label{eq:h_htil-a}
\widetilde{H}_k (g,\ell ) &:=& \int_0^{+\infty} \rmd z\, z^{g-1}\, \rme^{-\ell z}\, \log^k z = \frac{\partial^k}{\partial g^k} \frac{\Gamma(g)}{\ell^g}, \\
\label{eq:h_htil-b}
H_k (g,\ell) &:=& \int_0^{+\infty} \rmd z\, z^{g-1}\, \rme^{-\ell z}\, \log^k \left(-z\right) = \sum_{t=0}^k \binom{k}{t} \left(\rmi\pi\right)^{k-t} \widetilde{H}_t(g,\ell ).
\eea
\noindent
To arrive at the last equality, we have chosen the analytic continuation of the logarithm as: $\log z$ is real for $z$ real-positive, and $\log(-z)=\log z+\rmi\pi$ again for $z$ real-positive. The integrals in \eqref{eq:asy_series} then become
\bea
\int_0^{+\infty} \rmd w\, \rme^{-\ell \frac{A}{w}}\, \frac{\Phi_{(p+\ell ,p)}(w)}{w-x} &=& \sum_{g=0}^{+\infty} \sum_{h=0}^{+\infty} \sum_{k=0}^{p} \frac{x^g}{(-2)^k}\, u_{2h}^{(p+\ell ,p)[k]}\, \widetilde{H}_k \left( g-h-\beta^{(k)}_{(p+\ell ,p)}, \ell A \right), \\
\int_0^{-\infty} \rmd w\, \rme^{+\ell \frac{A}{w}}\, \frac{\Phi_{(p,p+\ell )}(w)}{w-x} &=& \sum_{g=0}^{+\infty} \sum_{h=0}^{+\infty} \sum_{k=0}^{p} \frac{x^g}{(-2)^k}\, (-1)^{g-h-\beta_{(p,p+\ell )}^{[k]}}\, u_{2h}^{(p,p+\ell )[k]} \times \\
&&
\hspace{120pt}
\times H_k \left( g-h-\beta^{(k)}_{(p,p+\ell )}, \ell A \right). \nonumber
\eea
\noindent
Using these integrals and translating $g\to g+n$ in the right-hand side of \eqref{eq:asy_series} allows us to finally obtain the main asymptotic relation for \textit{diagonal} transseries coefficients. This is:
\bea
u_{2g}^{(n,n)[0]} &\simeq& - \frac{1}{2\pi\rmi} \sum_{\ell =1}^{+\infty} \sum_{h=0}^{+\infty} \sum_{p=0}^{n} \sum_{k=0}^{p} \mathsf{S}_{(n,n)\to(p+\ell ,p)}\, (-2)^{-k}\, u_{2h}^{(p+\ell ,p)[k]}\, \widetilde{H}_k \left( g+n-h-\beta^{(k)}_{(p+\ell ,p)}, \ell A \right) - \nonumber \\
\label{eq:asym_main_before_back_forw}
&&
- \frac{1}{2\pi\rmi} \sum_{\ell =1}^{+\infty} \sum_{h=0}^{+\infty} \sum_{p=0}^{n} \sum_{k=0}^{p} \mathsf{S}_{(n,n)\to(p,p+\ell )}\, (-1)^{g-h+n-\beta_{(p,p+\ell )}^{[k]}}\, (-2)^{-k}\, u_{2h}^{(p,p+\ell )[k]} \times \\
&&
\hspace{220pt}
\times  H_k \left( g+n-h-\beta^{(k)}_{(p+s,p)}, \ell A \right). \nonumber
\eea
\noindent
All large-order relations which are necessary in order to compute Stokes data will be obtained from this relation. But there is one further simplification that makes explicit how our two Painlev\'e actions are symmetric: this is the \textit{backward-forward symmetry} to which we now turn.

\subsubsection*{Backward-Forward Symmetry}

The Painlev\'e transseries coefficients $u_{2g}^{(n,m)[k]}$ are iteratively constructed from recursion relations computed in \cite{asv11, sv13}. Among others, these recursion relations yield the properties \eqref{eq:properties_coefs} for the coefficients (essentially the same for \PI~and \PII). As we shall see now, these properties have a very relevant outcome. In fact, we may use them to reduce the amount of coefficients appearing in the right-hand side of \eqref{eq:asym_main_before_back_forw} solely to the set $u^{(p+\ell,p)[0]}_{2h}$, for $p,h\geq0$ and $\ell\geq1$. In this way, we obtain the (simpler) asymptotic relation\footnote{Having set the logarithm $\log \left(-1\right) = +\rmi\pi$, we now set the square-root $\left(-1\right)^{\frac{1}{2}} = +\rmi$.}
\bea
u_{2g}^{(n,n)[0]} &\simeq& - \frac{1}{2\pi\rmi} \sum_{\ell =1}^{+\infty} \sum_{h=0}^{+\infty} \sum_{p=0}^{n} \sum_{k=0}^{p} \frac{1}{k!} \left( \frac{\alpha}{2}\, \ell \right)^k u_{2h}^{(p+\ell-k,p-k)[0]} \times \\
&&
\times \left( \mathsf{S}_{(n,n)\to(p+\ell ,p)}\, \widetilde{H}_k \left( g+n-h-\beta_{(p+\ell ,p)}^{(k)}, \ell A \right) + \right. \nonumber \\
&&
\left. + (-\rmi)^\ell \left(-1\right)^{g+n+p}\, \mathsf{S}_{(n,n)\to(p,p+\ell )}\, H_k \left( g+n-h-\beta_{(p+\ell ,p)}^{(k)}, \ell A \right) \right). \nonumber
\eea
\noindent
The aforementioned symmetries of the coefficients further imply $u_{2(2g+1)}^{(n,n)[0]} \simeq  0$, in which case it follows
\bea
\label{eq:asym_back_forw_prepare}
0 &\simeq& - \frac{1}{2\pi\rmi} \sum_{\ell =1}^{+\infty} \sum_{h=0}^{+\infty} \sum_{p=0}^{n} \sum_{k=0}^{p} \frac{1}{k!} \left( \frac{\alpha}{2}\,\ell \right)^k u_{2h}^{(p+\ell-k,p-k)[0]} \times \\
&&
\times \left( \mathsf{S}_{(n,n)\to(p+\ell ,p)}\, \widetilde{H}_k \left( 2g+1+n-h-\beta_{(p+\ell-k,p-k)}^{(0)}, \ell A \right) - \right. \nonumber \\
&&
\left. - (-\rmi)^\ell  \left(-1\right)^{n+p}\, \mathsf{S}_{(n,n)\to(p,p+\ell )}\, H_k \left( 2g+1+n-h-\beta_{(p+\ell-k,p-k)}^{(0)}, \ell A \right) \right). \nonumber
\eea
\noindent
Now, the asymptotic behavior of the functions $H_k$, $\widetilde{H}_k$ is given by
\be
H_k \left(g,\ell A\right) \simeq \widetilde{H}_k \left(g,\ell A\right) \simeq \left(g-1\right)!\, \log^k g\, \left(\ell A\right)^{-g}.
\ee
\noindent
In order for \eqref{eq:asym_back_forw_prepare} to hold, we can look at the leading large-order behavior of its right-hand side and impose that the coefficient of the expansion vanishes; then look at the next-to-leading large-order behavior, and so on. In order to get necessary relations (which we will later prove to be also sufficient), we can drop the sum over $\ell$---as terms with growths $\left(\ell_1 A \right)^{-g}$ and $\left(\ell_2 A \right)^{-g}$ have different growths when $\ell_1\neq \ell_2$---and choose $h=0$---as the leading factorial contribution is obtained when the argument of $H_k$ and $\widetilde{H}_k$ is as large as possible. We finally use the inverse relation in \eqref{eq:h_htil-b}, between $H_k$ and $\widetilde{H}_k$, to obtain
\bea
0 &\simeq& \sum_{p=0}^{n} \sum_{k=0}^{p} \frac{1}{k!} \left( \frac{\alpha}{2}\,\ell \right)^k u_{0}^{(p+\ell-k,p-k)[0]} \times \\
&&
\times \left( \mathsf{S}_{(n,n)\to(p+\ell,p)} \sum_{t=0}^k \binom{k}{t} \left(-\rmi\pi\right)^{k-t} H_t \left( 2g+1+n-\beta_{(p+\ell-k,p-k)}^{(0)}, \ell A \right) - \right. \nonumber \\
&&
- (-\rmi)^\ell \left(-1\right)^{n+p}\, \mathsf{S}_{(n,n)\to(p,p+\ell)}\, H_k \left( 2g+1+n-\beta_{(p+\ell-k,p-k)}^{(0)}, \ell A \right) \Bigg). \nonumber
\eea
\noindent
Note how herein $\ell$ is no longer a summation variable---instead this relation now holds for every positive integer $\ell$. To extract the leading factorial, notice that $\beta_{(p+\ell-k,p-k)}^{[0]}$ is minimized when $p=k$ with value $\ell /2$. Then, the sum over $k$ can be dropped by fixing $k=p$, obtaining
\bea
0 &\simeq& \sum_{p=0}^{n} \frac{1}{p!} \left( \frac{\alpha}{2}\, \ell  \right)^p \left( \mathsf{S}_{(n,n)\to(p+\ell,p)} \sum_{t=0}^{p} \binom{p}{t} \left(-\rmi\pi\right)^{p-t}\, \frac{H_t \left( 2g+1+n-\frac{\ell}{2}, \ell A \right)}{H_0 \left( 2g+1+n-\frac{\ell}{2}, \ell A \right)} - \right. \nonumber \\
&&
\label{eq:last_for_back_forw}
\hspace{80pt}
\left. - (-\rmi)^\ell \left(-1\right)^{n+p}\, \mathsf{S}_{(n,n)\to(p,p+\ell)}\, \frac{H_p \left( 2g+1+n-\frac{\ell}{2}, \ell A \right)}{H_0 \left( 2g+1+n-\frac{\ell}{2}, \ell A \right)} \right).
\eea
\noindent
We have divided by an overall $H_0 (g,\ell)$ in order to cancel the factorial growth and obtain a power-series in $g^{-1}$ and $\log g$. One can now extract the backward-forward relation for Borel residues by simply considering the different $\log g$ growths in \eqref{eq:last_for_back_forw}.

First rewrite the above equation \eqref{eq:last_for_back_forw} as
\be
\label{eq:for_back_forw}
0 \simeq \sum_{p=0}^{n} d_p (n,\ell)\, \frac{H_p \left( 2g+1-n-\frac{\ell}{2}, \ell A \right)}{H_0 \left( 2g+1-n-\frac{\ell}{2}, \ell A \right)},
\ee
\noindent
with some coefficients $d_p (n,\ell)$ that contain the Borel residues. These coefficients must \textit{all} vanish in order for \eqref{eq:for_back_forw} to hold, as the $p$th term in the sum is asymptotic to $\log^p g$. The coefficients themselves can be easily obtained by direct computation,
\be
d_p (n,\ell) = \sum_{q=p}^{n} \frac1{q!} \left( \frac{\alpha}{2}\, \ell \right)^q \binom{q}{p} \left(-\rmi\pi\right)^{q-p}\, \mathsf{S}_{(n,n)\to(q+\ell,q)} - \frac{\rmi^\ell}{p!} \left( \frac{\alpha}{2}\, \ell\right)^p \left(-1\right)^{n+p}\, \mathsf{S}_{(n,n)\to(p,p+\ell)}.
\ee
\noindent
By setting them all to zero, we finally obtain the \textit{backward-forward relation}:
\be
\label{eq:back_forw}
\mathsf{S}_{(n,n)\to(p,p+\ell)} = (-\rmi)^\ell \left(-1\right)^{n+p+\ell}\, \sum_{q=p}^{n} \frac{1}{(q-p)!} \left( -\rmi\pi\, \frac{\alpha}{2}\, \ell \right)^{q-p} \mathsf{S}_{(n,n)\to(q+\ell,q)}.
\ee
\noindent
Through this relation, we are able to compute the set of backward Borel residues and Stokes data given the set of forward Borel residues and Stokes data.

This symmetry is necessary for property \eqref{eq:asym_back_forw_prepare} to hold. It can now be seen that it is also sufficient: insert \eqref{eq:back_forw} and \eqref{eq:h_htil-b} in \eqref{eq:asym_back_forw_prepare}, and after a long but straightforward calculation one finds the vanishing of the right-hand side in \eqref{eq:asym_back_forw_prepare}---hence the asymptotic equation holds.

One may now use this backward-forward symmetry to update \eqref{eq:asym_main_before_back_forw} in such a way as to only depend on forward Borel residues. Using the symmetry of coefficients \eqref{eq:properties_coefs}, relation \eqref{eq:h_htil-b} to eliminate the $H_k (g,s)$ functions, the backward-forward symmetry, and evaluating the expression at step $2g$ in order to only find non-zero coefficients, we get: 
\be
\label{eq:fin_as}
u_{4g}^{(n,n)[0]} \simeq - \frac{1}{\rmi\pi} \sum_{\ell=1}^{+\infty} \sum_{h=0}^{+\infty} \sum_{p=0}^{n} \sum_{k=0}^{p} \mathsf{S}_{(n,n)\to(p+\ell,p)}\, (-2)^{-k}\, u_{2h}^{(p+\ell,p)[k]}\, \widetilde{H}_k \left( 2g+n-h-\beta^{(k)}_{(p+\ell,p)}, \ell A \right).
\ee
\noindent
This is the \textit{fundamental relation} which is the basis for our large-order asymptotic analyses, and which will also be the basis of the next subsection where we propose an \textit{ansatz} to obtain closed-form results out from this formula (up to the determination of a number---see below).

\begin{figure}
\centering
\begin{subfigure}[t]{0.5\textwidth}
	\centering
	\includegraphics[height=1.7in]{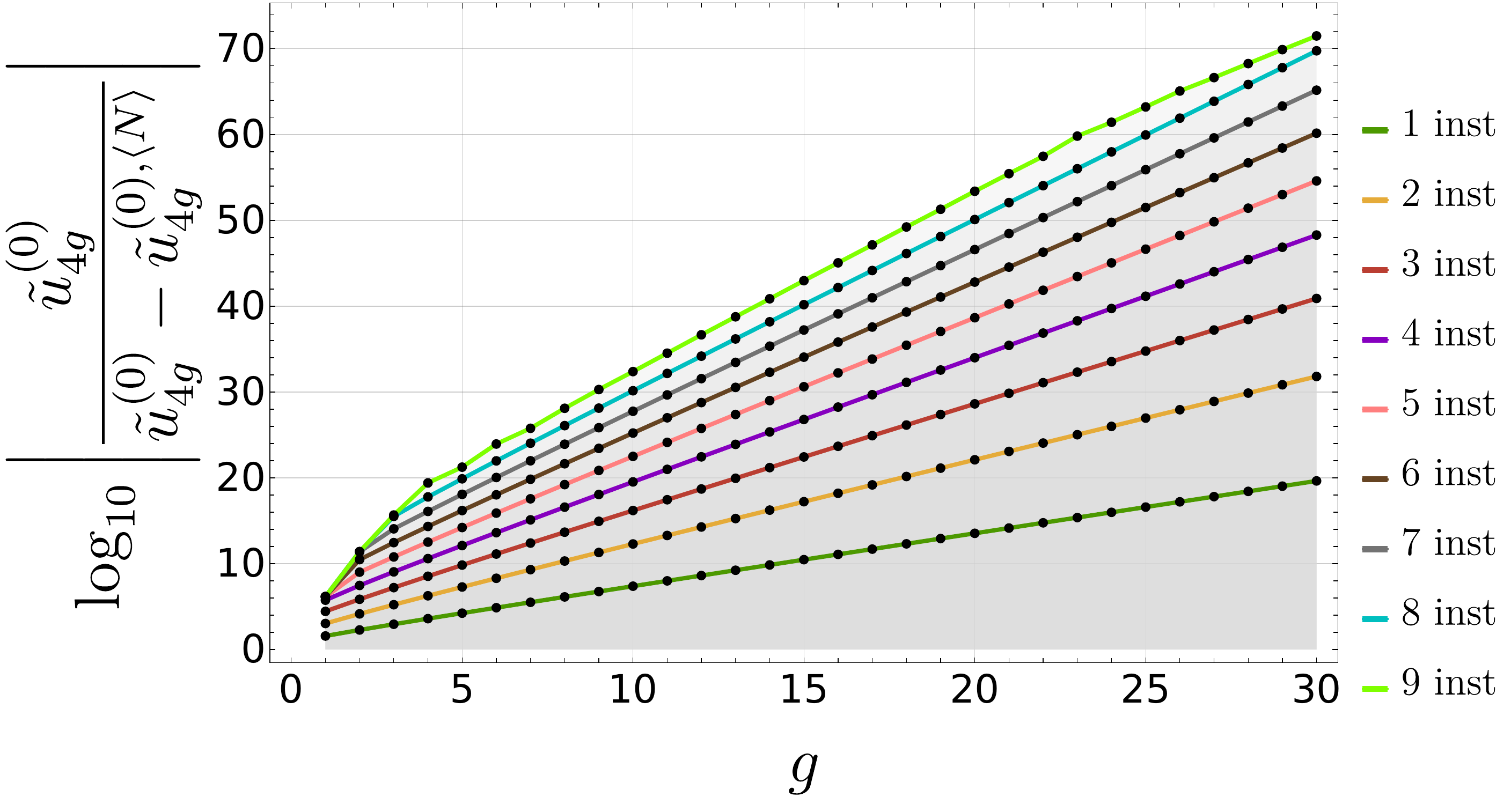}
	\caption{Analogue of computation in \cite{asv11}.}
\end{subfigure}%
~
\begin{subfigure}[t]{0.5\textwidth}
	\centering
	\includegraphics[height=1.7in]{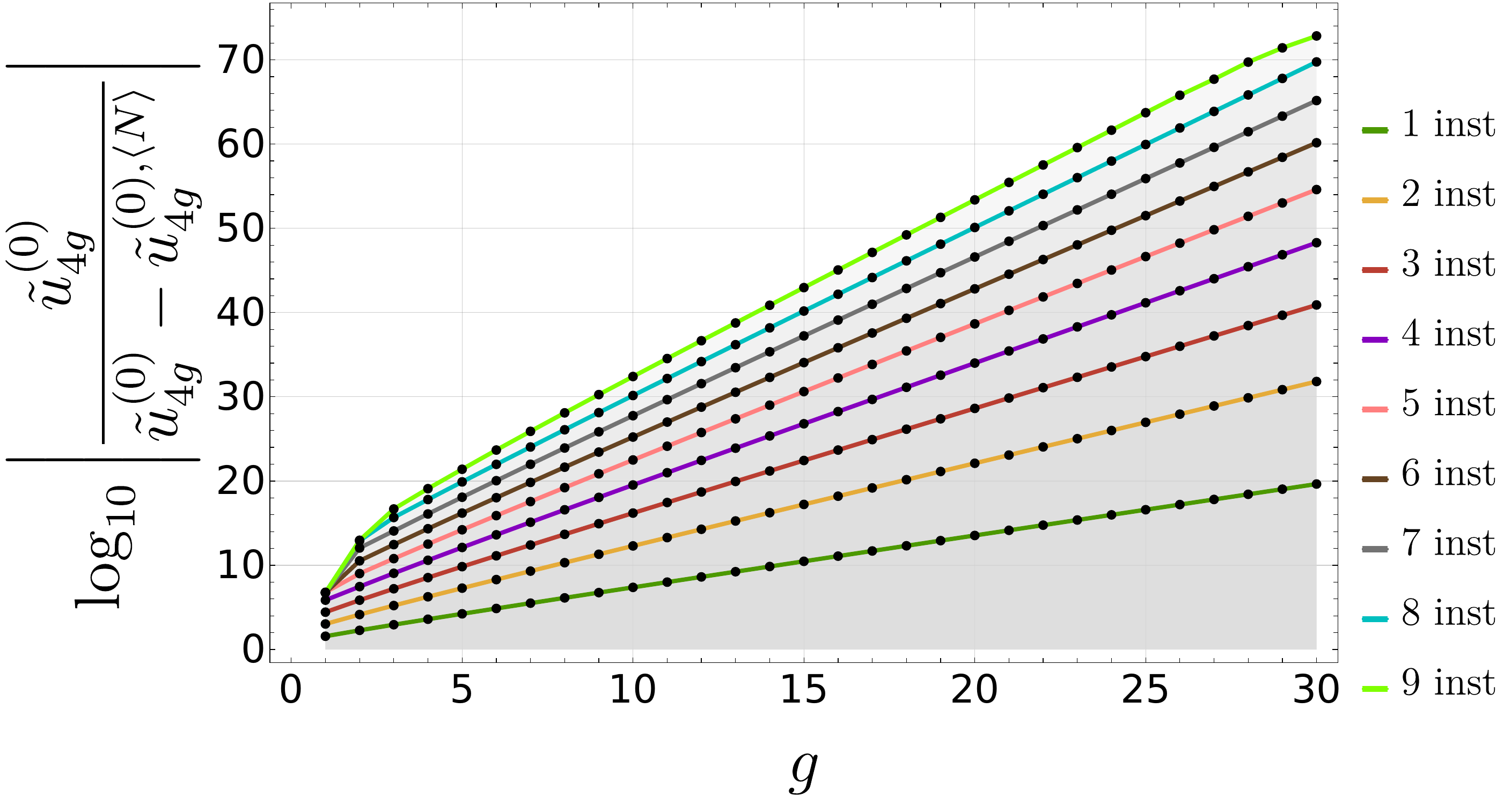}
	\caption{Upgrade with conformal transformations.}
\end{subfigure}
\caption{Precision of the large-order asymptotic formula for \PI. The figures plot numerical precision with increasing instanton corrections in the two approaches: a straightforward upgrade of \cite{asv11} and with the addition of conformal transformations (see subappendix~\ref{app:subsec:conformal}).}
\label{fig:PainleveIPrec}
\end{figure}

\begin{figure}
\centering
\begin{subfigure}[t]{0.5\textwidth}
	\centering
	\includegraphics[height=1.7in]{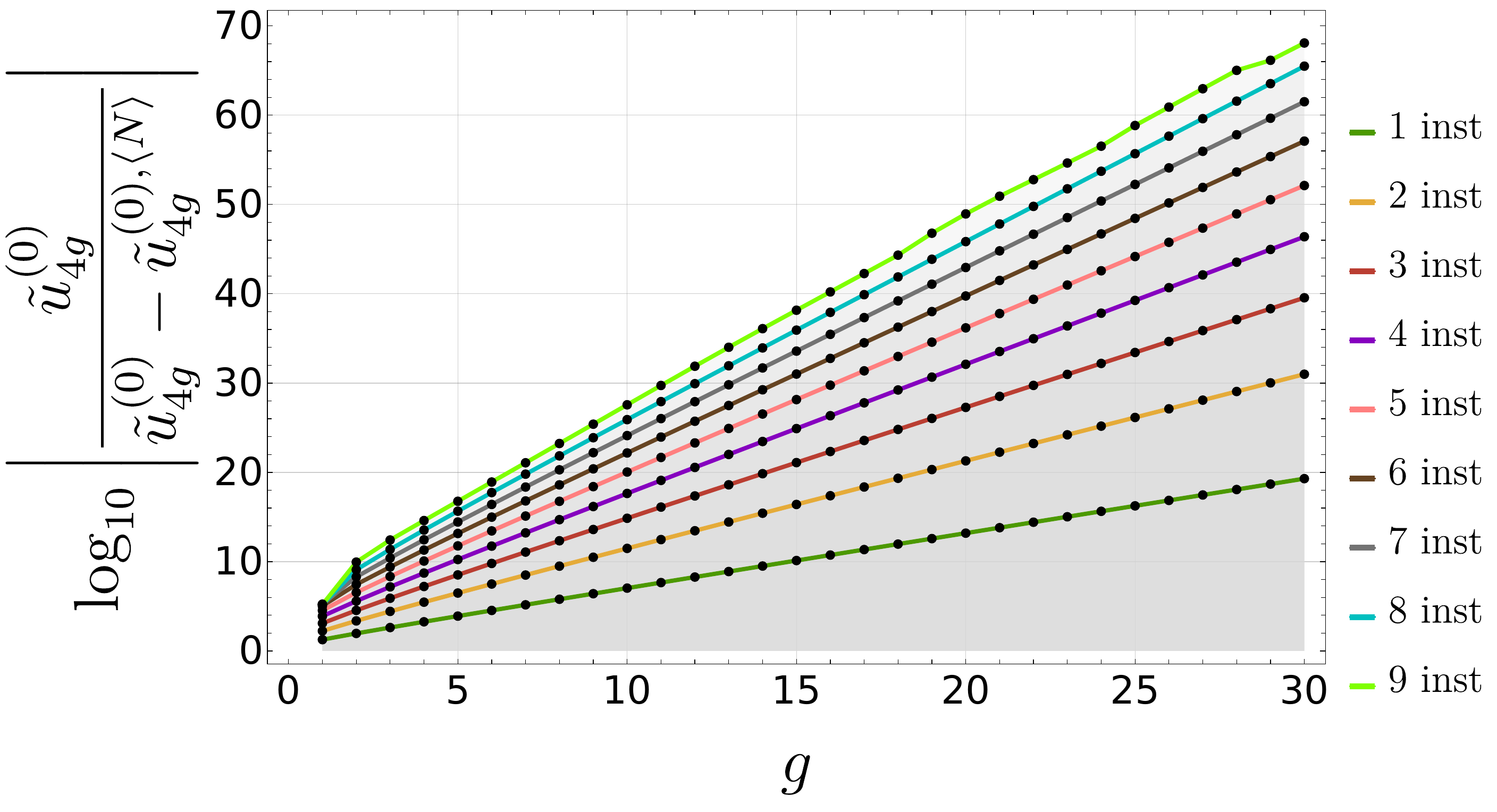}
	\caption{Analogue of computation in \cite{asv11}.}
\end{subfigure}%
~ 
\begin{subfigure}[t]{0.5\textwidth}
	\centering
	\includegraphics[height=1.7in]{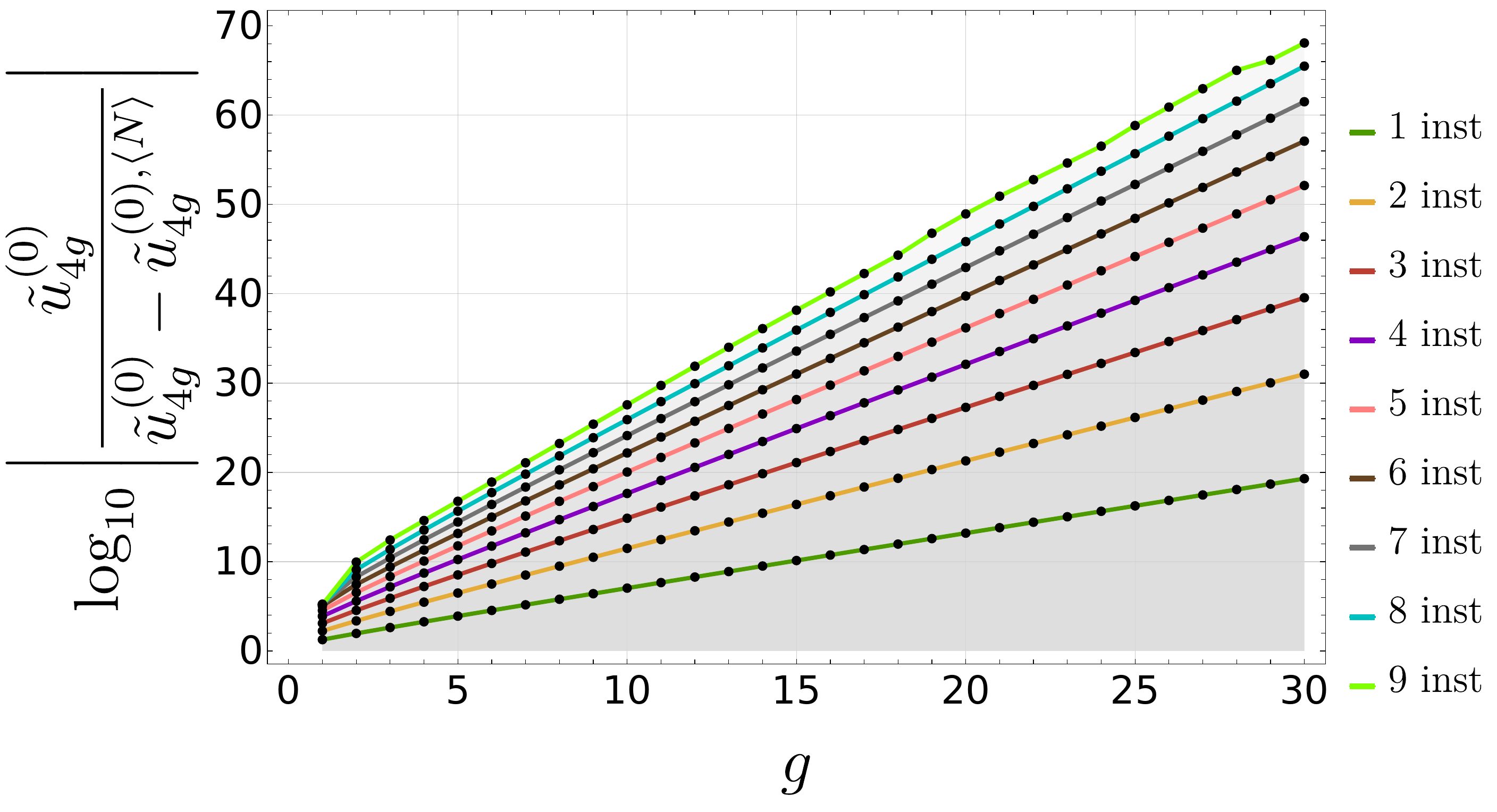}
	\caption{Upgrade with conformal transformations.}
\end{subfigure}
\caption{Precision of the large-order asymptotic formula for \PII. The figures plot numerical precision with increasing instanton corrections in the above two approaches.}
\label{fig:PainleveIIPrec}
\end{figure}

For the moment, let us run a couple of checks on this expression. This was already partially addressed in \cite{asv11}, but herein we have upgraded their equation (5.77) to better numerical precision and we have also explored the effect of conformal transformations\footnote{These will be discussed in further detail in subsection~\ref{subsec:conformal} and subappendix~\ref{app:subsec:conformal}---to where we refer the reader.}. Furthermore, apart from enlarging the precision for \PI, we have also carried this out for the first time in the case of \PII. We test \eqref{eq:fin_as} for the case $n=0$, which is a good test on the resurgent structure as the Stokes vector $\bm{S}_{(1,0)}$ (which is known analytically) is sufficient to build all Borel residues $\mathsf{S}_{(0,0)\to(\ell,0)}$. In order to work with quantities which have nicer behavior than factorial or exponential growth, we introduce as usual \cite{abs18}
\be
\label{eq:exacttildeu0}
\widetilde{u}_{4g}^{(0)} = \frac{\rmi\pi A^{2g-\frac{1}{2}}}{N_1^{(1)}\, \Gamma \left( 2g-\frac{1}{2} \right)}\, u_{4g}^{(0,0)[0]}.
\ee
\noindent
Using $\mathsf{S}_{(0,0)\to(\ell ,0)} =-  (N_1^{(1)})^\ell $, we also introduce
\be
\label{eq:Ninstapproxtotildeu0}
\widetilde{u}_{4g}^{(0), \langle N \rangle} = \sum_{\ell=1}^N \sum_{h=0}^{+\infty} (N_1^{(1)})^{\ell-1}\, u_{2h}^{(\ell,0)[0]}\, \frac{\Gamma \left( 2g-h-\frac{\ell}{2} \right)}{\Gamma \left( 2g-\frac{1}{2} \right)}\, \ell^{-2g+h+\frac{\ell}{2}}\, A^{h+\frac{\ell-1}{2}},
\ee
\noindent
which represents the first $N$ instanton contributions to the large-order behavior for the perturbative quantity $\widetilde{u}_{4g}^{(0)}$. Herein the $h$-sum is asymptotic and has to be evaluated via Borel--Pad\'e resummation; \textit{e.g.}, \cite{abs18}. We have detailed this method in a more general setting in subappendix~\ref{app:subsec:large-order}. For now, let us denote the resummed quantity with the same name as the asymptotic series. Focusing on the first $N$ instanton contributions implies one has, at the asymptotic level,
\be
\label{eq:test}
\widetilde{u}_{4g}^{(0)} \simeq \widetilde{u}_{4g}^{(0),\langle N \rangle} + o \left( \frac{1}{(N+1)^{2g}} \right).
\ee
\noindent
In order to check upon the resurgent structure, we have evaluated both $\widetilde{u}_{4g}^{(0)}$ exactly as in \eqref{eq:exacttildeu0} and $\widetilde{u}_{4g}^{(0),\langle N \rangle}$ numerically as in \eqref{eq:Ninstapproxtotildeu0}. For the latter, we have performed Borel--Pad\'e resummations up to orders $h = 300$, $280$, $260$, $240$, $220$, $200$, $180$, $160$, $140$ for the $\ell^{-2g} = 1^{-2g}$, $2^{-2g}$, $3^{-2g}$, $4^{-2g}$, $5^{-2g}$, $6^{-2g}$, $7^{-2g}$, $8^{-2g}$, $9^{-2g}$ contributions, respectively\footnote{Note that for each $\ell=1,\ldots,N$, we have kept different maximum $h$'s for each resummation. This means that in order to compute $\widetilde{u}_{4g}^{(0), \langle 4 \rangle}$, for example, we have resummed the $\ell =1$ sector up to $h=300$, $\ell =2$ up to $h=280$, $\ell =3$ up to $h=260$ and $\ell =4$ up to $h=240$.}. We may then compare both sides of \eqref{eq:test}, and the results are displayed in figures~\ref{fig:PainleveIPrec} and~\ref{fig:PainleveIIPrec}.

\subsection{Closed-Form Asymptotics and Stokes Data}\label{subsec:closedform-asymp}

We are finally ready to conjecture analytic equations for Stokes data, which we will be able to solve in order to find generating functions for all coefficients. We shall do this by focusing on the asymptotic relations \eqref{eq:fin_as}, and extracting all relevant information solely out of them. Let us also stress at this stage that the upcoming procedure is strongly supported by numerics and, although it allows for the conjecture of exact relations that determine Stokes data, the method presented herein is by no means a rigorous derivation. Supporting evidence possibly amounting to a fully rigorous proof of our conjectures will be later discussed in section~\ref{sec:stokes}.

\begin{figure}
\centering
\includegraphics[scale=0.35]{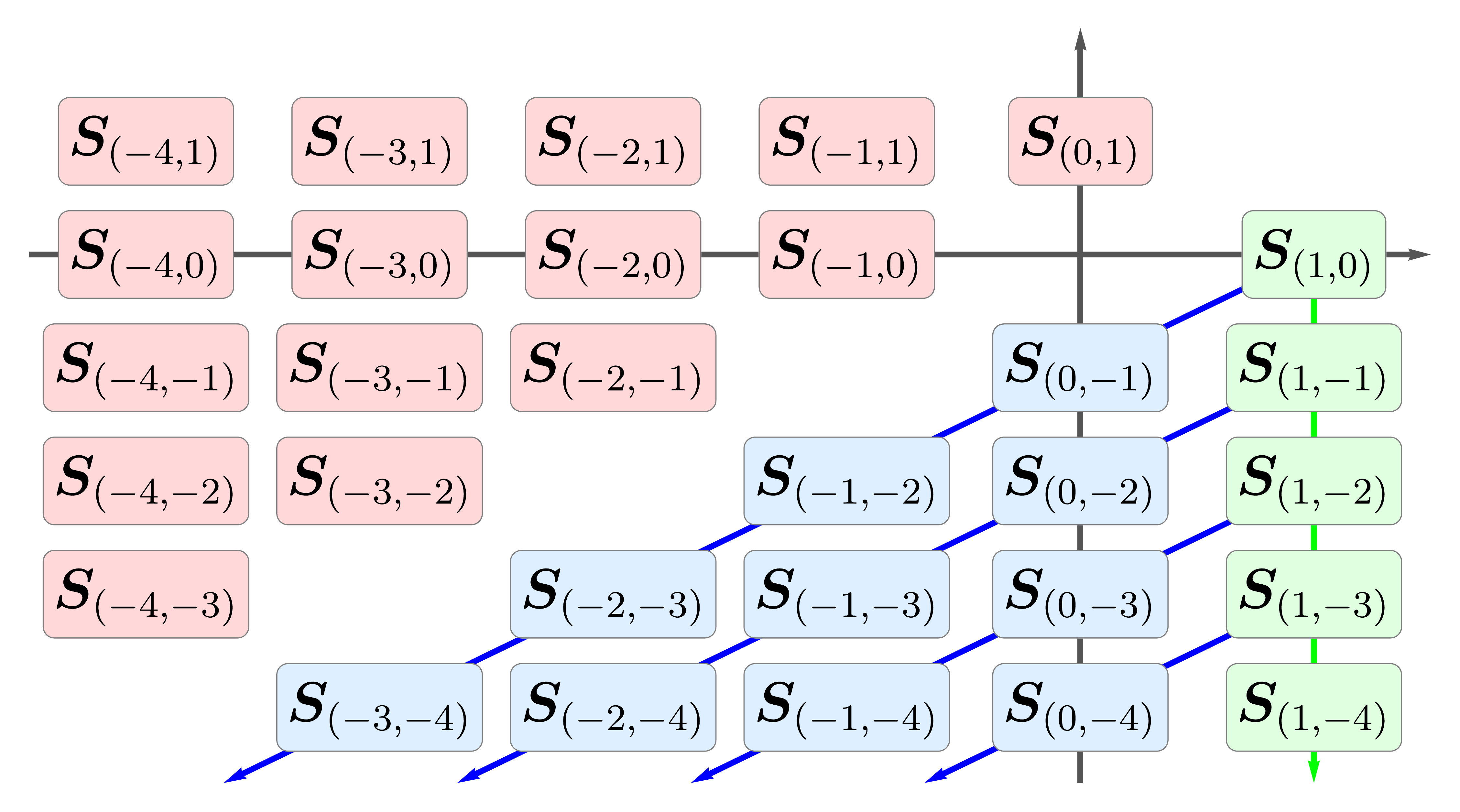}
\caption{The lattice of Stokes data revisited (recall figures~\ref{fig:lattice_stokes} and~\ref{fig:vec_struct}). It is sufficient to focus on the asymptotic relations for the forward Stokes data (in blue), as discussed in subsection~\ref{subsec:largeorder-asymp}. The edge or boundary of this blue region in the lattice is easily guessable (and we have hence changed its color to green). On the other hand, the numbers in the bulk of the blue region consist of sums of different terms and have a rather more non-trivial number content. In other words, they are much harder to guess without first properly understanding the logarithmic structure of the asymptotic relations \eqref{eq:fin_as}. In the following we will conjecture equations which allow us to compute all Stokes numbers along the diagonals (the blue arrows).}
\label{fig:lattice_stokes_guessing_game}
\end{figure}

Some (explicit) facts we do already have on \PI/\PII\, Stokes data. Firstly, we are dealing with a two-parameter resonant problem with logarithms---one could fear that these additional logarithmic terms could contribute to make the problem harder, but, as we shall see, and because of resonance, it turns out that at the end of the day they will make it simpler\footnote{Stokes data and connection formulae for the \PI~tau-function have been recently addressed in \cite{i19a}, following upon the exact WKB analysis in \cite{t95, t99, t00}; and in \cite{lr16}, following upon the gauge theoretic construction in \cite{blmst16}. Albeit completely different approaches from ours, in the end all results should match, \textit{i.e.}, in principle it should be possible to map those results to the highly non-trivial numbers we are computing \cite{i19b}. In the isomonodromic-like formulations used in the aforementioned papers, connection formulae have rather compact final expressions and logarithms or resonance seem to be hidden. Whereas resurgence relations and their associated asymptotics will always need all Stokes data we are computing, these results also imply that somehow our formulation of the problem should greatly simplify---on what concerns connection formulae---once we better understand the logarithmic structure present in our asymptotic relations. That this is the case will become fully clear in section~\ref{sec:stokes}, when we discuss how to compute connection formulae out of the bridge equations.}. Secondly, we are trying to compute vectors labelled on a lattice, figures~\ref{fig:lattice_stokes} or~\ref{fig:vec_struct}, and resonance is associated to the diagonal directions on this lattice---but it turns out that the boundaries of this Stokes lattice (depicted in figure \ref{fig:lattice_stokes_guessing_game}) have a particularly simple structure and are easy to guess\footnote{We will give a more complete exposition on how to guess closed-forms for numbers in subappendix~\ref{app:subsec:guess}.}. This structure was in fact already found in \cite{asv11} in the \PI~context (generalizable to \PII~in \cite{sv13}), and reads:
\be
N_1^{(\ell)}=\frac{\rmi^{\ell-1}}{\ell^2}\left(N_1^{(1)}\right)^{2-\ell}.
\ee
\noindent
There are two main reasons why these numbers were easily guessable from numerical results:
\begin{itemize}
\item The number content is basically dictated from $N_1^{(1)}$ alone, which is known analytically.
\item The are no sums of numbers---just products---which makes guessing such structures with computer code much easier and efficient.
\end{itemize}
\noindent
We are thus left with understanding the Stokes numbers located in the bulk of the lattice. These numbers are immediately harder to guess because the asymptotic relations suggest that on top of being sums of numbers, each additive term in these sums may have rather non-trivial number content. Understanding these asymptotic relations essentially means understanding the large $g$ limit in \eqref{eq:fin_as}. But, as already mentioned, such asymptotic behavior is also seemingly complicated by logarithms and we first need to understand their role. In order to achieve this, let us modify our asymptotic relations in the following three steps:
\begin{enumerate}
\item We will hide away everything which does not yield an immediate condition on the Borel residues, in different terms stored on the left-hand side of the asymptotic relation.
\item We then try to better understand the logarithmic structure in these resulting equations.
\item We finally deal with the asymptotic limit by making use of properties of the digamma functions, which we will see appear in these resulting equations.
\end{enumerate}
\noindent
At the end of these steps, we shall be able to conjecture analytic equations that determine Stokes data. As always, we find one may treat \PI/\PII~with the very same equations.

\paragraph{1. Simplifying the asymptotic relations:} Let us recast the asymptotic relations \eqref{eq:fin_as} derived in the previous subsection in a more convenient form. Start with \eqref{eq:fin_as} and incorporate the properties of the coefficients \eqref{eq:properties_coefs}. One obtains:
\bea
\label{eq:large-order-relations-closed-form}
u_{4g}^{(n,n)[0]} &\simeq& \\ 
&& 
\hspace{-50pt}
\simeq - \frac{1}{\rmi\pi} \sum_{\ell=1}^{+\infty} \sum_{h=0}^{+\infty} \sum_{p=0}^{n} \sum_{k=0}^{p} \mathsf{S}_{(n,n)\to(p+\ell,p)}\, \frac{1}{k!} \left( \frac{\alpha}{2}\ell \right)^{k} u_{2h}^{(p+\ell-k,p-k)[0]}\, \widetilde{H}_k \left( 2g+n-h-\beta^{(k)}_{(p+\ell,p)}, \ell A  \right). \nonumber
\eea
\noindent
Out of this expression, in order to lose subleading contributions, introduce the \textit{truncated} series $T_{g,\ell}^{(n,n)}$ in the obvious manner:
\bea
T_{g,\ell}^{(n,n)} &\simeq& \\ 
&& 
\hspace{-50pt}
\simeq - \frac{1}{\rmi\pi} \sum_{t=1}^{\ell-1} \sum_{h=0}^{+\infty} \sum_{p=0}^{n} \sum_{k=0}^{p} \mathsf{S}_{(n,n)\to(p+t,p)}\, \frac{1}{k!} \left( \frac{\alpha}{2} t \right)^{k} u_{2h}^{(p+t-k,p-k)[0]}\, \widetilde{H}_k \left( 2g+n-h-\beta^{(k)}_{(p+t,p)}, tA  \right). \nonumber
\eea
\noindent
Subtracting this truncated series from the original asymptotic expression \eqref{eq:large-order-relations-closed-form}, dividing by an appropriate quantity (which becomes clear in the following), and leaving out subleading terms, \textit{i.e.}, those associated to $h>0$, $k<p$, one finds:
\bea
\frac{u_{4g}^{(n,n)[0]} - T_{g,\ell}^{(n,n)}}{\widetilde{H}_0 \left( 2g+n-\frac{\ell }{2}, \ell A \right)} &\simeq& \\
&&
\hspace{-110pt}
\simeq - \frac{1}{\rmi\pi} \sum_{h=0}^{+\infty} \sum_{p=0}^{n} \sum_{k=0}^{p} \mathsf{S}_{(n,n)\to(p+\ell ,p)}\, \frac{1}{k!} \left( \frac{\alpha}{2}\ell \right)^{k} u_{2h}^{(p+\ell-k,p-k)[0]}\, \frac{\widetilde{H}_k \left( 2g+n-h-\beta^{(k)}_{(p+\ell,p)}, \ell A \right)}{\widetilde{H}_0 \left( 2g+n-\frac{\ell}{2}, \ell A \right)} + \cdots \nonumber \\
&&
\hspace{-110pt}
\simeq - \frac{1}{\rmi\pi} \sum_{p=0}^{n} \mathsf{S}_{(n,n)\to(p+\ell,p)}\, \frac{1}{p!} \left( \frac{\alpha}{2}\ell \right)^{p} u_{0}^{(\ell,0)[0]}\, \frac{\widetilde{H}_p \left( 2g+n-\frac{\ell}{2}, \ell A \right)}{\widetilde{H}_0 \left( 2g+n-\frac{\ell}{2}, \ell A \right)} + \cdots. \nonumber
\eea
\noindent
In our on-going quest to \textit{isolate} the unknown Borel residues, it is next natural to define the sequence $\widetilde{D}_{g,\ell}^{(n,n)}$ via
\be
\label{eq:PreAsymptoticRelation}
\underbrace{\frac{\rmi\pi}{u_{0}^{(\ell,0)[0]}}\, \frac{u_{4g}^{(n,n)[0]} - T_{g,\ell}^{(n,n)}}{\widetilde{H}_0 \left( 2g+n-\frac{\ell}{2}, \ell A \right)}}_{=:\, \left( n+1 \right) \widetilde{D}_{g,\ell }^{(n,n)}} \simeq - \sum_{p=0}^{n} \mathsf{S}_{(n,n)\to(p+\ell,p)}\, \frac{1}{p!} \left( \frac{\alpha}{2}\ell \right)^{p} \frac{\widetilde{H}_p \left( 2g+n-\frac{\ell}{2}, \ell A \right)}{\widetilde{H}_0 \left( 2g+n-\frac{\ell}{2}, \ell A \right)}.
\ee
\noindent
In particular, this expression implies that the large-$g$ behavior of the $\widetilde{D}_{g,\ell}^{(n,n)}$ \textit{will determine} the Borel residues. It is important to note that the $\widetilde{D}_{g,\ell}^{(n,n)}$ coefficients contain logarithms in $g$, matching those in the $\widetilde{H}_k$-functions. More explicitly, we can manipulate \eqref{eq:PreAsymptoticRelation} to explicitly give 
\bea
\left( n+1 \right) \widetilde{D}_{g,\ell}^{(n,n)} &\simeq& \sum_{p=0}^{n} \mathsf{S}_{(n,n)\to(p+\ell,p)}\, \frac{1}{p!} \left( \frac{\alpha}{2}\ell \right)^{p} \times \\ 
&&
\hspace{-40pt}
\times B_{p} \left( \psi^{(0)} \left( 2g+n-\frac{\ell}{2} \right) - \log \left(\ell A \right), \psi^{(1)} \left( 2g+n-\frac{\ell}{2} \right), \ldots, \psi^{(p-1)} \left( 2g+n-\frac{\ell}{2} \right) \right), \nonumber
\eea
\noindent
where $B_{p}$ denotes the complete Bell polynomial of order $p$ and $\psi^{(n)}$ the polygamma functions \cite{olbc10}. Given the known behavior of these polygamma functions \cite{olbc10}, it follows that all logarithmic growth originates from $\psi^{(0)}$ in the first entry of the Bell polynomial. Furthermore, by including the structure of the Stokes vectors \eqref{eq:vec_struct-a}-\eqref{eq:vec_struct-b} into equations \eqref{eq:rest_borel-a} or \eqref{eq:rest_borel-b} we may now obtain the relation
\be
\mathsf{S}_{(n,n) \to (p+\ell,p)} = -\left( n+1 \right) N_{\ell-n+p}^{(\ell)} + \mathsf{R}_{(p+\ell,p)}^{(n,n)},
\ee
\noindent
where, much as in \eqref{eq:rest_borel-a}, $\mathsf{R}_{(p+\ell,p)}^{(n,n)}$ contains products of the $N^{(\ell_i)}_{n_i}$'s such that for each term we have $\sum_i \ell_i=\ell$ and $\ell_i<\ell$ for any $i$. Furthermore, $\mathsf{R}_{\ell,p}^{(n,n)}=0$ for $\ell=1$. These conditions let us further constrict the above sum over $p$, as
\bea
\label{eq:AsymptoticRelationWithN}
D_{g,\ell}^{(n,n)} &\simeq& \sum_{p=0}^{n-\ell+1} N_{\ell-n+p}^{(\ell)}\, \frac{1}{p!} \left( \frac{\alpha}{2}\ell \right)^p \times \\
&&
\hspace{-40pt}
\times B_{p} \left( \psi^{(0)} \left( 2g+n-\frac{\ell}{2} \right) - \log \left(\ell A \right), \psi^{(1)} \left( 2g+n-\frac{\ell}{2} \right), \ldots, \psi^{(p-1)} \left( 2g+n-\frac{\ell}{2} \right) \right), \nonumber
\eea
\noindent
where we have absorbed the remainder terms $\mathsf{R}_{(p+\ell,p)}^{(n,n)}$ into the (new) $D_{g,\ell}^{(n,n)}$ (hence the small change in notation). This we do because $\mathsf{R}_{(p+\ell,p)}^{(n,n)}$ only contains Stokes data which can be determined from equations with lower $n$---and therefore we are not interested in it. Note also that $\mathsf{R}_{(p+\ell,p)}^{(n,n)}$ is zero for the first diagonal. The usual procedure dictated by large-order asymptotics would now be to understand the large-order behavior of the $D_{g,\ell}^{(n,n)}$. For large $g$, it takes the form
\begin{equation}
\label{eq:Dloggrowth}
D_{g,\ell}^{(n,n)} = \sum_{k=0}^{n} d_{\ell,k}^{(n,n)}(g)\, \log^k g + o \left(\frac{1}{g}\right).
\end{equation}
\noindent
In this way, knowledge of the coefficients $d_{\ell,k}^{(n,n)} (g)$ implies knowledge of all Stokes data. It turns out that understanding the $D_{g,\ell}^{(n,n)}$ is a much easier task than understanding ``bare'' Stokes data.

\paragraph{2. Dealing with the logarithmic structure:} As we have just seen, in the strict asymptotic limit equation \eqref{eq:AsymptoticRelationWithN} splits into different conditions, each one corresponding to a different logarithmic growth, as in \eqref{eq:Dloggrowth}. It turns out that only the lowest-order growth will give a non-trivial condition on Stokes data. What this means is that not all the numbers $d_{\ell,k}^{(n,n)} (g)$ encode relevant information; and it turns out that it is possible to find a reformulation of our problem which gets rid of all the logarithmic growth in equation \eqref{eq:AsymptoticRelationWithN}. We will use the fact that this equation has a very special structure dictated by the Bell polynomials. Let us introduce a new sequence in $g$ that we call
\begin{equation}
C_{\ell}(g)
\end{equation}
\noindent
which we leave unspecified for now. Considering the sum 
\be
\label{eq:BellsumationoverC}
\sum\limits_{p=0}^{n-\ell+1}\, \frac{1}{p!} \left(\frac{\alpha\,\ell\,C_{\ell}(g)}{2}\right)^p D_{g,\ell}^{(n-p,n-p)}
\ee
\noindent
is where the special structure of the Bell polynomials comes in. Using the explicit asymptotic relation \eqref{eq:AsymptoticRelationWithN} in the above sum \eqref{eq:BellsumationoverC}, it turns out that in the resulting right-hand side $C_\ell(g)$ will appear in the first argument of the Bell polynomials, but changing nothing else. Explicitly, we find
\bea
\label{eq:summedasymptoticrelationtoloosethelogs}
\sum\limits_{p=0}^{n-\ell+1}\, \frac{1}{p!} \left(\frac{\alpha\,\ell\,C_{\ell}(g)}{2}\right)^p D_{g,\ell}^{(n-p,n-p)} &\simeq& \sum_{p=0}^{n-\ell+1} N_{\ell-n+p}^{(\ell)}\, \frac{1}{p!} \left( \frac{\alpha}{2}\ell \right)^p \times \\
&&
\hspace{-180pt}
\times B_{p} \left( \psi^{(0)} \left( 2g+n-\frac{\ell}{2} \right) - \log \left(\ell A \right) + C_{\ell}(g), \psi^{(1)} \left( 2g+n-\frac{\ell}{2} \right), \ldots, \psi^{(p-1)} \left( 2g+n-\frac{\ell}{2} \right) \right). \nonumber
\eea
\noindent
Interestingly it turns out that if we now choose $C_{\ell }(g)$ such that it cancels the logarithmic growth of $\psi^{(0)} \left(2g+n-\frac{\ell}{2} \right)$, then the right-hand side of \eqref{eq:summedasymptoticrelationtoloosethelogs} becomes logarithmic-free. We then conclude that this also has to be the case for its left-hand side. This implies the choice of $C_\ell(g)$ which removes the logarithmic dependence of \eqref{eq:summedasymptoticrelationtoloosethelogs} is such that
\begin{equation}
\psi^{(0)} \left( 2g+n-\frac{\ell}{2} \right) - \log \left(\ell A \right)\,+\,C_{\ell}(g)\,=\,c_{\ell} + o \left(\frac{1}{g}\right),\quad c_{\ell}\,\in\,\mathbb{C},
\end{equation}
\noindent
where $c_\ell$ is constant in $g$. With this choice we now know that 
\be
\mathcal{D}^{n}_{g, \ell} := \sum\limits_{p=0}^{n-\ell+1} \frac{1}{p!} \left(\frac{\alpha\,\ell\,C_{\ell}(g)}{2}\right)^p D_{g,\ell}^{(n-p,n-p)}
\ee
\noindent
is logarithm-free, and we can spell out a new logarithm-free asymptotic relation as
\bea
\label{eq:summedasymptoticrelationnologs}
\mathcal{D}^{n}_{g,\ell} &\simeq& \sum_{p=0}^{n-\ell+1} N_{\ell-n+p}^{(\ell)}\, \frac{1}{p!} \left( \frac{\alpha}{2}\ell \right)^p \times \\
&&
\times B_{p} \left(c_{\ell}+ o \left(\frac{1}{g}\right), \psi^{(1)} \left( 2g+n-\frac{\ell}{2} \right), \ldots, \psi^{(p-1)} \left( 2g+n-\frac{\ell}{2} \right) \right). \nonumber
\eea
\noindent
Note how removing the logarithmic growth came at a price: we had to introduce the \textit{a priori} unknown constants $c_\ell$. The above resulting equation also has a much simpler structure than the original asymptotic relation \eqref{eq:fin_as} that we started-off with, but it is still an asymptotic relation. Let us finally try to understand this asymptotic limit in the next step.

\paragraph{3. Understanding the asymptotic limit:} As we have just seen, understanding Stokes data is tantamount to understanding the $\mathcal{D}^{n}_{g, \ell}$ set. Now, in order to make further progress along this analysis, one requires some intuition from our numerical explorations. As we will discuss later in subappendix~\ref{app:subsec:guess}, calculating numerical Stokes data and trying to match those numbers against possible closed-form analytic forms led us to find that they contain combinations of zeta-function values, essentially everywhere. At the same time, we have just seen in the previous paragraph how our main asymptotic equations contain polygamma functions, also essentially everywhere. These polygamma functions have the interesting property that, for integer arguments, they contain the zeta-function:
\be
\psi^{(n)}(x) =
\begin{cases}
\,-\gamma_{\text{E}} + \displaystyle\sum\limits_{k=1}^{x-1} \frac{1}{k}, & n = 0, \\
\,(-1)^{n+1}\, n!\, \zeta \left( n+1 \right) - (-1)^{n+1}\, n!\, \displaystyle\sum\limits_{k=1}^{x-1} \frac{1}{k^{n+1}}, & n\geq 1.
\end{cases}
\ee
\noindent
One first---very reasonable---assumption we can make is that all transcendental numbers appearing in Stokes data arise from the above polygamma functions; in particular from their $x$-independent contributions in the formulae above. In other words, we are assuming that no additional transcendental numbers may appear through the large-$g$ limit---which is in fact the simplest assumption one can make about such a limit, and further implies that the limit itself does not contain any non-trivial information. To do this let us try to group all the terms in \eqref{eq:summedasymptoticrelationnologs} that still contain a $g$ dependence on its left-hand side. We have already seen in \eqref{eq:summedasymptoticrelationtoloosethelogs} how we can modify the first entry of the Bell polynomials by considering clever summations of the asymptotic relation \eqref{eq:AsymptoticRelationWithN}. We can follow an analogous procedure to modify the other entries of the Bell polynomials in order to remove the $g$ dependence on its right-hand side. Let us start by spelling out the polygamma functions explicitly
\bea
\mathcal{D}^{n}_{g,\ell} &\simeq& \sum_{p=0}^{n-\ell+1} N_{\ell-n+p}^{(\ell)}\, \frac{1}{p!} \left( \frac{\alpha}{2}\ell \right)^p B_{p} \left( c_{\ell} + o \left(\frac{1}{g}\right), 2 \zeta \left(2\right) + 2 \sum\limits_{k=1}^{2g+n-1-\ell/2} \frac{1}{k^{n+1}},\, \ldots \right. \nonumber \\
&&
\left.\ldots\, , (-1)^{p}\, (p-1)!\, \zeta \left( p \right) - (-1)^{p}\, (p-1)!\, \sum\limits_{k=1}^{2g+n-1-\ell/2} \frac{1}{k^{p}} \right).
\label{eq:summedasymptoticrelationnologsgammasspelledout}
\eea
\noindent
Here, the parts which contain $g$-dependence are the sums that are generated by the polygamma functions inside the Bell polynomials. Now, as explained above, by picking good linear combinations of the asympotic relations \eqref{eq:summedasymptoticrelationnologsgammasspelledout} we can absorb those sums into the left-hand side. Let us call the resulting linear combination of the $\mathcal{D}^{n}_{g, \ell}$ by the name $d^{n}_{\ell}(g)$. We then arrive at
\be
\label{eq:summedasymptoticsnologsstillasymptotic}
d^{n}_{\ell}(g) \simeq \sum_{p=0}^{n-\ell+1} N_{\ell-n+p}^{(\ell)}\, \frac{1}{p!} \left( \frac{\alpha}{2}\ell \right)^p B_{p} \left( c_{\ell} + o \left(\frac{1}{g}\right), 2\zeta \left(2\right), \ldots, (-1)^{p}\, (p-1)!\, \zeta \left( p \right) \right).
\ee
\noindent
We note that the right-hand side of \eqref{eq:summedasymptoticsnologsstillasymptotic} goes to a finite value in the $g\to+\infty$ limit. For consistency reasons this also has to be the case for its left-hand side. We can therefore turn this asymptotic relation into an equality containing a limit:
\be
\label{eq:summedasymptoticsnologslimit}
\lim\limits_{g\to+\infty} d^{n}_{\ell}(g) = \sum_{p=0}^{n-\ell+1} N_{\ell-n+p}^{(\ell)}\, \frac{1}{p!} \left( \frac{\alpha}{2} \ell \right)^p B_{p} \left(c_{\ell}, \psi^{(1)}(1), \ldots, \psi^{(p-1)}(1) \right),
\ee
\noindent
where we have replaced the zeta expressions in the Bell polynomials back to polygamma functions evaluated at the finite value $x=1$ for conciseness. This final form no longer contains any asymptotic relation nor superfluous logarithms. All the non-triviality of the asymptotics is now contained in the limit $\lim\limits_{g\to+\infty}d^{n}_{\ell}(g)$ alone.

The last step in uniquely fixing Stokes data is now to understand this $\lim\limits_{g\to+\infty}d^{n}_{\ell}(g)$ limit. This is where the rewriting of the asymptotic relations that we did above pays-off, in the sense that we find simple expressions for the limits which we can calculate numerically. Let us start with the first diagonal, and thereafter try to generalize it to higher diagonals.
\begin{itemize}
\item \textbf{First diagonal, $\ell=1$:} \\
We have calculated a few of the desired numbers using our numerical methods (these will be discussed later). What we found\footnote{We explicitly checked this for \PI~with $n=0,\ldots,20$; and for \PII~with $n=0,\ldots,6$.}, for both \PI~and \PII, is that:
\bea
\lim\limits_{g\to+\infty}d^{0}_{1}(g) &=& N^{(1)}_{1}, \\
\lim\limits_{g\to+\infty}d^{n}_{1}(g) &=& 0, \qquad n > 0,
\eea
\noindent
with 
\begin{equation}
c_{1} =  \log \widetilde{A}_1 + \gamma_{\text{E}},
\end{equation}
\noindent
and where $\widetilde{A}_1$ can be calculated from $N^{(1)}_{0}$ alone. In light of this, one is immediately led to conjecture that \textit{all} $\lim\limits_{g\to+\infty}d^{n}_{1} (g)$ vanish for any $n>0$. As the calculation of $c_\ell$ needs $N^{(1)}_{0}$, this also implies that in order to fully determine the first diagonal we only need the first two numbers of said diagonal, \textit{i.e.}, $N^{(1)}_{0}$ and $N^{(1)}_{1}$.
\item \textbf{Higher diagonals, $\ell>1$:} \\
For higher diagonals the right-hand side in \eqref{eq:summedasymptoticsnologslimit} does not immediately yield such a simple form as for the first diagonal. Nonetheless, it is possible to perform further modifications to this equation so as to obtain the exact same structure for the (modified) $\lim\limits_{g\to+\infty}d^{n}_{\ell}(g)$ as for the first diagonal above (the decisive hint for this step once again came from our numerics, yet to be discussed). If we insert factors of $\ell$ in our asymptotic relations in the following way
\be
\label{eq:summedasymptoticsnologslimitinserteds}
\lim\limits_{g\to+\infty} d^{n}_{\ell}(g) = \sum_{p=0}^{n-\ell+1} N_{\ell-n+p}^{(\ell)}\, \frac{1}{p!} \left( \frac{\alpha}{2} \ell\right)^p B_{p} \left(c_{\ell}, \frac{1}{\ell}\psi^{(1)}(1), \ldots, \frac{1}{\ell^{p-1}}\psi^{(p-1)}(1) \right),
\ee
\noindent
we again arrive at a remarkably simple structure for the limiting procedure. Interestingly, this is something that would not change the $\ell=1$ case. We now find\footnote{We explicitly checked this for \PI~with $\ell=1,2,3$ and $n$ up to $20, 14, 5$, respectively; and for \PII~with $\ell=1,2$ and $n$ up to $6,4$, respectively.}, again for both \PI~and \PII, that
\bea
\lim\limits_{g\to+\infty}d^{0}_{\ell}(g) &=& N^{(\ell)}_{1}, \\
\lim\limits_{g\to+\infty}d^{n}_{\ell}(g) &=& 0, \qquad n > 0,
\eea
\noindent
with 
\begin{equation}
c_{\ell} =  \log \widetilde{A}_\ell + \gamma_{\text{E}},
\end{equation}
\noindent
and where $\widetilde{A}_\ell$ can be calculated from $N^{(\ell)}_{0}$ alone. 
As before, we now simply conjecture that \textit{all} $\lim\limits_{g\to+\infty}d^{n}_{\ell}(g)$ vanish for any $n>0$. Remarkably, it again suffices to know the first two non-trivial numbers in each diagonal, in order to construct the whole diagonal. Note that, in particular, this insertion of the $\ell$ factors in equation \eqref{eq:summedasymptoticsnologslimitinserteds} is necessary in order to reproduce the ``green factors'' appearing in the several formulae of figure~\ref{fig:patterns} (factors which were initially observed numerically).
\end{itemize}

We are now done. All cases discussed above may be assembled into a single set of recursive relations, which we conjecture uniquely determine \textit{full} Stokes data for the first two Painlev\'e equations. These relations are:
\begin{equation}
\label{eq:eq_for_N}
\sum_{p=0}^{n-\ell+1} N^{(s)}_{\ell-n+p}\, \frac{1}{p!} \left( \frac{\alpha}{2}\ell\right)^{p} B_{p} \left(c_\ell, \frac{1}{\ell}\, \psi^{(1)}(1), \ldots, \frac{1}{\ell^{p-1}}\, \psi^{(p-1)}(1) \right) = \lim\limits_{g\to+\infty}d^{n}_{\ell}(g),
\end{equation}
\noindent
where the limits are given by
\bea
\label{eq:lim_1}
\lim\limits_{g\to+\infty}d^{0}_{\ell}(g) &=& N^{(\ell)}_{1}, \\
\label{eq:lim_2}
\lim\limits_{g\to+\infty}d^{n}_{\ell}(g) &=& 0, \qquad n > 0,
\eea
\noindent
with
\begin{equation}
c_{\ell} =  \log \widetilde{A}_\ell - \gamma_{\text{E}},\label{eq:c_cons}
\end{equation}
\noindent
and where $\widetilde{A}_\ell$ is a number which may be calculated from $N^{(\ell)}_{0}$ alone. What this final result implies is that \textit{all} Stokes data may be specified once we know the \textit{first two numbers} associated to each diagonal. In other words, unfortunately, these first two numbers still need to be computed in some other way. Further, note that \eqref{eq:eq_for_N} is clearly problem-specific as we had to put-in information concerning the structure of the $\lim\limits_{g\to+\infty}d^{n}_{\ell}(g)$ (structure which we were able to conjecture after numerically exploring the first few Borel residues for the first two Painlev\'e equations). This part of the analysis would have to be readdressed if focusing on different (eventually non-Painlev\'e) problems---albeit in the exact same way. Finally, we assumed a very specific vector structure \eqref{eq:vec_struct-a}-\eqref{eq:vec_struct-b}, which allowed for the closure of the alien algebra \eqref{eq:585} (and is, as usual, strongly supported by numerical explorations). All in all, it seems very likely that for other resonant problems---as, \textit{e.g.}, those in the (m)KdV hierarchies \cite{gs21} mentioned in the introduction---very similar results and subsequent conjectures might hold. Analogous sequences $\lim\limits_{g\to+\infty}d^{n}_{\ell}(g)$ may be constructed for those problems, which, hopefully, will also enjoy a simpler structure than the ``bare'' original Stokes data, hence allowing for a similar solution to their full determination.

\section{Analytical Results for Borel Residues and Stokes Vectors}\label{sec:results}

We are now ready to present our complete results, for the first two Painlev\'e equations alongside their quantum gravity and minimal string incarnations. In other words, we will list Stokes data for both Painlev\'e \textit{solutions} and quantum-gravity/minimal-string \textit{free energies} (following up on the discussions in subsections~\ref{subsec:PI2dQG} and~\ref{subsec:PII2dSG}). Using the method of ``closed-form asymptotics'', just described in subsection~\ref{subsec:closedform-asymp}, we are able to obtain an \textit{ansatz} for the full set of forward Stokes data, which in sequence allows for the computation of all Stokes data---up to a single number. In the strict context of ``closed-form asymptotics'' this unknown number has to be guessed numerically---see such details in subappendix~\ref{app:subsec:guess}. Another goal of this section is to systematically assemble all Stokes data into adequate generating functions. Later, in section~\ref{sec:stokes}, we shall see how these generating functions are in fact the ones adequate to discuss Stokes automorphisms and connection formulae---as already partially discussed in subsection~\ref{subsec:stokes-trans}.

\subsection{Stokes Data for the Painlev\'e~I Equation}\label{subsec:PIresults}

We recall the structure of Stokes vectors that we described in subsection \ref{subsec:vec_struct}; see expressions \eqref{eq:vec_struct-a}-\eqref{eq:vec_struct-b}. The forward vectors $\bm{S}_{(1-p,1-p-\ell)}$ and the backward vectors $\bm{S}_{(1-p-\ell,1-p)}$ have the structure\footnote{In all equations that follow, $p$ is a non-negative integer and $\ell$ a strictly positive integer.}
\bea
\label{eq:struct-a}
\bm{S}_{(1-p,1-p-\ell)} &=& N^{(\ell)}_{1-p}\, \left[ \begin{array}{c}p+\ell \\ -p\end{array} \right], \\
\label{eq:struct-b}
\bm{S}_{(1-p-\ell,1-p)} &=& N^{(-\ell)}_{1-p}\, \left[ \begin{array}{c}-p \\ p+\ell\end{array} \right],
\eea
\noindent
which was illustrated in figure~\ref{fig:vec_struct}. How to find the (yet unknown) proportionality factors was just discussed at length in the previous section~\ref{sec:asymptotics}. Some things we do know, however. In particular, according to the analysis in \cite{as13}, the constants $N^{(\ell)}_{1-p}$ are purely imaginary. The constants $N^{(-\ell)}_{1-p}$, on the other hand, have both a real and imaginary part---albeit they are determined from the $N^{(\ell)}_{1-p}$ through the backward-forward relation \eqref{eq:back_forw}. We will first build a generating function for the $N^{(\ell)}_{1-p}$ data, and then use it to construct generating functions for the Stokes vectors themselves.

We claimed in subsection~\ref{subsec:closedform-asymp} that equations \eqref{eq:eq_for_N}-\eqref{eq:lim_1}-\eqref{eq:lim_2}-\eqref{eq:c_cons} are enough to compute all these unknowns. Let us now see how. For \PI, introduce the ratios
\be
\label{eq:stokes-R-ratios}
R^{(\ell)}_p := \frac{N^{(\ell)}_p}{N^{(\ell)}_1},
\ee
\noindent
which, for lack of a more imaginative name, we dub ``Stokes ratios''. Following upon the aforementioned results of subsection~\ref{subsec:closedform-asymp}, these ratios immediately fulfill
\be
\label{eq:closed_form_PI}
\sum_{q=0}^p R^{(\ell)}_{1-p+q}\, \frac{1}{q!} \left( \frac{2}{\sqrt3} \ell \right)^{q} B_{q} \left(c_\ell, \frac{1}{\ell}\, \psi^{(1)}(1), \ldots, \frac{1}{\ell^{q-1}}\, \psi^{(q-1)}(1) \right) = \delta_{p,0}.
\ee
\noindent
The finite sum on the left-hand side of this expression may be regarded as the coefficients for the Cauchy product\footnote{Given two formal series
\be
A(\uplambda) = \sum_{p=0}^{+\infty} a_p\uplambda^p \quad \text{ and } \quad B(\uplambda) = \sum_{p=0}^{+\infty} b_p\uplambda^p,
\ee
\noindent
the Cauchy product of the two series is the formal power-series defined by
\be
C(\uplambda) = \sum_{p=0}^{+\infty} c_p\uplambda^p \quad \text{ with } \quad c_p=\sum_{q=0}^p a_qb_{p-q}.
\ee
\noindent
In particular, given a formal power series $A(\uplambda)$ as above, with $a_0\neq0$, we say that the formal series $B(\uplambda)$ is a formal \textit{inverse} of $A(\uplambda)$ if their Cauchy product is the formal series $1$. This means that the coefficients obey the relation
\be
\sum_{q=0}^p a_q b_{p-q} = \delta_{p,0}.
\ee
\noindent
This relation can also be used to determine the coefficients of the inverse recursively.} between the two (so far formal) series:
\bea
\label{eq:Relllambda-genfun}
R^{(\ell)} (\uplambda) &=& \sum_{p=0}^{+\infty} R^{(\ell)}_{1-p}\, \uplambda^p, \\
B^{(\ell)} (\uplambda) &=& \sum_{p=0}^{+\infty} \frac{1}{p!} \left( \frac{2}{\sqrt3}\, \ell \right)^{p} B_{p} \left( c_\ell, \frac{1}{\ell}\, \psi^{(1)}(1), \ldots, \frac{1}{\ell^{p-1}}\, \psi^{(p-1)}(1) \right) \uplambda^p.\label{eq:bellsum}
\eea
\noindent
Furthermore, equation \eqref{eq:closed_form_PI} may be equivalently rewritten as
\be
R^{(\ell)} (\uplambda) = \frac{1}{B^{(\ell)} (\uplambda)},
\ee
\noindent
which immediately implies that finding a generating functional for Stokes data is equivalent to finding the closed-form of $B^{(\ell)} (\uplambda)$. This is actually quite simple. First, use well-known properties of Bell polynomials \cite{olbc10} to find\footnote{Recall Bell polynomials satisfy by the generating function
\be
\exp\left(\sum_{p=1}^{+\infty} x_p \frac{t^p}{p!} \right) = 1 + \sum_{k=1}^{+\infty} \frac{t^k}{k!}\, B_k(x_1,...,x_{k}).
\ee
\noindent
The right-hand side is exactly of the form \eqref{eq:bellsum}, with $x_1=c_\ell$, $x_n=\frac{1}{\ell^{n-1}}\psi^{(n-1)}(1)$ for $n>1$, and $t=\frac{2}{\sqrt{3}}\, \lambda\ell$.}
\bea
B^{(\ell)} (\uplambda) &=& \exp \left( \frac{2}{\sqrt{3}}\, \ell\, c_\ell\, \uplambda + \sum_{k=2}^{+\infty} \frac{1}{\ell^{k-1}\, k!}\, \psi^{(k-1)} (1) \left( \frac{2\ell}{\sqrt{3}}\, \uplambda \right)^{k} \right) = \nonumber \\
&=& \exp \left(\frac{2}{\sqrt{3}}\, \ell\, c_\ell\, \uplambda + \ell\, \sum_{k=2}^{+\infty} \frac{1}{k} \left( -\frac{2 \uplambda}{\sqrt{3}} \right)^k \zeta(k) \right),
\eea
\noindent
where we have used the explicit values already discussed in subsection~\ref{subsec:closedform-asymp}, \textit{i.e.}, $\psi^{(k)} (1) = \left(-1\right)^{k+1} k!\, \zeta(k+1)$. The remaining power-series is now readily evaluated in closed-form, leading to
\bea
B^{(\ell)} (\uplambda) &=& \exp \left\{ \frac{2}{\sqrt{3}}\, \ell\, c_\ell\, \uplambda + \ell \left( \log \Gamma \left( 1 + \frac{2 \uplambda}{\sqrt{3}} \right) + \frac{2 \gamma_{\text{E}}}{\sqrt{3}}\, \uplambda \right) \right\} = \nonumber \\
&=& \Gamma \left( 1 + \frac{2 \uplambda}{\sqrt{3}} \right)^{\ell} \exp \left\{ \frac{2\ell}{\sqrt{3}} \left( \gamma_{\text{E}} + c_\ell \right) \uplambda \right\}.
\label{eq:bell_pol}
\eea
\noindent
The corresponding inverse-function then yields $R^{(\ell)} (\uplambda)$, \textit{i.e.}, allows for a closed-form computation of \textit{all} Stokes ratios along the same fixed diagonal $\ell$, in terms of the first ratio $R^{(\ell)}_0$. To make this explicit, use the expression for $c_\ell$ \eqref{eq:c_cons} to obtain
\be
\label{eq:PIStokesratiosGF}
R^{(\ell)} (\uplambda) = \frac{\widetilde{A}_\ell^{\frac{2 \uplambda \ell}{\sqrt{3}}}}{\Gamma \left( 1 + \frac{2 \uplambda}{\sqrt{3}} \right)^{\ell}}.
\ee
\noindent
In particular, this expression further conveys the interesting property that $R^{(\ell)} (\uplambda) = \left( R^{(1)} (\uplambda) \right)^{\ell}$, unveiling how truly constrained this problem really is. One final step makes everything explicit for the Stokes vectors \eqref{eq:struct-a}. Using the conjecture in \cite{asv11} for the Stokes lattice boundary (recall figure~\ref{fig:lattice_stokes_guessing_game}), alongside the known analytic Stokes coefficient \eqref{eq:exampleStokesP1}, \textit{i.e.},
\be
\label{eq:n1sp1}
N_1^{(\ell)} = \frac{\rmi^{\ell-1}}{\ell^2} \left( N_1^{(1)} \right)^{2-\ell} \qquad \text{ and } \qquad  N_1^{(1)} = - \rmi\, \frac{\sqrt[4]{3}}{2\sqrt\pi},
\ee
\noindent
the explicit generating function of full \PI~Stokes data follows:
\bea
\label{eq:gen_fun-a}
N^{(\ell)} (\uplambda) &=& N^{(\ell)}_{1}\, R^{(\ell)} (\uplambda) = \frac{\rmi^{\ell-1}}{\ell^2} \left( N_1^{(1)} \right)^{2-\ell} \frac{\widetilde{A}_{\ell}^{\frac{2 \uplambda \ell}{\sqrt{3}}}}{\Gamma \left( 1 + \frac{2 \uplambda}{\sqrt{3}} \right)^{\ell}}, \\
\label{eq:gen_fun-b}
N_{1-p}^{(\ell)} &=&\frac{1}{p!} \left. \frac{\partial^{p}}{\partial \uplambda^{p}}\, N^{(\ell)} (\uplambda) \right|_{\uplambda=0}.
\eea

\begin{figure}
	\centering
	\begin{subfigure}[t]{0.5\textwidth}
		\centering
		\includegraphics[height=2in]{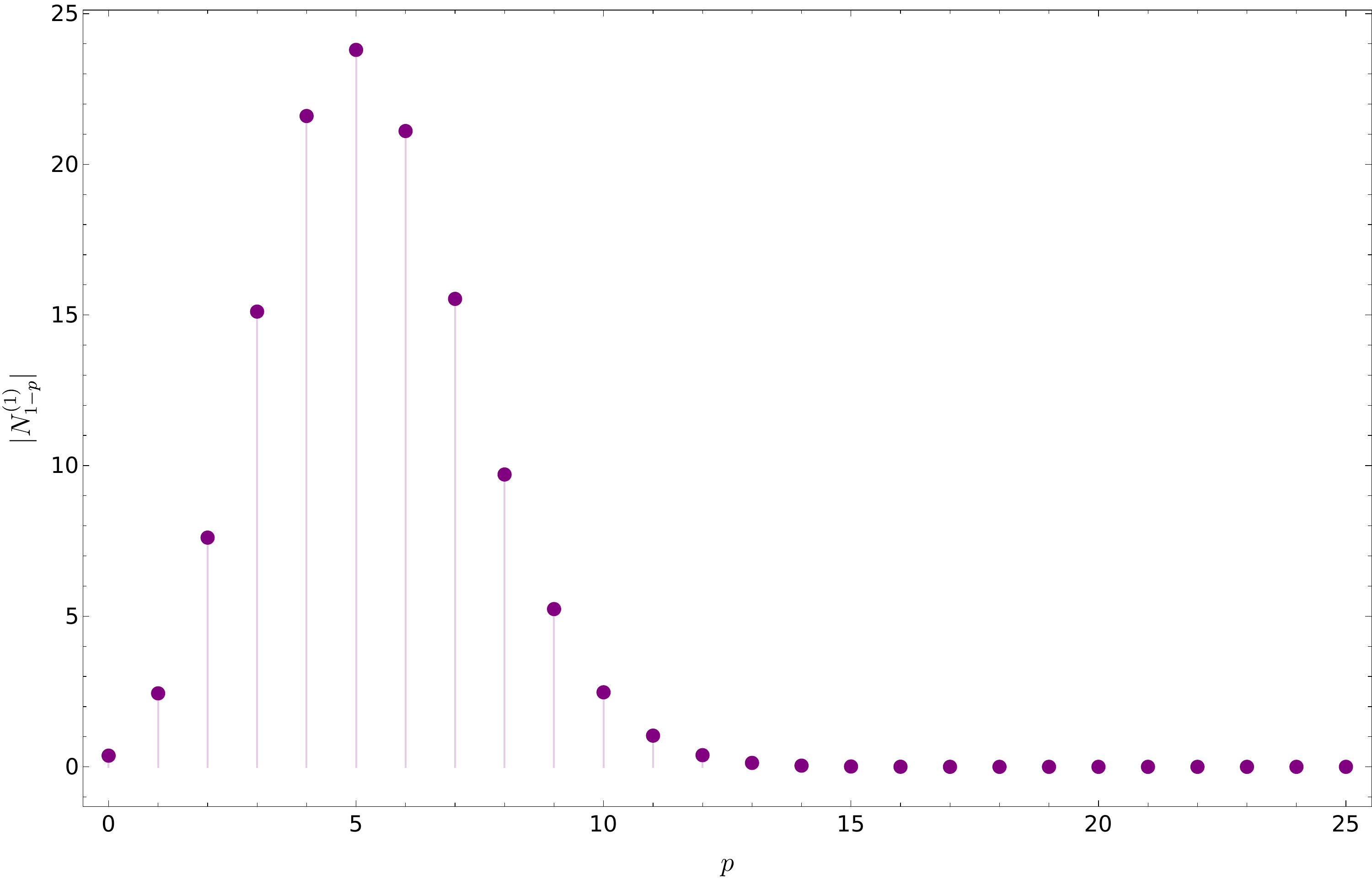}
		\caption{First diagonal $\ell=1$.}
	\end{subfigure}%
	~ 
	\begin{subfigure}[t]{0.5\textwidth}
		\centering
		\includegraphics[height=2in]{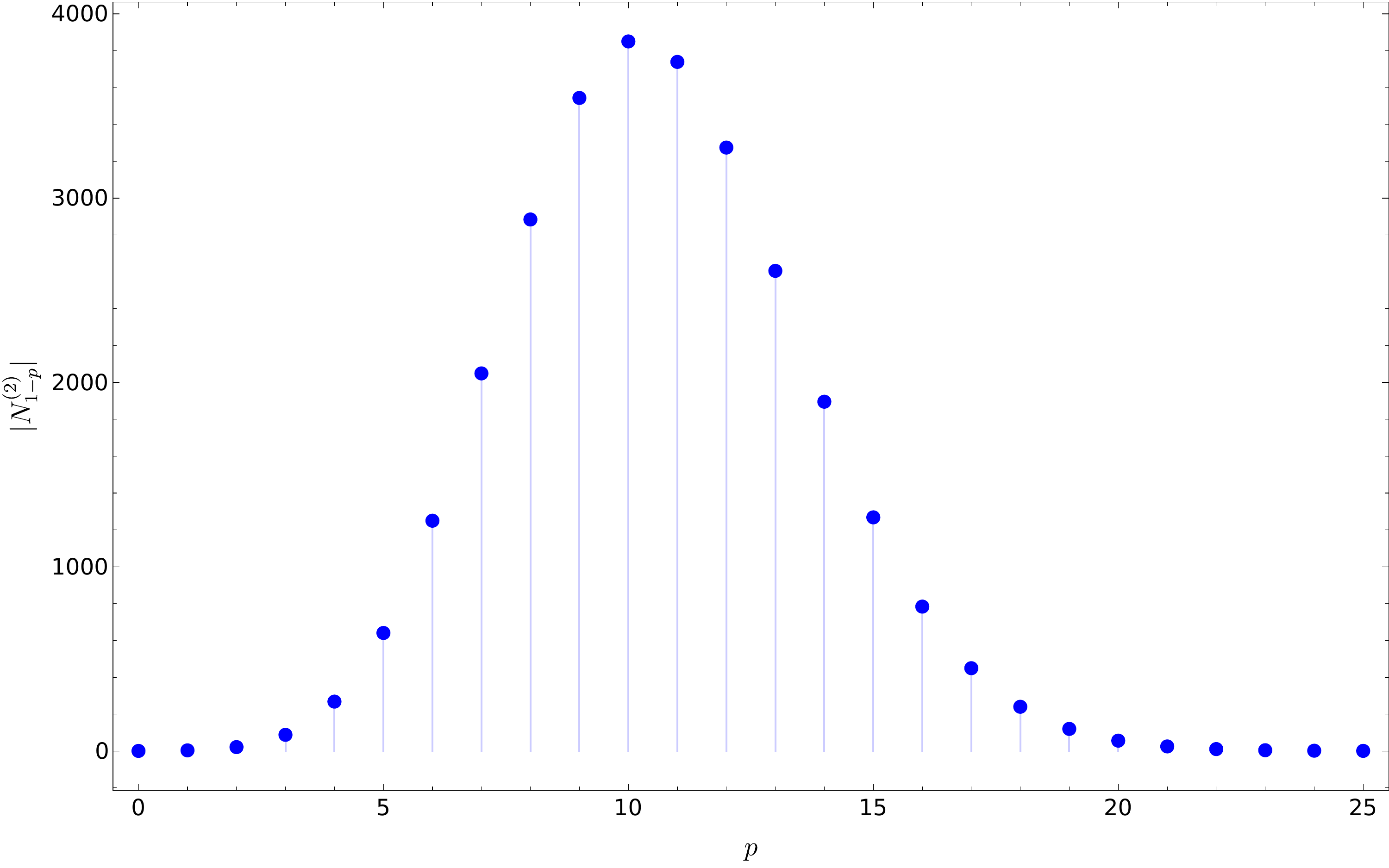}
		\caption{Second diagonal $\ell=2$.}
	\end{subfigure}
	\vspace{3mm}
	
	\begin{subfigure}[b]{0.5\textwidth}
		\centering
		\includegraphics[height=2in]{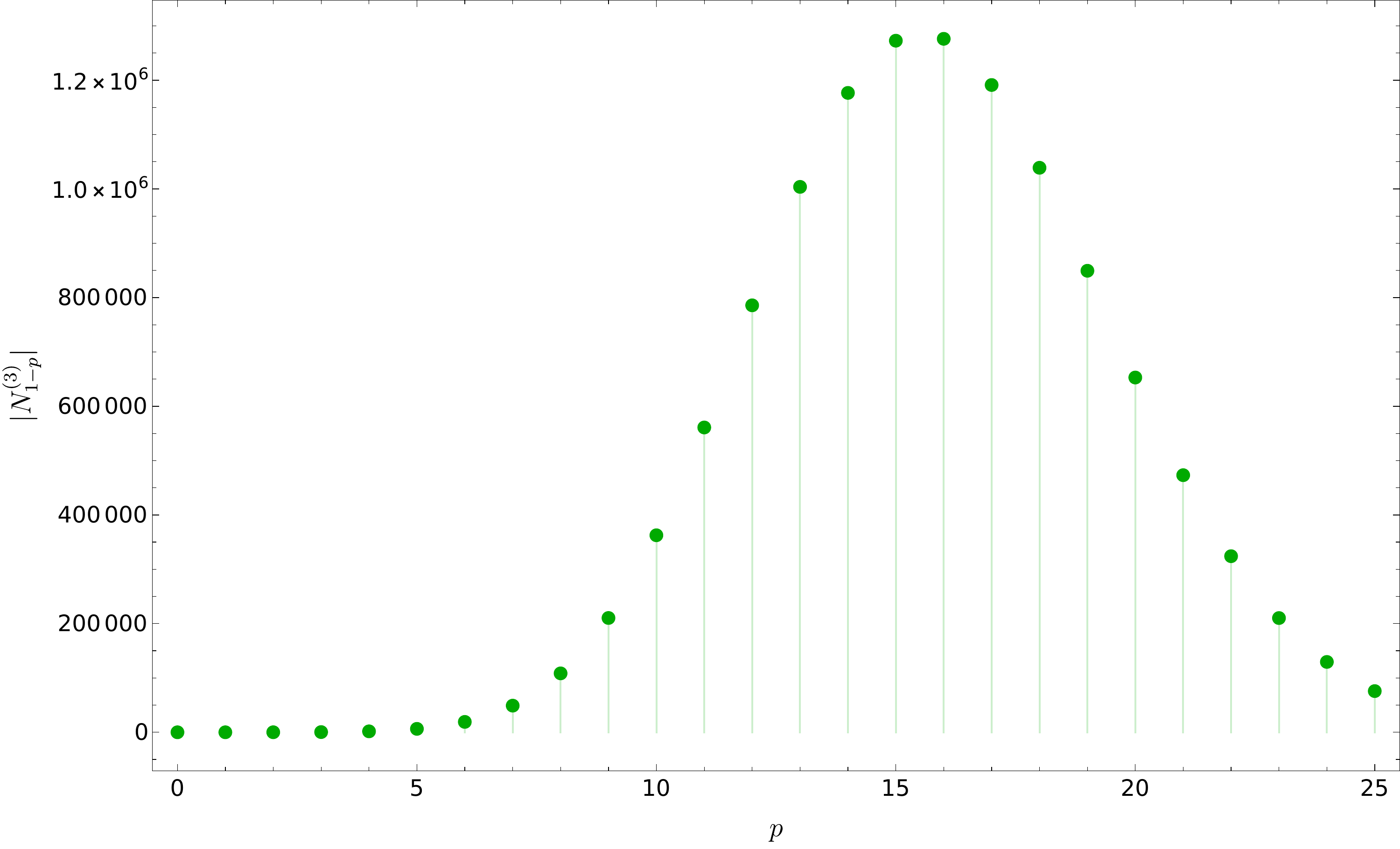}
		\caption{Third diagonal $\ell=3$.}
	\end{subfigure}
	\caption{Structure of the first few $N^{(\ell)}_{1-p}$ (in absolute value), for the first three diagonals.}
	\label{fig:Ns1mp}
\end{figure}

We have tested this expression thoroughly, which will be described in detail in section~\ref{sec:numerics}. In particular, setting
\be
\label{eq:Atildeell}
\widetilde{A}_\ell = 96 \sqrt{3}
\ee
\noindent
for \textit{all} $\ell$ and generating large sequences of the $N_{1-p}^{(\ell)}$ coefficients yields precise matches with our numerical results (at least up to very small numerical errors---see section~\ref{sec:numerics}). In particular, we have found that all $A_\ell$ are equal. As an example, the first coefficients of the first diagonal read\footnote{The ``single number'' we were missing to identify analytically in \cite{s19} was precisely the Euler--Mascheroni constant $\gamma_{\text{E}} \approx 0.5772156649...$, which immediately appears as we start moving down the first diagonal as in \eqref{eq:N10-examples}.}
\bea
\label{eq:N11-examples}
N^{(1)}_1 &=& - \rmi\, \frac{\sqrt[4]{3}}{2\sqrt{\pi}},\\
\label{eq:N10-examples}
N^{(1)}_0 &=& - \frac{2 \rmi}{\sqrt{3}}\, \frac{\sqrt[4]{3}}{2\sqrt{\pi}} \left( \gamma_{\text{E}} + \log \left( 96\sqrt{3} \right) \right), \\
\label{eq:N1-1-examples}
N^{(1)}_{-1} &=& - \frac{2 \rmi}{3}\, \frac{\sqrt[4]{3}}{2\sqrt{\pi}} \left( - \frac{\pi^2}{6} + \left( \gamma_{\text{E}} + \log \left(96\sqrt{3}\right) \right)^2 \right).
\eea
\noindent
In this way the \textit{complete} Stokes data is dictated by \textit{two} numbers: the ``canonical'' Stokes coefficient we already know, \eqref{eq:N11-examples}, and either, say, \eqref{eq:Atildeell} appearing in the generating function \eqref{eq:gen_fun-a}; or else, say, \eqref{eq:N10-examples} which explicitly appears in many of the patterns we shall illustrate below. Indeed, Stokes data displays many curious patterns. Some patterns are trivial, as the one along the Stokes lattice boundary in figure~\ref{fig:lattice_stokes_guessing_game} we already know
\be
\bm{S}_{(1,1-\ell)} = \frac{\rmi^{\ell-1}}{\ell} \left( N_1^{(1)} \right)^{2-\ell} \left[ \begin{array}{c} 1 \\ 0\end{array} \right].
\ee
\noindent
More intricate patterns may be found along each diagonal. Moving down each diagonal, the absolute values of the $N_{1-p}^{(\ell)}$ display a ``hump pattern'' as illustrated in figure~\ref{fig:Ns1mp} (for the first three diagonals $\ell=1,2,3$ and $p=0,1,\ldots,25$). The precise way in which these numbers actually behave displays even more interesting patterns, starting at the boundary of the Stokes lattice and moving deeper into the diagonal by the inclusion of more and more zeta-number combinations. Some such patterns are illustrated in figure~\ref{fig:patterns}, for both \PI~and \PII~cases (in fact, the only differences between these two cases are the values of $N^{(1)}_{1}$, $N_0^{(1)}$, and $\alpha$).

\begin{landscape}
\begin{figure}[hbt!]
	\centering
	\includegraphics[scale=0.5]{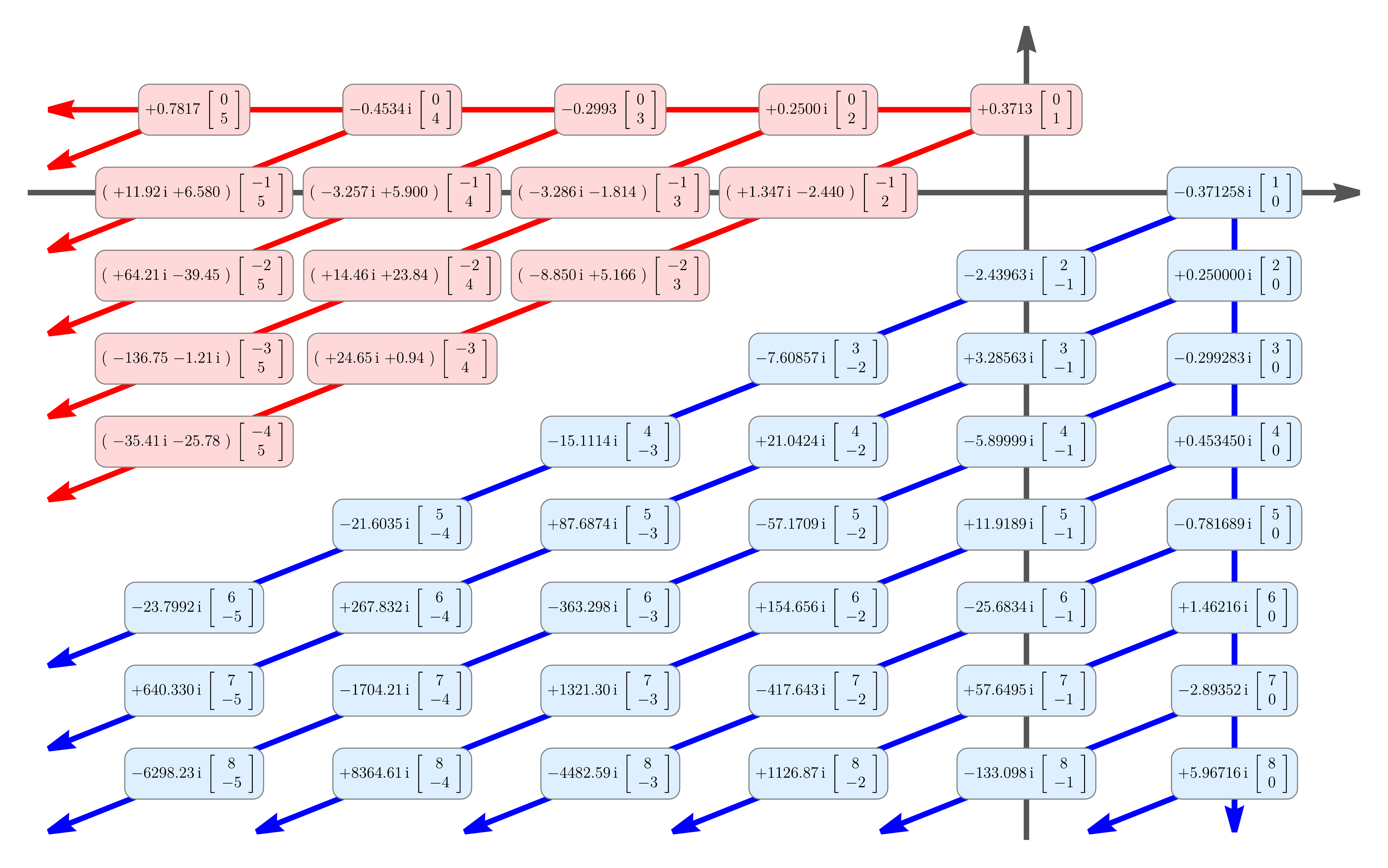}
	\caption{This plot illustrates (a subset of) Stokes data for \PI. All displayed numerical values are of course truncated, as they generically correspond to irrational quantities. Many of these numbers had been previously found in \cite{gikm10, asv11, as13}, and they are now all explainable with closed-form analytical expressions as discussed in the main text. They also encode many patterns and (at the end of the day, quite simple) structure, which is less evident from just staring at them. Comparing these numbers with the ones in \cite{asv17a, asv17b} we see that therein both lattice boundary and first diagonal were correct, but the ``bulk'' higher diagonals were not.}
	\label{fig:ComputedPainleveIStokesDataGridPatterns}
\end{figure}
\end{landscape}

Having obtained the complete forward Stokes data, the backward Stokes data immediately follows via the by-now familiar backward-forward symmetry. Start along the $\pm 1$ diagonals, where the relation between Stokes data \eqref{eq:alien-bridge-equation-NR} and Borel residues \eqref{eq:borel-sing-NR} is linear \cite{abs18}. Then the backward-forward equation \eqref{eq:back_forw} may be re-expressed at Stokes level quite simply,
\be
\label{eq:back_forw_explicit}
N^{(-1)}_{1-p} = \rmi \left(-1\right)^{p} \sum_{q=0}^{p} \left( - \frac{2\pi\rmi}{\sqrt{3}} \right)^{q} \frac{1}{q!}\, N^{(1)}_{1-(p-q)}.
\ee
\noindent
This expression may be again regarded as a Cauchy product,
\be
N^{(-1)}(\uplambda) =  \rmi \sum_{p=0}^{+\infty} \left(-\uplambda\right)^p \sum_{q=0}^p \left( - \frac{2\pi\rmi}{\sqrt{3}} \right)^{q} \frac{1}{q!}\, N^{(1)}_{1-(p-q)} = \rmi \sum_{p=0}^{+\infty} \frac{1}{p!}\, N^{(1)}_{1-p}\, \uplambda^p\, \sum_{q=0}^{+\infty} \frac{1}{q!} \left(\frac{2\pi\rmi}{\sqrt{3}}\, \uplambda \right)^q.
\ee
\noindent
The last expression yields a generating-function backward-forward relation
\be
\label{eq:back_forw_1}
N^{(-1)} (\uplambda) =  \rmi\, \rme^{\frac{2\pi\rmi}{\sqrt{3}}\, \uplambda}\, N^{(1)} (-\uplambda).
\ee
\noindent
To compute the generating function $N^{(-\ell)}(\uplambda)$, we first write it as $N^{(-\ell)}(\uplambda) = N^{(-\ell)}_1\, R^{(-\ell)}(\uplambda)$ and then impose the same property of  the ``forward'' generating function $R^{(\ell)}(\uplambda) = \left( R^{(1)}(\uplambda) \right)^\ell$, \textit{i.e.}, we impose $R^{(-\ell)}(\uplambda) = \left( R^{(-1)}(\uplambda) \right)^{\ell}$ and use the backward-forward relations for both $N_1^{(-1)}$ and $R^{(-1)}(\uplambda)$ to obtain
\be
N^{(-\ell)} (\uplambda) = \frac{N_1^{(-\ell)}}{\left( N_1^{(-1)} \right)^{\ell}} \left( N^{(-1)}(\uplambda) \right)^\ell.
\ee
\noindent
We can then use \eqref{eq:back_forw_1} and the lattice-boundary relation $N_1^{(-\ell)} = - (-\rmi)^\ell\, N_1^{(\ell)}$ (a consequence of the conjecture in \cite{asv11}) to write
\be
N^{(-\ell)}(\uplambda) = \frac{-(-\rmi)^\ell\, N_1^{(\ell)}}{\rmi^{\ell}  \left(N_1^{(1)}\right)^\ell} \left( \rmi\, \rme^{\frac{2\pi\rmi}{\sqrt{3}}\, \uplambda}\, N^{(1)}(-\uplambda) \right)^\ell.
\ee
\noindent
After simplification, the final generating-function backward-forward relation gets written as
\be
N^{(-\ell)}(\uplambda) = - (-\rmi)^\ell\, \rme^{\frac{2\pi\rmi}{\sqrt{3}}\, \uplambda\ell}\, N^{(\ell)}(-\uplambda).
\ee
\noindent
Again, we have tested this expression against our numerical results, and more details will be discussed in the upcoming section~\ref{sec:numerics}. We illustrate all these results for the Stokes-vectors lattice in figure~\ref{fig:ComputedPainleveIStokesDataGridPatterns}. These bare numerical values obviously hide the large number of patterns and structure we have just uncovered, but they are now \textit{all} analytically known---finally explaining all the data which had been previously found in \cite{gikm10, asv11, sv13, as13}.

\begin{landscape}
	\begin{figure}
	\begin{align*}
	N_{-1}^{(\textcolor{teal}{1})}=&\textcolor{violet}{\frac{1}{2!}\left(\frac{N^{(\textcolor{teal}{1})}_{0}}{N_1^{(\textcolor{teal}{1})}}\right)^1} \textcolor{blue}{N^{(\textcolor{teal}{1})}_{0}} {\color{violet}-}{\color{violet}\frac{1}{0!}\left(\frac{N^{(\textcolor{teal}{1})}_{0}}{N_1^{(\textcolor{teal}{1})}}\right)^0}\textcolor{blue}{\frac{\textcolor{teal}{1}}{2}\left(\frac{\alpha}{2}\right)^2 N^{(\textcolor{teal}{1})}_1{\color{red}\zeta(2)}},\\
	N_{-2}^{(\textcolor{teal}{1})}=&\textcolor{violet}{\frac{1}{3!}\left(\frac{N^{(\textcolor{teal}{1})}_{0}}{N_1^{(\textcolor{teal}{1})}}\right)^2}  \textcolor{blue}{ N^{(\textcolor{teal}{1})}_{0}} \textcolor{violet}{\color{violet}-}{\color{violet}\frac{1}{1!}\left(\frac{N^{(\textcolor{teal}{1})}_{0}}{N_1^{(\textcolor{teal}{1})}}\right)^1} \textcolor{blue}{\frac {\textcolor{teal}{1}}2\left(\frac{\alpha}{2}\right)^2 N^{(\textcolor{teal}{1})}_1{\color{red}\zeta(2)}}  \textcolor{violet}{+\left(\frac{N^{(\textcolor{teal}{1})}_{0}}{N_1^{(\textcolor{teal}{1})}}\right)^0}\textcolor{blue}{\frac {\textcolor{teal}{1}}3\left(\frac{\alpha}{2}\right)^3N^{(\textcolor{teal}{1})}_1}{\color{red}\zeta(3)},\\
	N_{-3}^{(\textcolor{teal}{1})}=&\textcolor{violet}{\frac{1}{4!}\left(\frac{N^{(\textcolor{teal}{1})}_{0}}{N_1^{(\textcolor{teal}{1})}}\right)^3}  \textcolor{blue}{N^{(\textcolor{teal}{1})}_{0}}  {\color{violet}-}{\color{violet}\frac{1}{2!}}{\color{violet}\left(\frac{N^{(\textcolor{teal}{1})}_{0}}{N_1^{(\textcolor{teal}{1})}}\right)^2}  \textcolor{blue}{\frac {\textcolor{teal}{1}}2\left(\frac{\alpha}{2}\right)^2 N^{(\textcolor{teal}{1})}_1{\color{red}\zeta(2)}}   +\textcolor{violet}{\left(\frac{N^{(\textcolor{teal}{1})}_{0}}{N_1^{(\textcolor{teal}{1})}}\right)^1} \textcolor{blue}{\frac {\textcolor{teal}{1}}3\left(\frac{\alpha}{2}\right)^3 N^{(\textcolor{teal}{1})}_1}{\color{red}\zeta(3)}{\color{violet}-}{\color{violet}\frac{1}{0!}\left(\frac{N^{(\textcolor{teal}{1})}_{0}}{N_1^{(\textcolor{teal}{1})}}\right)^0}{\color{blue}\frac {\textcolor{teal}{1}}4\left(\frac{\alpha}{2}\right)^4 N^{(\textcolor{teal}{1})}_1{\color{red}\left(\zeta(4)-\frac{\textcolor{teal}{1}}{2}\zeta(2)^2\right)}}.\\
	&\\
	N_{-1}^{(\textcolor{teal}{2})}=&\textcolor{violet}{\frac{1}{2!}\left(\frac{N^{(\textcolor{teal}{2})}_{0}}{N_1^{(\textcolor{teal}{2})}}\right)^1} \textcolor{blue}{N^{(\textcolor{teal}{2})}_{0}} {\color{violet}-}{\color{violet}\frac{1}{0!}\left(\frac{N^{(\textcolor{teal}{2})}_{0}}{N_1^{(\textcolor{teal}{2})}}\right)^0}\textcolor{blue}{\frac {\textcolor{teal}{2}}2\left(\frac{\alpha}{2}\right)^2 N^{(\textcolor{teal}{2})}_1{\color{red}\zeta(2)}},\\
	N_{-2}^{(\textcolor{teal}{2})}=&\textcolor{violet}{\frac{1}{3!}\left(\frac{N^{(\textcolor{teal}{2})}_{0}}{N_1^{(\textcolor{teal}{2})}}\right)^2}  \textcolor{blue}{ N^{(\textcolor{teal}{2})}_{0}} \textcolor{violet}{\color{violet}-}{\color{violet}\frac{1}{1!}\left(\frac{N^{(\textcolor{teal}{2})}_{0}}{N_1^{(\textcolor{teal}{2})}}\right)^1} \textcolor{blue}{\frac {\textcolor{teal}{2}}2\left(\frac{\alpha}{2}\right)^2 N^{(\textcolor{teal}{2})}_1{\color{red}\zeta(2)}}  \textcolor{violet}{+\left(\frac{N^{(\textcolor{teal}{2})}_{0}}{N_1^{(\textcolor{teal}{2})}}\right)^0}\textcolor{blue}{\frac {\textcolor{teal}{2}}3\left(\frac{\alpha}{2}\right)^3N^{(\textcolor{teal}{2})}_1}{\color{red}\zeta(3)},\\
	N_{-3}^{(\textcolor{teal}{2})}=&\textcolor{violet}{\frac{1}{4!}\left(\frac{N^{(\textcolor{teal}{2})}_{0}}{N_1^{(\textcolor{teal}{2})}}\right)^3}  \textcolor{blue}{N^{(\textcolor{teal}{2})}_{0}}  {\color{violet}-}{\color{violet}\frac{1}{2!}}{\color{violet}\left(\frac{N^{(\textcolor{teal}{2})}_{0}}{N_1^{(\textcolor{teal}{2})}}\right)^2}  \textcolor{blue}{\frac {\textcolor{teal}{2}}2\left(\frac{\alpha}{2}\right)^2 N^{(\textcolor{teal}{2})}_1{\color{red}\zeta(2)}}   +\textcolor{violet}{\left(\frac{N^{(\textcolor{teal}{2})}_{0}}{N_1^{(\textcolor{teal}{2})}}\right)^1} \textcolor{blue}{\frac {\textcolor{teal}{2}}3\left(\frac{\alpha}{2}\right)^3 N^{(\textcolor{teal}{2})}_1}{\color{red}\zeta(3)}{\color{violet}-}{\color{violet}\frac{1}{0!}\left(\frac{N^{(\textcolor{teal}{2})}_{0}}{N_1^{(\textcolor{teal}{2})}}\right)^0}{\color{blue}\frac {\textcolor{teal}{2}}4\left(\frac{\alpha}{2}\right)^4 N^{(\textcolor{teal}{2})}_1{\color{red}\left(\zeta(4)-\frac{\textcolor{teal}{2}}{2}\zeta(2)^2\right)}}.\\
	&\\
	N_{-1}^{(\textcolor{teal}{3})}=&\textcolor{violet}{\frac{1}{2!}\left(\frac{N^{(\textcolor{teal}{3})}_{0}}{N_1^{(\textcolor{teal}{3})}}\right)^1} \textcolor{blue}{N^{(\textcolor{teal}{3})}_{0}} {\color{violet}-}{\color{violet}\frac{1}{0!}\left(\frac{N^{(\textcolor{teal}{3})}_{0}}{N_1^{(\textcolor{teal}{3})}}\right)^0}\textcolor{blue}{\frac {\textcolor{teal}{3}}2\left(\frac{\alpha}{2}\right)^2 N^{(\textcolor{teal}{3})}_1{\color{red}\zeta(2)}},\\
	N_{-2}^{(\textcolor{teal}{3})}=&\textcolor{violet}{\frac{1}{3!}\left(\frac{N^{(\textcolor{teal}{3})}_{0}}{N_1^{(\textcolor{teal}{3})}}\right)^2}  \textcolor{blue}{ N^{(\textcolor{teal}{3})}_{0}} \textcolor{violet}{\color{violet}-}{\color{violet}\frac{1}{1!}\left(\frac{N^{(\textcolor{teal}{3})}_{0}}{N_1^{(\textcolor{teal}{3})}}\right)^1} \textcolor{blue}{\frac {\textcolor{teal}{3}}2\left(\frac{\alpha}{2}\right)^2 N^{(\textcolor{teal}{3})}_1{\color{red}\zeta(2)}}  \textcolor{violet}{+\left(\frac{N^{(\textcolor{teal}{3})}_{0}}{N_1^{(\textcolor{teal}{3})}}\right)^0}\textcolor{blue}{\frac {\textcolor{teal}{3}}3\left(\frac{\alpha}{2}\right)^3N^{(\textcolor{teal}{3})}_1}{\color{red}\zeta(3)},\\
	N_{-3}^{(\textcolor{teal}{3})}=&\textcolor{violet}{\frac{1}{4!}\left(\frac{N^{(\textcolor{teal}{3})}_{0}}{N_1^{(\textcolor{teal}{3})}}\right)^3}  \textcolor{blue}{N^{(\textcolor{teal}{3})}_{0}}  {\color{violet}-}{\color{violet}\frac{1}{2!}}{\color{violet}\left(\frac{N^{(\textcolor{teal}{3})}_{0}}{N_1^{(\textcolor{teal}{3})}}\right)^2}  \textcolor{blue}{\frac {\textcolor{teal}{3}}2\left(\frac{\alpha}{2}\right)^2 N^{(\textcolor{teal}{3})}_1{\color{red}\zeta(2)}}   +\textcolor{violet}{\left(\frac{N^{(\textcolor{teal}{3})}_{0}}{N_1^{(\textcolor{teal}{3})}}\right)^1} \textcolor{blue}{\frac {\textcolor{teal}{3}}3\left(\frac{\alpha}{2}\right)^3 N^{(\textcolor{teal}{3})}_1}{\color{red}\zeta(3)}{\color{violet}-}{\color{violet}\frac{1}{0!}\left(\frac{N^{(\textcolor{teal}{3})}_{0}}{N_1^{(\textcolor{teal}{3})}}\right)^0}{\color{blue}\frac {\textcolor{teal}{3}}4\left(\frac{\alpha}{2}\right)^4 N^{(\textcolor{teal}{3})}_1{\color{red}\left(\zeta(4)-\frac{\textcolor{teal}{3}}{2}\zeta(2)^2\right)}}.\\
	\end{align*}
	\caption{Explicit results going down the first three diagonals, $\ell=1,2,3$. These expressions hold for the Stokes data associated to both \PI~and \PII, where of course the values of $N^{(1)}_{1}$, $N_0^{(1)}$, and $\alpha$ differ for each case. We have used a color coding of the formulae in order to highlight key aspects of these patterns; \textit{e.g.}, in red we note the increasing amount of zeta-number combinations.}
	\label{fig:patterns}
	\end{figure}
\end{landscape}

The reader may now recall the generic discussion on bridge equations and Stokes transitions in subsection~\ref{subsec:stokes-trans}; in particular the (directional, pointed) bridge equation \eqref{eq:bridge-directional} and its associated (directional) Stokes vector field \eqref{eq:def_pointed} (where the flow of this vector field generated Stokes transitions as in, \textit{e.g.}, \eqref{eq:Stokes-flow-simplified}). One remark we had back then was that the Stokes vector field $\underline{\boldsymbol{S}}_{\theta} (\boldsymbol{\sigma})$ may also be regarded as a generating function for Stokes data. This may now be made precise, in light of our generating functions \eqref{eq:gen_fun-a}-\eqref{eq:gen_fun-b} for Stokes vectors \eqref{eq:struct-a}-\eqref{eq:struct-b}.

Start with the Stokes vector field $\underline{\boldsymbol{S}}_{\theta} (\boldsymbol{\sigma})$ \eqref{eq:def_pointed} which we rewrite herein,
\be
\label{eq:vec-fields}
\underline{\boldsymbol{S}}_{\theta} (\boldsymbol{\sigma}) = \sum_{\bm{\ell} \in \left\{ \bm{\ell}_{\theta} \right\}} \sum_{\bm{p} \in \ker \mathfrak{P}} \boldsymbol{\sigma}^{-\bm{\ell}-\bm{p}} \left[ \begin{array}{c} \sigma_1\, \bm{S}_{\bm{\ell}+\bm{p}}^{(1)} \\ \sigma_2\, \bm{S}_{\bm{\ell}+\bm{p}}^{(2)} \end{array} \right].
\ee
\noindent
What is left to do is to explicitly evaluate these sums---which may now be done in light of the data we have previously computed. First focus on the $\theta_{\text{S}} = 0$ Stokes line and insert the \eqref{eq:struct-a}-\eqref{eq:struct-b} vectorial structure in the expression above (after a shift in the $\bm{p}$-summation). The equation becomes
\be
\underline{\boldsymbol{S}}_{0} (\boldsymbol{\sigma}) = \sum_{\ell=1}^{+\infty} \sum_{p=0}^{+\infty} \sigma_1^{p-1} \sigma_2^{p-1+\ell}\, N^{(\ell)}_{1-p}\,  \left[ \begin{array}{c} \sigma_1 \left( p+\ell \right) \\ \sigma_2 \left( -p \right) \end{array} \right].
\ee
\noindent
The sums over the kernel are now straightforward. Denoting\footnote{The reader may also recall this variable has already appeared in diagonal-framing transseries \eqref{eq:PI/II-2PTS-diagonal}.} $\upmu \equiv \sigma_1 \sigma_2$, these are
\bea
\sum_{p=0}^{+\infty} \upmu^{p}\, N^{(\ell)}_{1-p} &=& N^{(\ell)} (\upmu), \\
\sum_{p=0}^{+\infty} p\, \upmu^{p}\, N^{(\ell)}_{1-p} &=& \upmu \frac{\rmd}{\rmd \upmu} N^{(\ell)} (\upmu),
\eea
\noindent
which we could sum immediately given our generating function of \PI~Stokes data \eqref{eq:gen_fun-a}. The Stokes vector field $\underline{\boldsymbol{S}}_{0} (\boldsymbol{\sigma})$ is hence rewritten as
\be
\underline{\boldsymbol{S}}_{0} (\boldsymbol{\sigma}) = \sum_{\ell=1}^{+\infty} \left[ \begin{array}{c} \sigma_2^{\ell-1} \left( \upmu \frac{\rmd}{\rmd \upmu} + \ell \right) N^{(\ell)} (\upmu) \\ - \sigma_2^{\ell+1}\, \frac{\rmd}{\rmd \upmu} N^{(\ell)} (\upmu) \end{array} \right].
\ee
\noindent
The final sum over all diagonals is also easily implemented. Observing that---via the above definitions and using a property inherited from the Stokes ratios---one has
\bea
N^{(\ell)} (\upmu) &=& - \frac{\rmi}{\ell^2} \left( N_1^{(1)} \right)^{2} \left( \rmi\, \frac{N^{(1)} (\upmu)}{\left( N_1^{(1)} \right)^2} \right)^{\ell} \\
&\Rightarrow& \qquad \frac{\rmd}{\rmd \upmu} N^{(\ell)} (\upmu) = - \frac{\rmi}{\ell} \left( N_1^{(1)} \right)^{2} \left( \rmi\, \frac{N^{(1)} (\upmu)}{\left( N_1^{(1)} \right)^2} \right)^{\ell} \frac{\rmd}{\rmd \upmu} \log N^{(1)} (\upmu),
\eea
\noindent
the final sums are evaluated to
\be
\label{eq:PIS0-final}
\underline{\boldsymbol{S}}_{0} (\boldsymbol{\sigma}) = \rmi \left( N_1^{(1)} \right)^{2} \left[ \begin{array}{c} \sigma_2^{-1} \left( 1 + \upmu \frac{\rmd}{\rmd \upmu} \log N^{(1)} (\upmu) \right) \\ - \sigma_2\, \frac{\rmd}{\rmd \upmu} \log N^{(1)} (\upmu) \end{array} \right] \log \left( 1 - \rmi \sigma_2\, \frac{N^{(1)} (\upmu)}{\left( N_1^{(1)} \right)^2} \right).
\ee
\noindent
Note how all the information is now solely encoded by the ``canonical'' Stokes coefficient, $N_1^{(1)}$, alongside the generating-function of the first diagonal, $N^{(1)} (\upmu)$. It is also interesting to note that this result has very explicit dependence on $\upmu = \sigma_1 \sigma_2$---which partially recovers the diagonal framing flavor of subsection~\ref{subsec:reson-frame}. As a consistency check on this expression, we may evaluate it when $\sigma_2=0$ (there is obviously no singular behavior). It follows
\be
\label{eq:PIS0-simple-1}
\underline{\boldsymbol{S}}_{0} \left(\sigma_1,0\right) = N_1^{(1)} \left[ \begin{array}{c} 1 \\ 0 \end{array} \right],
\ee
\noindent
which is exactly what one would expect in light of the results discussed in subsection~\ref{subsec:stokes-trans}. Likewise, albeit less trivial, the $\underline{\boldsymbol{S}}_{0} (\boldsymbol{\sigma})$ Stokes vector field may also be evaluated at $\sigma_1=0$. It now follows
\be
\label{eq:PIS0-simple-2}
\underline{\boldsymbol{S}}_{0} \left(0,\sigma_2\right) = \rmi N_1^{(1)} \left[ \begin{array}{c} \sigma_2^{-1}\, N_1^{(1)} \\ - \sigma_2\, N_0^{(1)} \end{array} \right] \log \left( 1 - \frac{\rmi \sigma_2}{N_1^{(1)}} \right).
\ee

Focusing on the $\theta_{\text{S}}=\pi$ Stokes line instead, a completely analogous procedure yields
\be
\label{eq:PISpi-final}
\underline{\boldsymbol{S}}_{\pi} (\boldsymbol{\sigma}) = \rmi \left( N_1^{(-1)} \right)^{2} \left[ \begin{array}{c} - \sigma_1\, \frac{\rmd}{\rmd \upmu} \log N^{(-1)} (\upmu) \\ \sigma_1^{-1} \left( 1 + \upmu \frac{\rmd}{\rmd \upmu} \log N^{(-1)} (\upmu) \right) \end{array} \right] \log \left( 1 - \rmi \sigma_1\, \frac{N^{(-1)} (\upmu)}{\left( N_1^{(-1)} \right)^2} \right).
\ee
\noindent
At either $\sigma_1=0$ or $\sigma_2=0$, this vector field simplifies as
\bea
\label{eq:PISpi-simple-2}
\underline{\boldsymbol{S}}_{\pi} \left(0,\sigma_2\right) &=& N_1^{(-1)} \left[ \begin{array}{c} 0 \\ 1 \end{array} \right], \\
\label{eq:PISpi-simple-1}
\underline{\boldsymbol{S}}_{\pi} \left(\sigma_1,0\right) &=& \rmi N_1^{(-1)} \left[ \begin{array}{c} - \sigma_1\, N_0^{(-1)} \\ \sigma_1^{-1}\, N_1^{(-1)} \end{array} \right] \log \left( 1 - \frac{\rmi \sigma_1}{N_1^{(-1)}} \right).
\eea

\subsection{Global Structure of 2D Quantum Gravity and Minimal Strings}\label{subsec:2DQGresults}

As discussed in subsection~\ref{subsec:PI2dQG}, \PI~solutions describe the specific-heat of the simplest ($k=2$ or $c=0$) multicritical model. Its free energy is given by \eqref{eq:PI-F/Z}, which we repeat in here
\be
F_{\text{I}}'' (z) = - u_{\text{I}} (z).
\ee
\noindent
This integration is performed formally, which means that out of the $u_{\text{I}} \left(z;\sigma_1,\sigma_2\right)$ transseries, in monomials $z^{-\frac{5}{4}}$ and transmonomials $\rme^{\pm z^{\frac{5}{4}}}$, one will obtain a $F_{\text{I}} \left(z;\sigma_1,\sigma_2\right)$ transseries in the same (trans)monomials---albeit it is also standard to use the string-coupling $g_{\text{s}} = z^{-\frac{5}{4}}$. This was explicitly carried through with many data in \cite{asv11}, and some brief examples were collected throughout section~\ref{sec:review}. What this further implies is that out of specific-heat Stokes data we will hence obtain \textit{free-energy Stokes data}---allowing for a global description of 2d quantum gravity. 

One may repeat the same analyses as in the previous subsection, starting out with a few of the free-energy Stokes-data (numerically) obtained in \cite{asv11}. The vectorial structure of free-energy Stokes data $\boldsymbol{S}_{\boldsymbol{n}}^{F}$ is unchanged as compared to their specific-heat counterparts \eqref{eq:struct-a}-\eqref{eq:struct-b}, \textit{i.e.},
\bea
\bm{S}_{(1-p,1-p-\ell)}^F &=& N^{(\ell)F}_{1-p}\, \left[ \begin{array}{c}p+\ell \\ -p\end{array} \right], \\
\bm{S}_{(1-p-\ell,1-p)}^F &=& N^{(-\ell)F}_{1-p}\, \left[ \begin{array}{c}-p \\ p+\ell\end{array} \right].
\eea
\noindent
It is also physically convenient \cite{d92, msw07, asv11} to rescale the free-energy transseries parameters as
\be
\sigma_1^F = \frac{\sigma_1}{N_1^{(1)}}, \qquad \sigma_2^F = \frac{\sigma_2}{N_1^{(-1)}}.
\ee
\noindent
In order to obtain the Stokes data for the free energy, we first argue that the Stokes data for a transseries and the Stokes data for its derivative with respect to $z$ are equal. This comes from the fact that the $\underline{\dot\Delta}_\theta$ operator and the $-\partial_z^2$ operator commute \cite{e81}, so applying the alien derivative operator before or after the derivatives gives the same result (this is in contrast with the discussion in \cite{asv11}). Due to our scaling of the transseries parameters, the Stokes data undergo the transformation
\bea
\label{eq:trans_1}
N_{1-p}^{(\ell)F} &=& \left(N_1^{(1)}\right)^{p-1} \left(N_1^{(-1)}\right)^{p+\ell-1} N_{1-p}^{(\ell)}, \\
\label{eq:trans_2}
N_{1-p}^{(-\ell)F} &=& \left(N_1^{(1)}\right)^{p+\ell-1} \left(N_1^{(-1)}\right)^{p-1} N_{1-p}^{(-\ell)}.
\eea
\noindent
From these relations, the generating functions for free-energy Stokes data immediately follow given \eqref{eq:gen_fun-a}
\bea
\label{eq:trans_1-GF}
N^{(\ell)F} (\uplambda) &=& \frac{\left(N_1^{(-1)}\right)^\ell}{N_1^{(1)}N_1^{(-1)}}\, N^{(\ell)} \left( N_1^{(1)} N_1^{(-1)} \uplambda \right), \\
\label{eq:trans_2-GF}
N^{(-\ell)F} (\uplambda) &=& \frac{\left(N_1^{(1)}\right)^\ell}{N_1^{(1)}N_1^{(-1)}}\, N^{(-\ell)} \left( N_1^{(1)} N_1^{(-1)} \uplambda \right).
\eea
\noindent
The backward-forward relation for free energies is
\be
N^{(-\ell)F}(\uplambda) = (-1)^{\ell-1}\, \rme^{\frac{1}{2}\, \uplambda\ell}\, N^{(\ell)F}(-\uplambda).
\ee
\noindent
In particular, all Stokes-lattice boundary factors $N_1^{(\ell)F}$ and $N_1^{(-\ell)F}$ turn out to be rational:
\be
\label{eq:boundary}
N_1^{(\ell)F} = \frac{(-1)^{\ell+1}}{\ell^2}, \qquad N_1^{(-\ell)F} = \frac{1}{\ell^2}.
\ee
\noindent
Explicitly, the forward-data generating-function is now given by
\be
N^{(\ell)F} (\uplambda) = \frac{(-1)^{\ell+1}}{\ell^2}\, \frac{\widetilde{A}^{\frac{\alpha_{F}}{2}\, \uplambda \ell}}{\Gamma \left( 1 + \frac{\alpha_F}{2}\, \uplambda \right)^{\ell}},
\ee
\noindent
with $\alpha_F = N_1^{(1)}N_1^{(-1)} \alpha = \frac{4}{\sqrt{3}}\, N_1^{(1)} N_1^{(-1)} = -\frac{\rmi}{\pi}$ and $\widetilde{A}$ defined in the previous subsection \eqref{eq:Atildeell}. Obtaining the corresponding backward-data generating-function is a straightforward exercise.

\begin{figure}
	\centering
	\begin{subfigure}[t]{0.5\textwidth}
		\centering
		\includegraphics[height=2in]{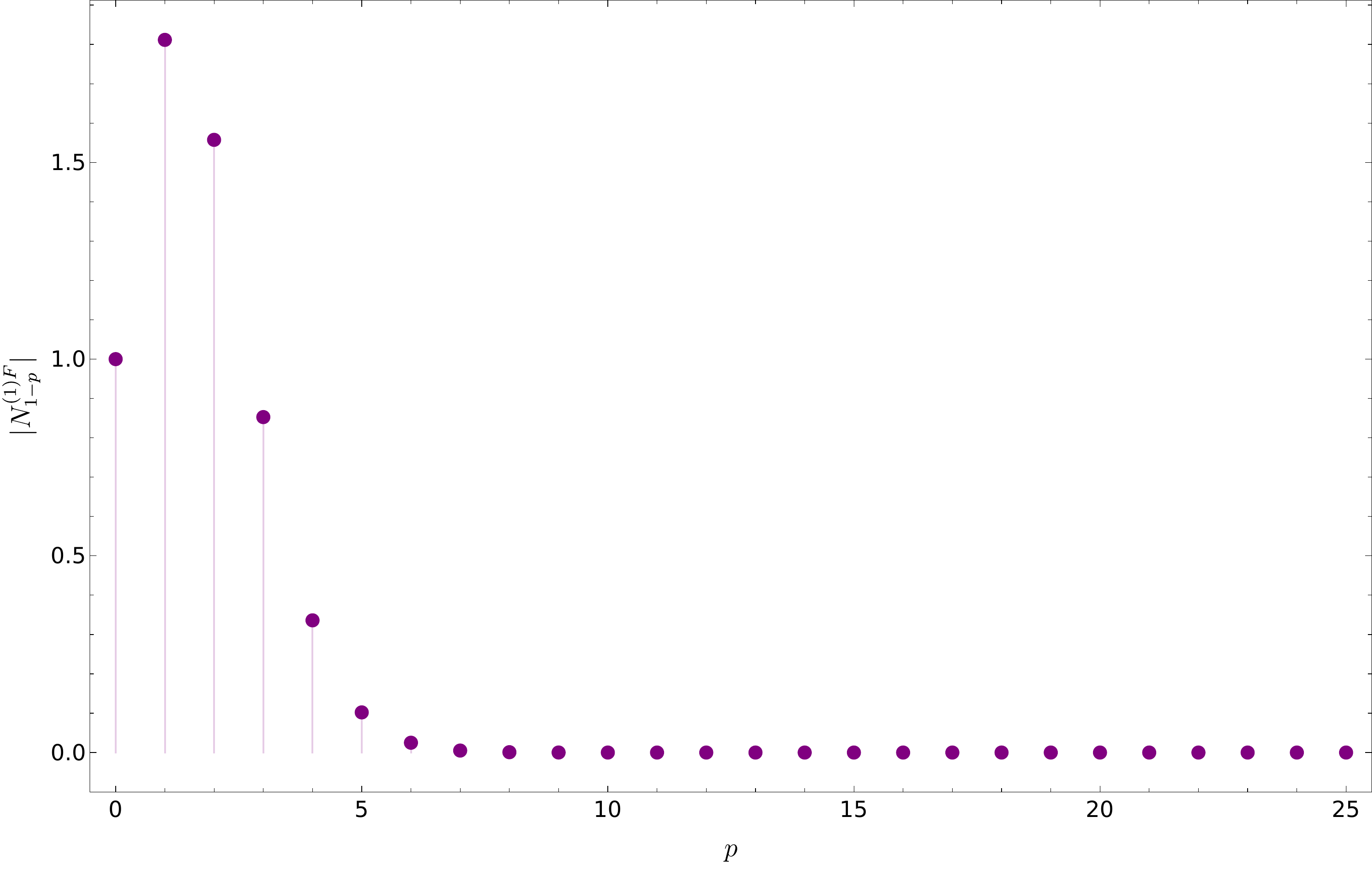}
		\caption{First diagonal $\ell=1$.}
	\end{subfigure}%
	~ 
	\begin{subfigure}[t]{0.5\textwidth}
		\centering
		\includegraphics[height=2in]{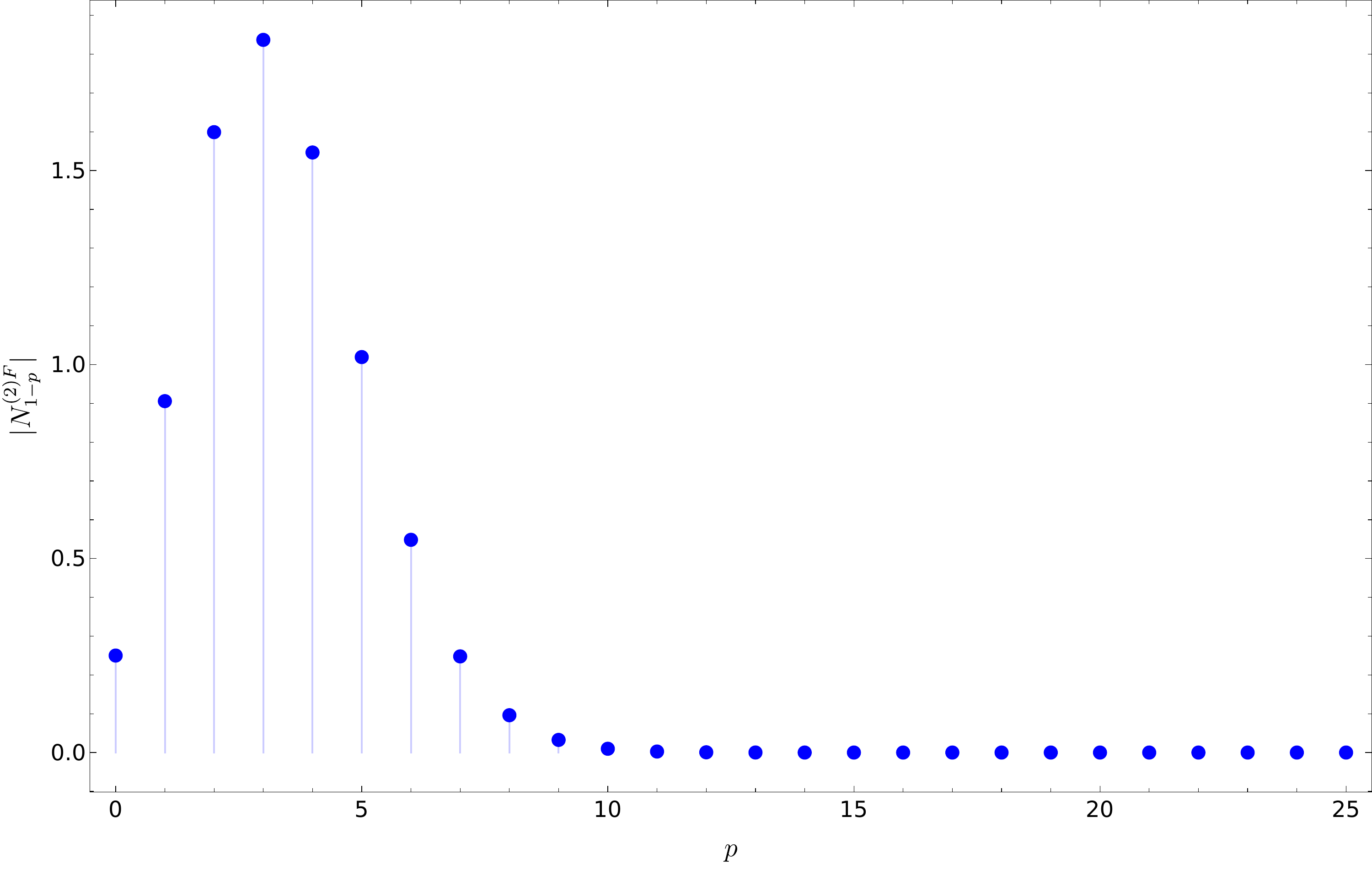}
		\caption{Second diagonal $\ell=2$.}
	\end{subfigure}
	\vspace{3mm}
	
	\begin{subfigure}[b]{0.5\textwidth}
		\centering
		\includegraphics[height=2in]{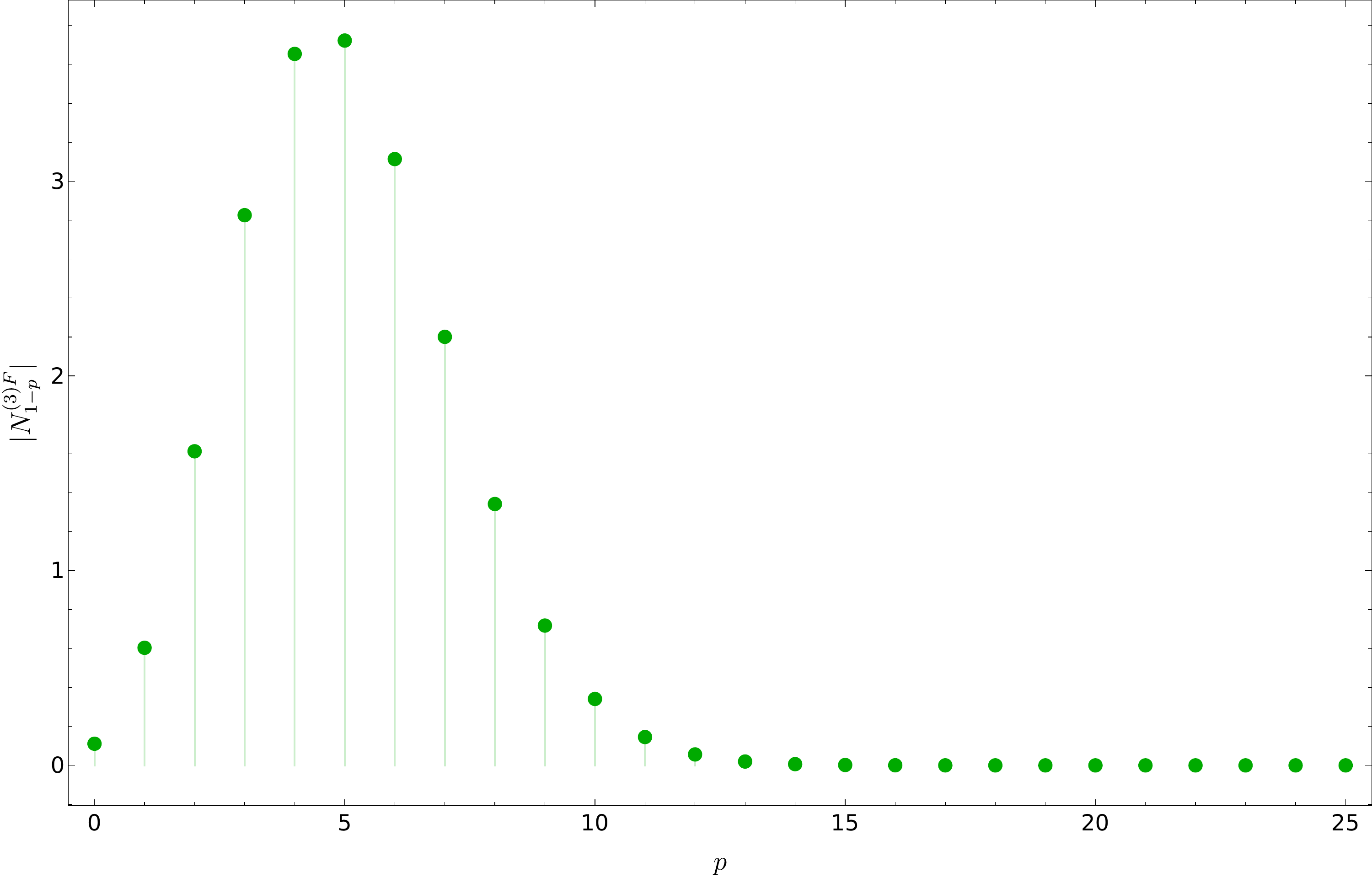}
		\caption{Third diagonal $\ell=3$.}
	\end{subfigure}
	\caption{Structure of the first few $N^{(\ell)F}_{1-p}$ (in absolute value), for the first three diagonals.}
	\label{fig:Ns1mpPIFE}
\end{figure}

\begin{landscape}
\begin{figure}
	\centering
	\includegraphics[scale=0.5]{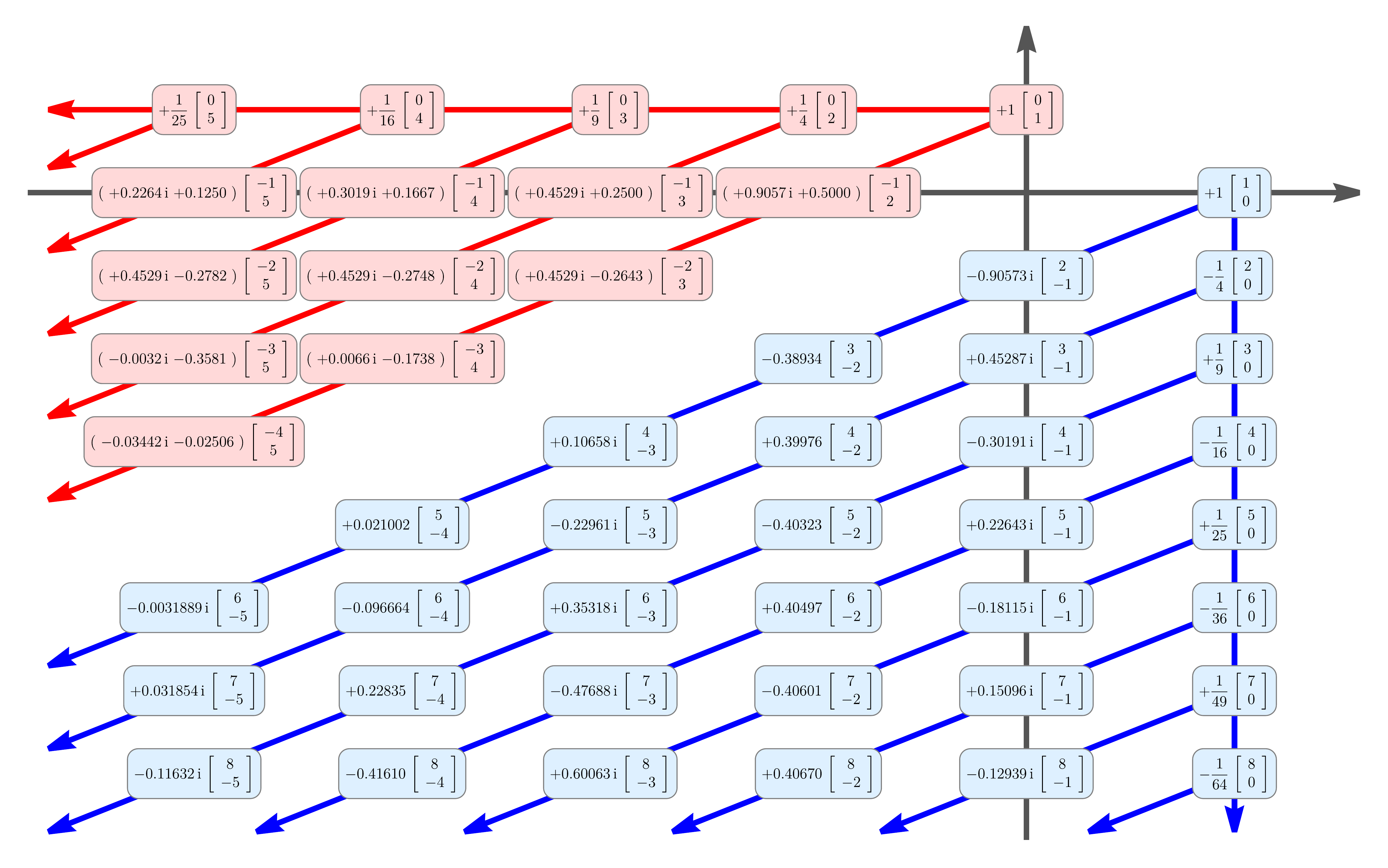}
	\caption{A subset of Stokes data for the free energy of the $k=2$ multicritical model. Similar to \PI, many numerical values are of course truncated irrational quantities. Yet, all these numbers are now fully understood with closed-form analytical expressions (see the main text). Again, these numbers encode many patterns, essentially inherited from those of \PI.}
	\label{fig:ComputedPainleveIStokesDataGridPatternsFreeEnergy}
\end{figure}
\end{landscape}

Just like for the specific-heat, also now Stokes data display the usual interesting patterns. There is a trivial pattern along the Stokes lattice boundary,
\be
\bm{S}_{(1,1-\ell)}^F = \frac{(-1)^{\ell+1}}{\ell}\, \left[ \begin{array}{c} 1 \\ 0 \end{array} \right].
\ee
\noindent
Then, moving down each diagonal, there are the ``hump patterns'' for the absolute values of the $N_{1-p}^{(\ell)F}$. These ``humps'' now occur earlier-on as compared to specific-heat data, as illustrated in figure~\ref{fig:Ns1mpPIFE} (for the first three diagonals $\ell=1,2,3$ and $p=0,1,\ldots,25$). Again, the precise way in which these numbers actually behave displays remarkable patterns---still seeing the general appearance of zeta-number combinations; compare figure~\ref{fig:patterns} with figure~\ref{fig:FEpatternsPI}. Bare numerical values for the free-energy Stokes-vectors lattice are finally illustrated in figure~\ref{fig:ComputedPainleveIStokesDataGridPatternsFreeEnergy}. 

\begin{figure}
	\begin{align*}
	N_{-1}^{(\textcolor{teal}{\ell})F}=&\textcolor{violet}{\frac{1}{2!}\left(\frac{N^{(\textcolor{teal}{\ell})F}_{0}}{N_1^{(\textcolor{teal}{\ell})F}}\right)^1} \textcolor{blue}{N^{(\textcolor{teal}{\ell})F}_{0}} {\color{violet}-}{\color{violet}\frac{1}{0!}\left(\frac{N^{(\textcolor{teal}{\ell})F}_{0}}{N_1^{(\textcolor{teal}{\ell})F}}\right)^0}\textcolor{blue}{\frac {\textcolor{teal}{\ell}}2\left(\frac{\alpha_F}{2}\right)^2 N^{(\textcolor{teal}{\ell})F}_1{\color{red}\zeta(2)}},\\
	N_{-2}^{(\textcolor{teal}{\ell})F}=&\textcolor{violet}{\frac{1}{3!}\left(\frac{N^{(\textcolor{teal}{\ell})F}_{0}}{N_1^{(\textcolor{teal}{\ell})F}}\right)^2}  \textcolor{blue}{ N^{(\textcolor{teal}{\ell})F}_{0}} \textcolor{violet}{\color{violet}-}{\color{violet}\frac{1}{1!}\left(\frac{N^{(\textcolor{teal}{\ell})F}_{0}}{N_1^{(\textcolor{teal}{\ell})F}}\right)^1} \textcolor{blue}{\frac {\textcolor{teal}{\ell}}2\left(\frac{\alpha_F}{2}\right)^2 N^{(\textcolor{teal}{\ell})F}_1{\color{red}\zeta(2)}}  \textcolor{violet}{+\left(\frac{N^{(\textcolor{teal}{\ell})F}_{0}}{N_1^{(\textcolor{teal}{\ell})F}}\right)^0}\textcolor{blue}{\frac {\textcolor{teal}{\ell}}3\left(\frac{\alpha_F}{2}\right)^3N^{(\textcolor{teal}{\ell})F}_1}{\color{red}\zeta(3)},\\
	N_{-3}^{(\textcolor{teal}{\ell})F}=&\textcolor{violet}{\frac{1}{4!}\left(\frac{N^{(\textcolor{teal}{\ell})F}_{0}}{N_1^{(\textcolor{teal}{\ell})F}}\right)^3}  \textcolor{blue}{N^{(\textcolor{teal}{\ell})F}_{0}}  {\color{violet}-}{\color{violet}\frac{1}{2!}}{\color{violet}\left(\frac{N^{(\textcolor{teal}{\ell})F}_{0}}{N_1^{(\textcolor{teal}{\ell})F}}\right)^2}  \textcolor{blue}{\frac {\textcolor{teal}{\ell}}2\left(\frac{\alpha_F}{2}\right)^2 N^{(\textcolor{teal}{\ell})F}_1{\color{red}\zeta(2)}}   +\textcolor{violet}{\left(\frac{N^{(\textcolor{teal}{\ell})F}_{0}}{N_1^{(\textcolor{teal}{\ell})F}}\right)^1} \textcolor{blue}{\frac {\textcolor{teal}{\ell}}3\left(\frac{\alpha_F}{2}\right)^3 N^{(\textcolor{teal}{\ell})F}_1}{\color{red}\zeta(3)}{\color{violet}}\\&{-\color{violet}\frac{1}{0!}\left(\frac{N^{(\textcolor{teal}{\ell})F}_{0}}{N_1^{(\textcolor{teal}{\ell})F}}\right)^0}{\color{blue}\frac {\textcolor{teal}{\ell}}4\left(\frac{\alpha_F}{2}\right)^4 N^{(\textcolor{teal}{\ell})F}_1{\color{red}\left(\zeta(4)-\frac{\textcolor{teal}{\ell}}{2}\zeta(2)^2\right)}}.\\
	\end{align*}
		\caption{Free-energy zeta-number patterns, going down any diagonal $\ell$. We have used the same color coding as before, in order to highlight key aspects of these patterns---again, in red, one notices the increasing amount of zeta-number combinations.}
		\label{fig:FEpatternsPI}
\end{figure}

Everything else follows in more-or-less the same lines as the discussion in the previous subsection. For example, the exact same backward-forward formulae for Stokes vectors hold. The Stokes vector field $\underline{\boldsymbol{S}}_{\theta}^F (\boldsymbol{\sigma})$ is also computed straightforwardly. Of course it is defined in the exact same (resurgence) manner, in \eqref{eq:vec-fields}, only now using free-energy Stokes data rather than specific-heat Stokes data (and further replacing specific-heat transseries parameters with their rescaled free-energy counterparts). Evaluating the sums with analogous reasoning now yields
\bea
\label{eq:PI-F-S0-final}
\underline{\boldsymbol{S}}_{0}^F \left( \bm{\sigma}^F \right) &=& \left[ \begin{array}{c} \left( \sigma_2^F \right)^{-1} \left( 1 + \upmu^F \frac{\rmd}{\rmd \upmu^F} \log N^{(1)F} (\upmu^F) \right) \\ - \sigma_2^F\, \frac{\rmd}{\rmd \upmu^F} \log N^{(1)F} (\upmu^F) \end{array} \right] \log \left( 1 + \sigma_2^F\, N^{(1)F} (\upmu^F) \right), \\
\label{eq:PI-F-Spi-final}
\underline{\boldsymbol{S}}_{\pi}^F \left( \boldsymbol{\sigma}^F \right) &=& - \left[ \begin{array}{c} -\sigma_1^F\, \frac{\rmd}{\rmd \upmu^F} \log N^{(-1)F} (\upmu^F) \\ \left( \sigma_1^F \right)^{-1} \left( 1 + \upmu^F \frac{\rmd}{\rmd \upmu^F} \log N^{(-1)F} (\upmu^F) \right) \end{array} \right] \log \left( 1- \sigma_1^F\, N^{(-1)F} (\upmu^F) \right),
\eea
\noindent
with the standard simplifications at either vanishing $\sigma_1^F$ or $\sigma_2^F$,
\bea
\label{eq:PI-F-S0-simple-1}
\underline{\boldsymbol{S}}_{0}^F \left(\sigma_1^F,0\right) &=& \left[ \begin{array}{c} 1 \\ 0 \end{array} \right], \\
\label{eq:PI-F-S0-simple-2}
\underline{\boldsymbol{S}}_{0}^F \left(0,\sigma_2^F\right) &=& \log \left( 1 + \sigma_2^F \right) \left[ \begin{array}{c} \left( \sigma_2^F \right)^{-1} \\ - \sigma_2^F\,  N_0^{(1)F} \end{array} \right], \\
\label{eq:PI-F-Spi-simple-2}
\underline{\boldsymbol{S}}_{\pi}^F \left(0,\sigma_2^F\right) &=& \left[ \begin{array}{c} 0 \\ 1 \end{array} \right], \\
\label{eq:PI-F-Spi-simple-1}
\underline{\boldsymbol{S}}_{\pi}^F \left(\sigma_1^F,0\right) &=& - \log \left( 1 - \sigma_1^F \right) \left[ \begin{array}{c} -\sigma_1^F\,  N_0^{(-1)F} \\ \left( \sigma_1^F \right)^{-1} \end{array} \right].
\eea

One natural question which arises after looking at the explicit numbers in figure~\ref{fig:ComputedPainleveIStokesDataGridPatternsFreeEnergy} is whether one may find a redefinition of the transseries parameters such that more (if not all) Stokes vectors would turn out to solely have integer entries (a question which was motivated to us by the recent but unrelated work \cite{ggm20a, ggm20b, gm21}). As argued in \cite{as13}, the most general transformation that preserves semiclassical decoding is a rescaling of transseries parameters by generally different factors---with a study of the transformation laws of the Stokes vectors under such rescaling. Whereas it is obvious that any one fixed Stokes vector may always be turned into an integer-valued vector by a specific transseries-parameters rescaling, this is much less evident for \textit{sets} of vectors. From the form of the transformation laws \eqref{eq:trans_1}-\eqref{eq:trans_2}, the parametrization with the greatest number of rational vectors turns out to be ours (up to rescalings that will keep the same number of integer vectors). This is because vectors which are not on the lattice-boundary have complicated relations between each other (\textit{e.g.}, figure~\ref{fig:patterns}), which cannot be reabsorbed by one of our transformation laws---as we can only pick two numbers to rescale $\boldsymbol{\sigma}^F$, we can only pick a parametrization that makes any two Stokes vectors rational (albeit the simplicity of \eqref{eq:boundary} ensures that if we decide to make the vectors $\boldsymbol{S}_{(1,0)}^{(F)}$ and $\boldsymbol{S}_{(0,1)}^{(F)}$ rational, then all vectors on the boundary will be rational).

Falling short of finding a ``purely rational'' parametrization of Stokes data, one could still ask whether that would be possible at the level of Borel residues. This is, however, immediately unlikely as Borel residues are computed out of Stokes vectors with the sole addition of rational-number factors and integer-valued inner-products \cite{abs18}. To be more explicit, Borel residues will have integer or rational real and imaginary components when they are obtained from Stokes vectors at the boundary of the lattice (in the adequate free-energy parametrization in figure~\ref{fig:ComputedPainleveIStokesDataGridPatternsFreeEnergy}), and none others. For example, for Borel residues of the form $\mathsf{S}^{F}_{(n,n)\to(n+\ell,n)}$ this can only occur if they only involve $\bm{S}^{F}_{(1,0)}$,
\be
\mathsf{S}_{(n,n)\to(n+\ell,n)} = - \frac{1}{\ell!}\, \prod_{j=1}^{\ell} \left(n+j\right) = - \binom{n+\ell}{\ell}.
\ee
\noindent
All other Borel residues must involve Stokes vectors in the ``bulk'' of the lattice, and hence cannot be integers. A couple of examples illustrating the cumbersome structure of Borel residues may be found in table~\ref{tab:P1Borel} for \PI, in table~\ref{tab:P1BorelF} for the rescaled version corresponding to the free energy, whereas in table~\ref{tab:P1BorelF-int} we display some Borel residues which are actually integers.

Let us make one last comment, concerning the $(2,3)$ minimal-string normalization of \PI~\eqref{eq:PI-(2,3)} and its free energy. In this context it is interesting to ask what is the role of Stokes data within the minimal string theory. On what concerns ``canonical'' Stokes data, there is a strong connection between ZZ-brane amplitudes and Stokes data which has been thoroughly established in the literature, \textit{e.g.}, \cite{ss03, akk03, kopss04, emms22}. It seems interesting to explore this connection further on what concerns the remaining ``non-canonical'' Stokes data, which we will discuss elsewhere. Further note that knowledge of \PI~Stokes data is enough to compute minimal-string Stokes data, given the discussion in subsection~\ref{subsec:PI2dQG}. For completeness of the discussion and illustration purposes, table~\ref{tab:PIresultsdiffconv} presents the relevant D-brane exponential factor alongside ``canonical'' Stokes coefficient (\textit{i.e.}, the invariant quantity $u^{(1,0)[0]}_1\,S^{(0)}_1$ \cite{msw07, asv11}) in the different relevant  \PI~normalizations.

Having computed generating functions for Stokes vectors, the next step would be to compute generating functions for Borel residues. This is a more complicated task as Borel residues are complicated expressions of Stokes data; hence a problem we shall later tackle in section~\ref{sec:stokes}.

\begin{landscape}
\begin{table}
\centering
$\begin{array}{c|c|c|c}
-\mathsf{S}_{(n,n)\to(\ell,0)} & \ell=1 & \ell=2 & \ell=3 \\ \hline
n=0 & -\frac{\rmi \sqrt[4]{3}}{2 \sqrt{\pi }} & -\frac{\sqrt{3}}{4 \pi } & \frac{\rmi 3^{3/4}}{8 \pi ^{3/2}} \\ \hline
n=1 & -\frac{\rmi (2 \gamma_{\text{E}} +2\log 96\sqrt3)}{\sqrt[4]{3} \sqrt{\pi }} & -\frac{6 \gamma_{\text{E}} -2 \rmi \pi +6 \log 96\sqrt3}{2 \pi } & \frac{3 \sqrt[4]{3} (2 \rmi \gamma_{\text{E}} +\pi +2\rmi \log 96\sqrt3)}{2 \pi ^{3/2}} \\ \hline
n=2 & -\frac{\rmi \left(3 (2 \gamma_{\text{E}} +4\log 96\sqrt3)^2-2 \pi ^2\right)}{4\ 3^{3/4} \sqrt{\pi }} & \frac{-12 \gamma_{\text{E}} ^2+6 \rmi \gamma_{\text{E}}  \pi +\pi ^2-12 (\log 96\sqrt3)^2+6 \rmi \pi  \log 96\sqrt3-36 \gamma_{\text{E}}  \log 96\sqrt3}{\sqrt{3} \pi } & \frac{2 \left(90 \rmi \gamma_{\text{E}} ^2+72 \gamma_{\text{E}}  \pi -13 \rmi \pi ^2\right)+180 \rmi (\log 96\sqrt3)^2+72 (2 \pi +5 \rmi \gamma_{\text{E}} ) \log 96\sqrt3}{8 \sqrt[4]{3} \pi ^{3/2}} \\
\end{array}$
\caption{Some Borel residues on the main diagonal, for the \PI~specific-heat. The complexity of these numbers increases with $n$ (more Stokes vectors from the ``bulk'' of the lattice may contribute) while $\ell$ has no apparent effect.}
	\label{tab:P1Borel}
\end{table}
\begin{table}
\centering
$\begin{array}{c|c|c|c}
-\mathsf S_{(n,n)\to(\ell,0)}^{F} & \ell=1 & \ell=2 & \ell=3 \\ \hline
n=0 & 1 & 1 & 1 \\ \hline
n=1 & -\frac{\rmi}{\pi}\left(\gamma_E+\log\left(96\sqrt3\right)\right) & -1-\frac{3\rmi}{\pi}\left(\gamma_E+\log\left(96\sqrt3\right)\right) & -3-\frac{6\rmi}{\pi}\left(\gamma_E+\log\left(96\sqrt3\right)\right)  \\
\end{array}$
\caption{The same Borel residues, now for the free energy of 2d quantum gravity. Residues starting at the $(0,0)$ perturbative sector are trivially $-1$, but all other residues have complicated expressions (albeit slightly less complicated than the previous case).}
	\label{tab:P1BorelF}
\end{table}
\begin{table}
\centering
$\begin{array}{c|c|c|c|c|c}
-\mathsf S^{F}_{(n,n)\to(n+\ell,n)} & \ell=1 & \ell=2 & \ell=3 & \ell=4 & \ell=5 \\ \hline
n=0 & 1 & 1 & 1 & 1 & 1 \\ \hline
n=1 & 2 & 3 & 4 & 5 & 6 \\ \hline
n=2 & 3 & 6 & 10 & 15 & 21 \\ \hline
n=3 & 4 & 10 & 20 & 35 & 56 \\ \hline
n=4 & 5 & 15 & 35 & 70 & 126
\end{array}$
\caption{A larger set of integer-valued Borel residues for the free energy of 2d quantum gravity. These are associated to transitions from a diagonal $(n,n)$-sector into an off-diagonal $(n+\ell,n)$-sector. The associated resurgence relations \cite{abs18} only involve the vector $\bm{S}^{F}_{(1,0)}$, which itself is integer-valued. Note that, in principle, other Borel residues could be integer- or rational-valued after cancellation of the transcendental parts in the Stokes data---but we did not find any such example after examining a large number of possibilities.}
	\label{tab:P1BorelF-int}
\end{table}
\end{landscape}

\begin{table}
	\begin{center}
		\begin{tabular}{c|c|c}
			\PI~Equation & $\displaystyle u^{(1,0)[0]}_1\, S^{(0)}_1\, g_{\text{s}}^{\beta_{g_{\text{s}}}}\, z^{\beta_z}\, \rme^{A(z,g_{\text{s}})}$ & Conformal Background\\ [0.5ex] 
			\hline &  & \\
			$\displaystyle u^2(z) - \frac{1}{6} u''(z) - z = 0$ & $\left( - \frac{3^{\frac{1}{4}}\rmi}{2\sqrt{\pi}} \right) \frac{1}{z^{\frac{1}{8}}}\, \rme^{-\frac{8\sqrt{3}}{5} z^{\frac{5}{4}}}$ & \\ \hline
			&  &  \\
			$\displaystyle u^2(z) - \frac{1}{3} u''(z) - z = 0$ & $\left( - \frac{3^{\frac{1}{4}}\rmi}{2^{\frac{3}{4}}\sqrt{\pi}} \right) \frac{1}{z^{\frac{1}{8}}}\, \rme^{-\frac{4\sqrt{6}}{5} z^{\frac{5}{4}}}$ & \\ \hline
			& & \\
			$\displaystyle u^2(z) - \frac{g_{\text{s}}^2}{3} u''(z) - z = 0$ & $\left( - \frac{3^{\frac{1}{4}}\rmi}{2^{\frac{3}{4}}} \sqrt{\frac{g_{\text{s}}}\pi} \right) \frac{1}{z^{\frac{1}{8}}}\, \rme^{-\frac{4\sqrt{6}}{5} \frac{z^{\frac{5}{4}}}{g_{\text{s}}}}$ & \\ \hline
			& & \\
			$\displaystyle u^2(z) - \frac{g_{\text{s}}^2}{3} u''(z) + \frac{4\sqrt{2}}{3} z = 0$ & $\left( - \frac{2^{\frac{3}{16}}\rmi}{3^{\frac{1}8}} \sqrt{\frac{g_{\text{s}}}{\pi}} \right) \frac{1}{(-z)^{\frac{1}{8}}}\, \rme^{-\frac{2^{\frac{25}{8}}3^{\frac{1}{4}}}{5} \frac{(-z)^{\frac{5}{4}}}{g_{\text{s}}}}$ & $\left( - \frac{\rmi}{3^{\frac{1}{4}}} \sqrt{\frac{2}{\pi}} \right) \sqrt{g_{\text{s}}}\, \rme^{-\frac{3\sqrt{3}}{5} \frac{1}{g_{\text{s}}}}$ 
		\end{tabular}
		\caption{Values of Stokes data alongside corresponding D-brane exponential weights (and starting powers of $z$ and $g_{\text{s}}$, when appropriate) of the standard $(1,0)$-instanton sector, for the different normalizations of \PI~discussed in subsection~\ref{subsec:PI2dQG}. For the $(2,3)$ minimal-string case we also give the result obtained when taking $z$ to the conformal background. \label{tab:PIresultsdiffconv}}
	\end{center}
\end{table}

\subsection{Stokes Data for the Painlev\'e~II Equation}\label{subsec:PIIresults}

Moving on to the second Painlev\'e equation is now almost straightforward. In fact the discussion in subsection~\ref{subsec:PII2dSG} completely parallels that of subsection~\ref{subsec:PI2dQG}, as was made clear in subsection~\ref{subsec:reson-frame}; \textit{i.e.}, from a resurgent transseries point-of-view, \PI~and \PII~essentially have the exact same structure. It is then only natural that \PII~Stokes data turns out to be obtained in very close---if not complete---analogy to what we did in our earlier subsection~\ref{subsec:PIresults}. Note that in both present and next subsections, all symbols refer to specific \PII~quantities (we will not be labelling any with a II).

The structure of Stokes vectors is always the same for these two-parameters transseries problems, either \eqref{eq:vec_struct-a}-\eqref{eq:vec_struct-b} or \eqref{eq:struct-a}-\eqref{eq:struct-b}; herein again
\bea
\bm{S}_{(1-p,1-p-\ell)} &=& N^{(\ell)}_{1-p}\, \left[ \begin{array}{c}p+\ell \\ -p\end{array} \right], \\
\bm{S}_{(1-p-\ell,1-p)} &=& N^{(-\ell)}_{1-p}\, \left[ \begin{array}{c}-p \\ p+\ell\end{array} \right].
\eea
\noindent
The corresponding \PII~Stokes ratios \eqref{eq:stokes-R-ratios} now fulfill (compare with \eqref{eq:closed_form_PI} for \PI)
\be
\sum_{q=0}^p R^{(\ell)}_{1-p+q}\, \frac{1}{q!} \left( 4 \ell \right)^{q} B_{q} \left( c_\ell, \frac{1}{\ell}\, \psi^{(1)}(1), \ldots, \frac{1}{\ell^{q-1}}\, \psi^{(q-1)}(1) \right) = \delta_{p,0},
\ee
\noindent
where $c_\ell$ still satisfies \eqref{eq:c_cons}. Following the same steps as earlier-on, from this equation follows the Stokes-ratios generating-function \eqref{eq:Relllambda-genfun} for \PII,
\be
\label{eq:PIIStokesratiosGF}
R^{(\ell)} (\uplambda) = \frac{\widetilde{A}_\ell^{4 \uplambda \ell}}{\Gamma \left( 1 + 4 \uplambda \right)^{\ell}}
\ee
\noindent
(compare with \eqref{eq:PIStokesratiosGF} for \PI). We can generalize the conjecture in \cite{asv11} for the \PI~Stokes lattice boundary \eqref{eq:n1sp1} to \PII---see as well \cite{sv13}. If we further join-in the known analytic Stokes coefficient \eqref{eq:exampleStokesP2}, alongside support from numerical explorations (see section~\ref{sec:numerics}), we have
\be
\label{eq:n1sp2}
N_1^{(\ell)} = \frac{(2\rmi)^{\ell-1}}{\ell^2} \left( N_1^{(1)} \right)^{2-\ell} \qquad \text{ and } \qquad N_1^{(1)} = -\rmi\, \frac{1}{\sqrt{2\pi}}.
\ee
\noindent
And this is now all the information we need in order to write down the explicit generating function of full \PII~Stokes data:
\bea
\label{eq:PII-gen_fun-a}
N^{(\ell)} (\uplambda) &=& \frac{(2\rmi)^{\ell-1}}{\ell^2} \left( N_1^{(1)} \right)^{2-\ell} \frac{\widetilde{A}_{\ell}^{4 \uplambda \ell}}{\Gamma \left( 1 + 4 \uplambda \right)^{\ell}}, \\
\label{eq:PII-gen_fun-b}
N_{1-p}^{(\ell)} &=&\frac{1}{p!} \left. \frac{\partial^{p}}{\partial \uplambda^{p}}\, N^{(\ell)} (\uplambda) \right|_{\uplambda=0}.
\eea

\begin{figure}
	\centering
	\begin{subfigure}[t]{0.5\textwidth}
		\centering
		\includegraphics[height=2in]{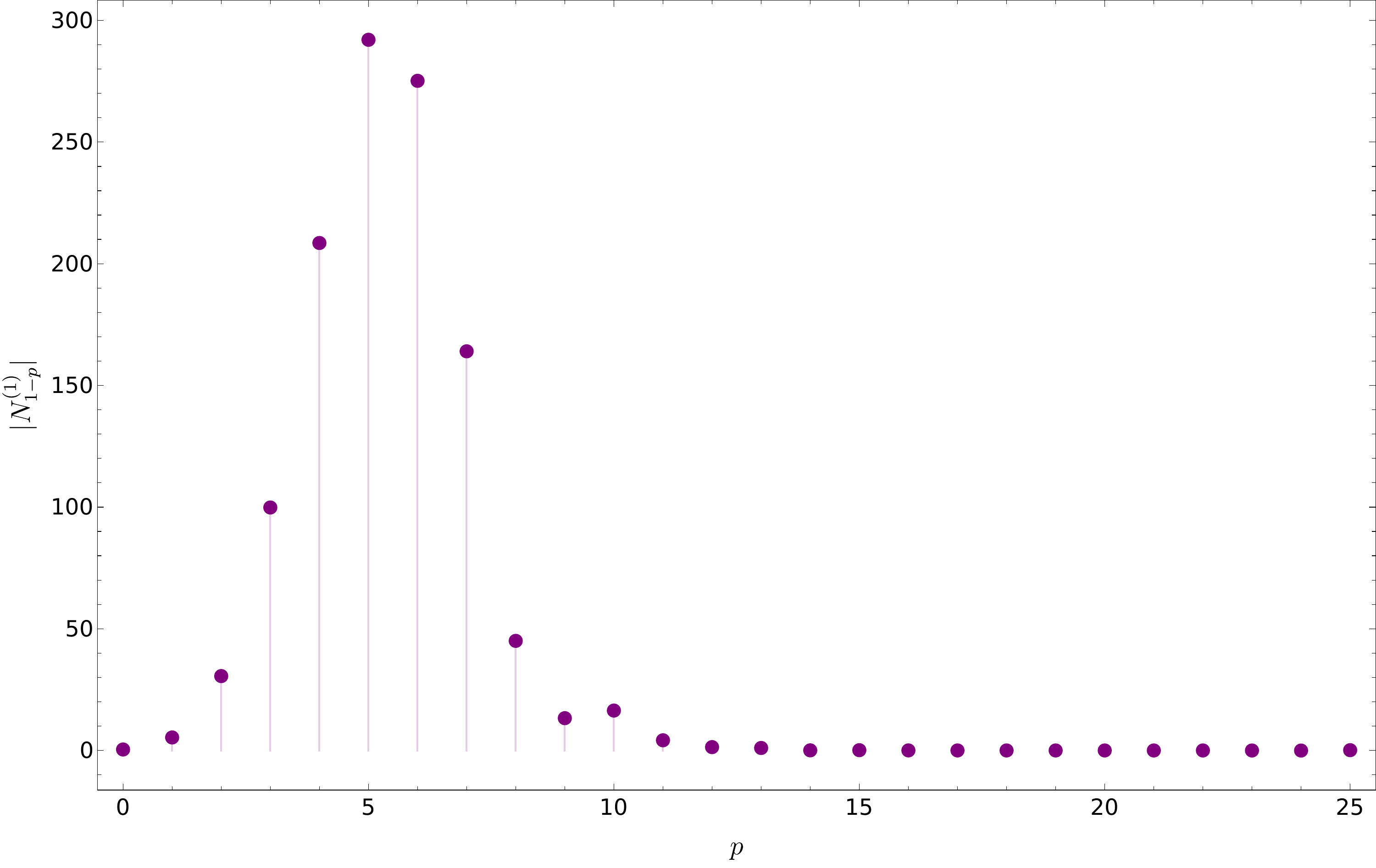}
		\caption{First diagonal $\ell=1$.}
	\end{subfigure}%
	~ 
	\begin{subfigure}[t]{0.5\textwidth}
		\centering
		\includegraphics[height=2in]{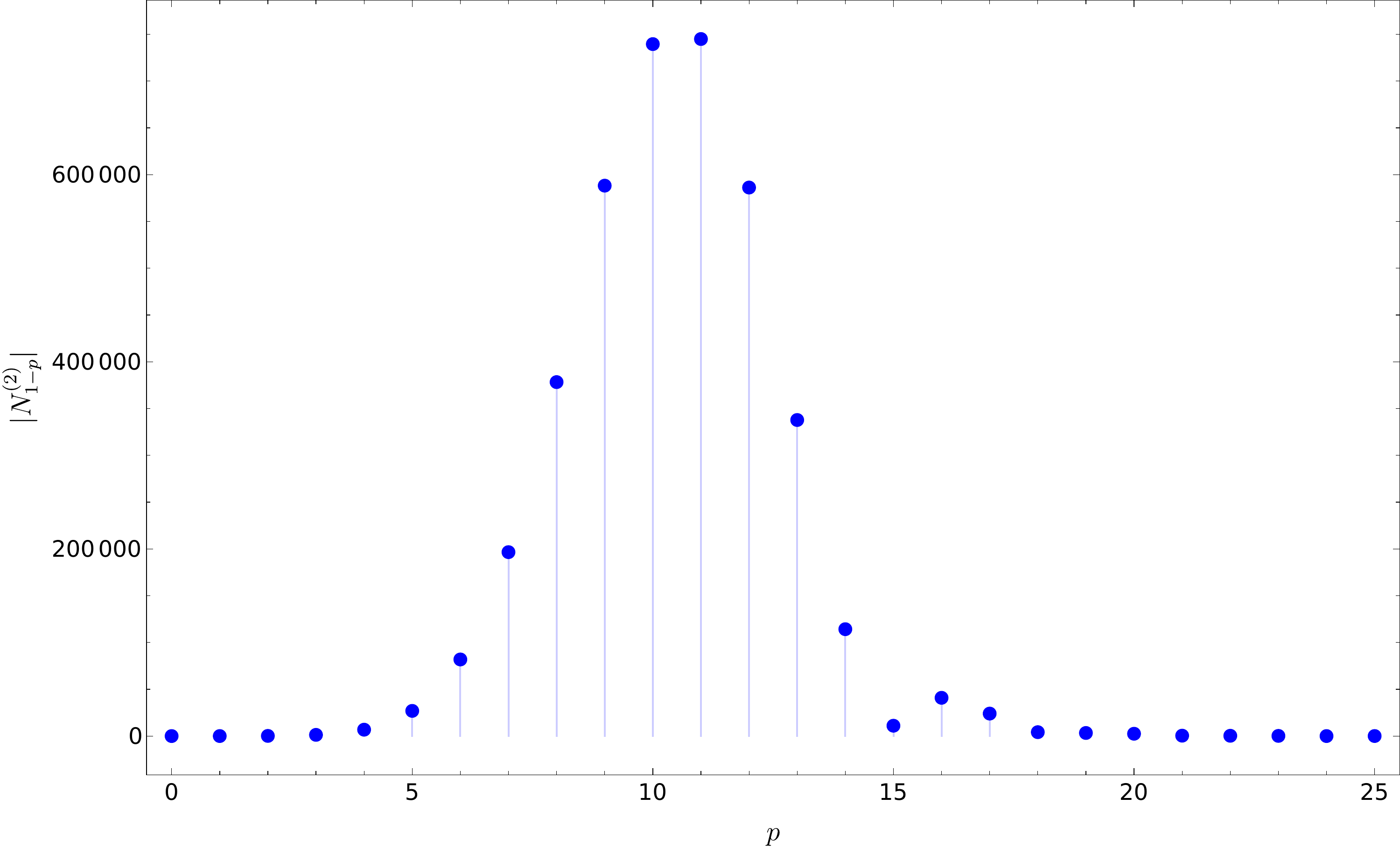}
		\caption{Second diagonal $\ell=2$.}
	\end{subfigure}
	\vspace{3mm}
	
	\begin{subfigure}[b]{0.5\textwidth}
		\centering
		\includegraphics[height=2in]{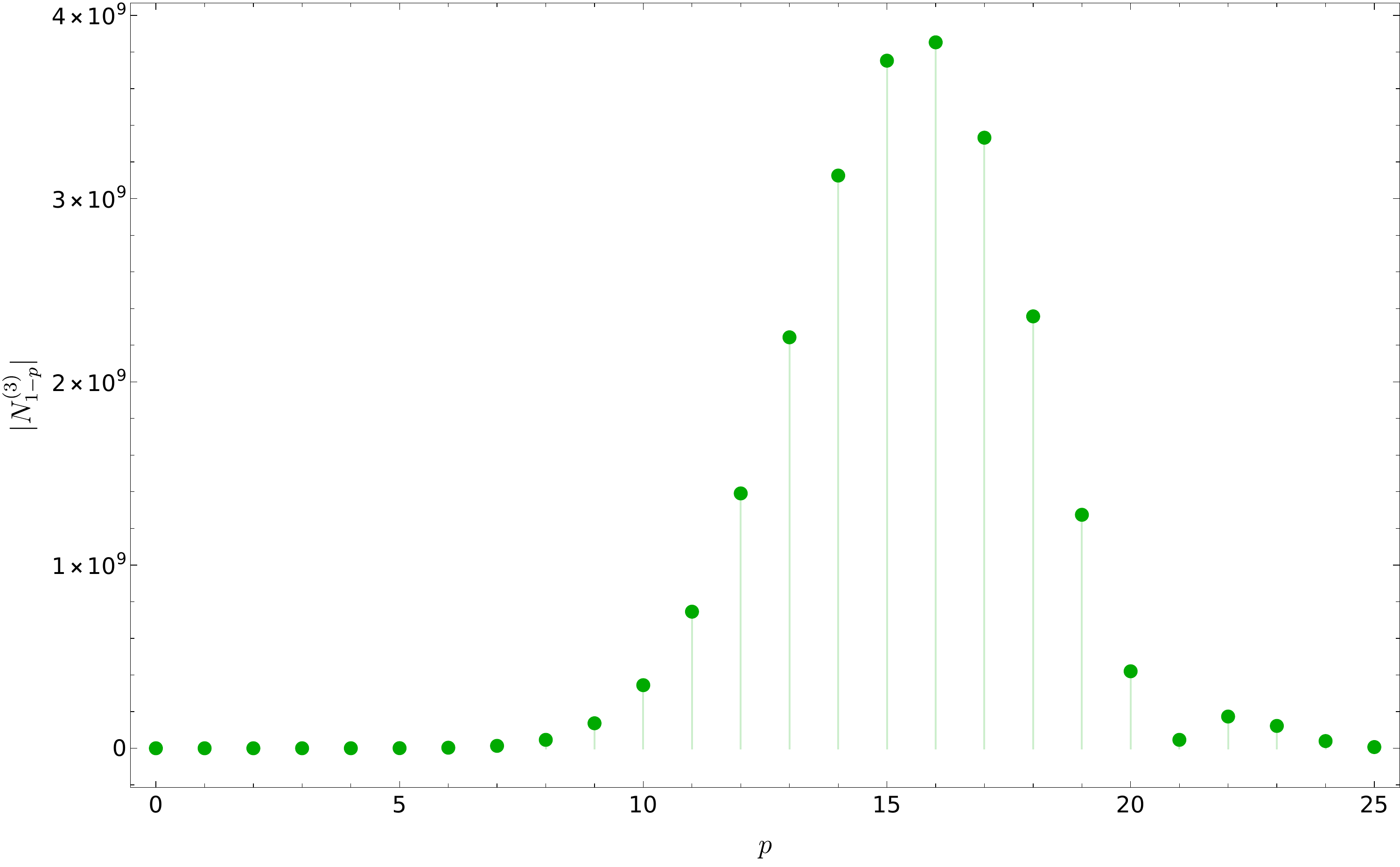}
		\caption{Third diagonal $\ell=3$.}
	\end{subfigure}
	\caption{Structure of the first few $N^{(\ell)}_{1-p}$ (in absolute value), for the first three diagonals.}
	\label{fig:Ns1mpPII}
\end{figure}

We have also tested this expression thoroughly, as will be described in section~\ref{sec:numerics}. This time around, it is setting
\be
\widetilde{A}_\ell = 16
\ee
\noindent
(again, $\widetilde{A}_\ell$ turns out not to depend on $\ell$) which yields precise matches with our numerics after generating large sequences of the $N^{(\ell)}_{1-p}$ coefficients---up to very small errors; see section~\ref{sec:numerics}. With all this at hand, the first couple of coefficients for the first diagonal read (compare with \eqref{eq:N11-examples}-\eqref{eq:N10-examples}-\eqref{eq:N1-1-examples} for \PI)
\bea
N^{(1)}_1 &=& - \frac{\rmi}{\sqrt{2\pi}},\\
N^{(1)}_0 &=& - 4\, \frac{\rmi}{\sqrt{2\pi}} \left( \gamma_{\text{E}} + \log \left(16\right) \right), \\
N^{(1)}_{-1} &=& \frac{8}{3}\, \frac{\rmi}{\sqrt{2\pi}} \left( \frac{\pi^2}{2} - 3 \left( \gamma_{\text{E}} + \log\left(16\right) \right)^2 \right).
\eea
\noindent
These \PII~Stokes data display the same curious patterns as for \PI~(which we already previewed in figure~\ref{fig:patterns}). Moving down each diagonal, the patterns of figure~\ref{fig:patterns} translate into the by-now familiar ``hump patterns'' for the absolute values of the $N_{1-p}^{(\ell)}$, as illustrated in figure~\ref{fig:Ns1mpPII}.

\begin{landscape}
\begin{figure}
	\includegraphics[scale=0.5]{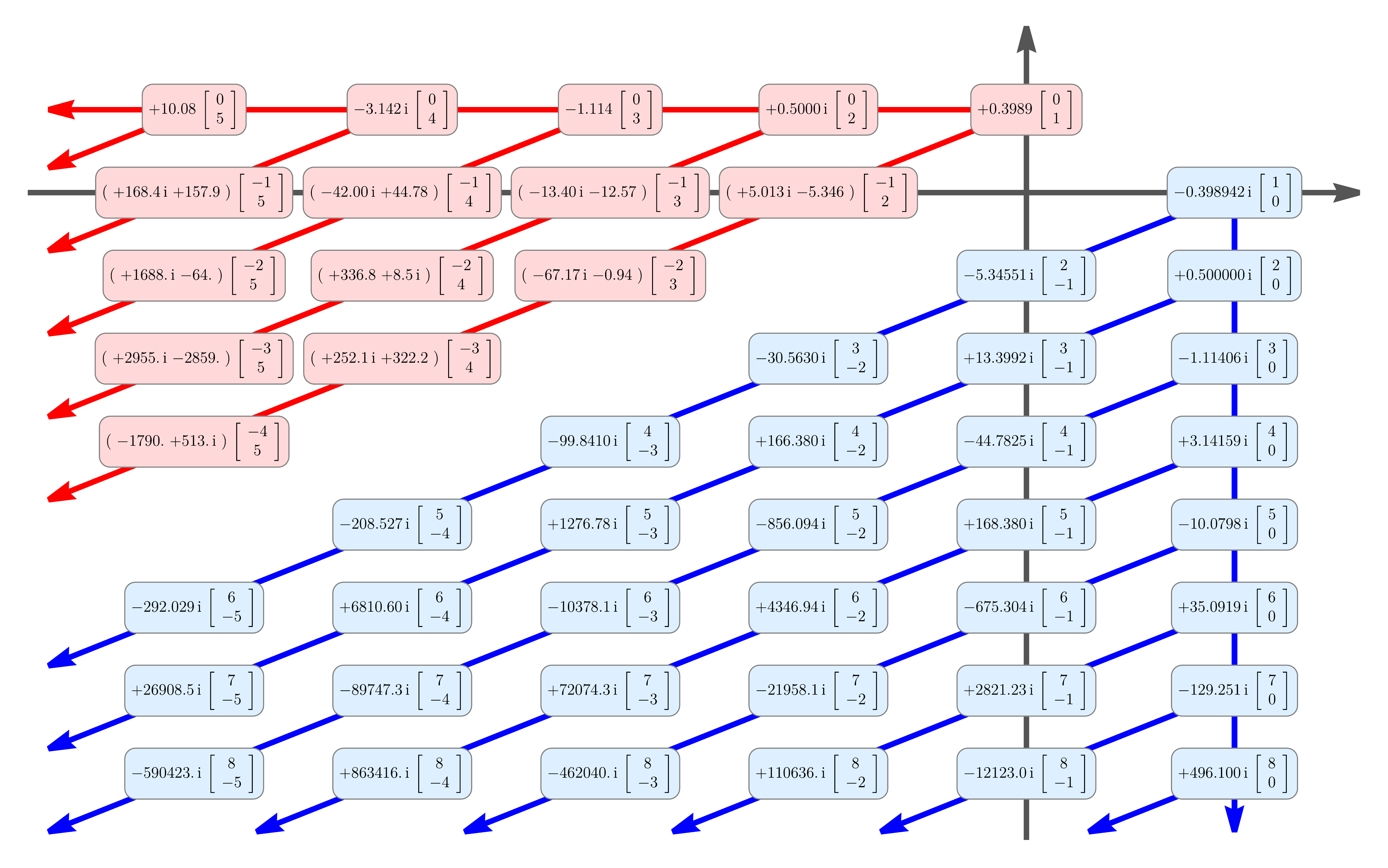}
	\caption{A subset of Stokes data for \PII. As usual, all displayed numerical values are of course truncations of irrational quantities. Some of these numbers had been previously found in \cite{sv13, as13} and they are now all understood with closed-form analytical expressions, as discussed in the text. These numbers encode many patterns (illustrated by the arrows), which are essentially the same as for \PI.}
	\label{fig:ComputedPainleveIIStokesDataGridPatterns}
\end{figure}
\end{landscape}

Backward Stokes data also follows in complete parallel with \PI. It is now encoded in the generating function
\be
N^{(-\ell)} (\uplambda) = - (-\rmi)^{\ell}\, \rme^{4\pi\rmi\, \uplambda\ell}\, N^{(\ell)}(-\uplambda).
\ee
\noindent
All these expressions are numerically testable, and those will be discussed in the upcoming section~\ref{sec:numerics}. We illustrate all these \PII~results for the Stokes-vectors lattice in figure~\ref{fig:ComputedPainleveIIStokesDataGridPatterns}.

Finally, we may address the $\underline{\boldsymbol{S}}_{\theta} (\boldsymbol{\sigma})$ Stokes vector field---which again reduces to its only non-trivial content, $\underline{\bm{S}}_{0} (\bm{\sigma})$ and $\underline{\bm{S}}_{\pi} (\bm{\sigma})$. Recalling the slight difference in the structure of $N_1^{(\ell)}$ data (compare the factor of $2$ in \eqref{eq:n1sp1} and \eqref{eq:n1sp2}), the present \PII~results match the earlier \PI~results in subsection~\ref{subsec:PIresults} up to factors of $2$. The computational steps of course remain the same. Hence, we now find (compare with \eqref{eq:PIS0-final}-\eqref{eq:PISpi-final} for \PI)
\bea
\label{eq:PIIS0-final}
\underline{\boldsymbol{S}}_{0} (\boldsymbol{\sigma}) &=& \frac{\rmi}{2} \left( N_1^{(1)} \right)^{2} \left[ \begin{array}{c} \sigma_2^{-1} \left( 1 + \upmu \frac{\rmd}{\rmd \upmu} \log N^{(1)} (\upmu) \right) \\ - \sigma_2\, \frac{\rmd}{\rmd \upmu} \log N^{(1)} (\upmu) \end{array} \right] \log \left( 1 - 2 \rmi \sigma_2\, \frac{N^{(1)}(\upmu)}{\left( N_1^{(1)}\right )^2} \right), \\
\label{eq:PIISpi-final}
\underline{\boldsymbol{S}}_{\pi} (\boldsymbol{\sigma}) &=& \frac{\rmi}{2} \left( N_1^{(-1)} \right)^{2} \left[ \begin{array}{c} - \sigma_1\, \frac{\rmd}{\rmd \upmu} \log N^{(-1)} (\upmu) \\ \sigma_1^{-1} \left( 1 + \upmu \frac{\rmd}{\rmd \upmu} \log N^{(-1)} (\upmu) \right) \end{array} \right] \log \left( 1 - 2 \rmi \sigma_1\, \frac{N^{(-1)}(\upmu)}{\left(N_1^{(-1)}\right)^2} \right),
\eea
\noindent
with the familiar simplifications at either vanishing $\sigma_1$ or $\sigma_2$ (compare with \eqref{eq:PIS0-simple-1}-\eqref{eq:PIS0-simple-2}-\eqref{eq:PISpi-simple-2}-\eqref{eq:PISpi-simple-1} for \PI)
\bea
\underline{\boldsymbol{S}}_{0} \left(\sigma_1,0\right) &=& N_1^{(1)} \left[ \begin{array}{c} 1 \\ 0 \end{array} \right], \\
\underline{\boldsymbol{S}}_{0} \left(0,\sigma_2\right) &=& \frac{\rmi}{2} N_1^{(1)} \left[ \begin{array}{c} \sigma_2^{-1}\, N_1^{(1)} \\ - \sigma_2\, N_0^{(1)} \end{array} \right] \log \left( 1 - \frac{2 \rmi \sigma_2}{N_1^{(1)}} \right), \\
\underline{\boldsymbol{S}}_{\pi} \left(0,\sigma_2\right) &=& N_1^{(-1)} \left[ \begin{array}{c} 0 \\ 1 \end{array} \right], \\
\underline{\boldsymbol{S}}_{\pi} \left(\sigma_1,0\right) &=& \frac{\rmi}{2} N_1^{(-1)} \left[ \begin{array}{c} - \sigma_1\, N_0^{(-1)} \\ \sigma_1^{-1}\, N_1^{(-1)} \end{array} \right] \log \left( 1 - \frac{2 \rmi \sigma_1}{N_1^{(-1)}} \right).
\eea

\subsection{Global Structure of 2D Quantum Supergravity}\label{subsec:2DQSGresults}

As discussed in subsection~\ref{subsec:PII2dSG}, \PII~solutions-\textit{squared} describe the specific-heat of the simplest ($k=1$) unitary multicritical model. Its free energy is given by \eqref{eq:PII-F/Z}, which we repeat in here
\be
F_{\text{II}}'' (z) = - u_{\text{II}} (z) = - \mu^2_{\text{II}} (z).
\ee
\noindent
As already discussed for \PI, the above integration is formal in which case out of the $\mu_{\text{II}} \left(x;\sigma_1,\sigma_2\right)$ transseries one hence obtains a $F_{\text{II}} \left(x;\sigma_1,\sigma_2\right)$ transseries in the same (trans)monomials. This was explicitly carried through with many data in \cite{sv13}, and some brief examples were collected throughout section~\ref{sec:review}. As for \PI, this further implies that out of the \PII~solution Stokes data we may hence obtain 2d supergravity \textit{free-energy Stokes data}.

\begin{figure}
	\centering
	\begin{subfigure}[t]{0.5\textwidth}
		\centering
		\includegraphics[height=2in]{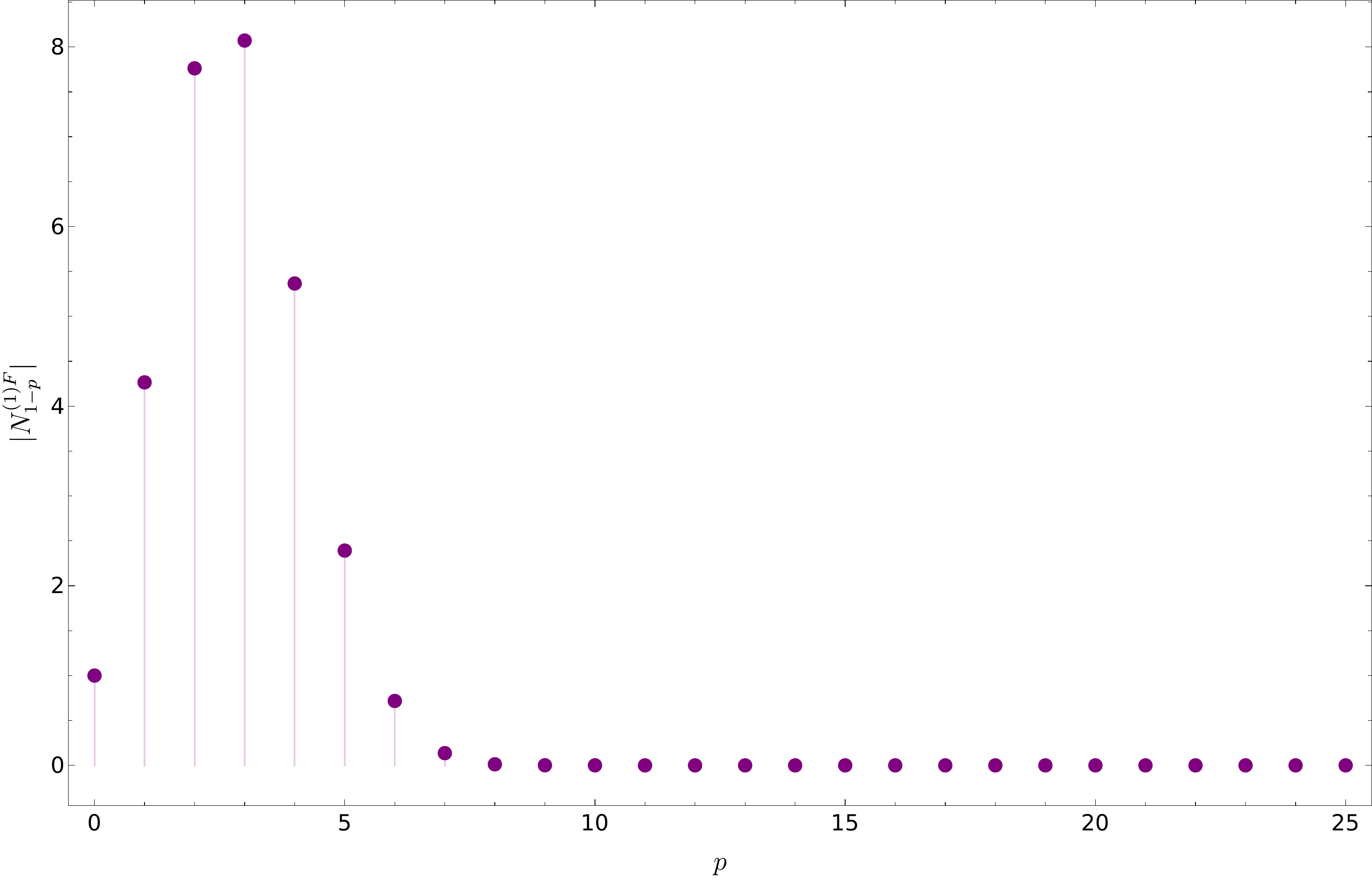}
		\caption{First diagonal $\ell=1$.}
	\end{subfigure}%
	~ 
	\begin{subfigure}[t]{0.5\textwidth}
		\centering
		\includegraphics[height=2in]{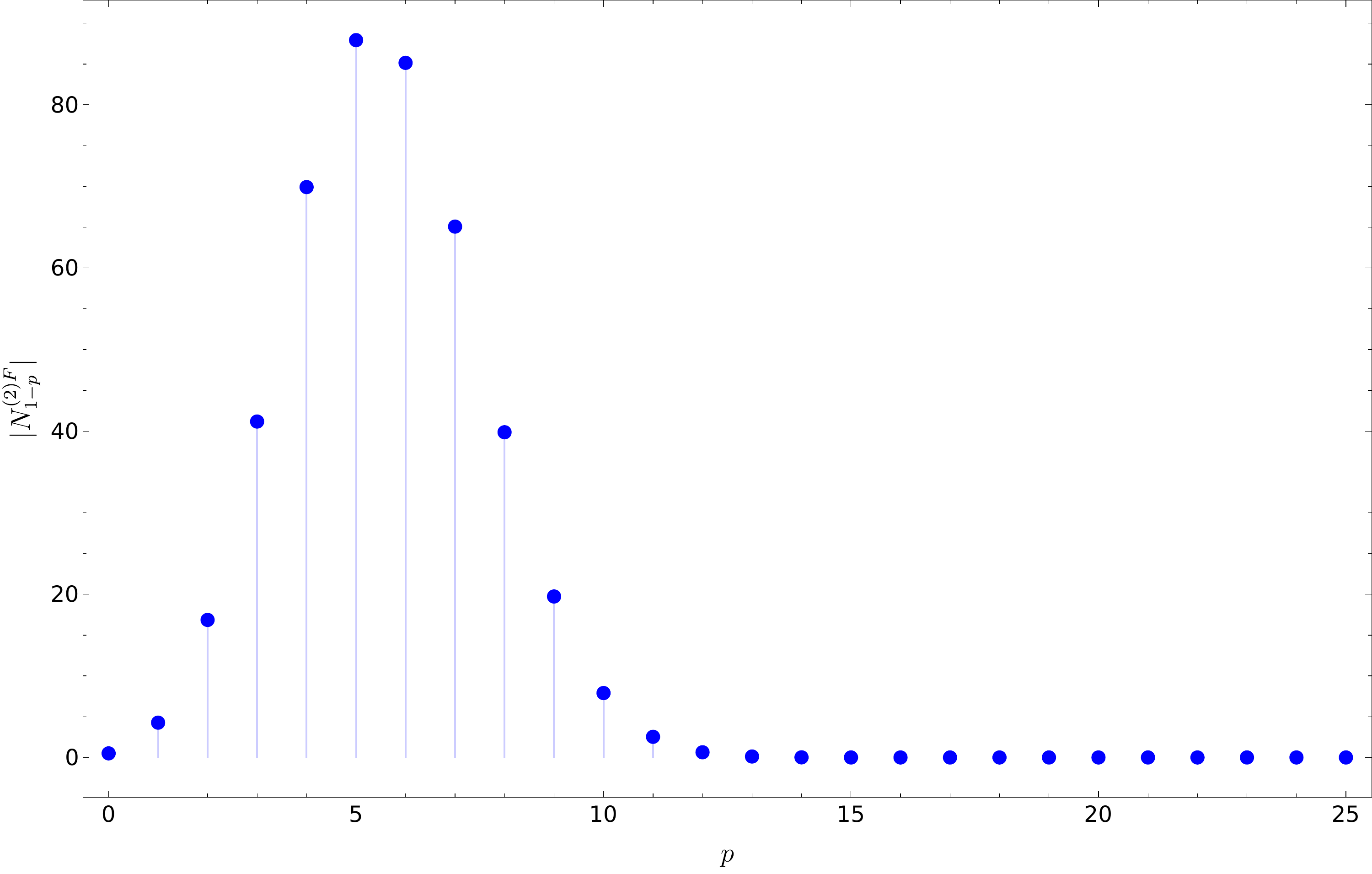}
		\caption{Second diagonal $\ell=2$.}
	\end{subfigure}
	\vspace{3mm}
	
	\begin{subfigure}[b]{0.5\textwidth}
		\centering
		\includegraphics[height=2in]{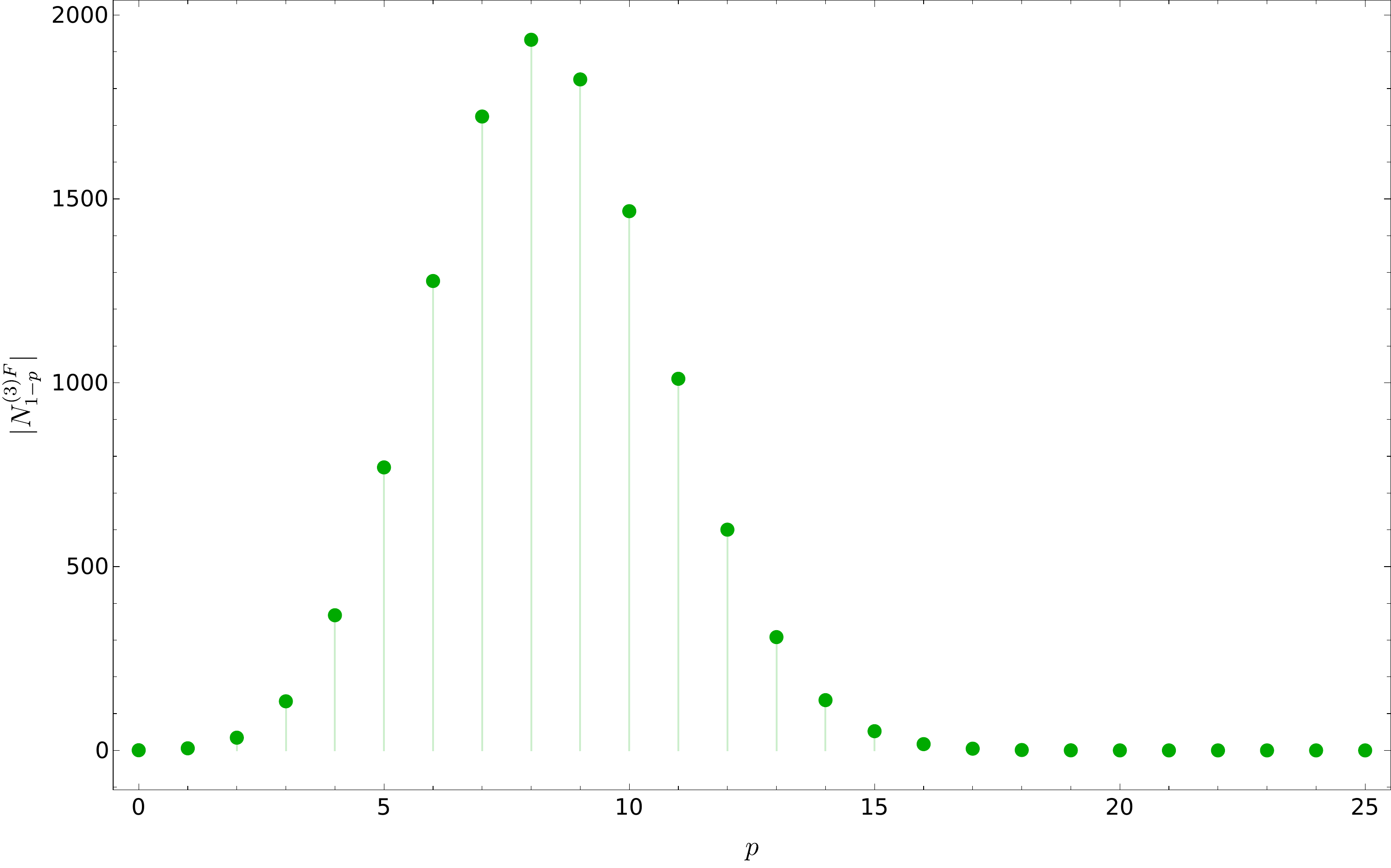}
		\caption{Third diagonal $\ell=3$.}
	\end{subfigure}
	\caption{Structure of the first few $N^{(\ell)F}_{1-p}$ (in absolute value), for the first three diagonals.}
	\label{fig:Ns1mpPIIFE}
\end{figure}

\begin{landscape}
\begin{figure}
	\centering
	\includegraphics[scale=0.5]{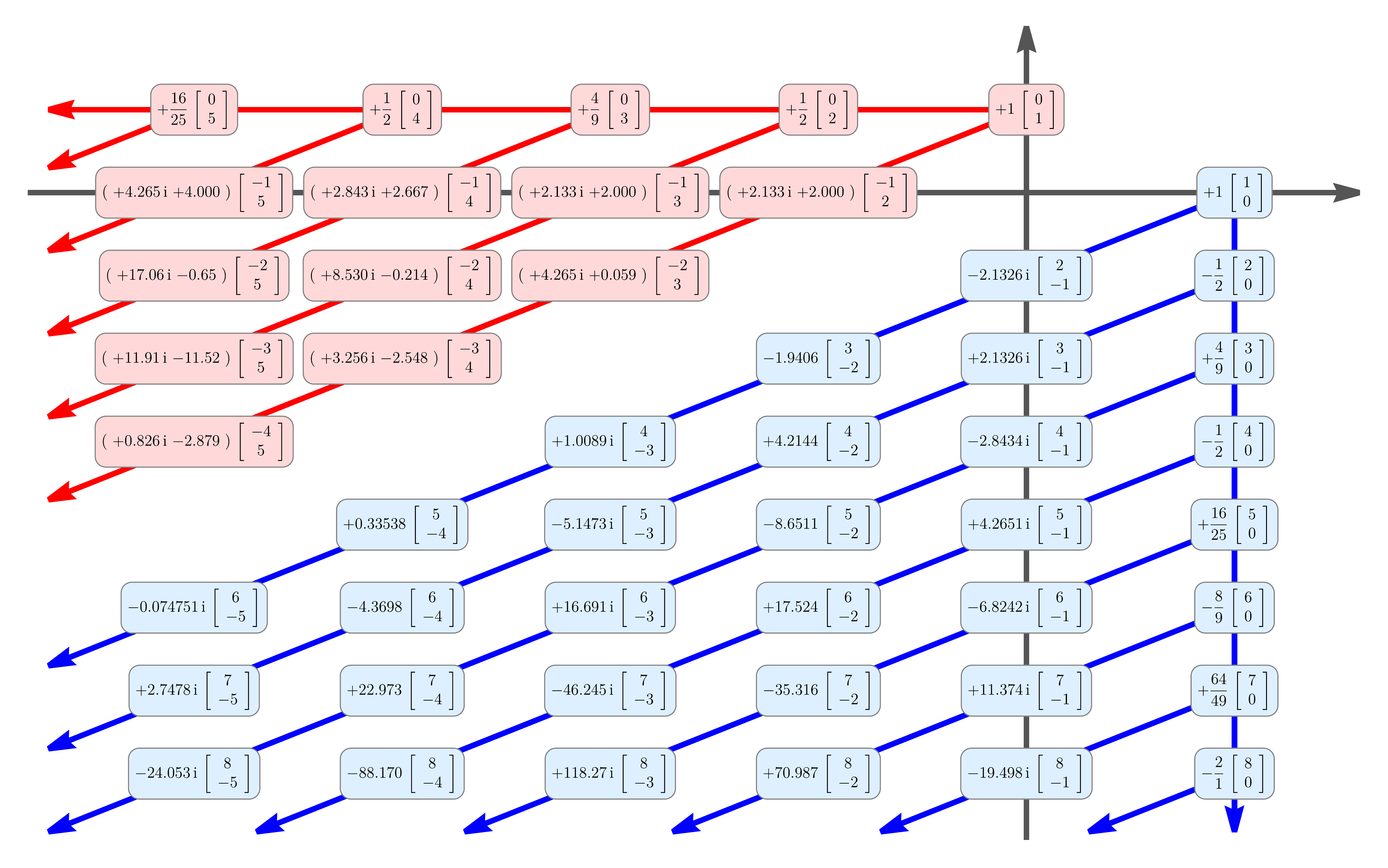}
	\caption{A subset of Stokes data for the free energy of the $k=1$ unitary multicritical model. As always, many numerical values are truncated irrational quantities. All these numbers are now fully understood, with closed-form analytical expressions. They encode many patterns (illustrated by the arrows), essentially inherited from those of \PII, and which end-up basically the same as for \PI.}
	\label{fig:ComputedPainleveIIStokesDataGridFreeEnergy}
\end{figure}
\end{landscape}

The whole calculational procedure follows in parallel with what was done in subsection \ref{subsec:2DQGresults}, and we will thus be brief in the present discussion. The vectorial structure of Stokes data $\bm{S}_{\bm{n}}^F$ is always the same, \textit{i.e.}, \eqref{eq:struct-a}-\eqref{eq:struct-b},
\bea
\bm{S}_{(1-p,1-p-\ell)}^F &=& N^{(\ell)F}_{1-p}\, \left[ \begin{array}{c}p+\ell \\ -p\end{array} \right], \\
\bm{S}_{(1-p-\ell,1-p)}^F &=& N^{(-\ell)F}_{1-p}\, \left[ \begin{array}{c}-p \\ p+\ell\end{array} \right].
\eea
\noindent
Rescaling the free-energy transseries parameters as
\be
\sigma_1^F = \frac{\sigma_1}{N_1^{(1)}}, \qquad \sigma_2^F = \frac{\sigma_2}{N_1^{(-1)}},
\ee
\noindent
then free-energy Stokes data scale as (same as \eqref{eq:trans_1}-\eqref{eq:trans_2} for \PI)
\bea
N_{1-p}^{(\ell)F} &=& \left(N_1^{(1)}\right)^{p-1} \left(N_1^{(-1)}\right)^{p+\ell-1} N_{1-p}^{(\ell)}, \\
N_{1-p}^{(-\ell)F} &=& \left(N_1^{(1)}\right)^{p+\ell-1} \left(N_1^{(-1)}\right)^{p-1} N_{1-p}^{(-\ell)}.
\eea
\noindent
This transformation law is obtained by combining the results of section \ref{subsec:2DQGresults} (double integration does not change the Stokes data), together with the result in \cite{sv13} (squaring a transseries does not change the Stokes data). Thus, the transformation law comes only from the rescaling of $\sigma_1$ and $\sigma_2$ in $\sigma_1^F$ and $\sigma_2^F$, just as in subsection \ref{subsec:2DQGresults}. From these relations the generating functions for free-energy Stokes data follow from \eqref{eq:PII-gen_fun-a} (same as \eqref{eq:trans_1-GF}-\eqref{eq:trans_2-GF} for \PI)
\bea
N^{(\ell)F} (\uplambda) &=& \frac{\left(N_1^{(-1)}\right)^\ell}{N_1^{(1)}N_1^{(-1)}}\, N^{(\ell)} \left( N_1^{(1)} N_1^{(-1)} \uplambda \right), \\
N^{(-\ell)F} (\uplambda) &=& \frac{\left(N_1^{(1)}\right)^\ell}{N_1^{(1)}N_1^{(-1)}}\, N^{(-\ell)} \left( N_1^{(1)} N_1^{(-1)} \uplambda \right).
\eea
\noindent
The forward-data generating-functions have the same form as in subsection \ref{subsec:2DQGresults},
\be
N^{(\ell)F} (\uplambda) = \frac{(-2)^{\ell-1}}{\ell^2}\, \frac{\widetilde{A}^{\frac{\alpha_{F}}{2}\, \uplambda \ell}}{\Gamma \left( 1 + \frac{\alpha_F}{2}\, \uplambda \right)^{\ell}}
\ee
\noindent
with $\alpha_F = N_1^{(1)} N_1^{(-1)} \alpha = 8\, N_1^{(1)} N_1^{(-1)} = -\frac{4\rmi}{\pi}$, and $\widetilde{A}$ as in the previous subsection. Backward data are generated by
\be
N^{(-\ell)F}(\uplambda) = (-1)^{\ell-1}\, \rme^{2\, \uplambda\ell}\, N^{(\ell)F}(-\uplambda).
\ee
\noindent
Further as before, the edge boundary factors turn out to be rational:
\be
N_1^{(\ell)F} = \frac{(-2)^{\ell-1}}{\ell^2}, \qquad N_1^{(-\ell)F} = \frac{2^{\ell-1}}{\ell^2}.
\ee
\noindent
The familiar Stokes-data patterns also repeat themselves. The trivial pattern along the Stokes-lattice boundary is fixed in the free-energy normalization,
\be
\bm{S}_{(1,1-\ell)}^F = \left[ \begin{array}{c} 1 \\ 0 \end{array} \right].
\ee
\noindent
Moving down each diagonal, the ``hump patterns'' (occurring earlier-on down the diagonal as typical of free-energy data) are illustrated in figure~\ref{fig:Ns1mpPIIFE} (for the first three diagonals $\ell=1,2,3$ and $p=0,1,\ldots,25$). Their actual zeta-number combinations may be then found in figure~\ref{fig:FEpatternsPII}, with corresponding bare numerical values illustrated in figure~\ref{fig:ComputedPainleveIIStokesDataGridFreeEnergy}.

\begin{figure}
	\begin{align*}
	N_{-1}^{(\textcolor{teal}{\ell})F}=&\textcolor{violet}{\frac{1}{2!}\left(\frac{N^{(\textcolor{teal}{\ell})F}_{0}}{N_1^{(\textcolor{teal}{\ell})F}}\right)^1} \textcolor{blue}{N^{(\textcolor{teal}{\ell})F}_{0}} {\color{violet}-}{\color{violet}\frac{1}{0!}\left(\frac{N^{(\textcolor{teal}{\ell})F}_{0}}{N_1^{(\textcolor{teal}{\ell})F}}\right)^0}\textcolor{blue}{\frac {\textcolor{teal}{\ell}}2\left(\frac{\alpha_F}{2}\right)^2 N^{(\textcolor{teal}{\ell})F}_1{\color{red}\zeta(2)}},\\
	N_{-2}^{(\textcolor{teal}{\ell})F}=&\textcolor{violet}{\frac{1}{3!}\left(\frac{N^{(\textcolor{teal}{\ell})F}_{0}}{N_1^{(\textcolor{teal}{\ell})F}}\right)^2}  \textcolor{blue}{ N^{(\textcolor{teal}{\ell})F}_{0}} \textcolor{violet}{\color{violet}-}{\color{violet}\frac{1}{1!}\left(\frac{N^{(\textcolor{teal}{\ell})F}_{0}}{N_1^{(\textcolor{teal}{\ell})F}}\right)^1} \textcolor{blue}{\frac {\textcolor{teal}{\ell}}2\left(\frac{\alpha_F}{2}\right)^2 N^{(\textcolor{teal}{\ell})F}_1{\color{red}\zeta(2)}}  \textcolor{violet}{+\left(\frac{N^{(\textcolor{teal}{\ell})F}_{0}}{N_1^{(\textcolor{teal}{\ell})F}}\right)^0}\textcolor{blue}{\frac {\textcolor{teal}{\ell}}3\left(\frac{\alpha_F}{2}\right)^3N^{(\textcolor{teal}{\ell})F}_1}{\color{red}\zeta(3)},\\
	N_{-3}^{(\textcolor{teal}{\ell})F}=&\textcolor{violet}{\frac{1}{4!}\left(\frac{N^{(\textcolor{teal}{\ell})F}_{0}}{N_1^{(\textcolor{teal}{\ell})F}}\right)^3}  \textcolor{blue}{N^{(\textcolor{teal}{\ell})F}_{0}}  {\color{violet}-}{\color{violet}\frac{1}{2!}}{\color{violet}\left(\frac{N^{(\textcolor{teal}{\ell})F}_{0}}{N_1^{(\textcolor{teal}{\ell})F}}\right)^2}  \textcolor{blue}{\frac {\textcolor{teal}{\ell}}2\left(\frac{\alpha_F}{2}\right)^2 N^{(\textcolor{teal}{\ell})F}_1{\color{red}\zeta(2)}}   +\textcolor{violet}{\left(\frac{N^{(\textcolor{teal}{\ell})F}_{0}}{N_1^{(\textcolor{teal}{\ell})F}}\right)^1} \textcolor{blue}{\frac {\textcolor{teal}{\ell}}3\left(\frac{\alpha_F}{2}\right)^3 N^{(\textcolor{teal}{\ell})F}_1}{\color{red}\zeta(3)}{\color{violet}}\\&{-\color{violet}\frac{1}{0!}\left(\frac{N^{(\textcolor{teal}{\ell})F}_{0}}{N_1^{(\textcolor{teal}{\ell})F}}\right)^0}{\color{blue}\frac {\textcolor{teal}{\ell}}4\left(\frac{\alpha_F}{2}\right)^4 N^{(\textcolor{teal}{\ell})F}_1{\color{red}\left(\zeta(4)-\frac{\textcolor{teal}{\ell}}{2}\zeta(2)^2\right)}}.
	\end{align*}
	\caption{Free-energy zeta-number patters, going down any diagonal $\ell$. We have used the same color coding as before, in order to highlight key aspects of these patterns---as usual, in red, one notices the increasing amount of zeta-number combinations.}
	\label{fig:FEpatternsPII}
\end{figure}

All we have left to address is the Stokes vector field $\underline{\boldsymbol{S}}_{\theta}^F (\boldsymbol{\sigma})$, which by-now is computed straightforwardly. Evaluating all relevant sums in the usual way yields (compare \eqref{eq:PI-F-S0-final}-\eqref{eq:PI-F-Spi-final})
\bea
\underline{\boldsymbol{S}}_{0}^F \left( \bm{\sigma}^F \right) &=& \frac{1}{2}  \left[ \begin{array}{c} \left( \sigma_2^F \right)^{-1} \left( 1 + \upmu^F \frac{\rmd}{\rmd \upmu^F} \log N^{(1)F} (\upmu^F) \right) \\ - \sigma_2^F\, \frac{\rmd}{\rmd \upmu^F} \log N^{(1)F} (\upmu^F) \end{array} \right] \log \left( 1 + 2 \sigma_2^F\, N^{(1)F} (\upmu^F)\right), \\
\underline{\boldsymbol{S}}_{\pi}^F \left( \boldsymbol{\sigma}^F \right) &=& - \frac{1}{2} \left[ \begin{array}{c} -\sigma_1^F\, \frac{\rmd}{\rmd \upmu^F} \log N^{(-1)F} (\upmu_F) \\ \left( \sigma_1^F \right)^{-1} \left( 1 + \upmu^F \frac{\rmd}{\rmd \upmu^F} \log N^{(-1)F} (\upmu^F) \right) \end{array} \right] \log \left( 1 - 2\sigma_1^F\, N^{(-1)F} (\upmu^F)\right),
\eea
\noindent
with the standard simplifications at either vanishing $\sigma_1^F$/$\sigma_2^F$ (compare \eqref{eq:PI-F-S0-simple-1}-\eqref{eq:PI-F-S0-simple-2}-\eqref{eq:PI-F-Spi-simple-2}-\eqref{eq:PI-F-Spi-simple-1}),
\bea
\underline{\boldsymbol{S}}_{0}^F \left(\sigma_1^F,0\right) &=& \left[ \begin{array}{c} 1 \\ 0 \end{array} \right], \\
\underline{\boldsymbol{S}}_{0}^F \left(0,\sigma_2^F\right) &=& \frac{1}{2} \log \left( 1 + 2\sigma_2^F \right) \left[ \begin{array}{c} \left( \sigma_2^F \right)^{-1} \\ - \sigma_2^F\, N_0^{(1)F} \end{array} \right], \\
\underline{\boldsymbol{S}}_{\pi}^F \left(0,\sigma_2^F\right) &=& \left[ \begin{array}{c} 0 \\ 1 \end{array} \right], \\
\underline{\boldsymbol{S}}_{\pi}^F \left(\sigma_1^F,0\right) &=& - \frac{1}{2} \log \left( 1  -2\sigma_1^F \right) \left[ \begin{array}{c} -\sigma_1^F\, N_0^{(-1)F} \\ \left( \sigma_1^F \right)^{-1} \end{array} \right].
\eea

Finally, we may recover the discussion we made at the end of subsection~\ref{subsec:2DQGresults}; asking if there might be better redefinitions of transseries parameters where more (if not all) Stokes vectors turn out to solely have integer entries? The answer is pretty much the same: our parametrization is the one in which the number of integer-valued Stokes vectors is maximized. Moving towards Borel residues, again we find similar results: the only such numbers (among the ones of the form $\mathsf{S}^{(F)}_{(n,n)\to(p+\ell,p)}$) that turn out to be (negative) integers are the Borel residues of the form $\mathsf{S}^{(F)}_{(n,n)\to(n+\ell,n)}$. A couple of examples illustrating the cumbersome structure of Borel residues may be found in table~\ref{tab:P2Borel} for \PII, in table~\ref{tab:P2BorelF} for the rescaled version corresponding to the free energy, whereas in table~\ref{tab:P2BorelF-int} we display some Borel residues which are actually integers. As in the previous \PI~case, we will present results regarding the generating functions for Borel residues in section \ref{sec:stokes}.

\begin{landscape}
\begin{table}
\centering
$\begin{array}{c|c|c|c}
-\mathsf{S}_{(n,n)\to(\ell,0)} & \ell=1 & \ell=2 & \ell=3 \\ \hline
n=0 & -\frac{\rmi}{\sqrt{2 \pi }} & -\frac{1}{2 \pi } & \frac{\rmi}{2 \sqrt{2} \pi ^{3/2}} \\ \hline
n=1 & -4 \rmi \sqrt{\frac{2}{\pi }} (\gamma_{\text{E}} +\log 16) & -\frac{2 (6 \gamma_{\text{E}} -\rmi \pi +6 \log 16)}{\pi } & \frac{3 \sqrt{2} (4 \rmi \gamma_{\text{E}} +\pi +4 \rmi \log 16)}{\pi ^{3/2}} \\ \hline
n=2 & -2 \rmi \sqrt{\frac{2}{\pi }} \left(6 (\gamma_{\text{E}} +\log 16)^2-\pi ^2\right) & \frac{8 \left(-12 \gamma_{\text{E}} ^2+3 \rmi \gamma_{\text{E}}  \pi +\pi ^2-12 (\log 16)^2-24 \gamma_{\text{E}}  \log 16+3 \rmi \pi  \log 16\right)}{\pi } & \frac{2\sqrt{2} \left(-7\rmi\pi^2+36\pi(\gamma_E+\log16)+90\rmi(\gamma_E+\log16)^2\right)}{	\pi ^{3/2}}
\end{array}$
\caption{Some Borel residues on the main diagonal, for the \PII~specific-heat. The complexity of these numbers increases with $n$ (more Stokes vectors from the ``bulk'' of the lattice may contribute) while $\ell$ has no apparent effect.}
	\label{tab:P2Borel}
\end{table}
\begin{table}
\centering
$\begin{array}{c|c|c|c}
-\mathsf{S}_{(n,n)\to(\ell,0)}^{F} & \ell=1 & \ell=2 & \ell=3 \\ \hline
n=0 & 1 & 1 & 1 \\ \hline
n=1 & -\frac{4\rmi}{\pi} (\gamma_{\text{E}} +\log 16) & -2-\frac{12\rmi}\pi (\gamma_{\text{E}}+\log 16) & -6-\frac{24\rmi}\pi (\gamma_{\text{E}}+\log 16)
\end{array}$
\caption{The same Borel residues, now for the free energy of 2d quantum supergravity. Residues starting at the $(0,0)$ perturbative sector are trivially $1$, but all other residues have complicated expressions (albeit slightly less complicated than the previous case).}
	\label{tab:P2BorelF}
\end{table}
\begin{table}
\centering
$\begin{array}{c|c|c|c|c|c}
-\mathsf{S}^{F}_{(n,n)\to(n+\ell,n)} & \ell=1 & \ell=2 & \ell=3 & \ell=4 & \ell=5 \\ \hline
n=0 & 1 & 1 & 1 & 1 & 1 \\ \hline
n=1 & 2 & 3 & 4 & 5 & 6 \\ \hline
n=2 & 3 & 6 & 10 & 15 & 21 \\ \hline
n=3 & 4 & 10 & 20 & 35 & 56 \\ \hline
n=4 & 5 & 15 & 35 & 70 & 126
\end{array}$
\caption{A larger set of integer-valued Borel residues for the free energy of 2d quantum supergravity. These are associated to transitions from a diagonal $(n,n)$-sector into an off-diagonal $(n+\ell,n)$-sector, and are exactly the same as the ones for 2d quantum gravity in table~\ref{tab:P1BorelF-int}---in fact, due to normalization, the only relevant Stokes vector is precisely the same in the two problems. As such, also the same considerations we had concerning table~\ref{tab:P1BorelF-int} hold herein.}
	\label{tab:P2BorelF-int}
\end{table}
\end{landscape}

\subsection{Alien Algebraic Structures from Stokes Data}\label{subsec:algebra}

Having computed full Stokes data for both \PI~and \PII, alongside their quantum gravity or minimal string incarnations, one may next ask what can one do with it. The one obvious answer, which we already started tackling back in subsection~\ref{subsec:stokes-trans}, is to compute Stokes transitions or connection formulae associated to the respective Stokes automorphisms. This we shall do in detail in upcoming section~\ref{sec:stokes}. But, as discussed in subsection~\ref{subsec:stokes-trans}, the road towards Stokes transitions starts with the alien derivative as in \eqref{eq:alien-bridge-equation-NR} or in \eqref{eq:alien_def_vectors} in the resonant case. Let us then address the computation of arbitrary alien derivatives in the present subsection.

One natural question---which we already alluded to in passing just before \eqref{eq:585}, and which in fact was pivotal to start our Stokes data calculation---is whether the commutator of two alien derivatives, say $\Delta_{\boldsymbol{n} \cdot \boldsymbol{A}}$ and $\Delta_{\boldsymbol{m} \cdot \boldsymbol{A}}$, with $\bm{m} \neq -\bm{n}$, will still result in an alien derivative, $\Delta_{\left( \boldsymbol{n}+\boldsymbol{m} \right) \cdot \boldsymbol{A}}$, or not. This was discussed in detail in \cite{abs18} and let us now briefly review it also to put \eqref{eq:585} in perspective. In analogy with the (non-resonant) alien derivative \eqref{eq:alien-bridge-equation-NR}---recall
\be
\label{eq:alien}
\Delta_{\boldsymbol{\ell} \cdot \boldsymbol{A}} \Phi_{\boldsymbol{n}} = \boldsymbol{S}_{\boldsymbol{\ell}} \cdot \left( \boldsymbol{n} + \boldsymbol{\ell} \right) \Phi_{\boldsymbol{n} + \boldsymbol{\ell}}
\ee
\noindent
---introduce the linear operator $G_{\bm{\ell}} \left(\bm{v}\right)$ (with $\bm{\ell} \in \BZ^2$ and $\bm{v} \in \BC^2$) acting on arbitrary transseries sectors through \cite{abs18}
\be
\label{eq:operator}
G_{\bm{\ell}} \left( \bm{v} \right) \Phi_{\boldsymbol{n}} = \bm{v} \cdot \left( \boldsymbol{n} + \boldsymbol{\ell} \right) \Phi_{\boldsymbol{n} + \boldsymbol{\ell}}.
\ee
\noindent
The linear operator $G_{\bm{\ell}} \left(\bm{v}\right)$ will match the alien derivative $\Delta_{\boldsymbol{\ell} \cdot \boldsymbol{A}}$ if $\bm{v} = \boldsymbol{S}_{\boldsymbol{\ell}}$ (plus the usual properties of Stokes data). The reason to introduce these more general operators is that they form a Lie algebra with standard commutator, as can be easily checked \cite{abs18}
\be
\left[ G_{\bm{n}} \left(\bm{v}\right), G_{\bm{m}} \left(\bm{u}\right) \right] = G_{\bm{n}+\bm{m}} \big( \left( \bm{v} \cdot \bm{m} \right) \bm{u} - \left( \bm{u} \cdot \bm{n} \right) \bm{v} \big).
\ee
\noindent
The above alien-derivation operators are certainly a subset of the $G_{\bm{\ell}} \left(\bm{v}\right)$ operators of the form $G_{\bm{\ell}} \left( \boldsymbol{S}_{\boldsymbol{\ell}} \right)$---but not necessarily form a Lie subalgebra. In fact, for these one finds \cite{abs18}
\be
\left[ G_{\bm{n}} \left( \boldsymbol{S}_{\boldsymbol{n}} \right), G_{\bm{m}} \left( \boldsymbol{S}_{\boldsymbol{m}} \right) \right] = G_{\bm{n}+\bm{m}} \big( \left( \boldsymbol{S}_{\boldsymbol{n}} \cdot \bm{m} \right) \boldsymbol{S}_{\boldsymbol{m}} - \left( \boldsymbol{S}_{\boldsymbol{m}} \cdot \bm{n} \right) \boldsymbol{S}_{\boldsymbol{n}} \big)
\ee
\noindent
and the right-hand side is an alien derivative if and only if \cite{abs18}
\be
\label{eq:non_res_algebra}
\left( \boldsymbol{S}_{\boldsymbol{n}} \cdot \bm{m} \right) \boldsymbol{S}_{\boldsymbol{m}} - \left( \boldsymbol{S}_{\boldsymbol{m}} \cdot \bm{n} \right) \boldsymbol{S}_{\boldsymbol{n}} \propto \boldsymbol{S}_{\boldsymbol{n}+\boldsymbol{m}}.
\ee

In the case where there is resonance, \eqref{eq:alien-bridge-equation-NR} or \eqref{eq:alien} need to be upgraded to \eqref{eq:alien_def_vectors}---recall
\be
\Delta_{\bm{\ell} \cdot \bm{A}} \Phi_{\bm{n}} = \sum_{\bm{p} \in \ker \mathfrak{P}} \bm{S}_{\bm{\ell}+\bm{p}} \cdot \left( \bm{n}+\bm{\ell}+\bm{p} \right) \Phi_{\bm{n}+\bm{\ell}+\bm{p}}.
\ee
\noindent
The condition for alien derivatives to form a Lie subalgebra is naturally now different. Computing the commutator
\bea
\label{eq:res-com-com}
\left[ \Delta_{\bm{n} \cdot \bm{A}}, \Delta_{\bm{m} \cdot \bm{A}} \right] \Phi_{\bm{k}} &=& \\
&=& \sum_{\bm{p},\bm{q} \in \ker\mathfrak{P}} \Big\{ \big( \boldsymbol{S}_{\boldsymbol{n}+\bm{p}} \cdot \left( \bm{m}+\bm{q} \right) \big)\, \boldsymbol{S}_{\boldsymbol{m}+\bm{q}} - \big( \boldsymbol{S}_{\boldsymbol{m}+\bm{q}} \cdot \left( \bm{n}+\bm{p} \right) \big)\, \boldsymbol{S}_{\boldsymbol{n}+\bm{p}} \Big\} \cdot \nonumber \\
&&
\label{eq:res-com-stnd}
\cdot \left( \bm{k} + \bm{n} + \bm{m} + \bm{p} + \bm{q} \right) \Phi_{\bm{k} + \bm{n} + \bm{m} + \bm{p} + \bm{q}} = \\
&&
\hspace{-60pt}
= \sum_{\bm{p} \in \ker\mathfrak{P}}\left( \sum_{\bm{q} \in \ker\mathfrak{P}} \Big\{ \big( \boldsymbol{S}_{\boldsymbol{n}+\bm{p}-\bm{q}} \cdot \left( \bm{m}+\bm{q} \right) \big)\, \boldsymbol{S}_{\boldsymbol{m}+\bm{q}} - \big( \boldsymbol{S}_{\boldsymbol{m}+\bm{q}} \cdot \left( \bm{n}+\bm{p}-\bm{q} \right) \big)\, \boldsymbol{S}_{\boldsymbol{n}+\bm{p}-\bm{q}} \Big\} \right) \cdot \nonumber \\
&&
\label{eq:res-com-ker}
\hspace{-60pt}
\cdot \left( \bm{k} + \bm{n} + \bm{m} + \bm{p} \right) \Phi_{\bm{k} + \bm{n} + \bm{m} + \bm{p}},
\eea
\noindent
and comparing it with
\be
\label{eq:alien-der-resonant-sum-over-kernel-explicit}
\Delta_{(\bm{n}+\bm{m}) \cdot \bm{A}} \Phi_{\bm{k}} = \sum_{\bm{p} \in \ker \mathfrak{P}} \bm{S}_{\bm{n}+\bm{m}+\bm{p}} \cdot \left( \bm{k}+\bm{n}+\bm{m}+\bm{p} \right) \Phi_{\bm{k}+\bm{n}+\bm{m}+\bm{p}},
\ee
\noindent
we see that that closure of the alien algebra in the resonant case now requires (compare with the simpler non-resonant \eqref{eq:non_res_algebra} above)
\be
\label{eq:algebra_res}
\sum_{\bm{q} \in \ker\mathfrak{P}} \Big\{ \big( \boldsymbol{S}_{\boldsymbol{n}+\bm{p}-\bm{q}} \cdot \left( \bm{m}+\bm{q} \right) \big)\, \boldsymbol{S}_{\boldsymbol{m}+\bm{q}} - \big( \boldsymbol{S}_{\boldsymbol{m}+\bm{q}} \cdot \left( \bm{n}+\bm{p}-\bm{q} \right) \big)\, \boldsymbol{S}_{\boldsymbol{n}+\bm{p}-\bm{q}} \Big\} \propto \boldsymbol{S}_{\boldsymbol{n}+\boldsymbol{m}+\bm{p}},
\ee
\noindent
where if to fit with \eqref{eq:alien-der-resonant-sum-over-kernel-explicit} the proportionality factor \textit{cannot} depend on $\bm{p} \in \ker \mathfrak{P}$. Further note that \eqref{eq:non_res_algebra} does not imply \eqref{eq:algebra_res} and vice-versa.
Such conditions, either \eqref{eq:non_res_algebra} or \eqref{eq:algebra_res}, of course need not be verified---and if they are this is pointing towards extra structure at the level of Stokes data which was not obvious from the starting equation itself. One can then simply check if such non-trivial conditions on Stokes data hold or not. In our context of \PI~and \PII~equations we surprisingly find that \eqref{eq:non_res_algebra} does hold, but \eqref{eq:algebra_res} does not (this is ``doubly surprising'' as we are finding the simpler condition one would expect for closure of a non-resonant alien algebra; not the one for closure of a resonant alien algebra). In particular, an algebra formed by operators $G_{\bm{\ell}} \left( \bm{S}_{\bm{\ell}} \right)$ in \eqref{eq:operator} will close. As explained in subsection~\ref{subsec:vec_struct}, this directly yields the structure of the Stokes vectors (together with the structure of the one-action vectors, that was found in \cite{asv11}). More specifically, what we have actually observed is that the following relation holds:
\be
\left( \boldsymbol{S}_{\boldsymbol{n}} \cdot \bm{m} \right) \boldsymbol{S}_{\boldsymbol{m}} - \left( \boldsymbol{S}_{\boldsymbol{m}} \cdot \bm{n} \right) \boldsymbol{S}_{\boldsymbol{n}} = \left( p\, \ell' - q\, \ell \right) \frac{N_{1-p}^{(\ell)}\, N_{1-q}^{(\ell')}}{N_{1-p-q}^{(\ell+\ell')}}\, \boldsymbol{S}_{\boldsymbol{n}+\boldsymbol{m}},
\ee
\noindent
with $\bm{n} = \left( 1-p,1-p-\ell \right)$ and $\bm{m} = \left( 1-q,1-q-\ell' \right)$ (and likewise for backward vectors), a somewhat unexpected result emerging from our data.

We have seen throughout this section that whilst individual Stokes vectors may look complicated and intricate, their (global) generating functions may be depicted quite easily. Hence perhaps the above discussion of looking at the individual alien derivatives is also being clouded with unnecessary complications, and we should be looking at directional (pointed) alien derivatives instead (the ones which are actually directly relevant for the Stokes automorphisms). Due to the bridge equation, recall \eqref{eq:bridge-original} or \eqref{eq:bridge-directional}, alien derivations may be rewritten as vector fields in $\bm{\sigma}$-space, in which case computing their commutator algebra becomes quite standard. The directional (pointed) alien derivative \eqref{eq:bridge-directional}
\be
\underline{\dot\Delta}_{\theta} \Phi = \underline{\boldsymbol{S}}_{\theta} (\boldsymbol{\sigma}) \cdot \frac{\partial \Phi}{\partial\boldsymbol{\sigma}}
\ee
\noindent
is non-trivial along the Stokes lines $\theta_{\text{S}} = 0, \pi$. There is hence one non-trivial commutator to compute, which is $[ \underline{\dot\Delta}_0, \underline{\dot\Delta}_\pi ]$. Using the results of this section, we can compute this commutator explicitly: it turns out it can always be written as
\be
[ \underline{\dot\Delta}_0, \underline{\dot\Delta}_\pi ] = a_0 (\sigma_1,\sigma_2)\, \underline{\dot\Delta}_0 + a_\pi (\sigma_1,\sigma_2)\, \underline{\dot\Delta}_\pi,
\ee
\noindent
but where the functions $a_0$ and $a_\pi$ turn out to be complicated functions of $\sigma_1$ and $\sigma_2$. As an example, we have in the case of the \PI~free energy\footnote{We chose this example simply because it leads to the most compact formulae. For ease of notation we herein use the functions $\rho_0$, $\rho_\pi$, to be introduced in \eqref{eqn:coord_chan-b}, \eqref{eqn:coord_chan-c}. These are just different coordinates on  $(\sigma_1,\sigma_2)$-space.}
\bea
a_0 &=& \left( \frac{\rho_0 \left(1+\rho_\pi\right)}{\rho_\pi \left(1+\rho_0\right)}\, \frac{1}{\log \left(1+\rho_0\right)} - \frac{1}{\rho_\pi} \right) \log \left( 1-\frac{\rho_\pi}{\rho_0} \right), \\
a_\pi &=& \frac{\log \left(1+\rho_0\right)}{\rho_\pi-\rho_0} \left( \frac{\rho_0}{\rho_\pi} - 1 + \frac{1+\rho_\pi}{\log \left(1-\frac{\rho_\pi}{\rho_0}\right)} \right).
\eea
\noindent
Unfortunately, the commutators we found are not the cleanest formulae. In principle one would now like to use the Stokes automorphism $\underline{\mathfrak{S}}_{\theta}$ in \eqref{eq:Stokes-auto-alien-der}, as
\be
\underline{\mathfrak{S}}_{\theta} = \exp \underline{\dot{\Delta}}_{\theta},
\ee
\noindent
and try to write some sort of wall-crossing formulae (\textit{e.g.}, via Baker--Campbell--Hausdorff); but due to the unfriendly nature of our above result we could not find any such expression.

\section{Numerical Checks of Stokes Data: Overview}\label{sec:numerics}

Having written down generating functions for the complete, resurgent Stokes data of \PI~and \PII, \eqref{eq:gen_fun-a} and \eqref{eq:PII-gen_fun-a}, respectively, it is now time to test the validity of our conjectures. Short of a fully rigorous proof (but see section~\ref{sec:stokes}) we will resort to numerics. As already mentioned in the previous section, our generating functions produce numbers which match against known numerical results \cite{gikm10, asv11, sv13, as13} to very high precision. Herein, we want to improve on those tests and checks, by presenting a complete and systematic approach to the numerical validation of analytical Stokes data (and which might be of broader use than just our Painlev\'e story). Two comments before proceeding. The first is that some of the following numerics actually \textit{started} our analysis: it was from numerics that some of the conjectures in the previous section were born. The second is that there is \textit{a lot} of numerics involved; hence, in order to keep readability, we present an overview of the numerics in the present section and we have moved the more technical details of the computations into appendix~\ref{app:numerics}. All methods were implemented in \textit{Wolfram Mathematica}.

\subsection{Asymptotics and Large-Order Analysis}\label{subsec:large-order}

Very much in the spirit of \cite{gikm10, asv11, sv13}, let us begin with numerical analysis of large-order asymptotics. In the present context, this is very much based upon the asymptotic expression \eqref{eq:fin_as}, dictating the (resurgent) large-order behavior of \textit{diagonal} transseries coefficients. As explained in section~\ref{sec:asymptotics}, out of this formula one may compute the minimal set of Borel residues which allow us to reconstruct arbitrary Stokes vectors. Let us next give an overview of its associated large-order analysis (more details may be found in subappendix~\ref{app:subsec:large-order}), which we shall later use to double-check results from our main numerical method (to be described in subsection~\ref{subsec:borel-plane}).

Let us start with the asymptotic relation \eqref{eq:fin_as} and let us turn it into an equation involving quantities which are easily evaluated numerically. If we divide both sides of \eqref{eq:fin_as} by $\widetilde{H}_{0} \left( 2g + n - \frac{\ell}{2}, \ell A \right)$, one may reorganize the result as
\bea
&&
\hspace{-10pt}
- \frac{\rmi\pi\, u_{4g}^{(n,n)[0]}}{\widetilde{H}_{0} \left( 2g + n - \frac{\ell}{2}, \ell A \right)} - \sum_{t=1}^{\ell-1} \sum_{h=0}^{+\infty} \sum_{p=0}^n \sum_{k=0}^p \mathsf{S}_{(n,n)\to(p+t,p)}\, \frac{u_{2h}^{(p+t,p)[k]}}{(-2)^k}\, \frac{\widetilde{H}_{k} \left( 2g + n - h - \beta_{(p+t,p)}^{(k)}, tA \right)}{\widetilde{H}_{0} \left( 2g + n - \frac{\ell}{2}, \ell A \right)} \simeq \nonumber \\
&&
\hspace{-10pt}
\label{eq:asy_div}
\simeq \sum_{t=\ell}^{+\infty} \sum_{h=0}^{+\infty} \sum_{p=0}^n \sum_{k=0}^p \mathsf{S}_{(n,n)\to(p+t,p)}\, \frac{u_{2h}^{(p+t,p)[k]}}{(-2)^k}\, \frac{\widetilde{H}_{k} \left( 2g + n - h - \beta_{(p+t,p)}^{(k)}, tA\right)}{\widetilde{H}_{0} \left( 2g + n - \frac{\ell}{2}, \ell A \right)}.
\eea
\noindent
Now consider the limit $g \to +\infty$ of this expression. The ratio of $\widetilde{H}_{k}$ functions may be expanded asymptotically as
\be
\label{eq:hratio}
\frac{\widetilde{H}_{k} \left( 2g + a, tA \right)}{\widetilde{H}_{0} \left( 2g + b, \ell A \right)} \simeq \left(\frac{\ell}{t} \right)^{2g} g^{a-b}\, \sum_{q=0}^k \sum_{h=0}^{+\infty}  c_{h,q}^{(k)} \left( a,b,t,\ell \right)\, \frac{\log^{q} g}{g^h},
\ee
\noindent
for some coefficients $c_{h,q}^{(k)}$ that depend on all other variables (subappendix~\ref{app:subsec:large-order} includes a recursive definition of these coefficients). This implies that, in the limit $g \to +\infty$, one may neglect all terms in the right-hand side of \eqref{eq:asy_div} with $t>\ell$, as they are suppressed. This turns out to be a quite relevant step which will later allow us to compute Borel residues inductively. We can then truncate the right-hand side of \eqref{eq:asy_div} by keeping only the $t=\ell$ term.

Next let us evaluate \eqref{eq:asy_div} at finite (but large) values of $g$. This presents a technical difficulty; namely, the sums
\be
\label{eq:ratio_sum_div}
\sum_{h=0}^{+\infty} \sum_{k=0}^p \frac{u_{2h}^{(p+t,p)[k]}}{(-2)^k}\, \frac{\widetilde{H}_{k} \left( 2g + n - h - \beta_{(p+t,p)}^{(k)}, tA \right)}{\widetilde{H}_{0} \left( 2g + n - \frac{\ell}{2},\ell A \right)}
\ee
\noindent
are in general divergent due to the factorial growth of the coefficients $u_{2h}^{(p+t,p)[k]}$. Using the expansion in \eqref{eq:hratio}, however, one can rewrite the above double-sum as
\be
\left( \frac{\ell}{t} \right)^{2g}\, \sum_{h=0}^{+\infty} \sum_{k=0}^p \sum_{q=0}^k \sum_{r=0}^{+\infty} \frac{u_{2h}^{(p+t,p)[k]}}{(-2)^k}\, g^{\frac{\ell}{2}-h-\beta^{(k)}_{(p+t,p)}}\, c_{r,q}^{(k)} \left( n - h - \beta^{(k)}_{(p+t,p)}, n - \frac{\ell}{2}, t, \ell \right) \frac{\log^q g}{g^r}.
\ee
\noindent
To this expression we perform the Cauchy product by substituting $h+r \to h$ in the exponent of $g$, and reshuffle the finite sums to obtain
\be
\left( \frac{\ell}{t} \right)^{2g}\, \sum_{h=0}^{+\infty} \sum_{q=0}^p \sum_{k=q}^p \frac{\log^q g}{g^{h+\beta_{(p+t,p)}^{(k)}-\frac{\ell}{2}}}\, \sum_{r=0}^h \frac{u_{2(h-r)}^{(p+t,p)[k]}}{(-2)^k}\, c_{r,q}^{(k)} \left( n - h - \beta^{(k)}_{(p+t,p)}, n - \frac{\ell}{2}, t, \ell \right).
\ee
\noindent
As such, the only problematic sum left to evaluate is now
\be
\sum_{k=q}^p \sum_{h=0}^{+\infty} \frac{1}{g^{h+\beta^{(k)}_{(p+t,p)}}}\, \sum_{r=0}^h u_{2(h-r)}^{(p+t,p)[k]}\, c^{(k)}_{r,q} \left( n - h - \beta^{(k)}_{(p+t,p)}, n -\frac{\ell}{2}, t, \ell \right).
\ee
\noindent
This is actually no longer a big deal, as it may simply be Borel resummed in order to yield finite results (a detailed description of this resummation may be found in subappendix~\ref{app:subsec:large-order}). Let $\Sigma_q(n,p,\ell,t,g)$ be this resulting Borel-resummed function. Then our initially problematic sum is now evaluated via the asymptotic equality
\be
\label{eq:ratio-B-resum}
\sum_{k=q}^p \sum_{h=0}^{+\infty} \frac{u_{2h}^{(p+t,p)[k]}}{(-2)^k}\, \frac{\widetilde{H}_{k} \left( 2g + n - h - \beta_{(p+t,p)}^{(k)}, tA \right)}{\widetilde{H}_{0} \left( 2g + n - \frac{\ell}{2}, \ell A \right)} \simeq \left( \frac{\ell}{t} \right)^{2g} g^{\frac{\ell}{2}}\, \sum_{q=0}^p \Sigma_q \left(n,p,\ell,t,g\right) \log^q g.
\ee
\noindent
Herein, the left-hand side is an asymptotic series whereas the right-hand side is a function (approximated through a numerical procedure) which may be evaluated at finite values of $g$.

We finally return to our starting asymptotic relation \eqref{eq:asy_div}, which we may now write in terms of finite quantities:
\bea
&&
- \frac{\rmi\pi u_{4g}^{(n,n)[0]}}{\widetilde{H}_{0} \left( 2g+n-\frac{\ell}{2},sA \right)} - \sum_{t=1}^{\ell-1} \sum_{p=0}^n \mathsf{S}_{(n,n)\to(p+t,p)} \left(\frac{\ell}{t}\right)^{2g} g^{\frac{\ell}{2}}\, \sum_{q=0}^p \Sigma_q \left(n,p,\ell,t,g\right) \log^q g = \nonumber \\
&&
\label{eq:asy_fin}
= \sum_{p=0}^n \mathsf{S}_{(n,n)\to(p+\ell,p)}\, g^{\frac{\ell}{2}}\, \sum_{q=0}^p \Sigma_q \left(n,p,\ell,\ell,g\right) \log^q g + o \left(\left(\frac{s}{s+1}\right)^g\right).
\eea
\noindent
We have used this expression in two ways: to compute Stokes data and compare them to our conjectures in section~\ref{sec:results}; and to test if the predicted Stokes data correctly describe the large-order behavior of the diagonal transseries coefficients $u_{4g}^{(n,n)[0]}$.

\subsubsection*{Computation of Stokes Data}

Let us then start by describing how to use \eqref{eq:asy_fin} to numerically compute Stokes data. In this process, we will first assume the vector structure \eqref{eq:vec_struct-a}-\eqref{eq:vec_struct-b}---we shall numerically check it later, using the method that we develop in subsection~\ref{subsec:borel-plane}. Note that even if in principle this vector structure can be checked via large-order analysis, we have instead decided to use our upcoming more powerful method to perform those checks, and just use large-order analysis to confirm results for the first diagonal. Suppose we wish to compute the vector $\boldsymbol{S}_{(1-p,1-p-\ell)}$, assuming that all Stokes vectors $\boldsymbol{S}_{(1-q,1-q-t)}$ with $q \leq p$ and $t<\ell$ are known, alongside the $\boldsymbol{S}_{(1-q,1-q-\ell)}$ for $q<p$. Further, construct this $\boldsymbol{S}_{(1-p,1-p-\ell)}$ using the Borel residue $\mathsf{S}_{(p,p)\to(\ell,0)}$ (with the Stokes vectors that we assume we already have, one can construct all Borel residues $\mathsf{S}_{(p,p)\to(q+t,q)}$ for $t<\ell$ and $q\leq p$, alongside the $\mathsf{S}_{(p,p)\to(q+\ell,q)}$ for $0<q\leq p$). To compute $\mathsf{S}_{(p,p)\to(\ell,0)}$, just write
\bea
&&
- \frac{1}{\Sigma_0 \left(n,0,\ell,\ell,g\right)}\, \frac{\rmi\pi g^{-\frac{\ell}{2}} u_{4g}^{(n,n)[0]}}{\widetilde{H}_{0} \left( 2g + n - \frac{\ell}{2}, \ell A \right)} - \sum_{t=1}^{\ell-1} \sum_{p=0}^n \mathsf{S}_{(n,n)\to(p+t,p)} \left(\frac{\ell}{t}\right)^{2g}\, \sum_{k=0}^q \frac{\Sigma_k \left(n,p,\ell,t,g\right)}{\Sigma_0 \left(n,0,\ell,\ell,g\right)}\, \log^k g - \nonumber \\
&&
\label{eq:asy_conv}
- \sum_{p=1}^n \mathsf{S}_{(n,n)\to(p+\ell,p)}\, g^{\frac{\ell}{2}}\, \sum_{k=0}^q \frac{\Sigma_k \left(n,p,\ell,\ell,g\right)}{\Sigma_0 \left(n,0,\ell,\ell,g\right)}\, \log^k g = \mathsf{S}_{(n,n)\to(\ell,0)} + o \left( \left(\frac{\ell}{\ell+1}\right)^g \right).
\eea
\noindent
The left-hand side above may in principle be computed for arbitrary values of $g$. As to the right-hand side, it only holds the unknown Borel residue. By taking $g$ large enough (eliminating the remainder factor, and improving the precision on the $\Sigma_k$ functions), this Borel residue follows. Further, employing numerical acceleration techniques---as for example  Richardson transforms (see subappendix~\ref{app:subsec:large-order} for details)---, we can obtain rather satisfactory answers for the Borel residue; and thus the Stokes vector.

Large-order asymptotics is, however, not without its own disadvantages. The remainder factor in \eqref{eq:asy_conv} becomes more and more significant as $\ell$ grows, making the extraction of higher-action Stokes data difficult. Furthermore, the computation of the coefficients in the asymptotic expansion of the $\Sigma_k$ function has a very-high computational cost. The method that we shall develop in the upcoming subsection~\ref{subsec:borel-plane} presents no such disadvantages and yields more precise results, hence the reason we already mentioned that we have used asymptotics only in order to check results for the first diagonals of \PI~and \PII~solutions. We report on the results of our computations, and a comparison with the data generated in section~\ref{sec:results}, in tables~\ref{tab:asymp1} and~\ref{tab:asymp2}.

\begin{table}
\centering
$
\begin{array}{c|c|c}
& \text{Computed value} & \text{Digits of agreement} \\ \hline
N_1^{(1)} & -0.371257624642845...\rmi & 45 \\ \hline
N_0^{(1)} & -2.439626908610028...\rmi & 39 \\ \hline
N_{-1}^{(1)} & -7.60857041604191...\rmi & 34 \\ \hline
N_{-2}^{(1)} & -15.1114015595599 ...\rmi & 28 \\ \hline
N_{-3}^{(1)} & -21.603496143998...\rmi & 22
\end{array}
$
\caption{Results from large-order analysis for \PI. We have computed the coefficients $N_{1-p}^{(1)}$ numerically and compared the results with the values predicted in section~\ref{sec:results}. Data have been obtained by selecting $g=248$ in \eqref{eq:asy_conv}, order $20$ Borel--Pad\'e for all resummations (but for $N_{-3}^{(1)}$, where we chose order $5$), and $2$ Richardson transforms. With these very low values, good results are obtained very quickly. In \cite{asv11}, part of these data have already been computed to a greater precision, using improved Borel--Pad\'e approximations. Via the map in \eqref{eq:old_new_not}, one finds that all data $S_1^{(p)}$ and $\widetilde{S}_1^{(p)}$ computed in \cite{asv11} may be reconstructed using the data in the present table.}
\label{tab:asymp1}
\end{table}

\begin{table}
\centering
$
\begin{array}{c|c|c}
& \text{Computed value} & \text{Digits of agreement} \\ \hline
N_1^{(1)} & -0.398942280401432...\rmi & 46 \\ \hline
N_0^{(1)} & -5.345514404419517...\rmi & 41 \\ \hline
N_{-1}^{(1)} & -30.562985224995706...\rmi & 35 \\ \hline
N_{-2}^{(1)} & -99.841024200127802...\rmi & 29
\end{array}
$
\caption{Results from large-order analysis for \PII. We have computed the coefficients $N_{1-p}^{(1)}$ numerically and compared the results with the values predicted in section~\ref{sec:results}. Data have been obtained by selecting $g=98$ in \eqref{eq:asy_conv}, order 20 Borel--Pad\'e for all resummations, and $5$ Richardson transforms. We are selecting a smaller value of $g$ than for \PI~because computing coefficients for \PII~has a very-high computational cost. With the data computed in the present table, we can reconstruct all data which was computed in \cite{sv13} (and in fact further predict new data with respect to what was found therein).}
\label{tab:asymp2}
\end{table}

\subsubsection*{Testing the Conjectures: Large-Order Growth}

Let us describe one other way to use large-order analysis in order to test our conjectures for \PI/\PII~Stokes data. Instead of extracting data from \eqref{eq:asy_conv}, we can use the following procedure.

In section~\ref{sec:asymptotics}, we tested the asymptotic (resurgence) relation \eqref{eq:fin_as} in a particular case, where knowledge of one particular Stokes vector (the one which is known analytically in literature) was sufficient to build all required Borel residues. Now, we can take the next step: assume \eqref{eq:fin_as} to hold as an asymptotic relation, and use it to validate Stokes data generated with our formulae. This is to say, if our Stokes data are indeed correct then the inclusion of further instanton contributions should improve the resulting precision of this asymptotic approximation.

The formula we shall use for the test is \eqref{eq:asy_fin} with $\ell=1$, and truncating the instanton sum at order $N$, \textit{i.e.},
\be
\label{eq:testasy}
\frac{\rmi\pi u_{4g}^{(n,n)[0]}}{\widetilde{H}_{0} \left( 2g + n - \frac{1}{2}, A \right)} \simeq - \sum_{t=1}^N \sum_{p=0}^n \mathsf{S}_{(n,n)\to(p+t,p)}\, t^{-2g}\, g^{\frac{1}{2}}\, \sum_{q=0}^p \Sigma_q \left(n,p,1,t,g\right) \log^q g + o \left( \left(N+1\right)^{-2g} \right).
\ee
\noindent
The remainder term gets less significant at increasing $N$, which means that higher instanton contributions may be used to improve the precision on the prediction for the quantity on the left-hand side. In order to keep notation consistent with the similar test which was done back in subsection~\ref{subsec:largeorder-asymp}, we will now use the following notation:
\bea
\widetilde{u}_{4g}^{(n)} &=& \frac{\rmi\pi}{N_1^{(1)}\, \widetilde{H}_{0} \left( 2g + n - \frac{1}{2}, A \right)}\, u_{4g}^{(n,n)[0]}, \\
\widetilde{u}_{4g}^{(n) \langle N\rangle} &=& - \sum_{t=1}^N \sum_{p=0}^n \frac{\mathsf{S}_{(n,n)\to(p+t,p)}}{N_1^{(1)}}\, t^{-2g}\, g^{\frac{1}{2}}\, \sum_{q=0}^p \Sigma_q \left(n,p,1,t,g\right) \log^q g.
\eea
\noindent
Equation \eqref{eq:testasy} is then rewritten as
\be
\widetilde{u}_{4g}^{(n)} = \widetilde{u}_{4g}^{(n) \langle N\rangle} + o \left( \left(N+1\right)^{-2g} \right).
\ee
\noindent
Due to the computational cost of computing coefficients for the above resummed functions, we resort to a procedure aimed at trying to keep the overall computational cost as low as possible. Every time we add an instanton contribution (\textit{i.e.}, raising the $N$ index in $\widetilde{u}_{4g}^{(n) \langle N\rangle}$), we first choose a low order for the Borel--Pad\'e approximation. We then raise this order, observing an improvement in the precision of the asymptotic prediction. This signals the fact that, at fixed $g$, the Borel--Pad\'e approximation is a larger source of error than the higher instanton contributions. Raising the order of the Borel--Pad\'e approximation will improve the precision up to some specific maximum order. After this, further raising the order will yield no improvement in the prediction for $\widetilde{u}_{4g}^{(n)}$. In order to increase its precision one then has to add another instanton, raising the $N$-index by $1$. Whenever this insertion of a new instanton further improves the precision, we conclude that the predicted Borel residue is correct (up to the specified precision).

\begin{figure}
\centering
\includegraphics[scale=0.35]{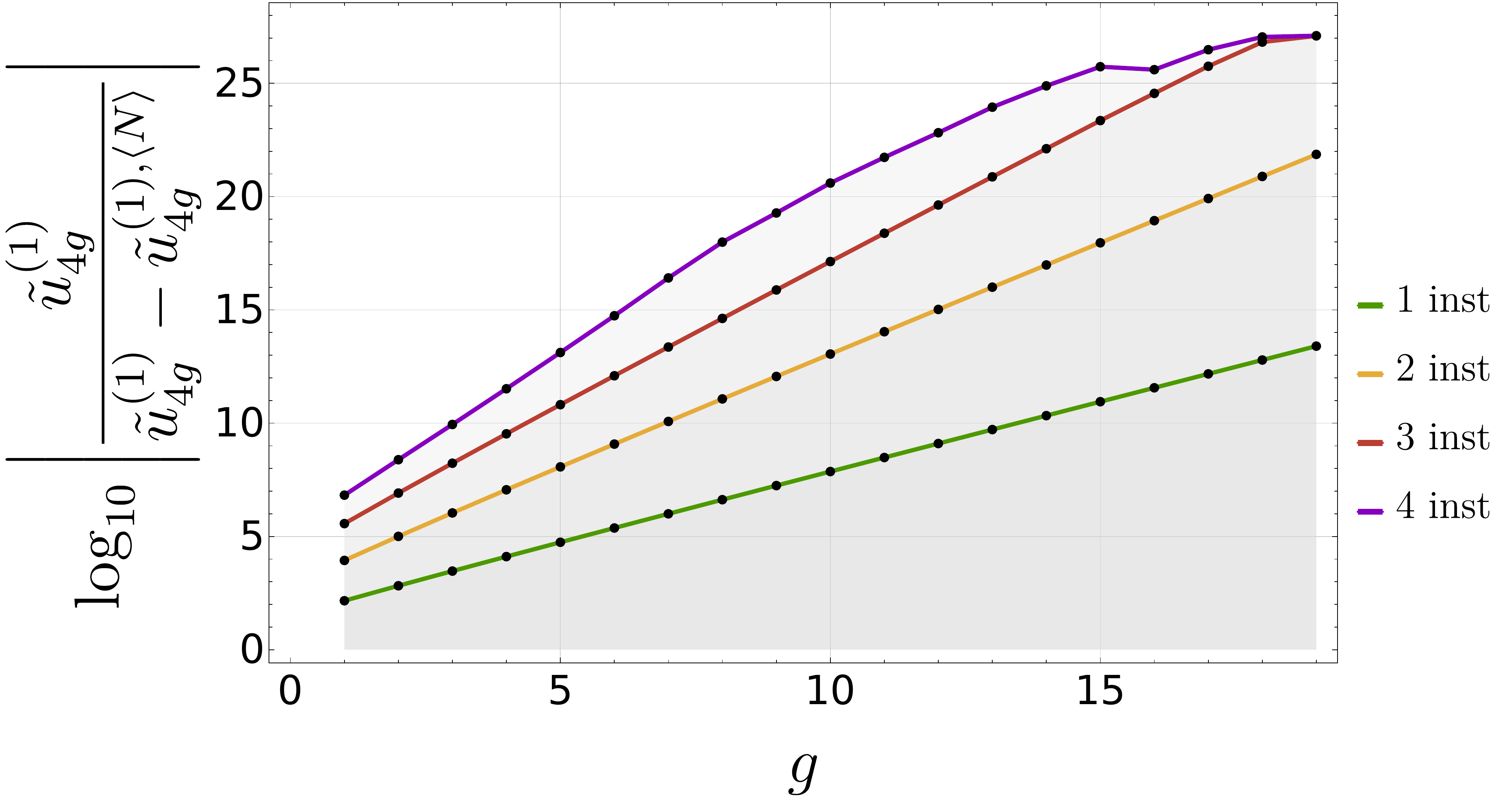}
\caption{Predicting the large-order growth of the coefficients $\widetilde{u}_{4g}^{(1)}$ for \PI. We have evaluated \eqref{eq:testasy} for $g=2,...,20$, with an order-$50$ Borel--Pad\'e for all resummations. The correction with one instanton already gives good precision, which is then improved by the correction with two instantons and so on. This test confirms that the data $N_1^{(1)}$, $N_0^{(1)}$, and $N_1^{(2)}$ are correct; up to our numerical precision. When $g$ becomes large the contribution from the fourth instanton ceases to improve the precision, because a better numerical precision in the integrals is then required.}
\label{fig:asy_testPI}
\end{figure}

\begin{figure}
\centering
\includegraphics[scale=0.35]{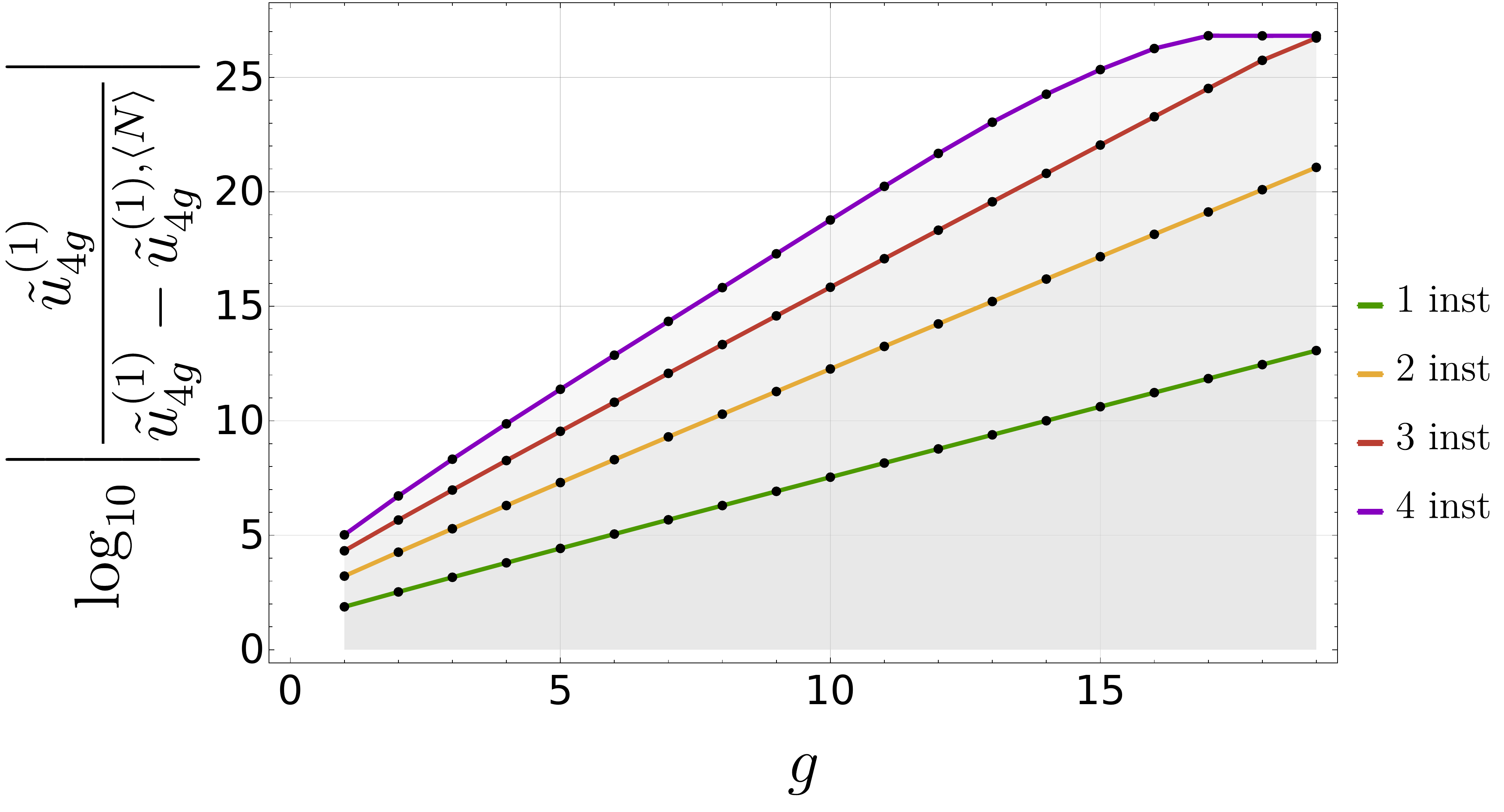}
\caption{Predicting the large-order growth of the coefficients $\widetilde{u}_{4g}^{(1)}$ for \PII. We have evaluated \eqref{eq:testasy} for $g=2,...,20$, with an order-$40$ Borel--Pad\'e for all resummations. Observation of the same trend as for \PI~leads to the same conclusion: the data $N_1^{(1)}$, $N_0^{(1)}$, and $N_1^{(2)}$ are correct; up to our numerical precision. The loss of precision at high $g$ has the same reason as before.}
\label{fig:asy_testPII}
\end{figure}

We report on this analysis for the $\Phi_{(1,1)}$ sectors in both \PI~and \PII~in figures~\ref{fig:asy_testPI} and~\ref{fig:asy_testPII}. We observe that for large values of $g$ our tests become more precise and require lower Borel--Pad\'e order for the resummations. Of course, there are disadvantages in raising $g$: we have to compute further coefficients $\widetilde{u}_{4g}^{(n)}$, and this has high computational cost. Furthermore, the numerical precision of the integrals has to be raised, as the corrections are becoming smaller---and this also adds to the overall computational cost. The true bottleneck of large-order asymptotics is the computation of coefficients for the functions which have to be resummed. The method we shall develop and present in the following subsection does not suffer from any of these issues, and will successfully produce many more Stokes data than the asymptotic method manages to.

\subsection{Stokes Data from Singularities on the Borel Plane}\label{subsec:borel-plane}

As already alluded to, if we are to see significant gains in numerical precision without unreasonable computational costs, then a more efficient numerical method is needed as compared to large-order asymptotics. The method we shall deploy is actually quite simple, basically extracting Stokes data straight from the (logarithmic) residues of the Borel singularities as in \eqref{eq:borel-sing-NR}. In order to understand its implementation we shall first explain it in the special case of a non-resonant system, where all its fundamentals may be neatly outlined. Then, we just upgrade it with the inclusion of resonance---which is needed for both \PI/\PII. The method then yields all our Borel residues, which of course encode the whole information for Stokes data.

\subsubsection*{The Setting for the Non-Resonant Case}

\begin{figure}[htbp]
	\centering
\begin{tikzpicture}[xscale=1]
\draw[thick, darkgray] (-0.8, 1.9)--(-0.5, 1.9);
\draw[thick, darkgray] (-0.5, 1.9)--(-0.5, 2.2);
\node at (-0.68, 2.1) {$s$};
\draw[->, ultra thick, darkgray] (-0.5,0) -- (13.5,0);
\draw[->, ultra thick, darkgray] (0,-1) -- (0,2);
\draw[violet, snake it, ultra thick] (3,0) -- (5.5,1);
\draw[violet, fill=violet] (3,0) circle (3pt);
\draw[blue, snake it, ultra thick] (6,0) -- (8.5,-1);
\draw[blue, fill=blue] (6,0) circle (3pt);
\draw[black!60!green, snake it, ultra thick] (9,0) -- (13,0.5);
\draw[black!60!green, fill=black!60!green] (9,0) circle (3pt);
\end{tikzpicture} 
	\caption{Schematic visualization of the first three branch-points (alongside their pictorial branch-cuts) on the complex Borel $s$-plane, for a simple resurgent function.}
	\label{fig:ResurgentFunctionOnBorelPlane}
\end{figure}
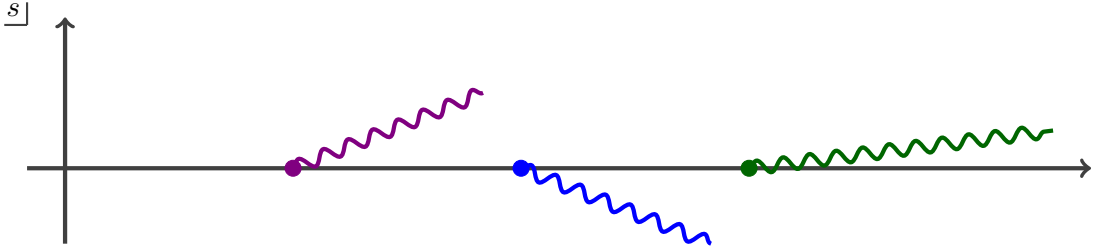

Let us start by introducing this method in the non-resonant case---which is structurally easier---, and let us focus on a case with a single Stokes line. On this line of singularities, let us then zoom-in on the first singularity, \textit{i.e.}, the one closest to the origin on the complex Borel plane (we shall later see how to access higher-action singularities). Further, choose conventions such that this Stokes line is located along the positive real line, associated with instanton action $A \in \mathbb{R}^+$. This setting is depicted in figure~\ref{fig:ResurgentFunctionOnBorelPlane}.

\subsubsection*{The First Singularity}

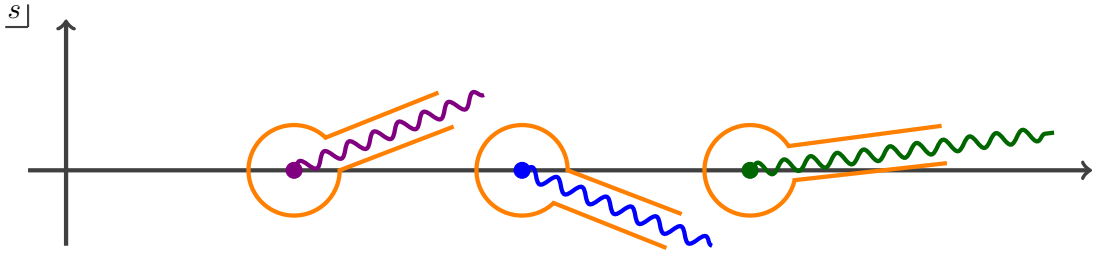
\begin{figure}
	\centering
\begin{tikzpicture}
\draw[thick, darkgray] (-0.8, 1.9)--(-0.5, 1.9);
\draw[thick, darkgray] (-0.5, 1.9)--(-0.5, 2.2);
\node at (-0.68, 2.1) {$s$};
\draw[->, ultra thick, darkgray] (-0.5,0) -- (13.5,0);
\draw[->, ultra thick, darkgray] (0,-1) -- (0,2);
\draw[violet, snake it, ultra thick] (3,0) -- (5.5,1);
\draw[violet, fill=violet] (3,0) circle (3pt);
\draw[blue, snake it, ultra thick] (6,0) -- (8.5,-1);
\draw[blue, fill=blue] (6,0) circle (3pt);
\draw[black!60!green, snake it, ultra thick] (9,0) -- (13,0.5);
\draw[black!60!green, fill=black!60!green] (9,0) circle (3pt);
\draw[ultra thick,orange] ([shift=(44.2:0.6cm)]3,0) arc (44.2:361:0.6cm);
\draw[ultra thick,orange] ([shift=(-1:0.6cm)]6,0) arc (-1:315:0.6cm);
\draw[ultra thick,orange] ([shift=(31:0.6cm)]9,0) arc (31:349:0.6cm);
\draw[orange, ultra thick] (3.415-2.586206*0.005,0.433-0.005) -- (4.9,1.03);
\draw[orange, ultra thick] (3.6-2.586206*0.0085,0-0.0085) -- (5.1,0.58);
\draw[orange, ultra thick] (6.6-2.586206*0.0085,0+0.0085) -- (8.1,-0.58);
\draw[orange, ultra thick] (6.415-2.586206*0.0075,-0.433+0.0075) -- (7.9,-1.03);
\draw[orange, ultra thick] (9.591-0.0285,-0.13-0.0285*0.11255) -- (11.59,0.095);
\draw[orange, ultra thick] (9.506-0.0188,0.321-0.0188*0.11255) -- (11.52,0.59);
\end{tikzpicture} 
	\caption{Visualization of possible Hankel contours for the branch-cuts depicted in figure~\ref{fig:ResurgentFunctionOnBorelPlane}.}
	\label{fig:ResurgentFunctionOnBorelPlaneWithHankelConours}
\end{figure}

As discussed in section~\ref{sec:stokes-resonance}, the Borel transform of any transseries sector $\Phi_{n}$, $\mathcal{B}[\Phi_{n}] (s)$, results in a germ which is holomorphic around the origin of the Borel complex $s$-plane. Its radius of convergence is dictated by $A$. Furthermore, this resulting function---the Borel transform---has the structure of a simple resurgent function \cite{e81, s07a, s07b, s14, abs18}. What this means is basically what we have already briefly discussed around \eqref{eq:borel-sing-NR}: the array of singularities along the Stokes line is labeled by an integer $\ell \in \mathbb{N}$, and each singularity essentially has the same structure which includes a simple pole\footnote{For simplicity we have omitted the pole in \eqref{eq:borel-sing-NR}, but herein it is convenient to be fully explicit. The $\phi$-\textit{coefficient} of the pole is associated to the so-called \textit{residual coefficient} of the asymptotic series (see, \textit{e.g.}, \cite{abs18} for more details).} and a logarithmic branch-cut,
\be
\label{eq:simpleResurgentFunctionOnBorelPlane}
\CB \left[ \Phi_{n} \right] (s) \Big|_{s = \ell A} \sim \mathsf{S}_{n \to n+\ell} \left( \frac{\phi_{n+\ell}}{2\pi\mathrm{i} \left(s - \ell A\right)} + \mathcal{B} \left[ \Phi_{n+\ell} \right] \left(s-\ell A\right)\, \frac{\log \left(s-\ell A\right)}{2\pi \mathrm{i}}\right)
\ee
\noindent
(recall that we are solely displaying the singular components of the Borel transform). Given the transseries coefficients, the above equation implies we know the full singular behavior of the Borel transform up to the Borel residues, $\mathsf{S}_{n \rightarrow n+\ell}$. These numbers are fixed by the unique analytic continuations of the power-series constructing the Borel transform $\mathcal{B}[\Phi_{n}] (s)$, by simply choosing appropriate Hankel contours and integrating both sides of \eqref{eq:simpleResurgentFunctionOnBorelPlane}---see figure~\ref{fig:ResurgentFunctionOnBorelPlaneWithHankelConours}. 

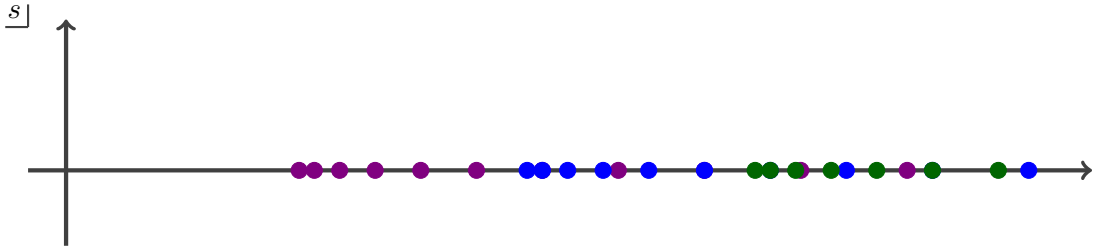
\begin{figure}
	\centering
	\begin{tikzpicture}[xscale=1]
\draw[thick, darkgray] (-0.8, 1.9)--(-0.5, 1.9);
\draw[thick, darkgray] (-0.5, 1.9)--(-0.5, 2.2);
\node at (-0.68, 2.1) {$s$};
\draw[->, ultra thick, darkgray] (-0.5,0) -- (13.5,0);
\draw[->, ultra thick, darkgray] (0,-1) -- (0,2);
	\foreach \n in {1,...,11}
	{
		\draw[violet, fill=violet] (3+\n^2/15,0) circle (3pt);
	};
	\foreach \n in {1,...,10}
	{
		\draw[blue, fill=blue] (6+\n^2/15,0) circle (3pt);
	};
	\foreach \n in {1,...,7}
	{
		\draw[black!60!green, fill=black!60!green] (9+\n^2/15,0) circle (3pt);
	}
	\end{tikzpicture} 
	\caption{Illustration of poles of the Pad\'e-approximant to a simple resurgent function, and their accumulation near its branch-points (further displaying tails of poles representing the branch-cuts). For the three branch-points of our example in figure~\ref{fig:ResurgentFunctionOnBorelPlane}, these numerical Pad\'e poles representing the branch-cuts end-up all falling on top of each other.}
	\label{fig:ResurgentFunctionPadeApproximation}
\end{figure}

Performing the aforementioned analytic continuations is, however, the main obstacle in calculating Borel residues analytically---since in most cases they cannot be explicitly constructed\footnote{As is the case for most nonlinear problems; and very much so in the present Painlev\'e contexts.}. One option is to then resort to some type of numerical approximation to these analytic continuations. Herein we will use a rather popular scheme: Pad\'e approximants. These will depict the Borel-transform logarithmic branch-cuts via accumulations of Pad\'e-poles on the complex plane; which is illustrated in figure~\ref{fig:ResurgentFunctionPadeApproximation}. Unfortunately, this procedure does not leave a lot of room to control exactly where the branch-cuts---\textit{i.e.}, the Pad\'e-poles representing them---will be located. Numerically, one observes that they end-up lying on top of each other while dispersing away from the origin. This simple fact prevents us from being able to isolate which poles represent which branch cut; hence from isolating Borel residues. This is also schematically illustrated in figure~\ref{fig:ResurgentFunctionPadeApproximation}, in the sense that, numerically, one cannot distinguish between the colors.

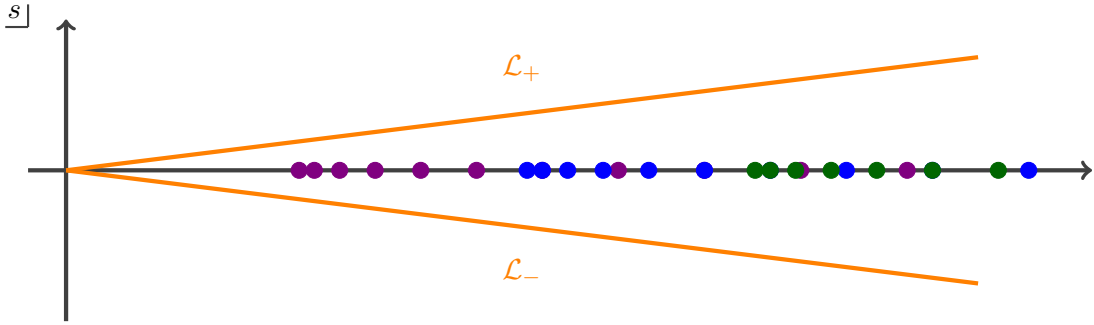
\begin{figure}
	\centering
	\begin{tikzpicture}[xscale=1]
	\draw[thick, darkgray] (-0.8, 1.9)--(-0.5, 1.9);
	\draw[thick, darkgray] (-0.5, 1.9)--(-0.5, 2.2);
	\node at (-0.68, 2.1) {$s$};
	\draw[->, ultra thick, darkgray] (-0.5,0) -- (13.5,0);
	\draw[->, ultra thick, darkgray] (0,-2) -- (0,2);
	\foreach \n in {1,...,11}
	{
		\draw[violet, fill=violet] (3+\n^2/15,0) circle (3pt);
	};
	\foreach \n in {1,...,10}
	{
		\draw[blue, fill=blue] (6+\n^2/15,0) circle (3pt);
	};
	\foreach \n in {1,...,7}
	{
		\draw[black!60!green, fill=black!60!green] (9+\n^2/15,0) circle (3pt);
	};
	\draw[orange, ultra thick] (0,0) -- (12, 1.5);
	\node[orange] at (6, 1.35) {$\mathcal L_{+}$};
	\draw[orange, ultra thick] (0,0) -- (12, -1.5);
	\node[orange] at (6, -1.35) {$\mathcal L_{-}$};
	\end{tikzpicture} 
	\caption{Starting from the numerical Pad\'e poles in figure~\ref{fig:ResurgentFunctionPadeApproximation}, the Hankel contour associated to the first singularity in figure~\ref{fig:ResurgentFunctionOnBorelPlaneWithHankelConours} is deformed to the two depicted rays, where the integration kernel is now weighted by the exponential from the Laplace transforms $\mathcal{L}_\pm$.}
	\label{fig:ResurgentFunctionLaplaceResummation}
\end{figure}

\begin{figure}
\centering
\begin{tikzpicture}
\node at (-1, 0.5) {\rotatebox{-25}{$s$}};
\draw[darkgray] (-1.15,0.3) -- (-0.9,0.3);
\draw[darkgray] (-0.9,0.3) -- (-0.6, 0.6);
\draw[->, ultra thick, darkgray] (-1,0) -- (13.5,0);
\draw[->, ultra thick, darkgray] (-0.5,-0.5) -- (0.5,0.5);
\draw[ultra thick, darkgray] (0,-0.7) -- (0,0.2);
\draw[dotted, ultra thick, darkgray] (0,0.1) -- (0,0.7);
\draw[dashed, violet, ultra thick] (3,-0.3) -- (3,3);
\draw[dashed, blue, ultra thick] (6,-0.3) -- (6,2.2);
\draw[dashed, black!60!green, ultra thick] (9,-0.3) -- (9,1.8);
\draw[->, ultra thick, darkgray] (-0.1, 0.8) -- (13.5, 0.8);
\draw[->, ultra thick, darkgray] (0,0.7) -- (0, 4);
\draw[ultra thick, darkgray] (-0.1,3) -- (0.1,3);
\node at (-0.3,0.8) {$0$};
\node at (-0.3,3) {$1$};
\node at (-0.6,3.9) {$\text{e}^{-{\color{brown}x} s}$};
\draw[brown, ultra thick] plot[domain=0:13,smooth,variable=\x] (\x,{0.8+2.2*2^(-\x*0.3)});
\draw[brown, ultra thick] plot[domain=0:13,smooth,variable=\x] (\x,{0.8+2.2*2^(-\x*1.5)});
\draw[brown, ultra thick] plot[domain=0:13,smooth,variable=\x] (\x,{0.8+2.2*2^(-\x*0.6)});
\draw[violet, snake it, ultra thick] (3,0) -- (13.5,0);
\draw[violet, fill=violet] (3,0) circle (3pt);
\draw[blue, snake it, ultra thick] (6,0) -- (13.5,0);
\draw[blue, fill=blue] (6,0) circle (3pt);
\draw[black!60!green, snake it, ultra thick] (9,0) -- (13.5,0);
\draw[black!60!green, fill=black!60!green] (9,0) circle (3pt);
\node at (13.5, 0.5) {$s$};
\end{tikzpicture}
\caption{Illustration of how higher-step branch-cuts are effectively damped by the kernel of the Laplace transformation, for different values of $x$. The lower-part of the figure represents the complex Borel plane, whereas the upper-part graphically displays distinct falling exponentials.}
\label{fig:ResurgentFunctionLaplaceWeightingFactor}
\end{figure}
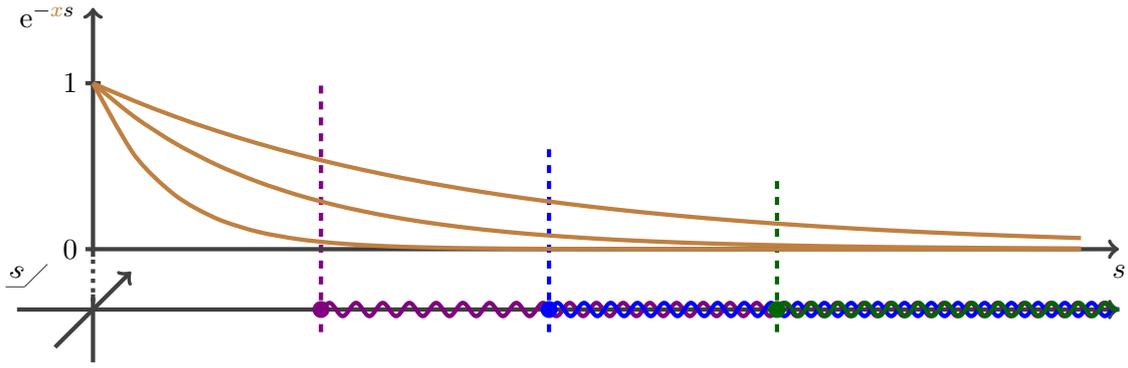

What we shall do is to calculate Borel residues via integration, as, clearly, the logarithmic singularities can only be captured using appropriate Hankel contours. But, from the above discussion, what this entails is the problem that it is hard to separate the different singularities from each other as the branch-cuts overlap. We then need to find a way of numerically isolating a given desired singularity. The solution we shall use is a method based on zooming-in on this desired singularity via the introduction of a weighting-factor before integrating over the Hankel contours. Additionally we need to make sure that the integrations are numerically stable. Interestingly, these two points above can be achieved by using Laplace transformations\footnote{Using Laplace transforms has the additional technical advantages that we do not have to concern ourselves with the $\beta$- or logarithm-factors. This is explained in subappendix~\ref{app:subsec:borel-plane}.}, which is illustrated in figures~\ref{fig:ResurgentFunctionLaplaceResummation} and~\ref{fig:ResurgentFunctionLaplaceWeightingFactor} for an evaluation point in the physical plane such that the higher-step branch-cuts are sufficiently damped. For the convenience of the reader let us here explicitly spell-out the relation between resummations and Laplace transforms: it is well known that one can rewrite the resummation in formula \eqref{eq:Borel-resum} in terms of a Laplace transform of the Borel transform as
\be
\CS_{\theta} \Phi (x) = \int_{0}^{\rme^{\rmi\theta} \infty} \rmd s\, \CB \left[ \Phi \right] (s)\, \rme^{-\frac{s}{x}} \equiv \mathcal{L}_{\theta} \left\{ \CB \left[ \Phi \right] (s)\right\} (x).
\ee
\noindent
For convenience, in the following discussion we shall denote $\mathcal{L}_{0^{\pm}}$ by $\mathcal{L}_{\pm}$ and $\CS_{0^{\pm}}$ by $\CS_{{\pm}}$.

Now being explicit, first focus on the Borel singularity closest to the origin and consider the corresponding Hankel contour integration via Borel resummation \eqref{eq:Borel-resum}. Since we cannot do these resummations over the real line, we slightly deform the integration contours to the lateral Borel resummations $\CS_{\pm}$. The difference between these resummations is given by 
\be
\label{eq:ResummedSimpleResurgentStructureFormula}
\left( \mathcal{S}_{+} - \mathcal{S}_{-} \right) \Phi_{n} (x) = \sum_{\ell \in \mathbb{N}} \mathsf{S}_{n \to n+\ell}\, \Big( \phi_{n+\ell} + \mathcal{S}_{+} \Phi_{n+\ell} (x) \Big)\, \rme^{-\ell \frac{A}{x}},
\ee
\noindent
where the sum in $\ell$ is over the singularities upon the real axis. As described above, we next want to pick a point $x$ such that the higher-step contributions are killed-off by the exponential damping present in the Laplace transformation. One is lead to pick
\begin{equation}
x = \check{x} \quad \text{small and real}.
\end{equation}
\noindent
At this point it follows
\be
\rme^{\frac{A}{\check{x}}} \left( \mathcal{S}_{+} - \mathcal{S}_{-} \right) \Phi_{n} (\check{x}) = \sum_{\ell\in\mathbb{N}} \mathsf{S}_{n \to n+\ell}\, \Big( \phi_{n+\ell} + \mathcal{S}_{+} \Phi_{n+\ell} (\check{x}) \Big)\, \rme^{-\left( \ell-1 \right) \frac{A}{\check{x}}}.
\ee
\noindent
It is quite simple to see that this procedure allows us to isolate the first branch-point up to high-precision, since contributions arising from the other branch-points are exponentially damped,
\be
\label{eq:DampingOutOfTheSecondBranchCuts}
\rme^{\frac{A}{\check{x}}} \left( \mathcal{S}_{+} - \mathcal{S}_{-} \right) \Phi_{n} (\check{x}) = \mathsf{S}_{n \to n+1}\, \Big( \phi_{n+1} + \mathcal{S}_{+} \Phi_{n+1} (\check{x}) \Big) + o \left( \rme^{- \frac{A}{\check{x}}} \right).
\ee
\noindent
This procedure is illustrated in figure~\ref{fig:ResurgentFunctionLaplaceWeightingFactor}.

This method has the great advantage that, at large $\check{x}$, the result one obtains via the above numerical resummation is very accurate. This is basically due to the zooming-in onto the origin of the complex Borel plane, where the Pad\'e approximation is most accurate. The above method may also have a different interpretation. Because we used Borel--\'Ecalle resummations the above is equivalent to imposing that the function obtained via resummation of the divergent series, $\mathcal{S}\Phi$, is continuous\footnote{Actually we will soon need to impose this condition also on the derivatives. This is consistent as it makes sense to ask for smoothness of the solution.} along a path crossing the Stokes line. This path precisely crosses the Stokes line at the chosen $\check{x}$. Formulated in this way one one can say that the set of Stokes data is exactly such that the function underlying the resurgence procedure is smooth across Stokes lines. In fact the smoothness property is enough to uniquely fix all Stokes data. This picture is equivalent to the one described above, and we could also have had presented the method purely from such point-of-view.

\subsubsection*{The Higher-Order Singularities}

Clearly the procedure outlined above cannot be identically copied for singularities which get overshadowed by branch-points closer to the origin. As an example let us consider the ``two-step'' singularity. In order to be able to go through with the above programme, we first need to find a way to remove the contribution from the first branch-point. One possibility would be to consider
\be
\left( \mathcal{S}_{+} - \mathcal{S}_{-} \right) \Phi_{n} (x) - \mathsf{S}_{n \to n+1}\, \Big( \phi_{n+1} + \mathcal{S}_{+} \Phi_{n+1} (x) \Big)\, \rme^{-\frac{A}{x}} = \sum_{\ell > 1} \mathsf{S}_{n \to n+\ell}\, \Big( \phi_{n+\ell} + \mathcal{S}_{+} \Phi_{n+\ell} (x) \Big)\, \rme^{-\ell \frac{A}{x}}.
\ee
\noindent
Multiplying by the appropriate exponential factor now yields
\bea
\label{eq:high_ord_sing}
\left\{ \left( \mathcal{S}_{+} - \mathcal{S}_{-} \right) \Phi_{n} (x) - \mathsf{S}_{n \to n+1}\, \Big( \phi_{n+1} + \mathcal{S}_{+} \Phi_{n+1} (x) \Big)\, \rme^{-\frac{A}{x}} \right\} \rme^{2 \frac{A}{x}} &=& \\
&&
\hspace{-100pt}
= \sum_{\ell > 1} \mathsf{S}_{n \to n+\ell}\, \Big( \phi_{n+\ell} + \mathcal{S}_{+} \Phi_{n+\ell} (x) \Big)\, \rme^{- \left(\ell-2\right) \frac{A}{x}}. \nonumber
\eea
\noindent
What we need to understand is how to pick $x$ in order to zoom-in precisely on the second branch-point. Equation \eqref{eq:DampingOutOfTheSecondBranchCuts} tells us that our knowledge of the first singularity is limited by $o \left( \rme^{- \frac{A}{\check{x}}} \right)$, which needs to be balanced by our present attempt of damping-out all higher-instanton singularities with $\ell > 2$. This implies we must first compute the Borel residue $\mathsf{S}_{n \to n+1}$ with some adequate damping $\check{x}$, and then choose a second damping for the computation of $\mathsf{S}_{n \to n+2}$ which is still adequate but still keeps itself
\begin{equation}
x = \check{\check{x}} \gg \check{x} \quad \text{small and real}.
\end{equation}
\noindent
Clearly this will only work if we have very-high precision on the first instanton, before computing the second one. In this way, going to higher and higher instanton contributions is a computationally hard endeavor. In this work we have carried-through this programme up to the \textit{fourth} instanton contribution. Details on this calculation may be found in subappendix~\ref{app:subsec:borel-plane}.

\subsubsection*{The Resonant Case}

Having outlined our main numerical method in a simplified example, it now just needs a slight upgrade in order to tackle the resonant Painlev\'e examples. Up to now, a single Borel residue was to be found at each particular singularity on the complex Borel plane. The difference to the resonant setting is that herein each such singularity will get associated to multiple Borel residues. This is quite clear from the general \textit{resonant} singular structure of simple resurgent functions in \eqref{eq:borel-sing-R}; which we repeat including the pole from the residual coefficient (recall \eqref{eq:simpleResurgentFunctionOnBorelPlane}):
\be
\label{eq:simpleresurgentfunction}
\CB \left[ \Phi_{\boldsymbol{n}} \right] (s) \Big|_{s = \boldsymbol{\ell} \cdot \boldsymbol{A}} \sim \sum_{\bm{p} \in \ker\mathfrak{P}} \mathsf{S}_{\boldsymbol{n} \to \boldsymbol{n} + \boldsymbol{\ell} + \bm{p}} \left( \frac{\phi_{\boldsymbol{n} + \boldsymbol{\ell} + \bm{p}}}{2\pi\rmi \left( s-\boldsymbol{\ell} \cdot \boldsymbol{A} \right)} + \CB \left[ \Phi_{\boldsymbol{n} + \boldsymbol{\ell} + \bm{p}} \right] (s-\boldsymbol{\ell} \cdot \boldsymbol{A})\, \frac{\log \left( s-\boldsymbol{\ell} \cdot \boldsymbol{A} \right)}{2\pi\rmi} \right).
\ee
\noindent
In our Painlev\'e problems singularities are still on the real axis (now both positive and negative axes). Let us focus on the positive real axis, where the picture is then basically the same as in figure~\ref{fig:ResurgentFunctionOnBorelPlane} only now to every branch-point singularity we will find multiple Borel residues attached to it. Let us denote this number of Borel residues at the $\ell$th singularity by $m$. The analogue to formula \eqref{eq:ResummedSimpleResurgentStructureFormula} above now reads 
\be
\label{eq:ResummedSimpleResurgentStructureResonantFormula}
\left( \mathcal{S}_{+} - \mathcal{S}_{-} \right) \Phi_{\boldsymbol{n}} (x) = \sum_{\ell\in\mathbb{N}} \sum_{\bm{p} \in \ker\mathfrak{P}} \mathsf{S}_{\boldsymbol{n} \to \boldsymbol{n} + \left(\ell,0\right) + \bm{p}}\, \Big( \phi_{\boldsymbol{n} + \left(\ell,0\right) + \bm{p}} + \mathcal{S}_{+} \Phi_{\boldsymbol{n} + \left(\ell,0\right) + \bm{p}} (x) \Big)\, \rme^{-\ell \frac{A}{x}}.
\ee
\noindent
The novelty is that when zooming-in on a singularity---as discussed above---we now still have to deal with the additional sum over the kernel. But the procedure we outlined in the non-resonant case yields a \textit{single} linear equation for the $m$ (unknown) Borel residues. This issue is solved by simply generating the needed ``missing'' equations via derivatives of \eqref{eq:ResummedSimpleResurgentStructureResonantFormula} with respect to $x$. This is consistent with the fact that our method can be seen as imposing smoothness when crossing Stokes lines. Then, all one has to do is to pick appropriate points $x$, allowing for the zoom-in on the desired contributions. This process results in a system of $m$ linear equations, which may be finally solved for the complete set of Borel residues at said singularity. This system of linear equations which allows us to deal with any resonant problem is:
\bea
\label{eq:ResummedSimpleResurgentStructureResonantFormulaWithDerivatives-a}
\left( \mathcal{S}_{+} - \mathcal{S}_{-} \right) \Phi_{\boldsymbol{n}} &=& \sum_{\ell\in\mathbb{N}} \sum_{\bm{p} \in \ker\mathfrak{P}} \mathsf{S}_{\boldsymbol{n} \to \boldsymbol{n} + \left(\ell,0\right) + \bm{p}}\, \Big( \phi_{\boldsymbol{n} + \left(\ell,0\right) + \bm{p}} + \mathcal{S}_{+} \Phi_{\boldsymbol{n} + \left(\ell,0\right) + \bm{p}} \Big)\, \rme^{-\ell \frac{A}{x}}, \\
\label{eq:ResummedSimpleResurgentStructureResonantFormulaWithDerivatives-b}
\partial_{x} \left( \mathcal{S}_{+} - \mathcal{S}_{-} \right) \Phi_{\boldsymbol{n}} &=& \partial_{x} \left\{ \sum_{\ell\in\mathbb{N}} \sum_{\bm{p} \in \ker\mathfrak{P}} \mathsf{S}_{\boldsymbol{n} \to \boldsymbol{n} + \left(\ell,0\right) + \bm{p}}\, \Big( \phi_{\boldsymbol{n} + \left(\ell,0\right) + \bm{p}} + \mathcal{S}_{+} \Phi_{\boldsymbol{n} + \left(\ell,0\right) + \bm{p}} \Big)\, \rme^{-\ell \frac{A}{x}} \right\}, \\
&\vdots& \nonumber \\
\partial_{x}^{m-1} \left( \mathcal{S}_{+} - \mathcal{S}_{-} \right) \Phi_{\boldsymbol{n}} &=& \partial_{x}^{m-1} \left\{ \sum_{\ell\in\mathbb{N}} \sum_{\bm{p} \in \ker\mathfrak{P}} \mathsf{S}_{\boldsymbol{n} \to \boldsymbol{n} + \left(\ell,0\right) + \bm{p}}\, \Big( \phi_{\boldsymbol{n} + \left(\ell,0\right) + \bm{p}} + \mathcal{S}_{+} \Phi_{\boldsymbol{n} + \left(\ell,0\right) + \bm{p}} \Big)\, \rme^{-\ell \frac{A}{x}} \right\}.
\eea

To finish this subsection we still need to describe how to estimate the precision of the numerical method outlined above. Let us describe the error estimate for the calculation of a first-step non-resonant Borel residue---the error estimates for (resonant-) higher-step Borel residues are analogous. Consider the setting shown in figures~\ref{fig:ResurgentFunctionLaplaceResummation} and~\ref{fig:ResurgentFunctionLaplaceWeightingFactor}. We start by understanding how to pick a damping $\check{x}$: it has to be sufficiently big so that we blend-out higher-order contributions coming from singularities that are further away from the origin---the first of which is the second-step contribution that is damped by $\rme^{-2\frac{A}{x}}$. We can therefore estimate the error $\epsilon_{\check{x}}$ coming from the damping via
\begin{equation}
\epsilon_{\check{x}} \sim \frac{\rme^{-2\frac{A}{\check{x}}}}{\rme^{-\frac{A}{\check{x}}}} \sim \rme^{-\frac{A}{\check{x}}}.
\end{equation}
\noindent
Having established a good damping value we can now go through the above calculation and estimate the error that we are making at each step. We start by performing a Borel--Pad\'e approximation of order $N$. In order to estimate the error for the whole procedure, we can compare the resummation of order $N+1$ with the resummation of order $N$. Now that we have all elements, we can propagate the errors through the elementary algebraic procedures to get the results of tables~\ref{tab:comparisonStokesDataWithBPErrorAnalysis} and~\ref{tab:comparisonStokesDataWithBPErrorAnalysisPII}.

\subsection{Results for Painlev\'e~I and~II}

As explained in detail in section~\ref{sec:stokes-resonance}, the properties of Stokes data and Borel residues allow for many simplifications in the calculations. For example, one such simplification originates from the vector structures in resonant Stokes data which, at the end of the day, implied that the full Stokes data could be reconstructed simply from the Borel residues starting off the main diagonal, $\mathsf{S}_{(n,n)\to(s,0)}$. Of course one thing we have done is to thoroughly test these numbers numerically. Yet, they assume specific properties which also need to be checked. We have also carefully checked the whole vector structure of Stokes data, by performing numerical checks where this structure was not assumed. Let us briefly outline the specific numerical results we obtained.

\begin{figure}
	\label{fig:ComputedPainleveStokesData}
	\centering
	\begin{subfigure}[t]{0.5\textwidth}
		\centering
		\includegraphics[height=3in]{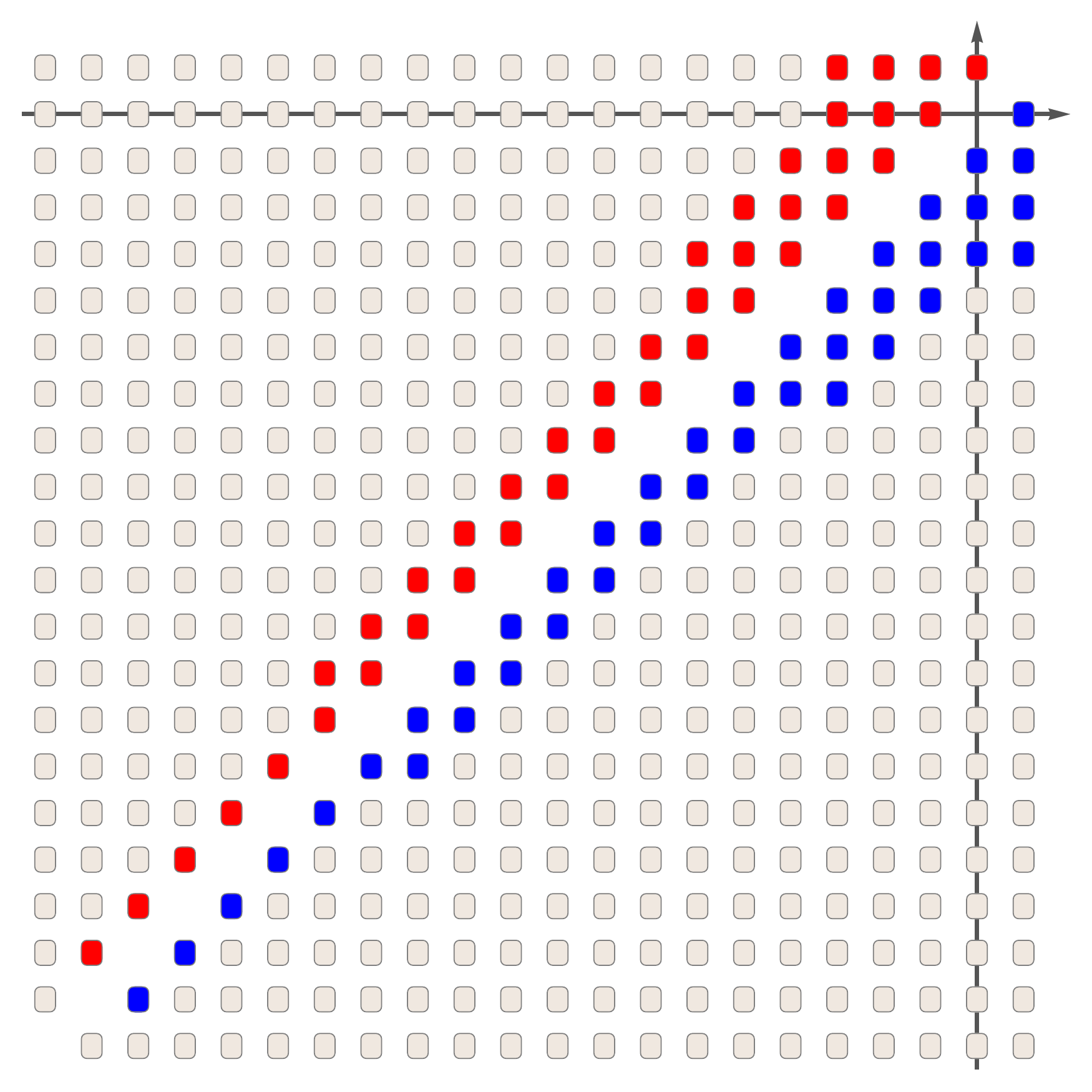}
		\caption{Painlev\'e~I.}
		\label{fig:ComputedPainleveIStokesData}
	\end{subfigure}%
	~ 
	\begin{subfigure}[t]{0.5\textwidth}
		\centering
		\includegraphics[height=3in]{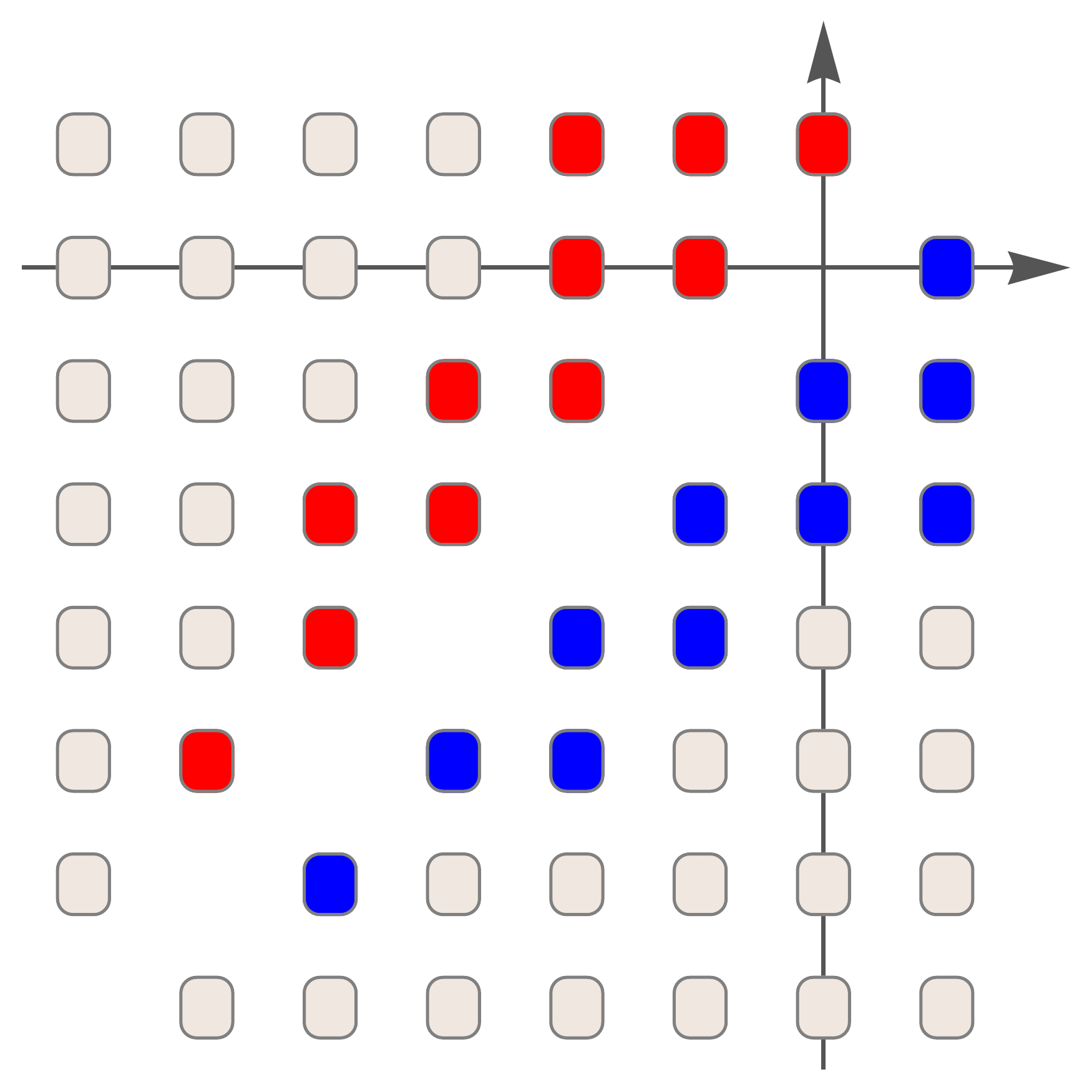}
		\caption{Painlev\'e~II.}
		\label{fig:ComputedPainleveIIStokesData}
	\end{subfigure}
	\caption{The lattice of Stokes data, with highlighted Stokes vectors which have been numerically checked in the cases of \PI~(left) and \PII~(right). The \PII~equation \eqref{eq:PII-eq-v2} has a cubic term, making the whole computation of data significantly slower as compared to the case of \PI.}
\end{figure}

We have numerically computed \PI~Stokes data as schematically displayed in figure~\ref{fig:ComputedPainleveIStokesData}, and as (very briefly) explicitly shown in table~\ref{tab:comparisonStokesDataWithAnalyticalCloseForm}. As we scan through the data in figure~\ref{fig:ComputedPainleveIStokesData}, its numerical precision varies. It is about 112-92, 50-37, and 20-12 digits for the first, second, and third diagonal, respectively; corresponding to the first 10, 8, and 4 vectors on each respective diagonal. A rough precision-chart may be found in table~\ref{tab:comparisonStokesDataWithBPErrorAnalysis}. In order to compute these numbers, we performed Borel transformations up to order $g=800$, with diagonal $(400,400)$-Pad\'e approximants, and chose damping factors of 0.011, 0.023, and 0.048, for the first, second, and third branch-points, respectively. To select these particular values one has to do some numerical exploration. Our general strategy was to search for the smallest damping factor allowed by the Borel--Pad\'e analysis, and then explore slightly larger ones in the search for the ``sweet-spot''. The remaining highlighted entries in figure~\ref{fig:ComputedPainleveIStokesData} which we have not discussed yet, correspond to lower-precision computations---which agree up to 42 and 11 digits for the first and second diagonals, respectively. These used order $g=200$ for the Borel transformation and $(100,100)$-Pad\'e approximants. See appendix~\ref{app:numerics} for further details.

In the case of Painlev\'e II, we have numerically computed data as schematically displayed in figure~\ref{fig:ComputedPainleveIIStokesData}. As we scan through these data, the ranges in precision are 93-84 for the first diagonal and 18-13 for the second diagonal and the first element of the third diagonal, as very briefly shown in table~\ref{tab:comparisonStokesDataWithAnalyticalCloseFormPII}. A rough precision-chart may be found in table~\ref{tab:comparisonStokesDataWithBPErrorAnalysisPII}. The Borel transforms went up to $g=600$, so that we used $(300,300)$-Pad\'e approximants with 0.004 and 0.006 as dampings. See appendix~\ref{app:numerics} for further details.

\begin{table}
	\centering
	\begin{tabular}{l|l}
		Painlev\'e~I & $N^{(\ell)}_{1}$ \\
		\hline
		Analytic: & $N^{(1)}_{1}=-0.37125762464284556748197742189264343969606293477341381894$\\ 
		& 9985892265068145661623767551055983672107532286489322969\textcolor{red}{271162524} $\rmi$  \\ 
		Numerical: & $N^{(1)}_{1}=-0.37125762464284556748197742189264343969606293477341381894$\\
		& 9985892265068145661623767551055983672107532286489322969\textcolor{red}{126049818} $\rmi$  \\ 
		\hline
		Analytic: & $N^{(2)}_{1}=0.250000000000000000000000000000000000000000000000000\textcolor{red}{000000000}$ $\rmi$\\ 
		Numerical: & $N^{(2)}_{1}=0.250000000000000000000000000000000000000000000000000$\textcolor{red}{418501898} $\rmi$\\ 
		\hline
		Analytic: & $N^{(3)}_{1}=-0.29928304157524408013\textcolor{red}{7573386121}$ $\rmi$\\ 
		Numerical: & $N^{(3)}_{1}=-0.29928304157524408013\textcolor{red}{8210639671} $ $\rmi$\\ 
	\end{tabular}
	\caption{Stokes data for \PI: illustrative comparison between our numerical results and our analytical closed-form expressions. The precision of numerical results validating analytical expressions is evident and highlighted in red. Note that these particular data have already been computed in \cite{asv11}, albeit to a smaller precision.}
	\label{tab:comparisonStokesDataWithAnalyticalCloseForm}
\end{table}

\begin{table}
	\centering
	\begin{tabular}{l|l}
		Painlev\'e~II & $N^{(\ell)}_{1}$ \\
		\hline
		Analytic: & $N^{(1)}_{1}=-0.39894228040143267793994605993438186847585863116493465766$\\ 
		& 5925829670657925899301838501252333907\textcolor{red}{3069364} $\rmi$  \\ 
		Numerical: & $N^{(1)}_{1}=-0.39894228040143267793994605993438186847585863116493465766$\\ 
		& 5925829670657925899301838501252333907\textcolor{red}{2839576} $\rmi$  \\ 
		\hline
		Analytic: & $N^{(2)}_{1}=0.50000000000000000\textcolor{red}{00000000}$ $\rmi$\\ 
		Numerical: & $N^{(2)}_{1}=0.50000000000000000\textcolor{red}{99465990}$ $\rmi$\\ 
		\hline
		Analytic: & $N^{(3)}_{1}=-1.1140570109471113{\color{red}3440700}$ $\rmi$\\ 
		Numerical: & $N^{(3)}_{1}=-1.1140570109471113{\color{red}4553594}$ $\rmi$\\ 
	\end{tabular}
	\caption{Stokes data for \PII: illustrative comparison between our numerical results and our analytical closed-form expressions. The precision of numerical results validating analytical expressions is evident and highlighted in red. Note that the first two of these particular data have already been computed in \cite{sv13}, albeit to a smaller precision.}
	\label{tab:comparisonStokesDataWithAnalyticalCloseFormPII}
\end{table}

\begin{table}
	\begin{center}
		\begin{tabular}{@{}c@{}}
			Precision (digits)\\
			\begin{tabular}{c|c|c|c}
				\hline
				$N^{(\ell)}_{1-p}$ & $\ell=1$ & $\ell=2$ & $\ell=3$  \\
				\hline
				$p=0$ & $\sim100$ & $\sim50$ & $\sim20$ \\
				\hline 
				$p=1$ & $\sim100$ & $\sim50$ & $\sim20$ \\
				\hline
				$p=2$ & $\sim100$ & $\sim50$ & $\sim10$ \\
				\hline
				$p=3$ & $\sim70$ & $\sim40$ &$\times$\\
				\hline
				$p=4$ & $\sim70$ & $\sim40$ &$\times$\\
				\hline 
				$p=5$ & $\sim70$ & $\sim40$ &$\times$\\
			\end{tabular}
		\end{tabular}
	\end{center}
	\caption{Rough, conservative estimates on the precision of \PI~numerical Stokes data, as based on the Borel--Pad\'e error analysis at the end of subsection~\ref{subsec:borel-plane}. The estimates we are displaying in the table correspond to the order-800 Borel transform with $(400,400)$-Pad\'e approximants.}
	\label{tab:comparisonStokesDataWithBPErrorAnalysis}
\end{table}

\begin{table}
	\begin{center}
		\begin{tabular}{@{}c@{}}
			Precision (digits)\\
			\begin{tabular}{c|c|c|c}
				\hline
				$N^{(\ell)}_{1-p}$ & $\ell=1$ & $\ell=2$ & $\ell=3$ \\
				\hline
				$p=0$ & $\sim90$ & $\sim20$  & $\sim20$\\
				\hline 
				$p=1$ & $\sim90$ & $\sim20$ &$\times$\\
				\hline
				$p=2$ & $\sim90$ & $\sim20$ &$\times$\\
				\hline
				$p=3$ & $\sim70$ & $\sim10$ &$\times$\\
				\hline
				$p=4$ & $\sim70$ & $\sim10$ &$\times$\\
				\hline 
				$p=5$ & $\sim70$ & $\sim10$ &$\times$\\
			\end{tabular}
		\end{tabular}
	\end{center}
	\caption{Rough, conservative estimates on the precision of \PII~numerical Stokes data, as based on the Borel--Pad\'e error analysis at the end of subsection~\ref{subsec:borel-plane}. The estimates we are displaying in the table correspond to the order-600 Borel transform with $(300,300)$-Pad\'e approximants.}
	\label{tab:comparisonStokesDataWithBPErrorAnalysisPII}
\end{table}

As mentioned above, the Stokes-data vector-structure \eqref{eq:vec_struct-a}-\eqref{eq:vec_struct-b} must also be tested. What we have done was to simply compute Borel residues using the numerical method described earlier in this section, without assuming any other relations in-between them but for standard resurgence relations. Once Stokes vectors get computed, we then compute the ratio of their two components which (if the vector structure we assumed is correct) must be a specific rational number as in \eqref{eq:structure_1}. Generically it should be
\be
\label{eq:structure_1-ell}
\boldsymbol{S}^{(2)}_{(1-p,1-p-\ell)} = - \frac{p}{p+\ell}\, \boldsymbol{S}^{(1)}_{(1-p,1-p-\ell)}.
\ee
\noindent
As usual, our results strongly validate this structure: we report on the digits of agreement between numerics and analytics in table~\ref{tab:pIstruct} for \PI, and in table~\ref{tab:pIIstruct} for \PII.

\begin{table}
	\centering
	\begin{tabular}{c|c|c|c|c|c|c|c|c|c|c}
		\diagbox{$\ell$}{$p$}&0&1&2&3&4&5&6&7&8&9\\\hline
		1&106&105&103&102&100&98&97&97&94&93\\\hline
		2&48&45&44&43&43&40&40&38&$\times$&$\times$\\\hline
		3&16&16&14&13&$\times$&$\times$&$\times$&$\times$&$\times$&$\times$
	\end{tabular}
	\caption{Testing the Stokes-data vector-structure for \PI. Each entry of the table refers to the number of decimal digits of agreement between the left-hand side and the right-hand side of equation \eqref{eq:structure_1-ell}. A cross indicates values of $\ell$ and $p$ for which we did not compute data without assuming the vector structure.}
	\label{tab:pIstruct}
\end{table}

\begin{table}
	\centering
	\begin{tabular}{c|c|c|c|c|c}
		\diagbox{$\ell$}{$p$}&0&1&2&3&4\\\hline
		1&93&93&90&89&87\\\hline
		2&16&14&14&$\times$&$\times$
	\end{tabular}
	\caption{Testing the Stokes-data vector-structure for \PII. Each entry of the table refers to the number of decimal digits of agreement between the left-hand side and the right-hand side of equation \eqref{eq:structure_1-ell}. A cross indicates values of $\ell$ and $p$ for which we did not compute data without assuming the vector structure. As usual, we have less data as compared to \PI~due to the higher computational cost of computing \PII~coefficients.}
	\label{tab:pIIstruct}
\end{table}

\subsection{Acceleration via Conformal Transformations}\label{subsec:conformal}

There is one last comment on numerics we would like to make. Whenever dealing with numerical methods based upon Pad\'e approximants to Borel transforms, significant gains in precision may be usually obtained by resorting to acceleration via conformal transformations; \textit{e.g.}, \cite{z98, cf99, cf00, z00, z10} (see subappendix~\ref{app:subsec:conformal} for the specific details). Adequate choices of conformal transformations on the Borel plane of the transseries sectors end-up replicating the underlying function (the Borel transform) with much better numerical precision. We have explored these techniques in subappendix~\ref{app:subsec:conformal}, as applied to both \PI/\PII, in particular asking how they might improve on the precision of the Borel residues we computed.

The punch-line of subappendix~\ref{app:subsec:conformal} is that acceleration via conformal transformations is not of significant help in our particular scenario. On the one hand, obviously, one can construct adequate conformal transformations which end-up increasing the precision of our calculations. On the other hand, however, the higher the number of coefficients in our Borel transform, the longer it takes to carry-through the method with the addition of conformal transformations. One thus has to make a decision which is problem dependent. This is, in a scenario where computing successive transseries coefficients is computationally heavy or even not easily accessible, conformal transformations are indeed a very valuable tool. In our case, however, computing coefficients is not computationally heavy \textit{in comparison with conformal transformations} once one goes to a high enough number of coefficients. As such, in our specific problems this acceleration is not very useful, and we have opted by leaving conformal transformations aside. Nevertheless, we point the reader to subappendix~\ref{app:subsec:conformal} for some extra details on this discussion.

\section{From Stokes Data to Connection Formulae}\label{sec:stokes}

Having conjectured formulae for the full Painlev\'e Stokes data (or Borel residues), and having tested and verified those formulae, it is now time to go back to our discussion in subsection~\ref{subsec:stokes-trans} and make the last step from \textit{local} to \textit{global} transseries solutions---\textit{i.e.}, integrate the flow of the Stokes vector field in order to obtain \textit{explicit} transition functions or connection formulae, as from \eqref{eq:diff_eq-a}-\eqref{eq:diff_eq-b} (see as well, \textit{e.g.}, \cite{jk92a, i96}). What we shall see is that, whereas integrating the flow in standard rectangular framing is quite intricate (hence also all the intricacies of the Painlev\'e generating functions for Stokes vectors or for Borel residues in this framing), it turns out that changing transseries variables to the diagonal framing of subsection~\ref{subsec:reson-frame} rewrites all these generating/transition functions almost trivially as simple \textit{shifts}. In other words, we shall obtain very simple and compact final formulae encoding all Stokes data (or Borel residues) and almost trivializing (to a shift) all Stokes transitions (albeit with some added subtleties we will discuss). 

Having set-up such connection formulae, we next want to proceed and address the \textit{direct} monodromy problem (\textit{i.e.}, directly at the level of Painlev\'e solutions). This is a harder question as in order to make progress one needs the full resummation of our two-parameter resonant transseries---a calculation which goes beyond the scope of this paper. We first outline the general chain of thought and set-up the direct monodromy calculation, investigating the necessary Stokes transitions. This includes a discussion of the transseries moduli space (in different rectangular- and diagonal-framing coordinates), also partially following the discussion in subsection~\ref{subsec:stokes-trans} (but stopping short of the geometrical discussion in \cite{g76, o79}). Direct monodromy is then explicitly shown albeit without fully addressing the required transseries resummations. Note that what this calculation also effectively amounts to is to a (partial) \textit{proof} of our conjectures---as at the end of the day this is essentially verifying the (defining) Painlev\'e property.

Two additional outcomes of this section are particularly relevant. Firstly, using the diagonal-framing formulation of the transseries, as in \eqref{eq:PI/II-2PTS-diagonal}, further allows for contact with related works in the literature; most notably the gauge-theoretic partition-function and exact WKB analyses in \cite{t95, t99, t00, lr16, i19a}. We will show how this takes place, following a suggestion first made in \cite{i19b}. Secondly, what this match between ``linear'' and ``nonlinear'' problems effectively amounts to is to \textit{another proof} (in addition to the aforementioned direct monodromy calculation) of our conjectures---at the end of the day, the aforementioned related works depend \textit{only} on the ``canonical'' Stokes data \eqref{eq:exampleStokesP1}-\eqref{eq:exampleStokesP2} which have been rigorously computed, \textit{e.g.}, via the indirect methods mentioned in the introduction \cite{k88, kk93, k94, ifk94, ik03, i03, k04a, k04b}. This also results in rather precise clues on how to generically obtain resurgent Stokes data efficiently, perhaps even for all string equations in the (m)KdV hierarchy---a topic to which we shall return in upcoming work.

\subsection{Computation of Stokes Transitions}

In section~\ref{sec:results} we computed closed-form Stokes vectors for \PI~and \PII. Our goal is now to find the corresponding Stokes transitions, $\underline{\pmb{\mathbb{S}}}_{\theta}^{(\uptau)} (\bm{\sigma})$, appearing in \eqref{eq:Stokes-flow-simplified-TAU}. In principle, these could be found via \eqref{eq:transitions} by writing closed-form expressions for Borel residues and then performing the sum. Alternatively, they can also be found by solving the differential system \eqref{eq:diff_eq-a}-\eqref{eq:diff_eq-b} and evaluating the solution at $\tau = 1$. This is the approach which we shall follow below.

We start with the \PI/\PII~vector fields, and initially consider the vector field $\boldsymbol{\underline S}_0(\boldsymbol{\sigma})$. For both equations, this is of the form (recall \eqref{eq:PIS0-final} and \eqref{eq:PIIS0-final})
\be
\underline{\boldsymbol{S}}_0 (\boldsymbol{\sigma}) = \underline{\text{P}}_0 \left( \sigma_2\, N^{(1)} (\upmu) \right) \left[ \begin{array}{c} \sigma_2^{-1} \left( 1 + \upmu \frac{\rmd}{\rmd \upmu} \log N^{(1)} (\upmu) \right) \\ - \sigma_2\, \frac{\rmd}{\rmd \upmu} \log N^{(1)} (\upmu) \end{array} \right],
\ee
\noindent
where the function $\underline{\text{P}}_0$ is given by\footnote{Recall that in \PI~and \PII~the values of $N_1^{(1)}$ and the forms of $N^{(1)}(\upmu)$ are obviously distinct; we are not going to distinguish between the two sets of data, as this is evident from the context.}
\be
\label{eq:P0-PI/PII-def}
\underline{\text{P}}_0(x) = 
\begin{cases}
\, \displaystyle {\rmi \left( N_1^{(1)} \right)^2 \log \left( 1 - \frac{\rmi x}{\left( N_1^{(1)} \right)^2} \right)}, \quad\quad & \text{\PI}, \\
\, \displaystyle \frac{\rmi}2{\left( N_1^{(1)} \right)^2\log \left( 1 - \frac{2\rmi x}{\left( N_1^{(1)} \right)^2} \right)}, \quad\quad & \text{\PII}.
\end{cases}
\ee
\noindent
Fully implementing diagonal-framing, we further make the change of variables $\left( \sigma_1, \sigma_2 \right) \mapsto \left( \upmu, \uprho_0 \right)$
\bea
\label{eqn:coord_chan-a}
\upmu &=& \sigma_1 \sigma_2, \\
\label{eqn:coord_chan-b}
\uprho_0 &=& \sigma_2\, N^{(1)} \left( \sigma_1 \sigma_2 \right),
\eea
\noindent
with inverse
\bea
\label{eqn:coord_inv-a}
\sigma_1 &=& \frac{\upmu\, N^{(1)} (\upmu)}{\uprho_0}, \\
\label{eqn:coord_inv-b}
\sigma_2 &=& \frac{\uprho_0}{N^{(1)} (\upmu)}.
\eea
\noindent
This change of variables makes solving \eqref{eq:diff_eq-a}-\eqref{eq:diff_eq-b} trivial: the alien vector-field $\underline{\dot\Delta}_0$ (in the sense as per the bridge equations \eqref{eq:bridge-directional}) is given in these new coordinates by
\bea
\underline{\dot\Delta}_0 \quad \leftrightarrow \quad \underline{\boldsymbol{S}}_{0} (\boldsymbol{\sigma}) \cdot \frac{\partial}{\partial\boldsymbol{\sigma}} = \underline{\text{P}}_0 \left(\uprho_0\right) \frac{\partial}{\partial\upmu},
\eea
\noindent
which shows how in the $\left(\upmu,\uprho_0\right)$ coordinates the alien vector-field is along the $\upmu$-direction, so that $\uprho_0$ becomes a constant of the flow; \textit{i.e.}, 
\bea
\frac{\rmd\upmu}{\rmd\uptau} &=& \underline{\text{P}}_0 \left(\uprho_0\right), \\
\frac{\rmd\uprho_0}{\rmd\uptau} &=& 0.
\eea
\noindent
Starting from an initial condition characterized by $\left( \sigma_1, \sigma_2 \right)$ or $\left( \upmu, \uprho_0 \right)$, we then have
\bea
\label{eq:shift-a}
\upmu \left( \uptau \right) &=& \sigma_1 \sigma_2 + \uptau\, \underline{\text{P}}_0 \left( \sigma_2\, N^{(1)} \left( \sigma_1 \sigma_2 \right) \right) \equiv \upmu + \uptau\, \underline{\text{P}}_0 \left(\uprho_0\right), \\
\label{eq:shift-b}
\uprho_0 \left( \uptau \right) &=& \sigma_2\, N^{(1)} \left( \sigma_1 \sigma_2 \right) \equiv \uprho_0.
\eea
\noindent
Remarkably, in these variables the Stokes transition $\underline{\pmb{\mathbb{S}}}_{0} \left( \upmu, \uprho_0 \right)$ is very simple: it keeps one coordinate \textit{constant} while simply \textit{shifting} the other coordinate. The effect on the original transseries parameters is more complicated---where transformation laws for $\sigma_1$ and $\sigma_2$ may be obtained using \eqref{eqn:coord_inv-a}-\eqref{eqn:coord_inv-b}---but the simple flow \eqref{eq:shift-a}-\eqref{eq:shift-b} is enough to fully characterize the Stokes jump.

Writing explicit formulae in the original variables, first perform the change of variables back to $\left( \sigma_1, \sigma_2 \right)$ to obtain
\be
\underline{\pmb{\mathbb{S}}}_{0}^{(\uptau)} \left( \bm{\sigma} \right) = \left[
\begin{array}{c}
\displaystyle {\frac{\left( \sigma_1 \sigma_2 + \uptau\, \underline{\text{P}}_0 \left( \sigma_2\, N^{(1)} \left( \sigma_1 \sigma_2 \right) \right) \right) N^{(1)} \left( \sigma_1 \sigma_2 + \uptau\, \underline{\text{P}}_0 \left( \sigma_2\, N^{(1)} \left( \sigma_1 \sigma_2 \right) \right) \right)}{\sigma_2\, N^{(1)} \left( \sigma_1 \sigma_2 \right)}} \\
\displaystyle {\frac{\sigma_2\, N^{(1)} \left( \sigma_1 \sigma_2 \right)}{N^{(1)} \left( \sigma_1 \sigma_2 + \uptau\, \underline{\text{P}}_0 \left( \sigma_2\, N^{(1)} \left( \sigma_1 \sigma_2 \right) \right) \right)}}
\end{array}
\right].
\ee
\noindent
Evaluating at $\uptau=1$, we finally get the transition function
\bea
\label{eq:transitionfunction0-a}
\underline{\pmb{\mathbb{S}}}_{0} \left( \bm{\sigma} \right) &=& \left[
\begin{array}{c}
\displaystyle {\frac{\left( \sigma_1 \sigma_2 + \underline{\text{P}}_0 \left( \sigma_2\, N^{(1)} \left( \sigma_1 \sigma_2 \right) \right) \right) N^{(1)} \left( \sigma_1 \sigma_2 + \underline{\text{P}}_0 \left( \sigma_2\, N^{(1)} \left( \sigma_1 \sigma_2 \right) \right) \right)}{\sigma_2\, N^{(1)} \left( \sigma_1 \sigma_2 \right)}} \\
\displaystyle {\frac{\sigma_2\, N^{(1)} \left( \sigma_1 \sigma_2 \right)}{N^{(1)} \left( \sigma_1 \sigma_2 + \underline{\text{P}}_0 \left( \sigma_2\, N^{(1)} \left( \sigma_1 \sigma_2 \right) \right) \right)}}
\end{array}
\right].
\eea
\noindent
This is a rather compact final expression. Further, interestingly, we can identify terms in the series expansion of this Stokes transition as the Borel residues. As an example, in the \PI~case, the first component of $\underline{\pmb{\mathbb{S}}}_{0} \left( \bm{\sigma} \right)$ is given by (compare with \eqref{eq:transitions})
\bea
&&
-\rmi \frac{3^{\frac{1}{4}}}{2\sqrt\pi} + \sigma_1 + \frac{2\pi\rmi - 2\gamma_{\text{E}} - 2 \log \left( 96\sqrt3 \right)}{4\pi}\, \sigma_2 - \rmi \frac{2\gamma_{\text{E}} + 2 \log \left( 96\sqrt{3} \right)}{3^{\frac{1}{4}}\sqrt{\pi}}\, \sigma_1\sigma_2 = \nonumber \\
&&
= - \mathsf{S}_{(0,0)\to(1,0)} + \sigma_1 - \mathsf{S}_{(0,1)\to(1,0)}\, \sigma_2 - \mathsf{S}_{(1,1)\to(1,0)}\, \sigma_1\sigma_2.
\eea
\noindent
In this way, the above function serves as a generating-function for Borel residues. In general, the correct way to identify the coefficients of $\left( \underline{\pmb{\mathbb{S}}}_{0} \left( \bm{\sigma} \right) \right)^{\bm{n}}$ as Borel residues is given by \eqref{eq:consistency}, evaluated at $\uptau=1$. In the case of our previous example, having expanded the first component of $\left( \underline{\pmb{\mathbb{S}}}_{0} \left( \bm{\sigma} \right) \right)^{\bm{n}}$, we then get a series expansion in which the coefficient of $\sigma_1^n \sigma_2^m$ equals $\mathsf{S}_{(n,m)\to(1,0)}$.

In the $\theta=\pi$ direction, the vector field to consider is now (recall \eqref{eq:PISpi-final} and \eqref{eq:PIISpi-final}) 
\be
\underline{\boldsymbol{S}}_\pi \left( \boldsymbol{\sigma} \right) = \underline{\text{P}}_\pi \left( \sigma_1\, N^{(-1)} (\mu) \right) \left[ \begin{array}{c} - \sigma_1\, \frac{\rmd}{\rmd \upmu} \log N^{(-1)} (\upmu) \\ \sigma_1^{-1} \left( 1 + \upmu \frac{\rmd}{\rmd \upmu} \log N^{(-1)} (\upmu) \right)\end{array} \right],
\ee
\noindent
where the $\underline{\text{P}}_\pi$ function can be obtained from $\underline{\text{P}}_0$ in \eqref{eq:P0-PI/PII-def} via backward-forward, obtaining $\underline{\text{P}}_\pi (x) \equiv - \underline{\text{P}}_0 (-x)$. The procedure to obtain the Stokes transitions is almost the same, with the difference that we now must perform the change of variables $\left( \sigma_1, \sigma_2 \right) \mapsto \left( \upmu, \uprho_\pi \right)$
\bea
\upmu &=& \sigma_1 \sigma_2, \\
\uprho_\pi &=& \sigma_1\, N^{(-1)} \left( \sigma_1 \sigma_2 \right).\label{eqn:coord_chan-c}
\eea
\noindent
Again, this change of variables trivializes the problem: the alien vector-field $\underline{\dot\Delta}_\pi$ (in the same sense as before) is given in these new coordinates by
\bea
\underline{\dot\Delta}_\pi \quad \leftrightarrow \quad \underline{\boldsymbol{S}}_{\pi} (\boldsymbol{\sigma}) \cdot \frac{\partial}{\partial\boldsymbol{\sigma}} = \underline{\text{P}}_\pi \left(\uprho_\pi\right) \frac{\partial}{\partial\upmu},
\eea
\noindent
with $\underline{\dot\Delta}_\pi$ again along the $\upmu$-direction, and where the flow equations are now
\bea
\frac{\rmd\upmu}{\rmd\uptau} &=& \underline{\text{P}}_\pi \left(\uprho_\pi\right), \\
\frac{\rmd\uprho_\pi}{\rmd\uptau} &=& 0.
\eea
\noindent
Analogously to the previous case, for an initial condition characterized by  $\left( \sigma_1,\sigma_2 \right)$ or $\left( \upmu,\uprho_\pi \right)$ one obtains
\bea
\label{eq:shift-a-II}
\upmu \left( \uptau \right) &=& \sigma_1 \sigma_2 + \uptau\, \underline{\text{P}}_\pi \left( \sigma_1\, N^{(-1)} \left( \sigma_1 \sigma_2 \right) \right) \equiv \upmu + \uptau\, \underline{\text{P}}_\pi \left(\uprho_\pi\right), \\
\label{eq:shift-b-II}
\uprho_\pi \left( \uptau \right) &=& \sigma_1\, N^{(-1)} \left( \sigma_1 \sigma_2 \right) \equiv \uprho_\pi,
\eea
\noindent
with Stokes jumps  $\underline{\pmb{\mathbb{S}}}_{\pi} \left( \upmu, \uprho_{\pi} \right)$ again trivializing to a simple \textit{shift}. The integration of the system is precisely the same; we now obtain back in the original $\left( \sigma_1, \sigma_2 \right)$ variables
\bea
\label{eq:transitionfunctionpi-a}
\underline{\pmb{\mathbb{S}}}_{\pi} \left( \bm{\sigma} \right) &=& \left[
\begin{array}{c}
\displaystyle {\frac{\sigma_1\, N^{(-1)} \left( \sigma_1 \sigma_2 \right)}{N^{(-1)} \left( \sigma_1\sigma_2 + \underline{\text{P}}_\pi \left( \sigma_1\, N^{(-1)} \left( \sigma_1 \sigma_2 \right) \right) \right)}}\\
\displaystyle {\frac{\left( \sigma_1 \sigma_2 + \underline{\text{P}}_\pi \left( \sigma_1\,  N^{(-1)} \left( \sigma_1 \sigma_2 \right) \right) \right) N^{(-1)} \left( \sigma_1\sigma_2 + \underline{\text{P}}_\pi \left( \sigma_1\, N^{(-1)} \left( \sigma_1 \sigma_2 \right) \right) \right)}{\sigma_1\, N^{(-1)} \left( \sigma_1 \sigma_2 \right)}}
\end{array}
\right].
\eea

For the free energy, one has to repeat the same procedure but now using the $\sigma_1^F$ and $\sigma_2^F$ variables. Define the function $\underline{\text{P}}_0^F (x)$ as
\be
\underline{\text{P}}^F_0 (x) =
\begin{cases}
\displaystyle {\log\left(1+x\right)}, \quad\quad & \text{\PI}, \\
\displaystyle {\frac12\log\left(1+2x\right)}, \quad\quad & \text{\PII},
\end{cases}
\ee
\noindent
and the function $\underline{\text{P}}^F_\pi$ which is given by $\underline{\text{P}}_\pi^F (x) \equiv - \underline{\text{P}}_0^F (-x)$. The Stokes transitions are finally given by\footnote{For clarity of presentation we are writing these formulae using $\upmu,\uprho$ as shorthand notation---but they are for $\bm{\sigma}$ coordinates. Likewise,  in these two formulae we dropped all $F$ superscripts---but they are implicitly there.}
\bea
\underline{\pmb{\mathbb{S}}}^F_{0} \left( \bm{\sigma}^F \right) &=& \left[
\begin{array}{c}
\displaystyle{\uprho_0^{-1} \left( \upmu + \underline{\text{P}}_0 \left( \uprho_0 \right) \right) N^{(1)} \left( \upmu + \underline{\text{P}}_0 \left( \uprho_0 \right) \right)} \\
\displaystyle{\frac{\uprho_0}{N^{(1)} \left( \upmu + \underline{\text{P}}_0 \left( \uprho_0 \right) \right)}}
\end{array}
\right],\\
\underline{\pmb{\mathbb{S}}}^F_{\pi} \left( \bm{\sigma}^F \right) &=& \left[
\begin{array}{c}
\displaystyle{\frac{\uprho_\pi}{N^{(-1)} \left( \upmu + \underline{\text{P}}_\pi \left( \uprho_\pi \right) \right)}} \\
\displaystyle{\uprho_\pi^{-1} \left( \upmu + \underline{\text{P}}_\pi \left( \uprho_\pi \right) \right) N^{(-1)} \left( \upmu + \underline{\text{P}}_\pi \left( \uprho_\pi \right) \right)}
\end{array}
\right].
\eea
\noindent
Having computed all, full two-parameter, Stokes transition functions, our next and final step is to put them to use in the analysis of the \PI/\PII~solution monodromy.

\subsection{A Preliminary Example: Monodromy of the Airy Function}\label{subsec:airymonodromy}

Before analyzing the harder cases of \PI~and \PII~it is pedagogical to start by recalling the simpler case of the Airy function; \textit{e.g.}, \cite{olbc10}. Following the analysis in \cite{as13}, we have that the transseries solution to the Airy equation $u''(z) - z\, u(z) = 0$ may be written as
\begin{equation}
u \left(z,\sigma_1,\sigma_2\right) = \sigma_1\, u_{\text{Ai}} \left(z\right) + \sigma_2\, u_{\text{Bi}} \left(z\right),
\end{equation}
\noindent
where the two transseries sectors are defined asymptotically in the $z \sim \infty$ region by
\bea
u_{\text{Ai}}(z) &=& \frac{1}{2\sqrt{\pi}\, z^{\frac{1}{4}}}\, \rme^{-\frac{1}{2} A  z^{\frac{3}{2}}}\, \Phi_{-1/2}(z),\\ 
u_{\text{Bi}}(z) &=& \frac{1}{2\sqrt{\pi}\, z^{\frac{1}{4}}}\, \rme^{+\frac{1}{2} A  z^{\frac{3}{2}}}\, \Phi_{+1/2}(z),
\eea
\noindent
with $A=\frac{4}{3}$, and the asymptotic series $\Phi_{\pm1/2}(z)$ given by 
\begin{equation}
	\Phi_{\pm1/2}(z)\simeq\sum_{n=0}^{+\infty}(\mp1)^na_n z^{-\frac32n}.
\end{equation}
\noindent
The coefficients $a_n$ can be recursively obtained by just plugging this expression into the Airy equation
\begin{equation}
u''(z) - z\, u (z) = 0.
\end{equation}
\noindent
The Stokes data for this case are $S_{\pm1} \equiv N^{(\pm1)}_1 = -\rmi$, and we have three Stokes directions upon the $z$-plane: $\arg z=0, \frac{2\pi}{3}, \frac{4\pi}{3}$. Note that when turning around once \textit{clockwise}\footnote{Turning around the anti-clockwise direction works in complete analogy to the calculation presented below.} on the $z$-plane one finds that the rays $\arg z = 0,-\frac{4\pi}3$ correspond to Borel-plane Stokes lines in the $\theta = 0,2\pi$ directions, respectively, while the ray $\arg z=-\frac{2\pi}3$ gets mapped into the Stokes line in the $\theta = \pi$ direction. Thus, in order to study the monodromy of the solution, one should find\footnote{The transition at $2\pi$ of the hypergeometric function gives the same Borel residue as that of the $0$ transition, so we have decided to denote $\underline{\mathfrak{S}}_{2\pi} \equiv \underline{\mathfrak{S}}_{0}$.} $\underline{\mathfrak{S}}_0^{-1} \circ \underline{\mathfrak{S}}_\pi^{-1} \circ \underline{\mathfrak{S}}_0^{-1}\, u \left( \rme^{-2\pi\rmi}\, z,\sigma_1,\sigma_2 \right) = u \left( z,\sigma_1,\sigma_2 \right)$ (an illustration of this path can be found in figure~\ref{fig:spiralPlanePII}, since the path that one has to follow for Airy is the same as for \PII). Indeed,
\bea
\underline{\mathfrak{S}}_0^{-1} \circ \underline{\mathfrak{S}}_\pi^{-1} \circ \underline{\mathfrak{S}}_0^{-1}\, u \left( \rme^{-2\pi\rmi}\, z,\sigma_1,\sigma_2 \right) &=& \underline{\mathfrak{S}}_0^{-1} \circ \underline{\mathfrak{S}}_\pi^{-1}\, u \left( \rme^{-2\pi\rmi}\, z,\sigma_1+S_1\sigma_2,\sigma_2 \right) \\
&=& \underline{\mathfrak{S}}_0^{-1}\, u \left( \rme^{-2\pi\rmi}\, z,\sigma_1+S_1\sigma_2,\sigma_2+S_{-1} \left(\sigma_1+S_1\sigma_2\right) \right) \\
&=& u \left( \rme^{-2\pi\rmi}\, z,\sigma_1+S_1\sigma_2+S_1S_{-1}\sigma_1,S_{-1}\sigma_1 \right) \\
&=& u \left( \rme^{-2\pi\rmi}\, z,-\rmi\sigma_2,-\rmi\sigma_1 \right),
\eea
\noindent
where we have used $S_1 S_{-1} = -1$. The last step is to now consider how the transseries sectors change whenever $z$ rotates once \textit{clockwise} on the complex plane, \textit{i.e.}, $z \mapsto \rme^{-2\pi\rmi}\, z$. In particular, we are interested in the quantities $z^{-\frac{1}{4}}$ and $z^{-\frac{3}{2}n}$ appearing in the transseries asymptotic series. Upon rotating $z$ once, these give 
\bea
z^{-\frac{1}{4}} &\mapsto& \rme^{\frac{1}{4}\, 2\pi\rmi}\, z^{-\frac{1}{4}} = \rmi\, z^{-\frac14}, \\
z^{-\frac{3}{2}n} &\mapsto& \rme^{\frac{3}{2} n\, 2\pi\rmi}\, z^{-\frac{3}{2}n} = (-1)^n\, z^{-\frac{3}{2}n}.
\eea
\noindent
What this implies is that the rotation yields both a global prefactor of $\rmi$, in each of the transseries sectors; as well as changing the sign of exponential weights associated to each of them. At the same time, the $(-1)^n$ factor swaps the asymptotic series as $\Phi_{\pm1/2} \left(\rme^{-2\pi\rmi}\, z\right) = \Phi_{\mp1/2} \left(z\right)$. As such, after one rotation one obtains
\bea
u_{\text{Ai}} \left( \rme^{-2\pi\rmi}\, z \right) &=& \rmi\, u_{\text{Bi}} \left(z\right), \\
u_{\text{Bi}} \left(\rme^{-2\pi\rmi}\, z \right) &=& \rmi\, u_{\text{Ai}} \left(z\right).
\eea
\noindent
This is in fact the last ingredient we needed to prove the (trivial) monodromy of the Airy solution, since now we have that:
\bea
\underline{\mathfrak{S}}_0^{-1} \circ \underline{\mathfrak{S}}_\pi^{-1} \circ \underline{\mathfrak{S}}_0^{-1}\, u \left( \rme^{-2\pi\rmi}\, z,\sigma_1,\sigma_2 \right) &=& u \left( \rme^{-2\pi\rmi}\, z,-\rmi\sigma_2,-\rmi\sigma_1 \right) = \\
&=& -\rmi \sigma_2\, u_{\text{Ai}} \left( \rme^{-2\pi\rmi}\, z \right) - \rmi \sigma_1\, u_{\text{Bi}} \left( \rme^{-2\pi\rmi}\, z \right) = \\
&=& \sigma_2\, u_{\text{Bi}} \left(z\right) + \sigma_1\, u_{\text{Ai}} \left(z\right) = \\
&=& u \left( z,\sigma_1,\sigma_2 \right).
\eea
\noindent
This basically shows (in a very simple example) how keeping track of both the Stokes automorphisms and the branch-cuts of the corresponding transseries (asymptotic series) monomials, one can easily recover the (trivial) monodromy of the Airy function---at the level of its full transseries. As we shall see next, playing the same game in the Painlev\'e context turns out naturally harder.

\subsection{Painlev\'e Direct Monodromy Calculation}\label{subsec:directmonodromy}

Let us now proceed analogously to the Airy example in the previous subsection, for our cases of \PI~and \PII---where multiple complications appear. First of all, in the Airy example we had a linear equation. This meant that the number of resurgent transseries sectors in the solution was finite. In our nonlinear cases of \PI~and \PII~this feature is immediately more complicated, as we now have infinitely many transseries sectors (including logarithmic transmonomials). What this could \textit{a priori} imply is that the solution direct-monodromy could actually be hidden away in the infinite sums of those sectors, and would not be directly translatable to the transseries parameters. This did not happen in the Airy example. In fact, secondly, in that example the monodromy statement was first set-up in the observation that the transseries parameters $\sigma_1$ and $\sigma_2$ returned to themselves after four rotations. This, however, was still \textit{not} the precise monodromy statement. Indeed, thirdly and finally, we still had to use properties of the Airy transseries sectors themselves in order to make the trivial, single-rotation monodromy explicit. For \PI~and \PII~these two features are also immediately more complicated. The geometry of moduli space is not flat $\BC \times \BC$ \cite{g76, o79} which implies that herein one can only assume the most general statement: even if the transseries parameters eventually do return to themselves, after a \textit{finite} amount of rotations, the following step is not immediately guaranteed in the sense that after one rotation around the complex $z$-plane---and taking into account all Stokes transitions and properties of the infinite Painlev\'e transseries sectors---we may still need to do a Borel--\'Ecalle transseries \textit{resummation} in order to explicitly find the return back to the starting point. This implies resummation of a two-parameter transseries which contains both enhanced and suppressed exponentials---a quite hard task in practice, for which we were unable to produce a sensible prescription. Further, making the geometry of moduli space precise \cite{g76, o79} is beyond the scope of this work. Nonetheless, in the following we shall try to understand it as much as possible in the attempt to make the trivial, single-rotation monodromy as explicit as we possible can.

Let us start by understanding the behavior of the transseries parameters under repeated Stokes transitions. Here, in contrast with the Airy example, an additional subtlety appears: the sectors for \PI~and \PII~contain logarithms and square-roots---as we outlined back in section~\ref{sec:review}---and these themselves already have non-trivial monodromy behavior. But this also implies that the action of the Stokes automorphism actually changes depending on the phase of $x$. In other words,
\begin{equation}
\label{eq:stokes-mismatch-0-pi}
\underline{\mathfrak S}_0\,\neq\,\underline{\mathfrak S}_{2\pi}.
\end{equation}
\noindent
Still, by understanding rotations of the transseries, we can relate the two automorphisms above. For the ensuing calculation it is beneficial to split-off the global square-root factor from the solution to \PI~and \PII,
\be
u(z) \equiv \sqrt{z}\, U(z).
\ee
\noindent
In this notation, let us repeat herein the transseries for illustrative purposes (with $x=z^{-\frac{5}{4}}$)
\be
\label{eq:transseries-repeat-U}
U \left(x;\sigma_1,\sigma_2\right) = \sum_{n=0}^{+\infty} \sum_{m=0}^{+\infty} \sigma_1^n\, \sigma_2^m\, \rme^{- \left( n-m \right) \frac{A}{x}}\, \sum_{k=0}^{k_{nm}} \left( \frac{\log x}{2} \right)^k\, \sum_{g=0}^{+\infty} u_{2g}^{(n,m)[k]}\, x^{g+\beta_{nm}^{[k]}}.
\ee
\noindent
Earlier in this paper, when we computed Stokes data, we only considered the first sheet of the above logarithms. Therefore, we now need to rotate the transseries by hand when applying the Stokes automorphisms at other phases of $x$. Interestingly, the change that the transseries undergoes when one rotates the variable $x$ can be absorbed by a change of its parameters $\boldsymbol{\sigma}$. This can be seen in a simple way by considering the logarithmic-resummed version of the transseries which we have already formulated in equation \eqref{eq:trans_form1-LOGSUM},
\be
U \left(x;\sigma_1,\sigma_2\right) =  \sum_{n=0}^{+\infty} \sum_{m=0}^{+\infty} \sigma_1^n\, \sigma_2^m\, \rme^{- \left( n-m \right) \frac{A}{x}}\, x^{- \frac{1}{2} \alpha \left( n-m \right) \sigma_1 \sigma_2}\, \sum_{g=0}^{+\infty} u_{2g}^{(n,m)[0]}\, x^{g+\beta_{nm}^{[0]}}.
\ee
\noindent
Let us investigate how $U$ changes under rotations. Recall that $\beta_{nm}^{[0]}$ is an integer when $n+m$ is even, while it is half-integer when $n+m$ is odd. Rotating $x$ by $\rme^{2\pi\rmi\, \rho}$ then yields
\bea
U \left( \rme^{2\pi\rmi\, \rho}\, x;\sigma_1,\sigma_2 \right) &=& \sum_{n=0}^{+\infty} \sum_{m=0}^{+\infty} \sigma_1^n\, \sigma_2^m\, \rme^{- \left( n-m \right) \frac{A}{x}}\, x^{- \frac{1}{2} \alpha \left( n-m \right) \sigma_1 \sigma_2}\, \rme^{- \rmi\pi\, \rho\, \alpha \left( n-m \right) \sigma_1 \sigma_2}\,\times \nonumber\\
&&
\times\, \sum_{g=0}^{+\infty} u_{2g}^{(n,m)[0]}\, x^{g+\beta_{nm}^{[0]}}\, \rme^{\rmi\pi\, \rho \left( n+m \right)}.
\eea
\noindent
Note how terms originating from the rotation of $x$ come nicely attached to powers of $n$ and $m$. We can then absorb these terms in $\sigma_1$ and $\sigma_2$---which also come with attached $n$ and $m$ powers. Explicitly, we will denote such a rotation on $\left( \sigma_1,\sigma_2 \right)$ by $\boldsymbol{R}$. Grouping together the above terms, it reads
\be
\boldsymbol{R}_\rho \left( \boldsymbol{\sigma} \right) = \rme^{\rmi\pi\, \rho} \left[ 
\begin{matrix}
\rme^{- \rmi\pi\, \rho \alpha\, \sigma_1 \sigma_2}\,\sigma_1\\
\rme^{\rmi\pi\, \rho \alpha\, \sigma_1 \sigma_2}\, \sigma_2
\end{matrix}
\right],
\ee
\noindent
where one full rotation corresponds to $\rho=1$. Then, on the full transseries, we have the rotation relation
\be
U \left( \rme^{2\pi\rmi \rho}\, x; \boldsymbol{\sigma} \right) = U \left(x; \boldsymbol{R}_\rho \left( \boldsymbol{\sigma} \right) \right).
\ee
\noindent
We may finally relate the two automorphisms above, in \eqref{eq:stokes-mismatch-0-pi}. One has:
\begin{equation}
\underline{\mathfrak{S}}_{2\pi} = \boldsymbol{R}_{1} \circ \underline{\mathfrak{S}}_0 \circ \boldsymbol{R}_{-1}.
\end{equation}
\noindent
Equipped with these tools, we may now spell-out iterations of the Stokes automorphism. It indeed turns out that the transseries parameters do not at first seem to return to themselves. Rather, depending on the starting point, one of them will grow large while the other one will compensate by decreasing at a similar pace. Within these changes, some structure does remain: the combination $\upmu = \sigma_1\sigma_2$ changes by a simple shift, as predicted in, \textit{e.g.}, \eqref{eq:shift-a}-\eqref{eq:shift-b}.

\begin{figure}
	\centering
	\begin{subfigure}[t]{0.5\textwidth}
		\centering
		\includegraphics[height=2.5in]{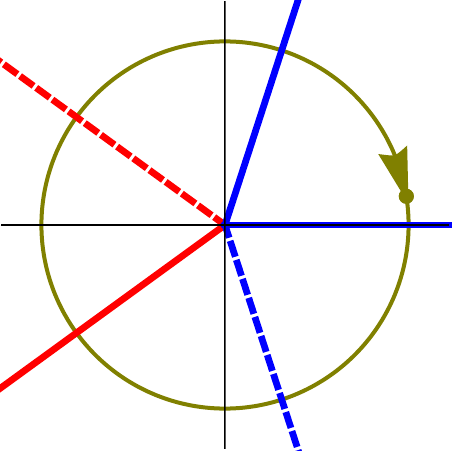}
		\caption{$z$-plane.}
	\end{subfigure}%
	~ 
	\begin{subfigure}[t]{0.5\textwidth}
		\centering
		\includegraphics[height=2.5in]{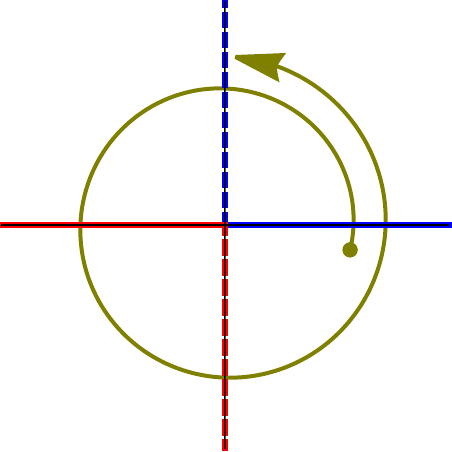}
		\caption{$x$-plane.}
	\end{subfigure}
	\caption{In image (a) we depict a visualization of rotating once on the complex $z$-plane for \PI. In image (b) we have plotted the same path but translated to the resurgent $x$-plane. Here, the Stokes lines that one needs to cross become clear (with the usual Stokes/anti-Stokes alternation also cleaner on the $x$-plane). Note that one has to carefully keep track of the phases of $x$ and $z$ in order to deal with the multi-sheeted nature of the transseries asymptotic sectors. In the plots, solid lines correspond to Stokes jumps while dashed ones represent anti-Stokes lines. In the same spirit, the blue lines are associated to the positive (real or imaginary) direction while the red ones represent the negative counterpart.}
	\label{fig:spiralPlanePI}
\end{figure}

\begin{figure}
\centering
		\begin{subfigure}[t]{0.5\textwidth}
		\centering
		\includegraphics[height=2.5in]{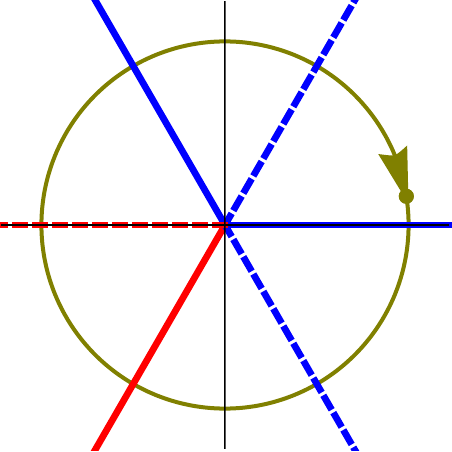}
		\caption{$z$-plane.}
	\end{subfigure}%
	~ 
	\begin{subfigure}[t]{0.5\textwidth}
		\centering
		\includegraphics[height=2.5in]{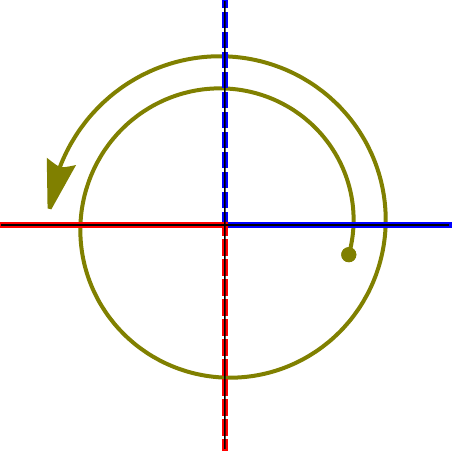}
		\caption{$x$-plane.}
	\end{subfigure}
	\caption{In image (a) we depict a visualization of rotating once on the complex $z$-plane for \PII~(or Airy). In image (b) we have plotted the same path but translated to the resurgent $x$-plane. Here, the Stokes lines that one needs to cross become clear. Note that one has to carefully keep track of the phases of $x$ and $z$ in order to deal with the multi-sheeted nature of the transseries asymptotic sectors. In the plots, solid lines correspond to Stokes jumps while dashed ones represent anti-Stokes lines. In the same spirit, the blue lines are associated to the positive (real or imaginary) direction while the red ones represent the negative counterpart.}
	\label{fig:spiralPlanePII}
\end{figure}

Having investigated the Stokes automorphisms involved in the direct monodromy calculation, let us now discuss the calculation\footnote{In the first \texttt{arXiv} version of the present paper this calculation had not been explicitly written due to a bug in its corresponding computer code. Once fixed, the code immediately yielded the results reported in the following.} itself. An explicit computation of the \PI~or \PII~monodromies would consist of walking a full circle on the complex $z$-plane and seeing whether the solution $u_{\text{I}}/\mu_{\text{II}}$ returns to itself. One first observes that, akin to the Airy case in subsection~\ref{subsec:airymonodromy}, our transseries parameters do not return to themselves after a \textit{single} rotation (even using the above rotation automorphism). What we should ask instead is whether they return to themselves after a \textit{finite} amount of rotations on the $z$-plane, with such number determined by inspection of the transmonomials in \eqref{eq:transseries-repeat-U}. Keeping in mind the map to the resurgent variable $x$ (which for \PI~reads $x=z^{-5/4}$ and for \PII~corresponds to $x=z^{-3/2}$), we can determine which Stokes lines are crossed and in which order. We have illustrated these transitions in figure~\ref{fig:spiralPlanePI} for \PI, whose transmonomials return to themselves after four rotations; and figure~\ref{fig:spiralPlanePII} for \PII, whose transmonomials return to themselves after two rotations.

\begin{figure}
\centering
\includegraphics[width=0.49\textwidth]{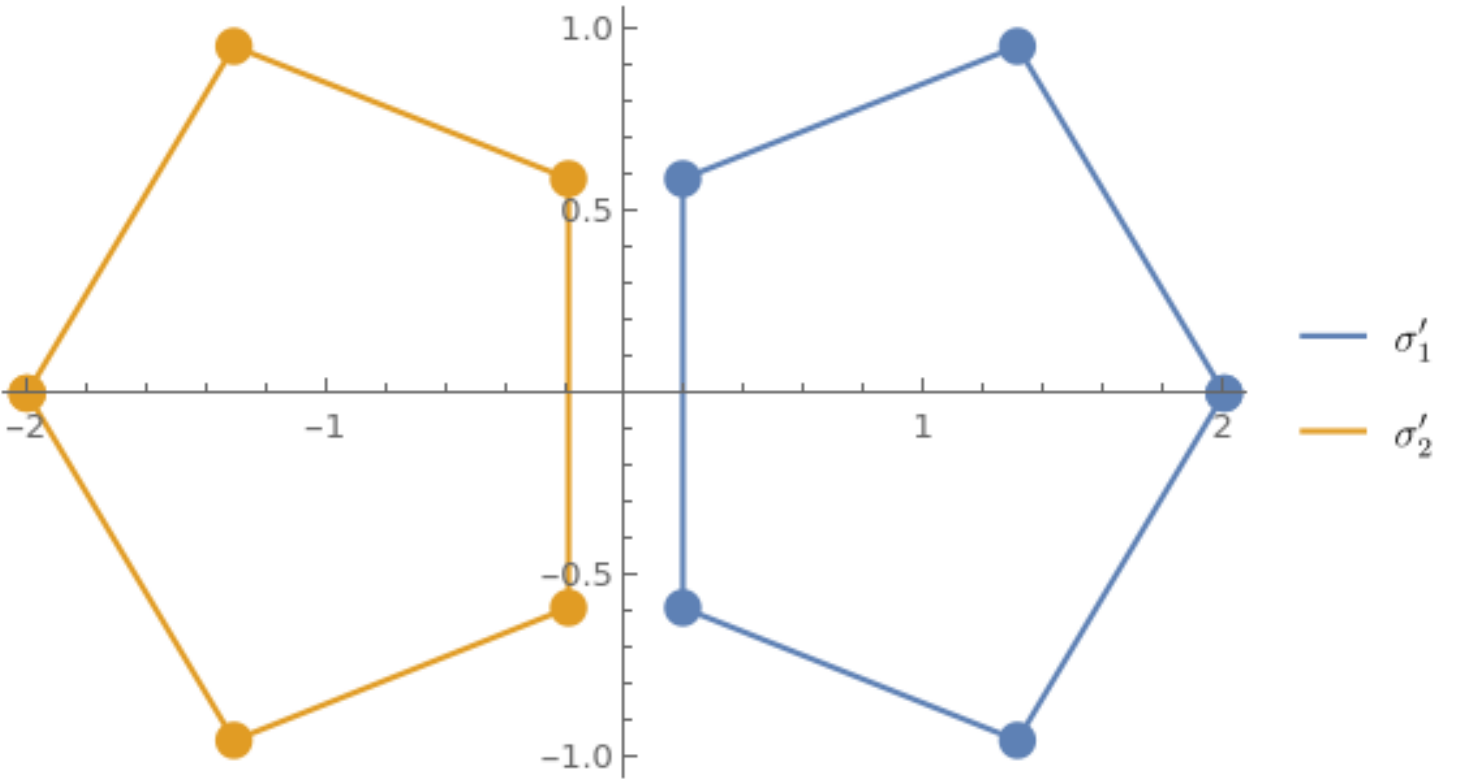}
\hfill
\includegraphics[width=0.49\textwidth]{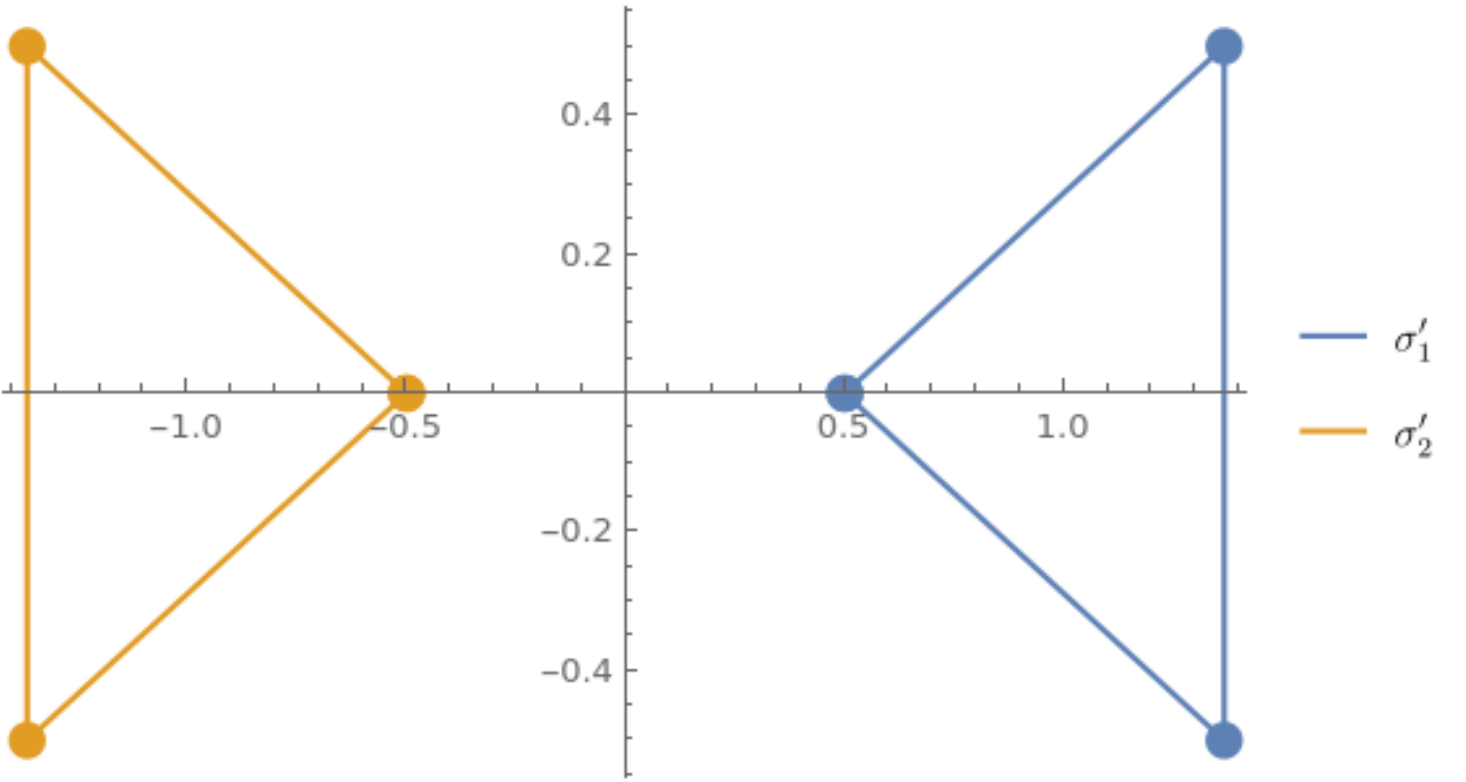}
\caption{Monodromy orbits in transseries-parameter moduli-space. These figures plot both $\sigma_1$ and $\sigma_2$ orbits on the same plane, and the numerical results have been polynomially mapped into natural regular shapes for ease of reading. The new variables $(\sigma_1',\sigma_2')$ are chosen to have a pentagonal shape in the case of \PI, and a triangular shape (rather than hexagonal) for \PII.}
\label{fig:automorphisms}
\end{figure}

Let us finally make the action of the combined automorphisms on the complex $z$-plane explicit. In the case of \PI, computing the monodromy after the aforementioned \textit{four} rotations amounts to computing the effect of the automorphism
\be
\boldsymbol{R}_{-5}\circ\underline{\mathfrak{S}}_{9\pi}\circ\underline{\mathfrak{S}}_{8\pi}\circ\underline{\mathfrak{S}}_{7\pi}\circ\underline{\mathfrak{S}}_{6\pi}\circ\underline{\mathfrak{S}}_{5\pi}\circ\underline{\mathfrak{S}}_{4\pi}\circ\underline{\mathfrak{S}}_{3\pi}\circ\underline{\mathfrak{S}}_{2\pi}\circ\underline{\mathfrak{S}}_\pi\circ\underline{\mathfrak{S}}_0
\ee
\noindent
on any arbitrary point of $(\sigma_1,\sigma_2)$ moduli space. While an explicit analytical computation is rather involved, we can numerically evaluate this map on different selected points and verify closure. We report one such example of our many results in table~\ref{tab:pi_free_en}. For all other cases which we have analyzed, points on the transseries-parameter moduli-space always return to themselves to very high precision (on the order of $\sim 2000$ significant digits). Turning to the case of \PII, computing the monodromy after the aforementioned \textit{two} rotations now amounts to computing
\be
\boldsymbol{R}_{-3}\circ\underline{\mathfrak{S}}_{5\pi}\circ\underline{\mathfrak{S}}_{4\pi}\circ\underline{\mathfrak{S}}_{3\pi}\circ\underline{\mathfrak{S}}_{2\pi}\circ\underline{\mathfrak{S}}_\pi\circ\underline{\mathfrak{S}}_0.
\ee
\noindent
Also herein we have numerically evaluated this map on many points in transseries-parameter moduli-space and we report one such example in table~\ref{tab:pii_free_en}. For all our many results, closure of the orbits is verified to very high precision (with the same precision as for \PI). We have further pictured these automorphisms (and their orbit closure) in figure~\ref{fig:automorphisms}.

\begin{table}
\centering
\begin{tabular}{c|c|c}
Monodromy Actions&$\sigma_1$&$\sigma_2$\\\hline
0&1&2\\\hline
1&-39.05-7.23$\rmi$&-0.03-0.03$\rmi$\\\hline
2&-3.75-0.85$\rmi$&0.53-0.01$\rmi$\\\hline
3&0.01+0.06$\rmi$&-22.15+11.76$\rmi$\\\hline
4&-0.03&27.31+48.3$\rmi$\\\hline
5&1&2
\end{tabular}
\caption{Iterated action of the \PI~automorphism $\boldsymbol{R}_{-n}\circ\prod_{j=0}^{n-1}\left(\underline{\mathfrak S}_{\pi(2j+1)}\circ\mathfrak S_{2\pi j}\right)$; with $n$ the number of distinct actions and with the product read as composition---where terms with lower $j$ act first. These are (truncated) numerical results on the example of $(\sigma_1,\sigma_2)=(1,2)$ for the \PI~free-energy parametrization (where intermediate points turn out to be dramatically simpler numbers as compared to the standard parametrization). Closure of the orbit is clear.}
\label{tab:pi_free_en}
\end{table}

\begin{table}
\centering
\begin{tabular}{c|c|c}
Monodromy Actions&$\sigma_1$&$\sigma_2$\\\hline
0&1&2\\\hline
1&118.27 + 1394.17 $\rmi$& 0.001+0.001$\rmi$\\\hline
2&-0.002+0.001$\rmi$& 1200.89 + 507.76 $\rmi$\\\hline
3&1&2
\end{tabular}
\caption{Iterated action of the \PII~automorphism (in the same setup as for table~\ref{tab:pi_free_en}). Again, these are (truncated) numerical results on the example of $(\sigma_1,\sigma_2)=(1,2)$ for the \PII~free-energy parametrization, and closure of the orbit is absolutely clear.}
\label{tab:pii_free_en}
\end{table}

It is interesting to observe that there are particular points on the transseries-parameter moduli-space which are \textit{singular} under the above automorphisms. For example, it is very easy to show that all points of the form $(\sigma,0)$ and $(0,\sigma)$ are singular, for any $\sigma \in \BC$ (equivalently, this is the $\upmu=0$ locus). It turns out that mapping those points to the coordinates of the isomonodromy approach (to be explained in detail in the upcoming subsection~\ref{subsec:monandisomon}), they precisely fall\footnote{At least in the case of \PI. In the case of \PII~the isomonodromic WKB approach is more involved and such subtleties will be explained in subsection~\ref{subsec:monandisomon}.} on the (adequate) singularity manifolds described in \cite{n13, n16}. Such singularity manifolds describe the special, truncated solutions of \PI~(which were earlier illustrated in figure~\ref{fig:P1-singular-Boutroux}) and this is in accord with the expectation that \textit{tronqu\'ee} Painlev\'e solutions may be expressed as one-parameter transseries, with the \textit{tritronqu\'ee} solution being the special zero-parameter case. In our language, it is hence the locus $\upmu=0$ which yields solutions with at least one sector free of poles. 

Having understood Stokes transitions and proved closure of orbits in transseries-parameter moduli-space, we should next, in principle, check monodromy of the \PI~or \PII~solutions (as earlier illustrated for Airy). For this, however, and as already mentioned, we would need a resummation of the two-parameter transseries---which unfortunately does not exist so far. As such, albeit we have not managed to prove the exact Painlev\'e property, due to difficulty with transseries resummations, we did manage to check the weaker property of orbit closure---which we still consider a partial proof of our Stokes-data results. Full (direct) proof of the  Painlev\'e property is an interesting open problem that we leave for future work.  Nonetheless, there is another (indirect) way of checking our results, by comparing to the isomonodromic deformations of \PI~and \PII. We shall perform such a comparison in the following subsection.

\subsection{Monodromy and Isomonodromy}\label{subsec:monandisomon}

Having discussed the implementation of the \textit{direct} monodromy approach, let us now take an alternative, \textit{indirect} route to the monodromy problem by comparing our results for \PI~to those of its isomonodromic deformation. More explicitly, we will compare our Stokes data to that in the work of \cite{i19a} (an analogous comparison could be employed for \PII, this time comparing with \cite{i11}). Let us start by laying out the main ideas. Stokes data for the associated isomonodromy systems are usually computed using WKB methods. In the following, we will refer to them as \textit{linear Stokes data}, in contrast to the \textit{nonlinear} Stokes data of \PI~and \PII~we have computed. For the comparison, we then need to answer the following question:
\begin{itemize}
\item How are the linear Stokes coefficients of the isomonodromic deformation related to the nonlinear Stokes data of \PI~and \PII?
\end{itemize}
\noindent
An answer to this question has been given in\footnote{We thank K.~Iwaki for sharing his unpublished ideas on this question with us.} \cite{t00, i19b}. Let us give a short review of this connection. We first need to introduce the isomonodromy system associated to \PI~(the following discussion can be analogously repeated for \PII)---which is done with \PI~as written in minimal-string conventions \eqref{eq:PI-(2,3)}. For that, we need to recall the most relevant quantities in the isomonodromy picture. Rescaling the \PI~solution $u_{(2,3)}$ and its variable $z$ from our conventions in subsection~\ref{subsec:PI2dQG} to the conventions used in \cite{i19a},
\begin{equation}
q = \left( -\frac{1}{2} \right)^{\frac{8}{5}} u_{(2,3)}, \qquad t = (-1)^{\frac{8}{5}}\, 2^{\frac{3}{10}}\, z, 
\end{equation}
\noindent
then the Hamiltonian system associated to \PI~may be written as (see, \textit{e.g.}, \cite{o80, o81})
\be
g_{\text{s}}\, \frac{\partial q}{\partial t} = \frac{\partial \CH}{\partial p}, \qquad g_{\text{s}}\, \frac{\partial p}{\partial t} = - \frac{\partial \CH}{\partial q},
\ee
\noindent
with the Hamiltonian
\begin{equation}
\CH = \frac{1}{2} p^2 - 2 q^3 - t q.
\end{equation}
\noindent
This Hamiltonian has the property
\begin{equation}
\frac{\partial}{\partial t} h(t) = -q (t), \qquad h(t) = \CH \left( t, q(t), p(t) \right).
\end{equation}
\noindent
The associated isomonodromy system then reads \cite{i19a}
\bea
\label{eq:L_I}
\left\{ g_{\text{s}}^2\, \frac{\partial^2}{\partial\xi^2} - \frac{g_{\text{s}}}{\xi - q} \left( g_{\text{s}}\, \frac{\partial}{\partial\xi} - p \right) - \left( 4 \xi^3 + 2 t \xi + 2 \CH \right) \right\} \Psi (\xi) &=& 0, \\
\label{eq:D_I}
\left\{ g_{\text{s}}\, \frac{\partial}{\partial t} - \frac{1}{2 \left(\xi - q\right)} \left( g_{\text{s}}\, \frac{\partial}{\partial \xi} - p\right) \right\} \Psi (\xi) &=& 0,
\eea
\noindent
where we have introduced the additional variable $\xi$. Finally, the partition function (or tau-function), in the sense of \eqref{eq:PI-F/Z}, associated to the minimal-string \PI~formulation is introduced via
\begin{equation}
Z (t) = \exp \left( \frac{1}{g_{\text{s}}^2}\, \int\rmd t\, h(t) \right).
\end{equation}

A (modified) two-parameter transseries \textit{ansatz} for this partition function was given in \cite{i19a}, which reads instead
\begin{equation}
\label{eq:taufunction}
Z \left( g_{\text{s}}; t,\nu,\rho\right) = \sum_{k\in\mathbb{Z}}\rme^{\frac{2\pi\rmi\, k\, \rho}{g_{\text{s}}}}\, \CZ \left( g_{\text{s}}; t, \nu + k g_{\text{s}} \right)
\end{equation}
\noindent
(with $\CZ \left( g_{\text{s}}; t, \nu \right)$ the topological-recursion partition-function for the \PI~spectral curve \cite{i19a}). It is interesting to note that the above (modified) transseries parameters $\rho$, $\nu$ are very akin to the transseries parameters in our diagonal-framing transseries formulation of formula \eqref{eq:PI/II-2PTS-diagonal}. Namely, we may schematically identify $\CZ \left( g_{\text{s}}; t, \nu \right)$ as the partition-function analogue of our diagonal-framing $\Phi^{(0)} (x,\upmu)$ in \eqref{eq:diagonalsummation}, and the generic $\CZ \left( g_{\text{s}}; t, \nu + k g_{\text{s}} \right)$ as the adequate analogues of our $\Phi^{(k)}_{+} (x,\upmu)$ and $\Phi^{(k)}_{-} (x,\upmu)$ in \eqref{eq:diagonalforwardsummation}-\eqref{eq:diagonalbackwardsummation}. Consequently we can further schematically identify
\bea
\rme^{\frac{\rho}{g_{\text{s}}}} \quad &\leftrightarrow& \quad \sigma_{1,2}\, \rme^{\pm\, \frac{A}{x}}, \\
\nu \quad &\leftrightarrow& \quad \upmu.
\eea
\noindent
This is also reflected in the Stokes transition-functions \eqref{eq:shift-a}-\eqref{eq:shift-b}, \eqref{eq:shift-a-II}-\eqref{eq:shift-b-II} as compared to \eqref{eq:transitionfunction0-a}, \eqref{eq:transitionfunctionpi-a}, which are much simpler in diagonal framing; and it is in these variables that the map to the isomonodromic problem will become clear as we shall make explicit next.

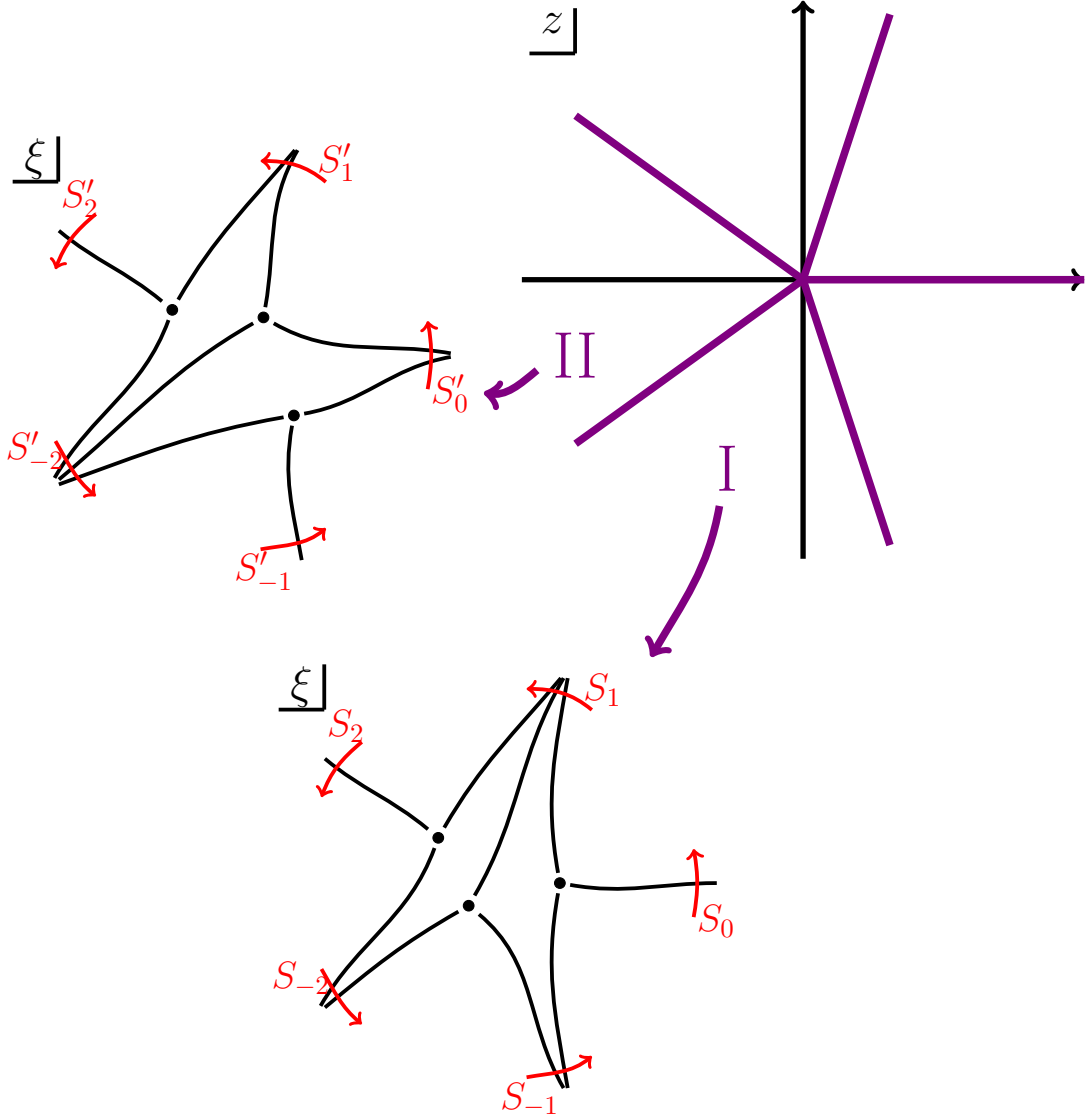
\begin{figure}
\centering
\begin{tikzpicture}[scale=1]
\draw[->, line width=0.07cm] (-3.7, 0) -- (3.7, 0);
\draw[->, line width=0.07cm] (0, -3.7) -- (0, 3.7);
\draw[line width=0.05cm] (-3.6, 3.0) -- (-3.0, 3.0);
\draw[line width=0.05cm] (-3.0, 3.0) -- (-3.0, 3.6);
\node at (-3.3, 3.4) {\LARGE $z$};
\draw[violet, line width=0.1cm] (0,0) -- (0:3.7);
\draw[violet, line width=0.1cm] (0,0) -- (360/5:3.7);
\draw[violet, line width=0.1cm] (0,0) -- (2*360/5:3.7);
\draw[violet, line width=0.1cm] (0,0) -- (3*360/5:3.7);
\draw[violet, line width=0.1cm] (0,0) -- (4*360/5:3.7);
\node[violet] at (-3, -1) {\Huge II};
\node[violet] at (-1, -2.5) {\Huge I};
\draw[violet, ->, line width=0.1cm] (-3.5, -1.2) to[out=220, in=-10] (-4.2, -1.5);
\draw[violet, ->, line width=0.1cm] (-1.1, -3) to[out=260, in=60] (-2, -5);
\def\PIx{-4};
\def\PIy{-8};
\node (s1) at ([shift=({0:3})](\PIx, \PIy) {};
\node (s2) at ([shift=({360/5:3})](\PIx, \PIy) {};
\node (s3) at ([shift=({2*360/5:3})](\PIx, \PIy) {};
\node (s4) at ([shift=({3*360/5:3})](\PIx, \PIy) {};
\node (s5) at ([shift=({4*360/5:3})](\PIx, \PIy) {};
\draw[line width=0.05cm] (\PIx-2.3, \PIy+2.9) -- (\PIx-2.3, \PIy+2.3);
\draw[line width=0.05cm] (\PIx-2.9, \PIy+2.3) -- (\PIx-2.3, \PIy+2.3);
\node at (\PIx-2.6, \PIy+2.6) {\LARGE $\xi$};
\node (tpI1) at (\PIx-0.8, \PIy+0.6) {};
\node (tpI2) at (\PIx-0.4, \PIy-0.3) {};
\node (tpI3) at (\PIx+0.8, \PIy) {};
\draw[fill=black] (tpI1) circle (2pt);
\draw[fill=black] (tpI2) circle (2pt);
\draw[fill=black] (tpI3) circle (2pt);
\draw[line width=0.05cm] (tpI3) to[out=-10, in=180] (s1);
\draw[line width=0.05cm] (tpI3) to[out=100, in=-100] (s2);
\draw[line width=0.05cm] (tpI3) to[out=-100, in=100] (s5);
\draw[line width=0.05cm] (tpI2) to[out=60, in=-120] (s2);
\draw[line width=0.05cm] (tpI2) to[out=-40, in=120] (s5);
\draw[line width=0.05cm] (tpI2) to[out=210, in=40] (s4);
\draw[line width=0.05cm] (tpI1) to[out=60, in=230] (s2);
\draw[line width=0.05cm] (tpI1) to[out=140, in=-40] (s3);
\draw[line width=0.05cm] (tpI1) to[out=250, in=60] (s4);
\draw[red, ->, line width=0.05cm] ([shift=({2*360/5-10:2.6})](\PIx, \PIy) to[out=220, in=70] ([shift=({2*360/5+10:2.6})](\PIx, \PIy);
\node[red] at ([shift=({2*360/5-10:2.9})](\PIx, \PIy) {\Large $S_{2}$};
\draw[red, ->, line width=0.05cm] ([shift=({3*360/5-10:2.6})](\PIx, \PIy) to[out=300, in=140] ([shift=({3*360/5+10:2.6})](\PIx, \PIy);
\node[red] at ([shift=({3*360/5-10:2.9})](\PIx, \PIy) {\Large $S_{-2}$};
\draw[red, ->, line width=0.05cm] ([shift=({4*360/5-10:2.6})](\PIx, \PIy) to[out=10, in=220] ([shift=({4*360/5+10:2.6})](\PIx, \PIy);
\node[red] at ([shift=({4*360/5-10:2.9})](\PIx, \PIy) {\Large $S_{-1}$};
\draw[red, ->, line width=0.05cm] ([shift=({5*360/5-10:2.6})](\PIx, \PIy) to[out=80, in=280] ([shift=({5*360/5+10:2.6})](\PIx, \PIy);
\node[red] at ([shift=({5*360/5-10:2.9})](\PIx, \PIy) {\Large $S_{0}$};
\draw[red, ->, line width=0.05cm] ([shift=({1*360/5-10:2.6})](\PIx, \PIy) to[out=140, in=360] ([shift=({1*360/5+10:2.6})](\PIx, \PIy);
\node[red] at ([shift=({1*360/5-10:2.9})](\PIx, \PIy) {\Large $S_{1}$};
\def\PIx{-7.5};
\def\PIy{-1};
\node (s1) at ([shift=({0:3})](\PIx, \PIy) {};
\node (s2) at ([shift=({360/5:3})](\PIx, \PIy) {};
\node (s3) at ([shift=({2*360/5:3})](\PIx, \PIy) {};
\node (s4) at ([shift=({3*360/5:3})](\PIx, \PIy) {};
\node (s5) at ([shift=({4*360/5:3})](\PIx, \PIy) {};
\draw[line width=0.05cm] (\PIx-2.3, \PIy+2.9) -- (\PIx-2.3, \PIy+2.3);
\draw[line width=0.05cm] (\PIx-2.9, \PIy+2.3) -- (\PIx-2.3, \PIy+2.3);
\node at (\PIx-2.6, \PIy+2.6) {\LARGE $\xi$};
\node (tpI1) at (\PIx-0.8, \PIy+0.6) {};
\node (tpI2) at (\PIx+0.4, \PIy+0.5) {};
\node (tpI3) at (\PIx+0.8, \PIy-0.8) {};
\draw[fill=black] (tpI1) circle (2pt);
\draw[fill=black] (tpI2) circle (2pt);
\draw[fill=black] (tpI3) circle (2pt);
\draw[line width=0.05cm] (tpI3) to[out=10, in=190] (s1);
\draw[line width=0.05cm] (tpI3) to[out=190, in=20] (s4);
\draw[line width=0.05cm] (tpI3) to[out=-100, in=100] (s5);
\draw[line width=0.05cm] (tpI2) to[out=80, in=-120] (s2);
\draw[line width=0.05cm] (tpI2) to[out=-30, in=170] (s1);
\draw[line width=0.05cm] (tpI2) to[out=210, in=40] (s4);
\draw[line width=0.05cm] (tpI1) to[out=60, in=230] (s2);
\draw[line width=0.05cm] (tpI1) to[out=140, in=-40] (s3);
\draw[line width=0.05cm] (tpI1) to[out=250, in=60] (s4);
\draw[red, ->, line width=0.05cm] ([shift=({2*360/5-10:2.6})](\PIx, \PIy) to[out=220, in=70] ([shift=({2*360/5+10:2.6})](\PIx, \PIy);
\node[red] at ([shift=({2*360/5-10:2.9})](\PIx, \PIy) {\Large $S_{2}^\prime$};
\draw[red, ->, line width=0.05cm] ([shift=({3*360/5-10:2.6})](\PIx, \PIy) to[out=300, in=140] ([shift=({3*360/5+10:2.6})](\PIx, \PIy);
\node[red] at ([shift=({3*360/5-10:2.9})](\PIx, \PIy) {\Large $S_{-2}^\prime$};
\draw[red, ->, line width=0.05cm] ([shift=({4*360/5-10:2.6})](\PIx, \PIy) to[out=10, in=220] ([shift=({4*360/5+10:2.6})](\PIx, \PIy);
\node[red] at ([shift=({4*360/5-10:2.9})](\PIx, \PIy) {\Large $S_{-1}^\prime$};
\draw[red, ->, line width=0.05cm] ([shift=({5*360/5-10:2.6})](\PIx, \PIy) to[out=80, in=280] ([shift=({5*360/5+10:2.6})](\PIx, \PIy);
\node[red] at ([shift=({5*360/5-10:2.9})](\PIx, \PIy) {\Large $S_{0}^\prime$};
\draw[red, ->, line width=0.05cm] ([shift=({1*360/5-10:2.6})](\PIx, \PIy) to[out=140, in=360] ([shift=({1*360/5+10:2.6})](\PIx, \PIy);
\node[red] at ([shift=({1*360/5-10:2.9})](\PIx, \PIy) {\Large $S_{1}^\prime$};
\end{tikzpicture}
\caption{Illustration of how the Stokes graph for the isomonodromy system changes depending on the wedge of the \PI~solution \cite{t00, i19b}. The original $z$-plane for the \PI~solution has five Stokes wedges (in purple; and as already illustrated earlier in both figures~\ref{fig:P1-singular-Boutroux} and~\ref{fig:spiralPlanePI}). Depending on the wedge under consideration, we will find different Stokes graphs for the isomonodromy system in the $\xi$-plane (as illustrated in two examples). From those graphs we can compute linear Stokes data, say, $S_i$ for region I and $S_i^{\prime}$ for region II (in red). But, because of the isomonodromy condition \eqref{eq:D_I}, the transitions are independent of $z$ at large $\xi$. Therefore, comparing the linear Stokes data of region I with that of region II determines all Stokes data.}
\label{fig:IsomonodromyNonLinearStokes}
\end{figure}

Having briefly introduced all our main players, we may now discuss the solution of the isomonodromy system \eqref{eq:L_I}-\eqref{eq:D_I} by use of WKB methods. In particular, we are interested in the linear Stokes data associated to the \PI~isomonodromy system. For this, we need to understand the Stokes graph of \eqref{eq:L_I}---but because \eqref{eq:L_I} depends on the \PI~solution $u_{(2,3)}$, the Stokes graph will change depending on which Stokes wedge of $u_{(2,3)}$ (as in figures~\ref{fig:P1-singular-Boutroux} or~\ref{fig:spiralPlanePI}) we are in. This is illustrated in figure~\ref{fig:IsomonodromyNonLinearStokes}. Now, as it turns out, equation \eqref{eq:D_I} in the isomonodromy system implies that at large $\xi$ the solution $\Psi$ of the isomonodromy system is independent of $t$ (and therefore also of $z$). This further implies that, for the same analytic solution of \PI~in two different wedges, the linear Stokes coefficients of the isomonodromy system must coincide---as they too have to be independent of $t$. But given that depending on which Stokes wedge one considers, the form of the linear Stokes multipliers changes, we can equate them and \textit{obtain conditions for the nonlinear Stokes data of} \PI. For the present isomonodromy system, the linear Stokes multipliers for \eqref{eq:L_I} have already been computed in \cite{i19a}; they are given by:
\begin{align}
S_0 &= \rmi\, \rme^{\frac{2\pi \rmi \nu}{g_{\text{s}}}}, & S_0^\prime &= \rmi\, \rme^{\frac{2\pi\rmi\nu'}{g_{\text{s}}}} \left( 1 - \rme^{\frac{2\pi \rmi \rho^\prime}{g_{\text{s}}}} \right),\label{eq:first_iso} \\
S_1 &= \rmi\, \rme^{-\frac{2\pi\rmi\nu}{g_{\text{s}}}} \left( 1 - \rme^{-\frac{2\pi \rmi \rho}{g_{\text{s}}}} + \rme^{-\frac{2\pi \rmi \left(\rho-\nu\right)}{g_{\text{s}}}} \right), & S_1^\prime &= \rmi\, \rme^{\frac{2\pi\rmi\rho'}{g_{\text{s}}}} \left( 1 - \rme^{-\frac{2\pi \rmi \nu^\prime}{g_{\text{s}}}} \right), \\
S_2 &= \rmi\, \rme^{\frac{2\pi \rmi \rho}{g_{\text{s}}}}, & S_2^\prime &= \rmi\, \rme^{\frac{2\pi \rmi \rho^\prime}{g_{\text{s}}}},\\
S_{-1} &= \rmi\, \rme^{\frac{2\pi\rmi\nu}{g_{\text{s}}}} \left( 1 - \rme^{\frac{2\pi \rmi \rho}{g_{\text{s}}}} \right), & S_{-1}^\prime &= \rmi\, \rme^{-\frac{2\pi \rmi \nu^\prime}{g_{\text{s}}}}, \\
S_{-2} &= \rmi\, \rme^{-\frac{2\pi\rmi\rho}{g_{\text{s}}}} \left( 1 - \rme^{\frac{2\pi \rmi \nu}{g_{\text{s}}}} \right), & S_{-2}^\prime &= \rmi\, \rme^{-\frac{2\pi\rmi\nu'}{g_{\text{s}}}} \left( 1 + \rme^{-\frac{2\pi \rmi \left(\rho^\prime-\nu'\right)}{g_{\text{s}}}} - \rme^{\frac{2\pi \rmi\rho^\prime}{g_{\text{s}}}} \right),\label{eq:last_iso}
\end{align}
\noindent
where the non-primed linear Stokes coefficients correspond to region I in figure~\ref{fig:IsomonodromyNonLinearStokes}, whereas the primed ones correspond to region II. The main idea is that it should be possible to obtain the nonlinear Stokes data out of the linear one, by simply equating the linear Stokes coefficients from two different wedges \cite{t00, i19b} (this is the isomonodromy property), \textit{i.e.},
\begin{equation}
\label{eq:stokesfromisomonodromy}
S_i \left(\nu, \rho\right) = S_i^\prime \left(\nu^\prime, \rho^\prime\right).
\end{equation}

One last required step in the comparison (and in actually matching against our Stokes data) is that the transseries given in \cite{i19a} needs to be rewritten in our conventions. A similar map has been proposed in \cite{i19b} between our standard rectangular-framing transseries \eqref{eq:PI/II-2PTS} and their diagonal-like formulation transseries \eqref{eq:taufunction} for the partition function. More specifically, the formulation of this map starts with the two-parameter transseries given in \cite{t00}, which we write herein \textit{schematically}\footnote{The reader should consult \cite{t00} for correct notation, precise expressions, and meaning of all quantities.} as
\be
\label{eq:takei_transseries}
u_{(2,3)} \left(g_s; t, \alpha,\beta \right) = u_{(2,3)}^{(0)} \left(g_s; t\right) + \sqrt{g_{\text{s}}}\, \sum_{g=0}^{+\infty} g_{\text{s}}^g\, \sum_{k=0}^{g+1} \uppsi^{(g/2)}_{g+1-2k} \left( t, \alpha,\beta \right)\, \rme^{\left( g+1-2k \right)\, \upphi \left( g_s; t, \alpha,\beta \right)},
\ee
\noindent
depending on transseries parameters $\alpha$ and $\beta$, where $\upphi$ contains the instanton action and more, and where the form of the prefactors $\uppsi^{(g/2)}_{g+1-2k}$ is given in \cite{t00}. This ``$\alpha,\beta$ transseries'' then relates to the partition-function transseries \eqref{eq:taufunction} via
\bea
\alpha &=& \frac{\sqrt{\pi}}{\Gamma\left(\frac{\nu}{g_{\text{s}}}\right)}\, 2^{\frac{\nu}{g_{\text{s}}}}\, \rme^{-2\pi\rmi\, \frac{\left( \nu + \rho \right)}{g_{\text{s}}}}, \\
\beta &=& \frac{\Gamma\left(1+\frac{\nu}{g_{\text{s}}}\right)}{\sqrt{\pi}}\, \frac{1}{2^{1+\frac{\nu}{g_{\text{s}}}}}\, \rme^{2\pi\rmi\, \frac{\left(\nu+\rho\right)}{g_{\text{s}}}}.
\eea
\noindent
We are now only left to link the transseries \eqref{eq:takei_transseries}, written in the ``$\alpha,\beta$ parametrization'', to our own \eqref{eq:PI-2PTS} or \eqref{eq:PI/II-2PTS}, written instead in the ``$\sigma_1,\sigma_2$ parametrization''---\textit{i.e.}, these transseries differ mostly by the way in which the transseries parameters were introduced. Computing a first few coefficients in the formulation of \cite{t00} (up to adequate analytic continuation),
\bea
u_{(2,3)} \left(g_s; t, \alpha,\beta \right) &\simeq& - \rmi\, \sqrt{\frac{t}{6}} + \alpha \beta\, g_{\text{s}}\, \sqrt{-\rmi} \left( \frac{3}{2} \right)^{\frac{1}{4}} t^{-\frac{3}{4}} - \\
&& 
- \alpha\, \sqrt{g_{\text{s}}} \left(-\rmi\right)^{\frac{7}{4}} t^{-\frac{1}{8}}\, 2^{- \frac{3}{8} + \frac{15}{2}\, \alpha \beta}\, 3^{- \frac{1}{8} + \frac{5}{2}\, \alpha \beta} \left( - \frac{\rmi\, t^{\frac{5}{2}}}{g_{\text{s}}^2} \right)^{\alpha \beta} \rme^{+ \sqrt{-\rmi}\, \frac{2^{\frac{11}{4}}\, 3^{\frac{1}{4}}}{5}\, \frac{t^{\frac{5}{4}}}{g_{\text{s}}}} - \nonumber \\
&& 
- \beta\, \sqrt{g_{\text{s}}} \left(-\rmi\right)^{\frac{7}{4}} t^{-\frac{1}{8}}\, 2^{- \frac{3}{8} - \frac{15}{2}\, \alpha \beta}\, 3^{- \frac{1}{8} - \frac{5}{2}\, \alpha \beta} \left( - \frac{\rmi\, t^{\frac{5}{2}}}{g_{\text{s}}^2} \right)^{-\alpha \beta} \rme^{- \sqrt{-\rmi}\, \frac{2^{\frac{11}{4}}\, 3^{\frac{1}{4}}}{5}\, \frac{t^{\frac{5}{4}}}{g_{\text{s}}}} + \cdots, \nonumber
\eea
\noindent
one can immediately see, upon the appropriate variable transformations, that such relation is given by\footnote{We have also checked that this relation is consistent with the first few coefficents of the $(2,0)$ and $(0,2)$ sectors. Given the homogeneity property of the coefficients stated in \cite{t00}, we can ensure that this map is exact.}
\bea
\sigma_1 &=& \rmi \beta\, 3^{\frac{1}{4}-3\alpha\beta}\, 2^{-8\alpha\beta} = \rmi\, \frac{\Gamma \left( 1 + \frac{\nu}{g_{\text{s}}} \right)}{\sqrt{\pi}}\, \frac{3^{\frac{1}{4} - \frac{3 \nu}{2 g_{\text{s}}}}}{2^{1 + \frac{5\nu}{g_{\text{s}}}}}\, \rme^{2\pi\rmi\, \frac{\left(\nu+\rho\right)}{g_{\text{s}}}}, \\
\sigma_2 &=& \rmi \alpha\, 3^{\frac{1}{4}+3\alpha\beta}\, 2^{+8\alpha\beta} = \rmi\, \frac{\sqrt{\pi}}{\Gamma \left( \frac{\nu}{g_{\text{s}}} \right)}\, 3^{\frac{1}{4} + \frac{3 \nu}{2 g_{\text{s}}}}\, 2^{\frac{5\nu}{g_{\text{s}}}}\, \rme^{-2\pi\rmi\, \frac{\left( \nu + \rho \right)}{g_{\text{s}}}}.
\eea
\noindent
In practice this map gets expanded order-by-order in $\alpha\beta$ to explicitly match Stokes data.

We can finally plug-in the transition functions for nonlinear Stokes data into the above isomonodromic-formula for Stokes data \eqref{eq:stokesfromisomonodromy}. Once this is done, obtaining nonlinear Stokes data is a straightforward exercise of Taylor-expanding both sides. For example, starting with the purely perturbative solution ($\sigma_1=\sigma_2=0$) in region I, one will undergo an inverse backwards Stokes transition by entering region II. Then, one can calculate the first nonlinear backwards Stokes coefficients from that transition. In particular, under the inverse Stokes transition the $\left(\alpha,\beta\right)$ parameters jump from $\left(0,0\right)$ to $\left(-\frac\rmi{2\sqrt\pi},0\right)$; consequently, the $\left(\sigma_1,\sigma_2\right)$ jump from $\left(0,0\right)$ to $\left(0,\frac{3^{\frac14}}{2\sqrt\pi}\right)=\left(0,-N_1^{(-1)}\right)$. This is indeed exactly what we expect from the limits of the transition functions which we have computed. In a similar fashion, one may easily calculate other nonlinear Stokes data and compare with our previous results in this paper. One such simple albeit rather non-trival example is obtaining the transition function for $\upmu$ in the $3\pi$ direction; \textit{i.e.}, the one we are considering in the isomonodromy formulation of figure~\ref{fig:IsomonodromyNonLinearStokes}. Using the definitions in \cite{t00}
\be
\mathsf{E} \equiv - 8 \alpha \beta = \frac{8}{\sqrt{3}}\, \sigma_1\sigma_2, \qquad \qquad \upchi (z) \equiv \sqrt{\pi}\, \frac{2^{1 + \frac{z}{4}}}{\Gamma \left( 1 + \frac{z}{4} \right)},
\ee
\noindent
the isomonodromy condition $S_0=S'_0$ in \eqref{eq:stokesfromisomonodromy} reads
\be
\rmi\, \rme^{-\frac{\rmi\pi}{2}\, \mathsf{E}} = \rmi\, \rme^{-\frac{\rmi\pi}{2}\, \mathsf{E}^{\prime}} - \rmi\beta'\, \upchi (-\mathsf{E}^{\prime}).
\ee
\noindent
Using our map between $\alpha,\beta$ and $\sigma_1,\sigma_2$, one can now obtain the transition function for $\upmu$ in terms of $\upmu'$ and $\sigma_1'$, or, equivalently  $\upmu'$ and $\uprho_{3\pi}'$; \textit{i.e.}, we can find our earlier\footnote{It is important to point out that the  transition from regions II to I corresponds to Stokes automorphism $\underline{\mathfrak{ S}}_{3\pi}$.} generating function. First, the isomonodromy condition condition above is immediately rewritten as
\be
\rme^{-\frac{\rmi\pi}{2} \left( \mathsf{E}-\mathsf{E}^{\prime} \right)} = 1 - \beta^{\prime}\, \upchi (-\mathsf{E}^{\prime})\, \rme^{\frac{\rmi\pi}{2}\, \mathsf{E}^{\prime}} \qquad \Leftrightarrow \qquad \mathsf{E} = \mathsf{E}^{\prime} - \frac{2}{\rmi\pi}\, \log \left( 1 - \rme^{\frac{\rmi\pi}{2}\, \mathsf{E}^{\prime}}\, \beta^{\prime}\, \upchi(-\mathsf{E}^{\prime}) \right).
\ee
\noindent
Translating this relation back to our diagonal-framing coordinates, it reads
\be
\upmu = \upmu^{\prime} - \rmi \left( N_1^{(1)} \right)^2 \log \left( 1 - \rme^{\frac{4\pi\rmi}{\sqrt{3}}\, \upmu^{\prime}} \left( - \rmi\sigma'_1\, 2^{-\frac{8}{\sqrt{3}}\, \upmu}\, 3^{-\frac{1}{4} - \sqrt{3}\, \upmu^{\prime}}\, \sqrt{\pi}\, \frac{2^{1-\frac{2}{\sqrt{3}}\, \upmu^{\prime}}}{\Gamma \left( 1 - \frac{2}{\sqrt{3}}\, \upmu^{\prime} \right)} \right) \right).
\ee
\noindent
All we have left to do is to identify the different ingredients in terms of earlier familiar quantities. First identify the ``canonical'' Stokes coefficient,
\be
\upmu = \upmu^{\prime} - \rmi \left( N_1^{(1)} \right)^2 \log \left( 1 + \frac{\rmi\sigma'_1}{\left( N^{(1)}_1 \right)^2}\, N^{(-1)}_1\, \rme^{\frac{4\pi\rmi}{\sqrt{3}}\, \upmu^{\prime}}\, \frac{\left( 96\sqrt{3} \right)^{-\frac{2}{\sqrt{3}}\, \upmu^{\prime}}}{\Gamma \left( 1 - \frac{2}{\sqrt{3}}\, \upmu^{\prime} \right)} \right),
\ee
\noindent
and finally identify the backward generating function,
\be
\label{eq:iso_fin}
\upmu = \upmu^{\prime} - \rmi \left( N_1^{(1)} \right)^2 \log \left( 1 + \frac{\rmi\sigma'_1}{\left( N_1^{(1)} \right)^2}\, \rme^{\frac{2\pi\rmi}{\sqrt{3}}\, \upmu'}\, N^{(-1)} \left(\upmu^{\prime}\right) \right).
\ee
\noindent
This final result is precisely \eqref{eq:shift-a-II} at\footnote{Note that $\uprho_\pi$ back from \eqref{eq:shift-a-II} appears in \eqref{eq:iso_fin} together with an additional exponential. This additional exponential is to be associated to the fact that the rotations examined in \cite{i19a,t00} correspond to our backward transitions evaluated at 3$\pi$. The constant of motion in this case will be $\rme^{\frac{2\pi\rmi}{\sqrt{3}}\, \upmu}\uprho_\pi$, which may be identified as $\uprho_{3\pi}$.} $3\pi$. As such, essentially equipped with the above relations, the bridge between the isomonodromy calculations of linear Stokes data and our present results for nonlinear Stokes data has been successfully established. Indeed, by performing the above calculation we have verified that our Stokes data are fully consistent with all results obtained in the isomonodromy systems literature---more specifically, the results obtained in \cite{i19a, t00}.

Having checked consistency of our nonlinear Stokes data with their linear counterparts, let us finish by returning to our initial question of monodromy. Indeed the linear Stokes multipliers fulfill the following cyclic relation \cite{i19a}
\begin{equation}
\begin{bmatrix}
1 & 0\\
s_2 & 1
\end{bmatrix}\cdot
\begin{bmatrix}
1 & s_1\\
0 & 1
\end{bmatrix}\cdot
\begin{bmatrix}
1 & 0\\
s_0 & 1
\end{bmatrix}\cdot
\begin{bmatrix}
1 & s_{-1}\\
0 & 1
\end{bmatrix}\cdot
\begin{bmatrix}
1 & 0\\
s_{-2} & 1
\end{bmatrix} = 
\begin{bmatrix}
0 & \rmi\\
\rmi & 0
\end{bmatrix},
\end{equation}
\noindent
which is the defining relation in monodromy space. This hence finally \textit{proves} our conjectures regarding the complete nonlinear Stokes data of the \PI~problem.

\begin{figure}
    \centering
	\begin{tikzpicture}[scale=0.42]
		\draw[black, ->, line width=1pt] (-8, 0) -- (8,0);
		\draw[black, ->, line width=1pt] (0, -8) -- (0,8);
		\draw[violet, line width=0.1cm,dashed] (0,0) -- (7,0);
		\draw[violet, line width=0.1cm] (0,0) -- (-7,0);
		\draw[violet, line width=0.05cm] (0,0) -- (3.5, 6.06218);
		\draw[violet, line width=0.05cm] (0,0) -- (3.5, -6.06218);
		\draw[violet, line width=0.05cm,dashed] (0,0) -- (-3.5, 6.06218);
		\draw[violet, line width=0.05cm,dashed] (0,0) -- (-3.5, -6.06218);
		\draw[red, fill=red, line width=0.05cm] (0,0) circle (.7ex);
	\end{tikzpicture}
\hfill
\begin{tikzpicture}[scale=0.42]
		\draw[black, ->, line width=1pt] (-8, 0) -- (8,0);
		\draw[black, ->, line width=1pt] (0, -8) -- (0,8);
		\draw[violet, line width=0.05cm,dashed] (5/2,1.85/2*5) -- (1.25, 6.79006);
		\draw[violet, line width=0.05cm,dashed] (5/2,1.85/2*5) -- (-7.28148, 2.54588);
		\draw[violet, line width=0.05cm,dashed] (5/2,1.85/2*5) to[out=260, in=90] (2,0);
		\draw[violet, line width=0.05cm] (2,0) to[out=270, in=120] (4.27283, -6.48286);
		\draw[violet, line width=0.05cm,dashed] (5/2,1.85/2*5) -- (4.78386, 3.60816);
		\draw[violet, line width=0.05cm,dashed] (5/2,1.85/2*5) -- (4.17283, 6.48286);
		\draw[red, fill=red, line width=0.05cm] (5/2,1.85/2*5) circle (.7ex);
		\draw[violet, line width=0.05cm,dashed] (5/2,-1.85/2*5) -- (4.17283, -6.48286);
		\draw[violet, line width=0.05cm,dashed] (5/2,-1.85/2*5) -- (4.78386, -3.60816);
		\draw[violet, line width=0.05cm,dashed] (5/2,-1.85/2*5) -- (1.25, -6.79006);
		\draw[violet, line width=0.05cm,dashed] (5/2,-1.85/2*5) -- (-7.28148, -2.54588);
		\draw[violet, line width=0.05cm,dashed] (5/2,-1.85/2*5) to[out=100, in=270] (1.5,0);
		\draw[violet, line width=0.05cm] (1.5,0) to[out=90, in=240] (4.27283, 6.48286);
		\draw[red, fill=red, line width=0.05cm] (5/2,-1.85/2*5) circle (.7ex);
		\draw[violet, line width=0.1cm] (-5,0) -- (-7.5,0);
		\draw[violet, line width=0.05cm,dashed] (-5,0) -- (-5.77254, -2.37764);
		\draw[violet, line width=0.05cm] (-5,0) -- (3.89919, 6.46564);
		\draw[violet, line width=0.05cm] (-5,0) -- (3.89919, -6.46564);
		\draw[violet, line width=0.05cm,dashed] (-5,0) -- (-5.77254, 2.37764);
		\draw[red, fill=red, line width=0.05cm] (-5,0) circle (.7ex);
		\draw[red,style={decorate, decoration=snake}, line width=0.05cm] (-5,0)--(8,0);
	\end{tikzpicture}
	\caption{The turning points of homogeneous (left) and inhomogeneous (right) \PII~equations, with a schematic representation of both Stokes and anti-Stokes lines. The turning point of homogeneous \PII~is not simple, and six Stokes and anti-Stokes lines emanate from it. Considering the inhomogeneous equation instead, the triple turning-point splits into three simple turning-points, where the behavior of the solution around each simple turning-point can be described in terms of \PI. In the limit where $\alpha_{\text{II}}$ vanishes, these three simple turning-points merge, and the Stokes network on the right image mutates into the configuration on the left image.}
\label{fig:turnings}
\end{figure}
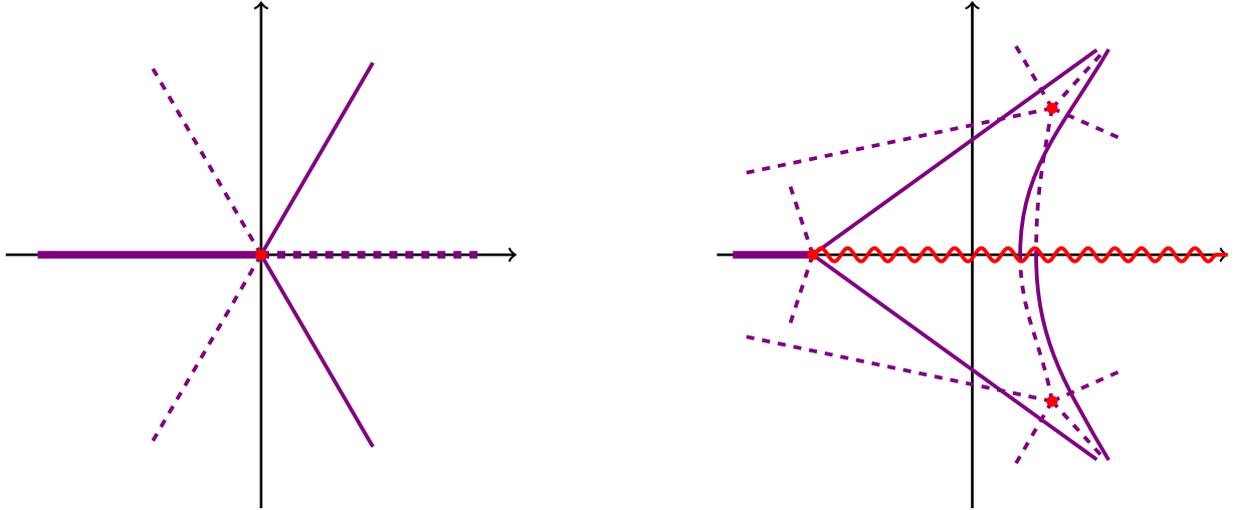

The case of \PII~turns out to be more complicated. The isomonodromic approach for \PI~worked out particularly well as its WKB turning point is simple (recall figure~\ref{fig:IsomonodromyNonLinearStokes}), but this is not quite the case for \PII. To better understand this issue let us first recall the definition of Painlev\'e turning points in, \textit{e.g.}, \cite{t02}. For that, slightly rewrite our two Painlev\'e equations, \eqref{eq:PI-eq-v2} and \eqref{eq:PII-eq-v2}, as
\be
u_{\text{I}}'' (z) = 6 \left( u_{\text{I}}^2(z) - z \right), \qquad \mu_{\text{II}}'' (z) = 2 \left( \mu_{\text{II}}(z)^3 - z\, \mu_{\text{II}}(z) \right).
\ee
\noindent
Turning points on the $z$-plane are now obtained by imposing vanishing of the right-hand side of the above Painlev\'e equations, alongside vanishing of their first functional-derivatives. In particular, \PI~has a simple turning point defined by
\be
\begin{cases}
\, u_{\text{I}}^2 (z) - z = 0, \\
\, u_{\text{I}} (z) = 0.
\end{cases}
\ee
\noindent
In fact, solving this system indeed yields $z=0$ as the simple turning point. For \PII, an equivalent computation results in the system
\be
\begin{cases}
\, \mu_{\text{II}}^3 (z) - z\, \mu_{\text{II}} (z) = 0, \\
\, 3 \mu_{\text{II}} (z) = z.
\end{cases}
\ee
\noindent
As for \PI, there is a turning point at $z=0$. The difference is that this time around the turning point is not simple---it is a \textit{triple} turning point. This implies the standard WKB approach used for \PI~cannot be taken, as it relies on all turning points being simple (see, \textit{e.g.}, \cite{t02}).

Happily, there is a clear way out of this difficulty. This entails turning to the \text{inhomogeneous} \PII~equation, with the introduction of an additional complex parameter $\alpha_{\text{II}}$,
\be
\mu_{\text{II}}'' (z) = 2 \left( \mu_{\text{II}}(z)^3 - z\, \mu_{\text{II}}(z) + \alpha_{\text{II}} \right).
\ee
\noindent
The system determining the turning points of inhomogeneous \PII~reads
\be
\begin{cases}
\, \mu_{\text{II}}^3 (z) - z\, \mu_{\text{II}} (z) + \alpha_{\text{II}} = 0, \\
\, 3 \mu_{\text{II}} (z) = z.
\end{cases}
\ee
\noindent
While this system now has three distinct turning points, they are all \textit{simple}---coalescing in the above \PII~triple turning-point in the limit $\alpha_{\text{II}} \to 0$. With non-vanishing $\alpha_{\text{II}}$, and solely in the presence of simple turning-points, one may now use the theorem\footnote{See \cite{t02} for the statement and proof of the theorem.} of local reduction to \PI, in order to relate the behavior of inhomogeneous \PII~solutions to the behavior of \PI~solutions. We schematically illustrate how the reduction back from inhomogeneous to homogeneous \PII~results would take place in figure~\ref{fig:turnings}. However, a complete study of the resurgent Stokes data for the inhomogeneous \PII~equation goes beyond the scope of the present paper---it will appear in \cite{v22}.

\acknowledgments
We would like to thank
In\^es Aniceto,
Ovidiu Costin, 
Fr\'ed\'eric Fauvet,
M\'ario Figueiredo,
Stavros Garoufalidis,
Kohei Iwaki,
Marcos Mari\~no,
Jean-Pierre Ramis,
David Sauzin,
Valerio Toledano-Laredo,
Marcel Vonk,
for useful discussions, comments and/or correspondence. RS would further like to thank In\^es Aniceto and Marcel Vonk for earlier collaboration on the present topics, and to thank Kohei Iwaki for sharing his unpublished results. We would further like to thank Alexander van Spaendonck and Marcel Vonk for pointing out a relevant typo\footnote{This typo seems to have some history. It is associated to the ambiguous expression $S_1^{(0)} = \left( -1 \right)^{\frac{1}{2}} \widetilde{S}_{-1}^{(0)}$ first appearing as formula (5.47) in \cite{asv11} for \PI, then appearing (unambiguously) as formula (5.46) in \cite{sv13} for \PII, and then again ambiguously as formula (A.59) in \cite{as13} now for both \PI~and \PII. As it turns out, the explicit (\textit{i.e.}, unambiguous) determinations of \cite{asv11} and \cite{sv13} are opposite to each other, leading to some backward-forward incompatibility. The explicit determinations in our present paper are now fully compatible between \PI~and \PII.} in an earlier version of our draft. RS would further like to thank CERN TH-Division and the University of Geneva for extended hospitality, where parts of this work were conducted. SB is supported by the FCT-Portugal scholarship SFRH/BD/130088/2017. MS is supported by the LisMath Doctoral program and FCT-Portugal scholarship SFRH/PD/BD/135525/ 2018. RV is supported by the LisMath Doctoral program and FCT-Portugal scholarship SFRH/ PD/BD/135514/2018. This research was supported in part by CAMGSD/IST-ID and via the FCT-Portugal grants UIDB/04459/2020, UIDP/04459/2020, PTDC/MAT-OUT/28784/2017. This paper is partly a result of the ERC-SyG project, Recursive and Exact New Quantum Theory (ReNewQuantum) funded by the European Research Council (ERC) under the European Union's Horizon 2020 research and innovation programme, grant agreement 810573.

\newpage

\appendix

\section{Numerical Checks of Stokes Data: Details}\label{app:numerics}

This appendix includes many technical details concerning the numerical checks on our conjectures for \PI/\PII~Stokes data; essentially complementing the overview described in section~\ref{sec:numerics}.

\subsection{Asymptotics and Large-Order Analysis}\label{app:subsec:large-order}

Let us begin by discussing certain assumptions and give further details on the numerical procedures which were used in subsection~\ref{subsec:large-order}. We split this discussion into three parts. First, we prove one assumption which we used, and obtain a set of coefficients needed in a resummation procedure. Second, we explain the procedure to perform Borel resummations of the required asymptotic series. Third, we detail the method of Richardson transforms and their application to our problem.

\subsubsection*{Structure of the Ratio of $\widetilde{H}_k$ Functions}

Equation \eqref{eq:hratio} uses the asymptotic expansion
\be
\frac{\widetilde{H}_{k} \left( 2g + a, tA \right)}{\widetilde{H}_{0} \left( 2g + b, \ell A \right)} \simeq \left(\frac{\ell}{t} \right)^{2g} g^{a-b}\, \sum_{q=0}^k \sum_{h=0}^{+\infty}  c_{h,q}^{(k)} \left( a,b,t,\ell \right)\, \frac{\log^{q} g}{g^h}.
\label{eq:ratio_form}
\ee
\noindent
The goal is to find the coefficients $c_{h,q}^{(k)} \left( a,b,t,\ell \right)$. Starting from the definition of these functions, \eqref{eq:h_htil-a}, it is possible to convert it into a recursive definition of the $\widetilde{H}_k$ as
\be
\widetilde{H}_{k+1} \left(2g+a,tA\right) = \frac{1}{2}\, \frac{\partial}{\partial g} \widetilde{H}_k \left(2g+a,tA\right), \qquad \widetilde{H}_0 \left(2g+a,tA\right) = \frac{\Gamma \left(2g+a\right)}{\left(tA\right)^{2g+a}}.
\ee
\noindent
Then, for the ratio of these functions, one finds
\be
\label{eq:recursive}
\frac{\partial}{\partial g} \frac{\widetilde{H}_k \left(2g+a,tA\right)}{\widetilde{H}_0 \left(2g+b,\ell A\right)} = 2\, \frac{\widetilde{H}_{k+1} \left(2g+a,tA\right)}{\widetilde{H}_0 \left(2g+b,\ell A\right)} - 2\, \frac{\widetilde{H}_k \left(2g+a,tA\right)}{\widetilde{H}_0 \left(2g+b,\ell A\right)}\, \frac{\widetilde{H}_1 \left(2g+b,\ell A\right)}{\widetilde{H}_0 \left(2g+b,\ell A\right)}.
\ee
\noindent
The coefficients of equation \eqref{eq:ratio_form} may now be computed via induction on $k$: the case $k=0$ is trivial, where the coefficients are given by the well-known coefficients of the ratio of gamma functions \cite{olbc10}
\be
c^{(0)}_{h,q} \left(a,b,t,\ell\right) = \delta_{q,0}\, 2^{a-b-h}\, \frac{\left( \ell A \right)^a}{\left( tA \right)^b}\, \binom{a-b}{h}\, B_h^{(a-b+1)}(a),
\ee
\noindent
where $B^{(a)}_h(b)$ is a generalized Bernoulli polynomial (also known in the literature as a N\o rlund polynomial) \cite{olbc10}. Going forward, we see from \eqref{eq:recursive} that the $c_{h,q}^{(1)} \left(b,b,s,s\right)$ coefficients are also required. But these may be obtained by definition: with $\psi^{(0)}(x)$ the ordinary polygamma function, we have
\be
\frac{\widetilde{H}_1 \left(2g+b,\ell A\right)}{\widetilde{H}_0 \left(2g+b,sA\right)} = \psi^{(0)} \left(2g+b\right) - \log \ell A.
\ee
\noindent
Power-expansion of the polygamma function then yields $c^{(1)}_{h,1} \left(b,b,s,s\right) = \delta_{h,0}$ and
\be
c^{(1)}_{h,0} \left(b,b,\ell,\ell\right) = \begin{cases}
\displaystyle{\log 2 - \log \ell A}, \quad & h=0, \\
\displaystyle{- \frac{1}{4} + \frac{b}{2}}, & h=1, \\
\displaystyle{- \frac{1}{2^h} \left( \sum_{k=2}^h \frac{b^{h-k}}{k}\, B_k\, \binom{-k}{h-k} + \frac{1}{2}\, b^{h-1}\, \binom{-1}{h-1} + \frac{\left(-b\right)^h}{h}\right)}, & h\geq2
\end{cases}
\ee
\noindent
(here the $B_k$ are the Bernoulli numbers). Finally, we obtain the recursion relation that defines the coefficients for the series-expansion,
\bea
c^{(k)}_{h,q} \left(a,b,t,\ell\right) &=& c_{h,q}^{(k-1)} \left(a,b,t,\ell\right) \log \frac{\ell}{t} + \frac{a-b-h+1}{2}\, c_{h-1,q}^{(k-1)} \left(a,b,t,\ell\right) + \\
&&
+ \frac{q+1}{2}\, c_{h-1,q+1}^{(k-1)} \left(a,b,t,\ell\right) + \sum_{r=0}^h \sum_{d=0}^1 c^{(k-1)}_{h-r,q-d} \left(a,b,t,\ell\right)\, c_{r,d}^{(1)}\left(b,b,\ell,\ell\right). \nonumber
\eea
\noindent
Herein we have implicitly assumed that all $c_{h,q}^{(k)} \left(a,b,t,\ell\right)$ vanish for $h<0$ or $q<0$ or $q>k$.

\subsubsection*{Evaluation of Asymptotic Sums}

One issue we had to deal with in \eqref{eq:ratio_sum_div} concerned divergent sums. Formally, \eqref{eq:ratio_sum_div} defines a formal power series of the form
\be
\label{eq:asy_sum}
R_q^{(k)} \left(n,p,t,\ell\right) \left(g\right) \simeq \sum_{h=0}^{+\infty} \frac{1}{g^h}\, R_{h,q}^{(k)} \left(n,p,t,\ell\right),
\ee
\noindent
which is asymptotic due to the factorially divergent nature of the $R_{h,q}^{(k)} \left(n,p,t,\ell\right)$ coefficients---themselves given by
\be
R_{h,q}^{(k)} \left(n,p,t,\ell\right) = \sum_{r=0}^{h} \frac{u_{2(h-r)}^{(p+t,p)[k]}}{\left(-2\right)^k}\, c^{(k)}_{r,q} \left(n-h+r-\beta^{(k)}_{(p+t,p)},n-\frac{\ell}{2},t,\ell\right),
\ee
\noindent
where the factorial divergence arises from the transseries coefficients $u^{(p+t,p)[k]}_{2h}$. In order to extract finite numerical values for large-order formulae, we hence need to build some function which will be asymptotic to \eqref{eq:asy_sum} in the $g\to+\infty$ limit. According to the conventions of subsection~\ref{subsec:large-order}, we denote such a function by $\Sigma_q^{(k)}(n,p,t,\ell)(g)$. The general theory of Borel resummation via resurgence is of course well-known; \textit{e.g.}, \cite{abs18}. What we wish to discuss herein is the practical numerical implementation of such resummations. Let us see how.

\begin{figure}
\centering
\includegraphics[width=\columnwidth]{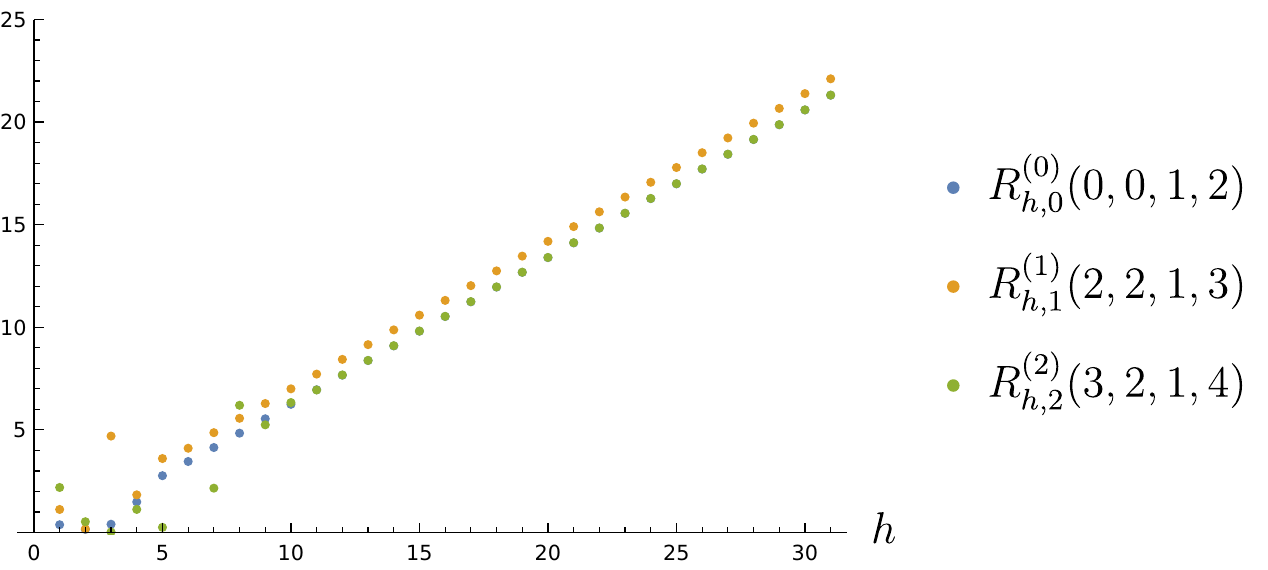}
\caption{Asymptotic growth of the coefficients in the sums \eqref{eq:asy_sum}. Their factorial growth \eqref{eq:factorial_growth} naturally leads to considering the test-ratios of these coefficients as $\left|R_{h+1}/R_h\right|$, which are asymptotic to lines of slope $2/A$  independently of any further details concerning their remaining indices. We illustrate this with the explicit coefficients for the asymptotic series: $R_{h,0}^{(0)} \left(0,0,1,2\right)$ in green; $R_{h,1}^{(1)} \left(2,2,1,3\right)$ in orange; and $R_{h,2}^{(2)} \left(3,2,1,4\right)$ in red. At high $h$, the different ratios become very difficult to distinguish, as they all just depend on $A$.}
\label{fig:growth}
\end{figure}

For brevity of exposition, let us temporarily drop all indices in \eqref{eq:asy_sum}. Thus consider
\be
\label{eq:to_resum}
R(g) \simeq \sum_{h=0}^{+\infty} \frac{1}{g^h}\, R_{h},
\ee
\noindent
with the $R_h$ real coefficients asymptotically growing as
\be
\label{eq:factorial_growth}
R_h \sim C\, \frac{h!}{2^{-h} A^{h}},
\ee
\noindent
with $C \in \BR$, $A$ the (real, positive) instanton action, and the factor of $2$ arises from the contributions of the forward and backward series. We denote by $\Sigma(g)$ the function that is asymptotic to $R(g)$ obtained through Borel--Pad\'e resummation. Due to the above growth, the sum in \eqref{eq:to_resum} cannot be evaluated---all that expression yields are asymptotic properties of the (yet unknown) function $\Sigma(g)$. Such a simplified setting includes all our cases \eqref{eq:asy_sum}, and this is briefly illustrated in figure~\ref{fig:growth} where we plot a few ratios of the form $R_{h+1}/R_h$---which, in the $h\to{+\infty}$ limit are asymptotic to straight lines with slope $2/A$---for some explicit examples of \eqref{eq:asy_sum}.

The evaluation of the function $\Sigma(g)$ must then be made via Borel resummation. Its practical, numerical implementation proceeds as follows. The Borel transform of the asymptotic sum is
\be
\label{eq:bor_an}
\CB \left[R\right] (s) = \sum_{h=0}^{+\infty} \frac{R_{h+1}}{h!}\, s^h,
\ee
\noindent
and it does not include the residual coefficient $R_0$; it will have to be reinserted at the end of the computation. Once analytically continued throughout the complex $s$-plane, beyond its radius of convergence $\abs{A}$, Borel $\theta$-resummation \eqref{eq:Borel-resum} follows (set $\theta=-\arg g$ to ensure convergence of the integral)
\be
\Sigma(g) \equiv \CS_{\theta} R (g) = R_0 + \int_{0}^{\rme^{\rmi\theta} \infty} \rmd s\, \CB \left[ R \right] (s)\, \rme^{- s g}.
\ee
\noindent
This function is asymptotic to the series $R(g)$, taking finite values for every $g$ with $\arg g \neq 0,\pi$.

Now this procedure relies on knowledge of all coefficients $R_h$, on explicit implementation of the aforementioned analytic continuation, and on evaluation of the above Laplace transform. In our present numerical setting, we start by having but a finite number of coefficients $R_h$, hence start by needing some way to approximate the whole procedure. Note that a truncated Borel transform 
\be
\label{eq:bor_num}
\CB_{N} \left[R\right] (s) = \sum_{h=0}^{N} \frac{R_{h+1}}{h!}\, s^h,
\ee
\noindent
with whatever finite number of coefficients one might have, will not be enough: whereas the approximation might be good in the convergent region $\abs{s} < \abs{A}$, it will see no singularities outside of this disk. But it is this structure of singularities which is absolutely crucial in order to obtain a realistic Borel resummation. Hence this truncation requires some sort of upgrading. The natural upgrading is to consider a Pad\'e approximant to the truncated Borel transform, see, \textit{e.g.}, \cite{abs18},
\be
\label{eq:borel_pade}
\text{BP}_N \left[R\right] (s) = \frac{\sum_{h=0}^{N} b_h\, s^h}{\sum_{h=0}^{N} c_h\, s^h},
\ee
\noindent
where the $b_h$ and $c_h$ coefficients are defined by imposing $c_0=1$ for normalization and that the difference between the Taylor series at $s=0$ of \eqref{eq:borel_pade} and \eqref{eq:bor_num} is of order $s^{2N}$. Then, the $b_h$ and $c_h$ are computed in terms of the $R_h$. This Borel--Pad\'e approximant reduces to the numerical Borel transform for $\abs{s} < \abs{A}$, but importantly has poles accumulating near the branch-points which appear in the analytic continuation of the exact Borel transform (poles ``approximating the branch-cuts''). We can then define the numerical Borel resummation procedure by\footnote{In our computations we have used $\theta = - \arg g$ for faster convergence, but other choices are possible.}
\be
\label{eq:resum_num}
\CS_{\theta}^{\ev{N}} R(g) = R_0 + \int_{0}^{\rme^{\rmi\theta} \infty} \rmd s\, \text{BP}_{N} \left[ R \right] (s)\, \rme^{- s g}.
\ee
\noindent
This function tends to $\CS_{\theta} R (g)$ as $N\to{+\infty}$. Restoring all indices, our strategy for the resummation is then to evaluate
\be
\Sigma_q^{(k)\ev{N}} \left(n,p,t,s\right) \left(g\right)= \CS_{\theta}^{\ev{N}} R_q^{(k)} \left(n,p,t,s\right) \left(g\right),
\ee
\noindent
so that the functions $\Sigma_q(n,p,\ell,t,g)$ appearing in \eqref{eq:ratio-B-resum} are then computed in terms of Borel-resummed functions as
\be
\Sigma_q \left(n,p,t,\ell,g\right) \mapsto \Sigma_q^{\ev{N}} \left(n,p,t,\ell,g\right) = \sum_{k=0}^p \frac{1}{g^{\beta^{(k)}_{(p+t,p)}}}\, \CS_{\theta}^{\ev{N}} R_q^{(k)} \left(n,p,t,\ell\right) \left(g\right).
\ee

One last practical point is the following: we cannot straight-off evaluate  \eqref{eq:resum_num} for $g$ real, due to the structure of singularities on the real axis. The simplest way to proceed is to resort to the usual lateral resummations, \textit{i.e.}, substitute $g \to g + \rmi \epsilon$ in \eqref{eq:resum_num}, with $g$ real, positive, and large; and $\epsilon$ real, positive, and small. In practice this introduces small imaginary components in the sums, which, at the end of the day, translate to small real parts in the numerical results for the Stokes data (and these may further be used as first indicators on the precision of our numerical approximations). This replacement $g \to g + \rmi \epsilon$ and the cancellation of the real parts in the Stokes data is discussed in \cite{asv11} for the \PI~example, and in \cite{abs18} in more general settings.

\subsubsection*{Convergence Acceleration: Richardson Transforms}

The large-order analysis of subsection~\ref{subsec:large-order} obtains Borel residues by focusing on the asymptotic behavior of specific transseries coefficients. In particular, by setting-up certain ratios of these coefficients which yield convergent sequences $\BS_g$ whose limits are the particular Stokes data we want to find; see, \textit{e.g.}, \cite{msw07, gikm10, asv11, sv13}. An approximation to the limit $\lim_{g \to + \infty} \BS_g \equiv \BS_{\infty}$ is of course obtained by choosing $g_\star$ large-enough so that $\BS_{\infty} \approx \BS_{g_\star}$; the larger $g_\star$ the better the approximation. However in practice this amounts to the need of computing transseries coefficients $u^{(n,m)[k]}_{2g}$ for very large values of $g$, a recursive procedure which very quickly achieves very-high computational cost. In order to obtain good estimates on $\BS_{\infty}$ without computing additional terms in its sequence, one accelerates convergence via the method of Richardson transforms; see, \textit{e.g.}, \cite{msw07, gikm10, asv11, sv13, abs18}. For example, our relevant sequences (\textit{e.g.}, obtained from \eqref{eq:asy_conv}) are of the form
\be
\BS_g \simeq \BS_{\infty} + \frac{\BS_{-1}}{g} + \frac{\BS_{-2}}{g^2} + \cdots.
\ee
\noindent
Approximating $\BS_{\infty}$ (say, whatever Borel residue we are looking for) by evaluating the sequence at large $g_\star$ yields an error of order $g_\star^{-1}$. But the \textit{Richardson transformed} sequence
\be
\text{RT}_1 \BS_g = \BS_{g+1} + g \left( \BS_{g+1} - \BS_g \right),
\ee
\noindent
behaves asymptotically as
\be
\label{eq:asy_form}
\text{RT}_1 \BS_g \simeq \BS_{\infty} - \frac{\BS_{-2}}{g^2} + \cdots,
\ee
\noindent
which means that at large $g_\star$ the error is now smaller, of order $g_\star^{-2}$. This procedure can of course be iterated, where the $N$th Richardson transform of $\BS_g$ is given by
\be
\text{RT}_N \BS_g = \sum_{h=0}^{N} \left(-1\right)^{h+N} \frac{\left(g+h\right)^N}{h! \left(N-h\right)!}\, \BS_{g+h}.
\ee
\noindent
This procedure accelerates convergence, albeit at the cost that every iteration requires one coefficient more, which in practice translates to being evaluated at smaller $\check{g}_\star < g_\star$. More precisely, if we start-off knowing $g_\star$ terms in the sequence, after $N$ Richardson transforms we will be left with only $g_\star - N$ terms in the ``accelerated'' sequence. We have thus traded an error of order $g_\star^{-1}$ by an error of order $\left( g_\star-N \right)^{-N-1}$. One must thus for each example decide which is the adequate order of the Richardson transform producing accurate results. Additionally, precise cancellation of the logarithmic divergences is required in order for this method to work.

\subsection{Singularity Structure on the Borel Plane}\label{app:subsec:borel-plane}

The main method we ended-up using, as explained in subsection~\ref{subsec:borel-plane}, was to extract Stokes data straight from the singularity structure of the Borel transforms on the complex plane. Concerning this method, let us discuss in a bit more detail: the logarithms arising from resonance and their associated starting-powers in the asymptotic series; how to improve the precision and speed of the algorithm making use of the map between Stokes vectors and Borel residues; and a quick word on the choice of dampings and their relation to the degree of the Borel--Pad\'e approximation.

For both \PI/\PII, their two-parameter transseries \eqref{eq:PI/II-2PTS} has nonperturbative sectors associated with logarithms and asymptotic series with $\beta$-starting-powers and logarithms, as in
\be
\Phi_{(n,m)} (x) \simeq \sum_{k=0}^{k_{nm}} \left( \frac{\log x}{2} \right)^k\, \sum_{g=0}^{+\infty} u_{2g}^{(n,m)[k]}\, x^{g+\beta_{nm}^{[k]}}.
\ee
\noindent
To implement our numerical method, we are interested in the resummation of such sectors. One may consistently take these factors to the Borel plane---and carry through our analysis---but this would not be numerically very efficient. Alternatively observe that we can use different ways of resumming the same function whilst still arriving at the same result. For example, we have
\begin{equation}
\CS_{\pm\theta} \left[ \log(x)\, \Phi^{[1]}_{(n,m)}(x) \right] = \log(x)\, \CS_{\pm\theta} \left[ \Phi^{[1]}_{(n,m)}(x) \right].
\end{equation}
\noindent
Here we have taken the logarithm out of the resummation and kept it as a prefactor as it is already a finite quantity that does not need to be resummed. In other words we are taking $x \to \check{x}$ (with $\check{x}$ the damping factor) in terms that we want to leave as prefactors, and then we multiply them by the Borel--Pad\'e resummation of the remaining ones. In other words if we take into account that we resum correctly, the explicit way in which we evaluate the Borel--Pad\'e approximant to the $\Phi_{(n,m)} (x)$ sectors reduces to: 
\be
\text{BP}_{N} \left[ \Phi_{(n,m)} \right] (s) := \sum_{k=0}^{k_{nm}} \left( \frac{\log \check{x}}{2} \right)^k \check{x}^{\beta_{nm}^{[k]}}\, \text{BP}_{N} \left[ \sum_{g=0}^{+\infty} u_{2g}^{(n,m)[k]}\, x^{g} \right] (s).
\ee
\noindent
In this way, the (damped) lateral Borel resummations are numerically expressed as
\be
\CS_{\pm\theta} \Phi_{(n,m)} (\check{x}) = \int_{0}^{\rme^{\rmi\theta} \Lambda} \rmd s\, \text{BP}_{N} \left[ \Phi_{(n,m)} \right] (s)\, \rme^{-\frac{s}{\check{x}}},
\ee
\noindent
with $\Lambda$ a numerical cutoff for the integral which we normally choose to be $50$, and $\theta$ a small angle for the lateral resummations which we took to be $0.01$.

As discussed in subsection~\ref{subsec:stok_to_borel}, by means of the map between Borel residues and Stokes vectors \cite{abs18}, the sufficient set of Borel residues that one needs to determine in order to obtain all Stokes vectors is given in formula \eqref{eq:sufficientsetofborels}. This means that, via such map, one should be able to compute \textit{every other} ``non-sufficient'' Borel residue from previously computed ones. This is expectable from the fact that Borel residues are degenerate with respect to Stokes vectors: whereas both sets include the very same information, we do have a considerably larger number of Borel residues than Stokes vectors. The construction of the ``non-sufficient'' Borel residues from previous data helps increasing the precision of our algorithm as follows. Assume that at each step we do not build the Borel residues out of previously computed ones. Then, we would need to consider more derivatives in the calculation to access these ``unknown'' resonant Borel residues. Eventually, this would lead to the inversion of large matrices\footnote{Since one would have to solve a large system of equations instead of a $2\times2$ matrix. For example, the Borel transform of the $(n,n)$ sector has, at the first singularity located at $s=A$, $n+1$ contributions coming from the $(p+1,p)$, $p=0,\ldots,n$, sectors. Thus, if we consider \textit{all} the Borel residues to be unknown, we would need $n+1$ equations to determine all such unknowns. These equations are obtained from taking derivatives, as explained earlier. But we would also need to construct this system and solve it. This is computationally very heavy. If one considers all Borel residues associated to $(p+1,p)$, $p=2,\ldots,n$, to be known (and one can compute them from previous results) then we would have only two unknowns, $\mathsf{S}_{(n,n)\to(1,0)}$ and $\mathsf{S}_{(n,n)\to(2,1)}$, thus needing just two equations to determine them. This means inverting a $2\times2$ matrix instead of a $(n+1)\times(n+1)$ one.}, increasing the number of required operations, and contributing to the rapid loss of precision. On the other hand, if at each step we reconstruct Stokes vectors, then we end up just needing to invert a $2 \times 2$ matrix---with much smaller computational cost. Of course, numerical errors will still appear as the computation of Stokes vectors out from Borel residues evaluates sums of products of Borel residues, and these operations increase errors.

Finally, a note on the choice of damping factors. According to numerical analyses, one should be able to obtain a proper ``optimal'' damping depending on the order of the Pad\'e approximant. Our approach was a little bit more pragmatical. We tried to choose the lowest damping (largest suppression) such that the results were consistent, and then numerically explore the precision of each attempt by comparing with the Borel residues that were analytically known, \textit{i.e.}, trying to achieve the highest precision when computing $\mathsf{S}_{(0,0)\to(\ell,0)}=-\left(N_1^{(1)}\right)^\ell$ or $\mathsf{S}_{(1,1)\to(2,1)}=-2N_1^{(1)}$.

\subsection{Acceleration via Conformal Transformations}\label{app:subsec:conformal}

As briefly mentioned in subsection~\ref{subsec:conformal}, in this appendix we will explore the role of conformal acceleration in the context of our numerical methods. As we will see, one can obtain significant gains in precision by supplementing the implementation of Pad\'e approximants to Borel transforms with adequate conformal transformations on the Borel plane. This works as follows: for each problem, one would like to find an appropriate transformation of the complex Borel plane variable such that the Pad\'e approximation gets improved. In both examples we shall analyze below, these suitable transformations will then map the negative (and positive) real line(s) into the unit circle. Thus, these transformations will effectively set \textit{all singularities} at equal distance from the origin. Since the information of the Borel transform is encoded mostly in its singularity structure, we thus expect the approximations encoded in these conformally-transformed Borel sectors to be better than the original ones. Of course, the conformal transformations themselves need to be invertible such that one may go back to this original set-up. These methods have been extensively discussed in the literature \cite{cmrsj07, cf99, cf00, js00, cd19, cd20} and we will review them in here. In particular, we will be interested in the effect of conformal transformations in the following two scenarios: an infinite set of branch-points along a single direction (illustrated with an example involving the Riccati equation), and an infinite set of branch-points along two directions (as occurs in our Painlev\'e problems, and illustrated with \PI). Of special interest to our discussion is whether these techniques of conformal acceleration may be good tools to extract information from \textit{far away} branch-points\footnote{Since these transformations are supposed to give somehow ``similar'' importance to different singularities, by bringing them all into the unit circle, we expect this method to be specially effective when dealing with the singularities that lie the furthest away. Following this reasoning, we do not expect to see a significant improvement in the information from the first singularity.}; \textit{i.e.}, if they can improve the numerical precision in the computation of Borel residues which are associated to (eventually very) subleading singularities.

\subsubsection*{The Riccati Case}

Consider the Riccati equation written as (we briefly build on the example in \cite{c14})
\be
u'(z) = u(z) - \frac{1}{z}\, u^2(z) - \frac{1}{z},
\ee
\noindent
which is a first-order nonlinear ODE. Its corresponding one-parameter transseries solution is
\be
u \left(z;\sigma\right) = \sum_{n=0}^{+\infty} \sigma^n\, \rme^{-nAz}\, \Phi_n (z),
\ee
\noindent
where $\Phi_n(z)$ are asymptotic series given by
\be
\Phi_n (z) \simeq \sum_{g=0}^{+\infty} a_g^{(n)}\, z^{-g-\beta_n},
\ee
\noindent
with factorially-growing coefficients $a_g^{(n)} \sim g!$, and with $\beta_n$ the starting powers. The (single) instanton action $A=-1$ is straightforward to compute; as are the recursion-relations for the coefficients $a_g^{(n)}$, a few of which are \cite{c14}
\bea
\Phi_0 (z) &=& \frac{1}{z} - \frac{1}{z^2} + \frac{3}{z^3} - \frac{11}{z^4} + \frac{51}{z^5} + \cdots, \\
\Phi_1 (z) &=& 1 + \frac{2}{z} + \frac{1}{z^2} + \frac{4}{3 z^3} - \frac{7}{3 z^4} + \cdots, \\
\Phi_2 (z) &=& - \frac{1}{z} - \frac{5}{z^2} - \frac{14}{z^3} - \frac{122}{3 z^4} - \frac{421}{3 z^5} + \cdots.
\eea

As stated at the beginning of this subappendix, the idea behind conformal transformations is to bring singularities to the same distance from the origin on the Borel plane, such that the information they encode gets better reflected in the subsequent approximations. In the case of an infinite set of branch-points along a single direction (starting from a singularity located at $A$), the natural conformal transformation that maps, say, the negative real axis $(-\infty,A)$ into the unit circle\footnote{Sending $A$ to $-1$, $2A$ to $\pm\rmi$, and going clockwise and anticlockwise for the different determinations of the square-root along the negative direction. Note that the map degenerates due to the presence of square-roots.} as $\mathfrak{u} : \BR^{-} \to \BS^1$ alongside its inverse $\mathfrak{z} : \BS^1 \to \BR^{-}$ are\footnote{We chose to label the conformal transformations $\mathfrak{u}$, $\mathfrak{z}$ to make the analogy with the original $u$, $z$ transformations common in the literature, respectively, but not to confuse with the $u$, $z$ used everywhere else in this paper.} (from \cite{js00})
\bea
\mathfrak{u} (s) &=& \frac{\sqrt{A-s}-\sqrt{A}}{\sqrt{A}+\sqrt{A-s}}, \\
\mathfrak{z} (s) &=& - \frac{4sA}{\left(s-1\right)^2}.
\eea
\noindent
Now, in order to Borel--Pad\'e approximate a sector $\Phi_n$, we have to compute (breaking-down into steps the Borel--Pad\'e earlier defined in subappendix~\ref{app:subsec:large-order}) the $(n,m)$-Pad\'e approximant to the corresponding (truncated) Borel transform
\begin{equation}
\mathrm{P}_{n,m} \left\{ \mathcal{B} \left[\Phi_n\right] \left(\mathfrak z\left(s\right)\right) \right\} \Big|_{s\to \mathfrak{u} \left(s\right)},
\end{equation}
\noindent
where the last step $s \mapsto \mathfrak{u} (s)$ recovers the original set-up by performing the inverse transformation.

\begin{figure}
	\centering
	\begin{subfigure}[t]{0.5\textwidth}
		\centering
		\includegraphics[height=2in]{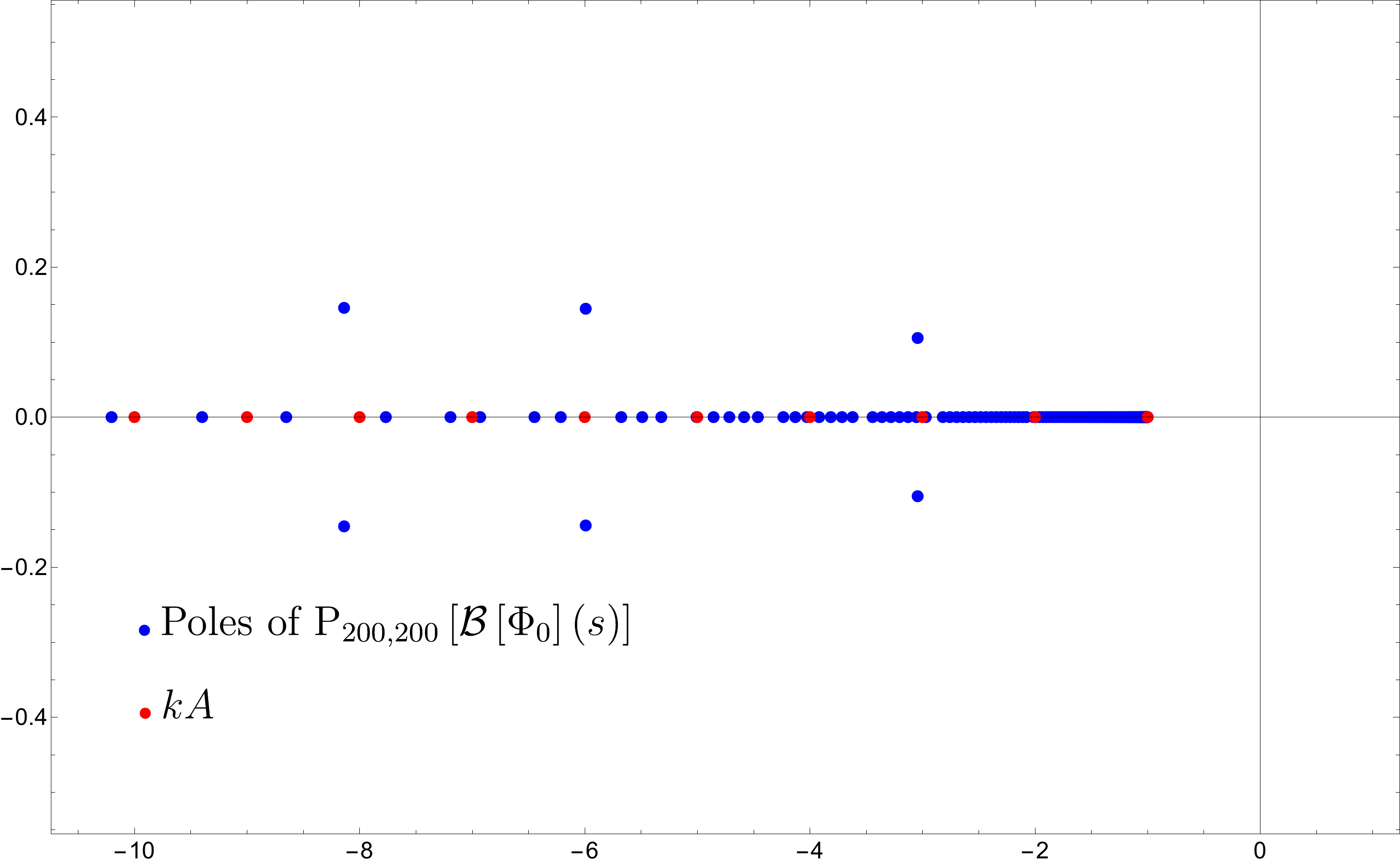}
		\caption{Riccati Borel plane.}
		\label{fig:poles1branch}
	\end{subfigure}%
	~ 
	\begin{subfigure}[t]{0.5\textwidth}
		\centering
		\includegraphics[height=2in]{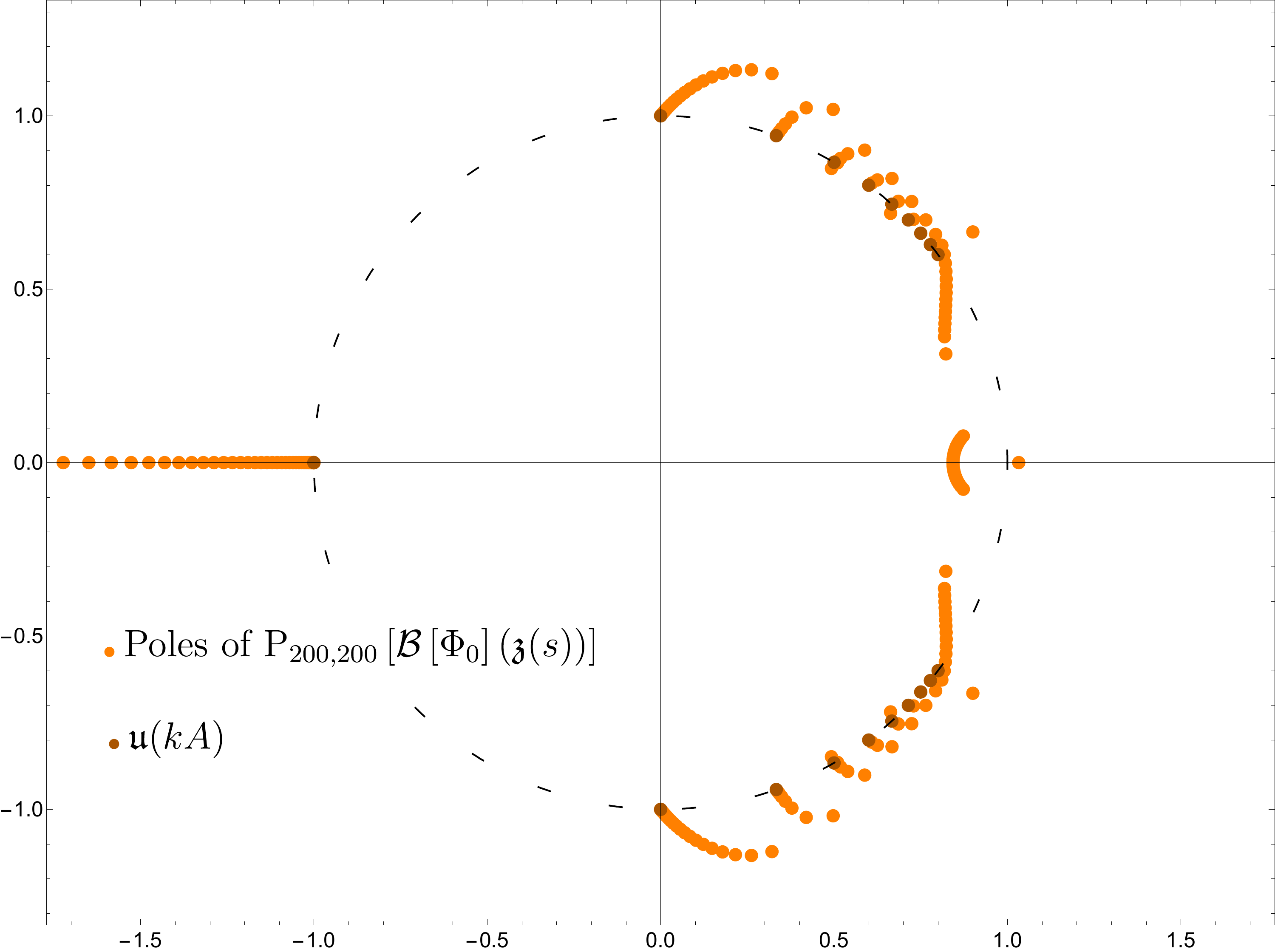}
		\caption{Riccati conformal Borel plane.}
		\label{fig:polesconftrans1branch}
	\end{subfigure}
	\caption{Illustration of our conformal transformation on the perturbative sector of the Riccati transseries. The left plot corresponds to the standard Borel plane, whereas the right one shows the effect of the conformal transformation on the distribution of Borel singularities.
	\label{fig:polesnormalandconftrans1branch}}
\end{figure}

One intuitive visual description behind this procedure is to place all branch-points at the same distance to the origin---say, along the unit circle---so that the Pad\'e approximant gives them all the same ``weight''. In figure~\ref{fig:polesnormalandconftrans1branch}, we plot the Pad\'e-poles for both $\text{P}_{200,200}\left[ \mathcal{B}[\Phi_0](s)\right]$ and $\text{P}_{200,200}\left[\mathcal{B}[\Phi_0](\mathfrak z(s))\right]$ to illustrate the effect of these transformations on the locations of the poles and why they should, in principle, improve the accuracy in the approximation of the chosen sector. This also reflects how the conformal transformation yields much better precision on the further-away singularities, since they receive increased weight in the approximation by being brought closer to the point of expansion, \textit{i.e.}, to the origin.

Having said this, we now have two options to computing Borel residues: our standard method from subsection~\ref{subsec:borel-plane} or subappendix~\ref{app:subsec:borel-plane}, and its corresponding conformal acceleration just described. We compare the precisions one may obtain with these two methods, in the present context, in table~\ref{tab:1branchconfvsnorm}, for the Borel residues $\mathsf{S}_{0\to k}$, for $k=1,2,3$ (addressed analytically in \cite{tsds03, c14}).

\begin{table}
	\centering
	\begin{tabular}{c|c|c|c}
		Borel Residue & Analytical &  Precision non-CT & Precision CT  \\
		\hline
		$\mathsf{S}_{0\to 1}$ & $-2\rmi \sinh \pi$ & 79 & 84\\ 
		\hline
		$\mathsf{S}_{0\to 2}$ & $-(2\rmi \sinh \pi)^2$ & 61 & 83\\
		\hline
		$\mathsf{S}_{0\to 3}$ &  $-(2\rmi \sinh \pi)^3$ & 22 & 39\\
	\end{tabular}
	\caption{A few Borel residues $\mathsf{S}_{0\to k}$ for the Riccati transseries (their exact value computed from asymptotics \cite{c14} or from analytics \cite{tsds03}). We display the precision our methods achieve in computing these Borel residues, using both the conformally transformed Borel--Pad\'e method, and our regular procedure.  The dampings needed in both cases were different, in which case we are displaying the comparison at fixed $(150,150)$-Pad\'e approximants.}
	\label{tab:1branchconfvsnorm}
\end{table}

\subsubsection*{The Painlev\'e~I Case}

\begin{figure}
	\centering
	\begin{subfigure}[t]{0.5\textwidth}
		\centering
		\includegraphics[height=1.8in]{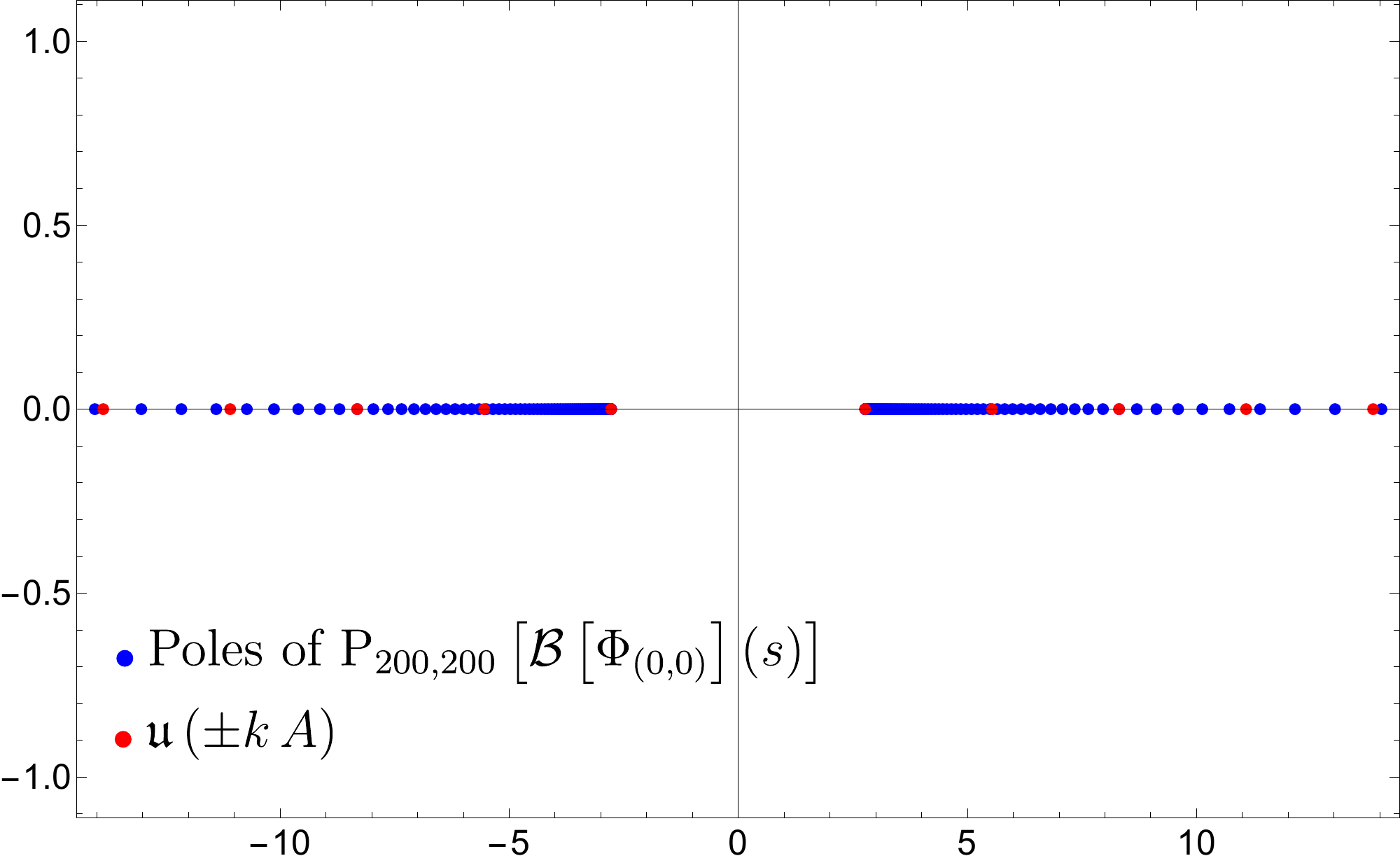}
		\caption{\PI~Borel plane.}
		\label{fig:poles2branches}
	\end{subfigure}%
	~ 
	\begin{subfigure}[t]{0.5\textwidth}
		\centering
		\includegraphics[height=1.8in]{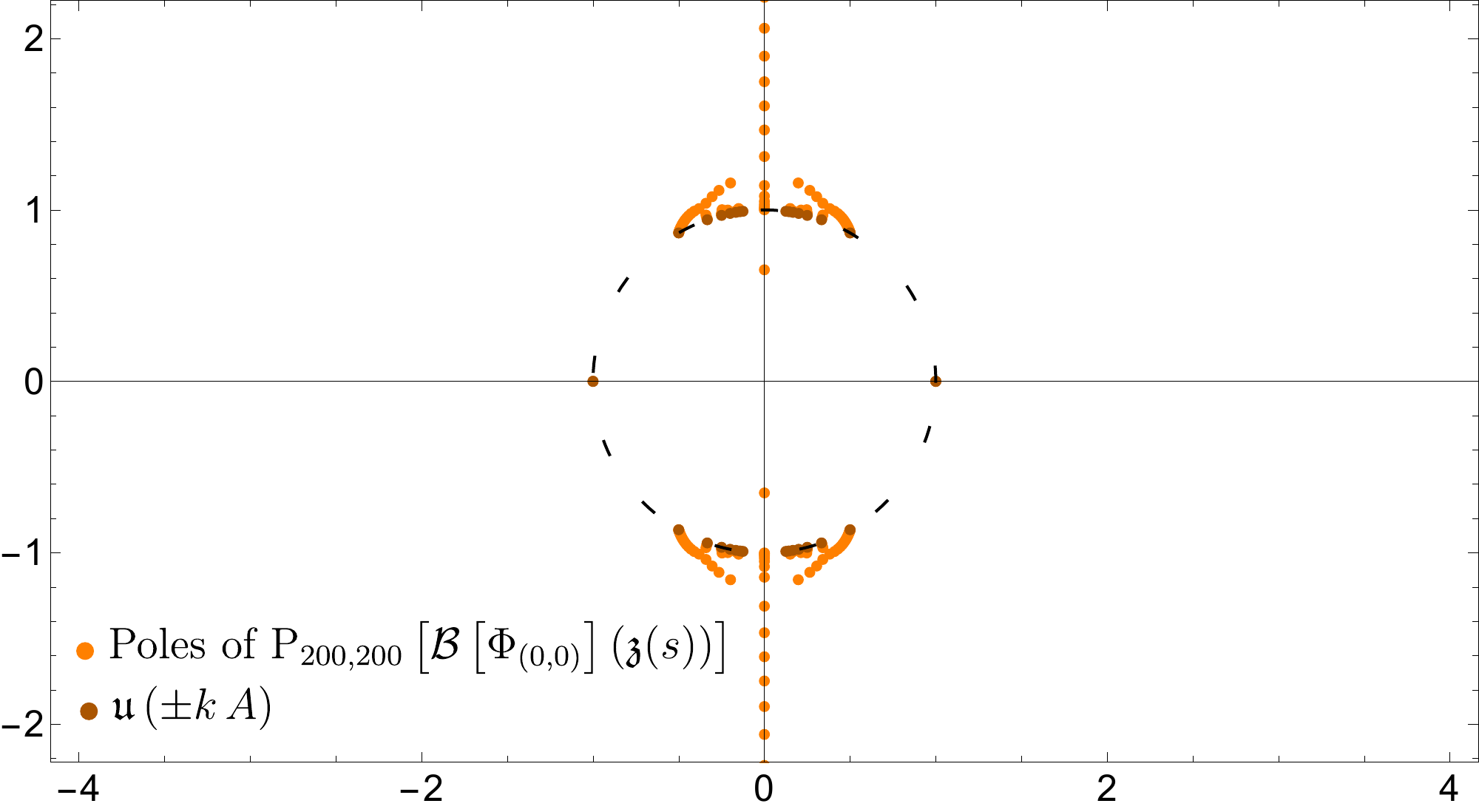}
		\caption{\PI~conformal Borel plane.}
		\label{fig:polesconftrans2branches}
	\end{subfigure}
	\caption{Poles of the Borel--Pad\'e approximation to the \PI~transseries perturbative-sector, shown both without (left) and with (right) conformal transformation improvement.}
	\label{fig:polesnormalandconftrans2branches}
\end{figure}

The discussion we just had for Riccati essentially holds very similarly in the Painlev\'e case; which we shall illustrate herein with \PI. Basically the only novelty are different conformal transformations. Now, branch-points are originally located along positive and negative real axes, which get mapped to upper and lower halves\footnote{Due to the square-roots in the transformation, each axis gets degenerated into two complex conjugate lines.} of the unit circle via the transformations (from \cite{cf99})
\bea
\mathfrak{u} (s) &=& \frac{\sqrt{A+s}-\sqrt{A-s}}{\sqrt{A+s}+\sqrt{A-s}}, \\
\mathfrak{z} (s) &=& \frac{2sA}{s^2-1} .
\eea
\noindent
The Pad\'e poles associated to both coordinates are displayed in figure~\ref{fig:polesnormalandconftrans2branches}. Comparing both sets of Pad\'e poles in the two plots, one might expect, in principle, that the present ``two-direction'' transformation could be less effective than the earlier ``one-direction'' case. Nonetheless, as we shall see, the results are pretty similar in both cases.

We have numerically evaluated a few Borel residues $\mathsf{S}_{(0,0)\to (k,0)}$ of the \PI~perturbative sector. Our results always indicate an increase in precision whenever using conformal transformations, as can be seen in table~\ref{tab:2branchconfvsnorm}. However, as already mentioned in the main text, do note that the conformal transformation implies extra computational time. The question then arises of whether it makes sense, in the case of the two lines of branch points, to invest this extra time in order to gain a few extra digits. This becomes a more pressing question once we realize that the higher the number of digits we may wish for, the higher the degrees of the polynomials appearing in the Pad\'e approximants, the more time it will take to execute the conformal transformations. It so happens that in our cases in the main body of the text, the computation of extra transseries coefficients is not terribly time-consuming. As such, we decided to avoid the use of conformal transformations. Nonetheless, we would like to mention that in cases where the amount of available data might be scarce, or when it might be computationally heavy to obtain, then conformal transformations are indeed a great and valuable tool.

\begin{table}
	\centering
	\begin{tabular}{c|c|c}
		Borel Residue & Precision non-CT & Precision CT \\
		\hline
		$\mathsf{S}_{(0,0)\to(1,0)}$ & 73 & 85\\ 
		\hline
		$\mathsf{S}_{(0,0)\to(2,0)}$ & 30 & 42\\
		\hline
		$\mathsf{S}_{(0,0)\to(3,0)}$ & 16 & 21\\
	\end{tabular}
	\caption{Borel residues for \PI~(see the main text). We display the precision our methods achieve in computing these Borel residues, using both the conformally transformed Borel--Pad\'e method, and our regular procedure. The dampings needed in both cases were different, in which case we are displaying the comparison at fixed $(150,150)$-Pad\'e approximants.}
	\label{tab:2branchconfvsnorm}
\end{table}

\subsection{Guessing the One Undetermined Number}\label{app:subsec:guess}

The reader will recall from our results throughout section~\ref{sec:results}, for instance figure~\ref{fig:patterns}, that we generically need two Stokes data, the well-known canonical $N_1^{(1)}$, and $N_0^{(1)}$; and then their higher-diagonal counterparts, to compute all others. Whereas $N_1^{(\ell)}$ are well-known (\textit{e.g.}, \eqref{eq:n1sp1} for \PI), we still need an analytic expression for $N_0^{(\ell)}$. This does not follow unequivocally from asymptotics as in section~\ref{sec:asymptotics} (albeit see how it might follow otherwise in section~\ref{sec:stokes}). So, in the asymptotics context, it requires a bit of guess-work. As for $N_1^{(\ell)}$, also for $N_0^{(\ell)}$ all we really need to know is $N_0^{(1)}$ and we shall next focus on this number. There are two ways to obtain it. The easiest is to actually see how far asymptotics takes us (following on closed-form asymptotics from subsection~\ref{subsec:closedform-asymp}). This method takes us as far as to show that
\begin{equation}
N^{(1)}_0 = \frac{\alpha}{2}\, N^{(1)}_1 \left( \gamma_{\text{E}} + \log \widetilde{A} \right).
\end{equation}
\noindent
The remaining guess-work is then quite straightforward. Just take our numerical results for $N^{(1)}_0$, with a precision of $\sim 10^3$ digits, and obtain numerical estimates for $\widetilde{A}$. Then plugging such number into any online numerical guesser (we used \textit{WolframAlpha}) one easily finds
\begin{equation}
\widetilde{A} = \left\{\begin{array}{ll}
96\sqrt 3 & \quad \text{(\PI)} \\
16 & \quad \text{(\PII)}
\end{array}\right..
\end{equation}

\begin{figure}
	\centering
		\includegraphics[width=4.5in]{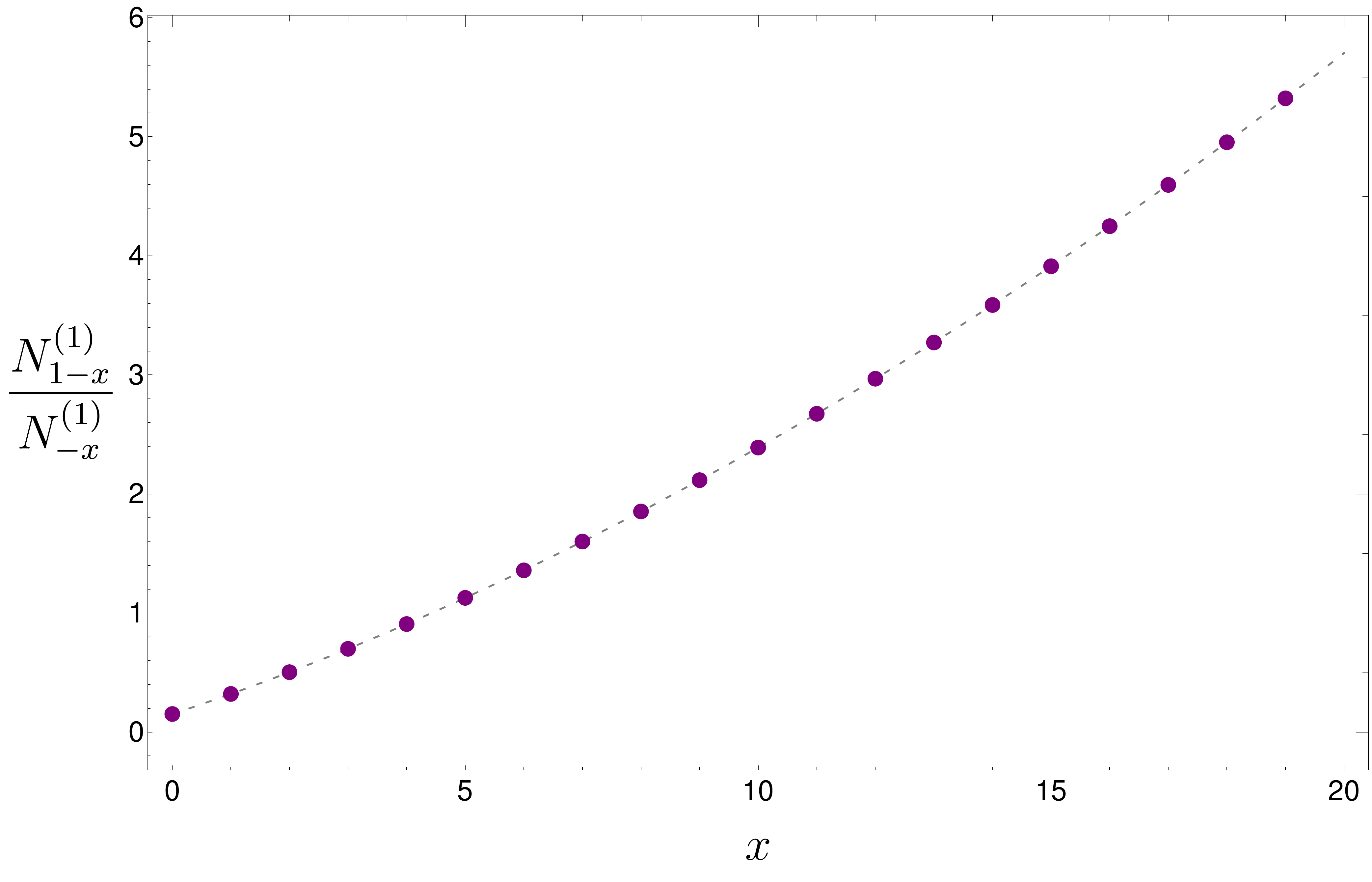}
		\caption{Numerical plot of the ratio $\frac{N^{(1)}_{1-x}}{N^{(1)}_{-x}}$, as a function of $x$.}
	\label{fig:ratioN1s}
\end{figure}

Having obtained this result rather swiftly, we believe it might also be of interest to the reader to present an alternative approach to this same result---the reason being that the following method, albeit more intricate, may be of use in other cases say for the problems in the (m)KdV hierarchy. Let us assume that we have no knowledge about closed-form asymptotics, and all we have is access to the numerical data. One empirical observation that follows is
\begin{equation}
\frac{N^{(1)}_{1-x}}{N^{(1)}_{-x}} \approx c \cdot x,
\end{equation}
\noindent
where $c$ is a constant. This approximate estimate is numerically tested in figure~\ref{fig:ratioN1s}. It then follows one can estimate
\begin{equation}
N^{(1)}_{1-x} \approx \frac{1}{x!}\, c^{x}\, N^{(1)}_{1},
\end{equation}
\noindent
and, increasing accuracy in the numerical tests, it further follows that the natural completion of our initial approximation is to write
\begin{equation}
N^{(1)}_{1-x} = \frac{1}{x!}\,c^{x}\,N^{(1)}_{1} + \frac{1}{(x-1)!}\,c^{x-1}\,a_{2} + \cdots + a_{x+1}.
\end{equation}
\noindent
We now need to describe the $a_{i}$'s. A clue on what they might be is already present in the general resurgence literature: mould calculus \cite{s07b} very concretely points in the direction that Stokes data should equal sums over values of multi-zeta functions. Now this may sound way too broad to be of any use, but it is actually not. We first do the reasonable assumption that all these sums are finite sums; and then conjecture that all $a_{i}$ are given by such sums of multi-zeta values. In this context, the simplest first-guess is to just use values of the zeta-function itself. In other words, a very reasonable \textit{ansatz} is
\begin{equation}
N^{(1)}_{-1} = \frac{1}{2}\,c^{2}\,N^{(1)}_{1} + \frac{b_{1}}{b_{2}}\,\zeta(b_{3})\,N^{(1)}_{1},
\end{equation}
\noindent
where we take $c=\frac{N^{(1)}_{0}}{N^{(1)}_{1}}$. Given that we know the numerical value for $N^{(1)}_{-1}$, we can now try values for the $b_{i}$ and see if there is a match. In the particular case above, the search-space happens to be sufficiently small that we could almost try values by hand. For numbers further down the diagonal (\textit{i.e.}, larger $x$), this space gets rather big very fast. Therefore, we wrote a small code that, given a test-function, tries integer values in a given range. Then it compares the result to the desired number. It finally outputs the combinations whose error is smaller than a certain precision. For the example of \PI, one very quickly finds
\begin{equation}
b_{1} = 2, \qquad b_{2} = 3, \qquad b_{3} = 2.
\end{equation}
\noindent
In this way, by making similar \textit{ans\"atze} as before, we can guess analytic forms for the numbers listed below (recall figure~\ref{fig:patterns}, where these are identical results for \PI~and \PII)
\bea
N_{-1}^{(1)} &=& \textcolor{violet}{\frac{1}{2!}\left(\frac{N^{(1)}_{0}}{N_1^{(1)}}\right)^1} \textcolor{blue}{N^{(1)}_{0}} {\color{violet}-}{\color{violet}\frac{1}{0!}\left(\frac{N^{(1)}_{0}}{N_1^{(1)}}\right)^0}\textcolor{blue}{\frac 12\left(\frac{\alpha}{2}\right)^2 N^{(1)}_1{\color{red}\zeta(2)}}, \\
N_{-2}^{(1)} &=& \textcolor{violet}{\frac{1}{3!}\left(\frac{N^{(1)}_{0}}{N_1^{(1)}}\right)^2}  \textcolor{blue}{ N^{(1)}_{0}} \textcolor{violet}{\color{violet}-}{\color{violet}\frac{1}{1!}\left(\frac{N^{(1)}_{0}}{N_1^{(1)}}\right)^1} \textcolor{blue}{\frac 12\left(\frac{\alpha}{2}\right)^2 N^{(1)}_1{\color{red}\zeta(2)}}  \textcolor{violet}{-\left(\frac{N^{(1)}_{0}}{N_1^{(1)}}\right)^0}\textcolor{blue}{\frac 13\left(\frac{\alpha}{2}\right)^3N^{(1)}_1}{\color{red}\zeta(3)}, \\
N_{-3}^{(1)} &=& \textcolor{violet}{\frac{1}{4!}\left(\frac{N^{(1)}_{0}}{N_1^{(1)}}\right)^3}  \textcolor{blue}{N^{(1)}_{0}}  {\color{violet}-}{\color{violet}\frac{1}{2!}}{\color{violet}\left(\frac{N^{(1)}_{0}}{N_1^{(1)}}\right)^2}  \textcolor{blue}{\frac 12\left(\frac{\alpha}{2}\right)^2 N^{(1)}_1{\color{red}\zeta(2)}}   -\textcolor{violet}{\left(\frac{N^{(1)}_{0}}{N_1^{(1)}}\right)^1} \textcolor{blue}{\frac 13\left(\frac{\alpha}{2}\right)^3 N^{(1)}_1}{\color{red}\zeta(3)}- \nonumber\\
&&
{\color{violet}-}{\color{violet}\frac{1}{0!}\left(\frac{N^{(1)}_{0}}{N_1^{(1)}}\right)^0}{\color{blue}\frac 14\left(\frac{\alpha}{2}\right)^4 N^{(1)}_1{\color{red}\left(\zeta(4)-\frac{1}{2}\zeta(2)^2\right)}}.
\eea
\noindent
Our ``brute force'' numerical analysis ensemble with our ``educated guess'' resurgence analysis have thus produced a very clean pattern for all these Stokes data, up to the determination of $N^{(1)}_0$. Two interesting properties stand out in this pattern:
\begin{itemize}
\item The ``new'' contribution appearing in $N^{(1)}_{1-x}$ is always a combination of products of zeta functions $\prod_i\zeta(x_i)$ such that $\sum_i x_i=x$.
\item This contribution is preceded by a factor of $\frac{\alpha}{2}$ to some power.
\end{itemize}

With these observations in mind it is natural to guess that $N^{(1)}_0$ would be related to $\zeta(1)$, whose most common normalization is given by the Euler--Mascheroni constant $\gamma_{\text{E}} \approx 0.5772156649...$. One is thus very naturally led to conjecture
\begin{equation}
N^{(1)}_0 = \frac{\alpha}{2}\, N^{(1)}_1 \left(\gamma_{\text{E}}+\widetilde{B}\right).
\end{equation}
\noindent
Obtaining $\widetilde{B}$ numerically, and again introducing the result into \textit{WolframAlpha}, one immediately finds the expected answer,
\begin{equation}
\widetilde{B} = \left\{\begin{array}{ll}
\log\left(96\sqrt 3\right) & \quad \text{(\PI)} \\
\log\left(16\right) & \quad \text{(\PII)}\\
\end{array}\right..
\end{equation}

\newpage

\bibliographystyle{plain}

\end{document}